\documentclass[aip,amsmath,amssymb,reprint]{revtex4-1}

\usepackage{graphicx}% Include figure files
\usepackage{bm}% bold math
\usepackage[utf8]{inputenc}
\usepackage[T1]{fontenc}
\usepackage{mathptmx}
\usepackage{etoolbox}

\usepackage{subcaption} %% to combine two or more subfigures in one figure
\usepackage{bbm} %% for imaginary unit

\makeatletter
\def\@email#1#2{%
 \endgroup
 \patchcmd{\titleblock@produce}
  {\frontmatter@RRAPformat}
  {\frontmatter@RRAPformat{\produce@RRAP{*#1\href{mailto:#2}{#2}}}\frontmatter@RRAPformat}
  {}{}
}%
\makeatother
\begin{document}

\preprint{AIP/123-QED}

\title[Eigenvalues of regular symmetric Hall-plates]{Eigenvalues of regular symmetric Hall-plates}
% Force line breaks with \\
\author{Udo Ausserlechner}
\email{udo.ausserlechner@infineon.com}
\affiliation{Infineon Technologies Austria AG, Siemensstrasse 2, A-9500 Villach, Austria.
ORCID ID: https://orcid.org/0000-0002-8229-9143 }
% \altaffiliation[Also at ]{Physics Department, XYZ University.}%Lines break automatically or can be forced with \\
%\author{B. Author}%
% \email{Second.Author@institution.edu.}
%\affiliation{ 
%Authors' institution and/or address%\\This line break forced with \textbackslash\textbackslash
%}%

%\author{C. Author}
% \homepage{http://www.Second.institution.edu/~Charlie.Author.}
%\affiliation{%
%Second institution and/or address%\\This line break forced% with \\
%}%

\date{\today}% It is always \today, today,
             %  but any date may be explicitly specified

\begin{abstract}
I discuss uniform, isotropic, plane, singly connected, electrically linear, regular symmetric Hall-plates with an arbitrary number of $N$ peripheral contacts exposed to a uniform perpendicular magnetic field of arbitrary strength. In practice, the regular symmetry is the most common one. If the Hall-plates are mapped conformally to the unit disk, regular symmetry means that all contacts are equally large and all contacts spacings are equally large, yet the contacts spacings may have a different size than the contacts. Such Hall-plates do not change when they are rotated by $360$°$/N$. Their indefinite conductance matrices are circulant matrices, whose complex eigenvalues are computable in closed form. These eigenvalues are used to discuss the Hall-output voltage, the maximum noise-efficiency, and Van-der-Pauw's method for measuring sheet resistances. For practical use, I report simple approximations for Hall-plates with four contacts and $90$° symmetry with popular shapes like disks, rectangles, octagons, squares, and Greek crosses with and without rounded corners. 
\end{abstract}

\maketitle

\section{Introduction}
\label{sec:intro}

\noindent Hall-plates are thin flat pieces of (semi-)conductors with a large mobility of the majority charge carriers. If a magnetic field acts on the carriers the Lorentz force diverts the current streamlines, and this builds up a Hall-electric field $\bm{E}_H$. The exact solution of the electric field problem in closed analytical form is not trivial, all the more if the Hall-plates have a non-symmetric geometry with extended contacts. Even if the potential in response to a single input current is known everywhere in the Hall-plate, it is still much work to compute the conductance matrix, which relates the voltages at the contacts of the plate to the currents into all contacts. Luckily, all shapes, which are equivalent by conformal transformation, have the same conductance matrix \cite{Wick1954}. Therefore, it is sufficient to study simple geometries, like circular disk Hall-plates, because from Riemann's mapping theorem \cite{Riemann} we know that there always exists a conformal transformation that maps the disk to any other singly connected domain. Recently, Homentcovschi and Murray found a general method to compute the conductance matrix of cicular disk Hall-plates without conformal transformation \cite{Homentcovschi2019}. For singly-connected 2D-domains with $N$ extended peripheral contacts, they defined two matrices, whose entries are numerical integrals that depend on the locations of the vertices of the contacts and on the Hall-angle. One of the matrices needs to be inverted and multiplied by the other one to get the resistance matrix normalized by the sheet resistance. The method is very general, but it does not readily give closed formulae for the resistance matrix of Hall-plates. Yet, for \emph{symmetric} Hall-plates with $N$ equal contacts the equations can be simplified and one gets closed formulae. This has been done for \emph{strictly regular symmetric} Hall-plates \cite{ArXiv2023}---they have the highest possible degree of symmetry (see Fig. \ref{fig:regular-vs-strictly-regular-Hall-plates}). Instead of computing the resistance matrix directly, it turned out to be simpler to compute the eigenvalues of the indefinite conductance matrix. In this work I apply the same techniques from Ref.~\onlinecite{ArXiv2023} to less symmetric Hall-plates --- \emph{regular symmetric} instead of strictly regular symmetric Hall-plates.

This paper starts with the definitions of the indefinite conductance matrix, Ohm's law, and the stream function. The regular symmetry leads to a symmetry in the indefinite conductance matrix---it is a circulant matrix, whose eigenvalues and eigenvectors are computed in closed form in Section \ref{sec:regular-sym}. In Section \ref{sec:iG} I apply these findings to the theory of Homentcovschi and Murray, to get a closed formula for the complex eigenvalues. Section \ref{sec:igamma} discusses some basic properties of these eigenvalues. Section \ref{sec:R} shows how to compute the resistance matrix from the eigenvalues. Section \ref{sec:4C-Hall} computes the output voltage of regular symmetric Hall-plates with four contacts. It also gives a very good approximation for the Hall-geometry factor of this most common type of Hall-plate and compares it to formulae given in the literature. Section \ref{sec:noise} compares the maximum noise efficiency of regular symmetric versus strictly regular symmetric Hall-plates, and Section \ref{sec:VdP} explains how to generalize van-der-Pauw's method for regular symmetric Hall-plates with four contacts with or without applied magnetic field. The appendix specifies how Hall-plates with popular shapes (Greek crosses, octagons, and rectangles) are equivalent to disk shaped Hall-plates.

\section{Definitions}
\label{sec:fundamentals}

\noindent In electrically linear Hall-plates the currents into the terminals are linear combinations of the potentials at the terminals. Suppose that the Hall-plate has $N$ terminals. Let us group all $N$ currents to a vector $\bm{I}$ and all $N$ voltages to a vector $\bm{V}$. Then we can express the linear combination as a matrix multiplication $\bm{I} = {^i}\!\bm{G}\bm{V}$, with the \emph{indefinite} conductance matrix ${^i}\!\bm{G}$. Hereby I use the historic nomenclature in circuit theory\cite{Shekel,Haykin,Balabanian,Simonyi}, where 'indefinite' denotes \emph{undefined} reference potential (ground node). In a mathematical sense the indefinite conductance matrix is not indefinite, but positive semi-definite. Both indefinite and semi-definite matrices have zero determinants, and therefore ${^i}\!\bm{G}$ cannot be inverted. 
%Thereby, the reference potential (ground node) is arbitrary -- none of the terminals needs to be grounded. 
If we ground the $\ell$-th terminal, we write $\bm{I}=\bm{G}\bm{V}$, where we delete the $\ell$-th current and voltage in $\bm{I},\bm{V}$, respectively, and we delete the $\ell$-th row and column in ${^i}\!\bm{G}$ to get $\bm{G}$. 
%The indefinite conductance matrix is positive semi-definite -- its determinant vanishes and therefore it cannot be inverted. 
The definite conductance matrix of any passive (= dissipative) system is positive definite, its determinant is positive, and an inverse exists: $\bm{V}=\bm{R}\bm{I}$ with the resistance matrix $\bm{R}=\bm{G}^{-1}$. 

In the presence of the Hall-effect Ohm's law is \cite{Hall-Ohm-law}
\begin{equation}\label{eq:RMFoCD2}
\bm{E}-\bm{E}_H=\rho\bm{J} \quad\text{with } \bm{E}_H = -\rho\mu_H\bm{J}\times (B_z\bm{n}_z) , 
\end{equation}
with the electric field, $\bm{E}=-\nabla\phi$, being the negative gradient of the electrostatic potential $\phi$, with the current density $\bm{J}$, the vector product $\times$, the unit vector $\bm{n}_z$ orthogonal to the plane Hall-plate, and with the $z$-component of the magnetic flux density $B_z$. $\rho$ is the homogeneous specific resistivity and $\mu_H$ is the Hall-mobility. 
In (\ref{eq:RMFoCD2}) I use the traditional sign convention for $\mu_H$, which means \emph{negative} Hall-mobility for \emph{electrons} as majority carriers, and I neglect minority carriers. According to (\ref{eq:RMFoCD2}) $\bm{E}$ and $\bm{J}$ are not colinear -- there is the Hall-angle $\theta_H$ in-between, with $\tan(\theta_H)=\mu_H B_z$. 
I use the following abbreviations, 
\begin{equation}\label{eq:RMFoCD3}
R_\mathrm{sheet} = \frac{\rho}{t_H}, \quad R_\mathrm{sq} = \frac{R_\mathrm{sheet}}{\cos(\theta_H)} ,
\end{equation}
whereby $t_H$ is the thickness of the Hall-plate. For a square plate with two contacts fully covering two opposite edges, the resistance between these contacts is the sheet resistance $R_\mathrm{sheet}$ at zero magnetic field, and it is the square resistance $R_\mathrm{sq}$ in the presence of a magnetic field \cite{Lippmann1958}.  

In two-dimensional space we can describe the current density $\bm{J}$ by a stream function $\psi$ via 
\begin{equation}\label{eq:RMFoCD1a}
\bm{J} = \frac{-1}{\rho} \nabla\times (\psi\bm{n}_z) = \frac{-1}{\rho}\frac{\partial\psi}{\partial y}\bm{n}_x + \frac{1}{\rho}\frac{\partial\psi}{\partial x}\bm{n}_y ,
\end{equation}
with the nabla operator $\nabla$ and the Cartesian unit vectors $\bm{n}_x,\bm{n}_y,\bm{n}_z$, the last one being orthogonal to the plane Hall-plate. Equation (\ref{eq:RMFoCD1a}) holds only, if current enters and exits the Hall-plate merely through contacts on the single boundary \cite{Ausserlechner2019b,Ausserlechner2019c}. Then the continuity equation $\nabla\cdot\bm{J}=0$ follows immediately from (\ref{eq:RMFoCD1a}) (the dot denotes the scalar product of two vectors). $\bm{J}$ is orthogonal to the gradient of $\psi$, 
\begin{equation}\label{eq:RMFoCD1b}
\bm{J}\cdot\nabla\psi = \frac{1}{\rho} \left(\begin{array}{r} -\partial\psi/\partial y\\ \partial\psi/\partial x \end{array} \right) \cdot \left(\begin{array}{c} \partial\psi/\partial x\\ \partial\psi/\partial y \end{array} \right) = 0 ,
\end{equation}
which explains the name 'stream function': it is constant along current streamlines. Along the insulating segments of the boundary there are also specific current streamlines. Therefore, $\psi$ is constant along each insulating segment.  Thus, $\psi$ behaves on the insulating boundary segments just like $\phi$ behaves on the conducting boundary segments.  
In the stationary condition Maxwell's second equation is 
\begin{equation}\label{eq:RMFoCD1c}\begin{split}
& \nabla\times\bm{E} = \bm{0} = \rho\nabla\times\bm{J} - \rho\tan(\theta_H)\nabla\times(\bm{J}\times\bm{n}_z), \\
& \qquad\nabla\times(\bm{J}\times\bm{n}_z) = (\bm{n}_z\cdot\nabla)\bm{J}-\bm{n}_z(\nabla\cdot\bm{J})=\bm{0}, \\
& \qquad\rho\nabla\times\bm{J} = -\nabla\times(\nabla\times\psi\bm{n}_z) \\
& \qquad\qquad\qquad=-\nabla(\nabla\cdot\bm{n}_z\psi)+\nabla^2\psi\bm{n}_z , \\
& \Rightarrow \nabla^2\psi = 0.
\end{split}\end{equation}
In the first line of (\ref{eq:RMFoCD1c}) I used (\ref{eq:RMFoCD2}), in the second line I used $\bm{n}_z\cdot\nabla=0$ and $\nabla\cdot\bm{J}=0$, in the third line I used (\ref{eq:RMFoCD1a}), and in the fourth line I used $\nabla\cdot\bm{n}_z\psi=0$. Equation (\ref{eq:RMFoCD1c}) says that the stream function fulfils the Laplace equation inside the Hall-plate---just like the electric potential.
Finally, let us compute the current $I_{12}$ flowing from left to right across any contour (extruded into thickness direction) with start-point 1 and end-point 2, 
\begin{equation}\label{eq:RMFoCD1d}\begin{split}
I_{12} & = t_H\int_1^2 J_n\,\mathrm{d}s = t_H\bm{n}_z\cdot\int_1^2 (\bm{J}\times\bm{t})\,\mathrm{d}s \\ 
& = \frac{-t_H}{\rho} \int_1^2 \left(\frac{\partial\psi}{\partial x}\,\mathrm{d}x+\frac{\partial\psi}{\partial y}\,\mathrm{d}y\right) \\
& = \frac{-t_H}{\rho} \int_1^2 \mathrm{d}\psi = -\frac{\psi_2-\psi_1}{R_\mathrm{sheet}} ,
\end{split}\end{equation}
with $J_n=\bm{J}\cdot\bm{n}=\bm{J}\cdot(\bm{t}\times\bm{n}_z)=\bm{n}_z\cdot(\bm{J}\times\bm{t})$.The unit vector $\bm{n}$ is normal to the contour pointing away from the Hall region, and $\bm{t}$ is the unit vector tangential to the contour pointing from 1 to 2, $\bm{t}=\mathrm{d}x\,\bm{n}_x+\mathrm{d}y\,\bm{n}_y$. 

\section{Symmetry Properties of Regular Symmetric Hall-plates}
\label{sec:regular-sym}

\begin{figure}[t]
%\vspace{1mm}
  \centering
                \includegraphics[width=0.34\textwidth]{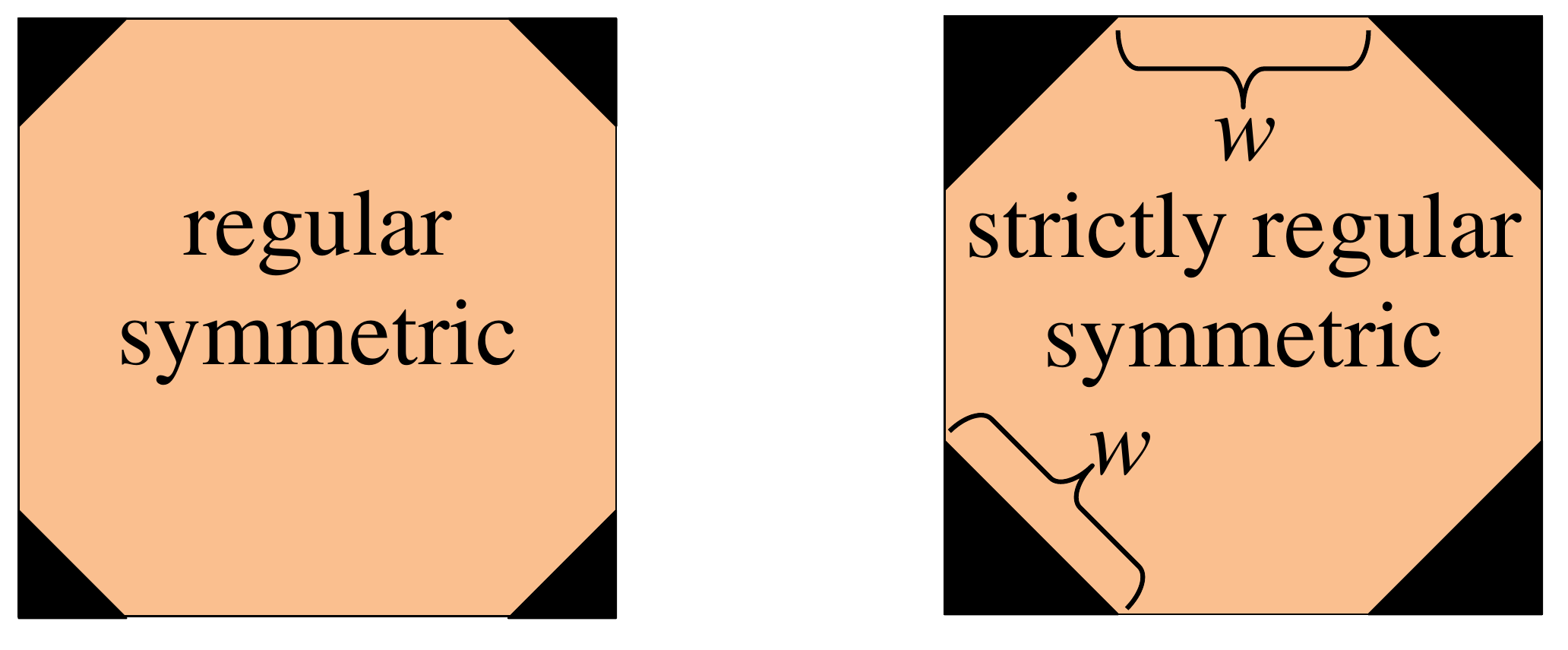}
    \caption{ Examples of regular symmetric and strictly regular symmetric Hall-plates with four contacts. The perimeters are squares, the dark rectangles are perfectly conductive contacts, and the orange octagons are domains where (\ref{eq:RMFoCD2}) holds. }
   \label{fig:regular-vs-strictly-regular-Hall-plates}
\end{figure}

\noindent Suppose we have a regular symmetric boundary with $N$ contacts of equal size, the spacings between the contacts are all equal, yet they are not necessarily identical to the size of the contacts (see Fig. \ref{fig:regular-vs-strictly-regular-Hall-plates}). Suppose we connect an external circuit to this Hall-plate. If we keep the Hall-plate fixed and move the external circuit by one terminal, say clock-wise, then the resulting potentials and currents into the external circuit must remain the same, because the Hall-plate is regular symmetric in shape and the applied magnetic field is parallel to the rotation axis.This means that the entries in the current and voltage vectors move one instance downwards, and the entries in the indefinite conductance matrix move one row down and one column to the right. Both the original and the new ${^i}\!\bm{G}$ matrices must be identical. We can repeat this procedure $N$ times. The conclusion is that the elements along diagonals parallel to the main diagonal are identical, which gives the following patterns for ${^i}\!\bm{G}$ and $\bm{G}$, 
%Then the symmetry of the indefinite and definite conductance matrices, ${^i}\!\bm{G}$, and $\bm{G}$, respectively, was derived in Ref.~\onlinecite{Ausserlechner2023a} by the simple fact that shifting the indices of the contacts by $m$ instances means shifting all rows and columns in ${^i}\!\bm{G}$ cyclically. For $N=5$ this gives the following patterns  
\begin{equation}\label{G-symmetry}
{^i}\!\bm{G} = \left(\!\!\begin{array}{ccccc} A&B&C&D&E\\ E&A&B&C&D\\ D&E&A&B&C\\ C&D&E&A&B\\ B&C&D&E&A \end{array} \!\!\right) \,\Rightarrow\,
\bm{G} = \left(\!\!\begin{array}{cccc} A&B&C&D\\ E&A&B&C\\ D&E&A&B\\ C&D&E&A \end{array} \!\!\right) .
\end{equation}
${^i}\!\bm{G}$ is an $N\times N$ matrix, $\bm{G}$ is an $(N-1)\times (N-1)$ matrix. We get $\bm{G}$ from ${^i}\!\bm{G}$ by deleting the row and column which corresponds to the grounded contact. $\bm{G}$ and ${^i}\!\bm{G}$ are Toeplitz matrices, which have the property that $G_{\ell ,m}=g_{\ell-m}$, which depends only on $\ell-m$. Moreover, ${^i}\!\bm{G}$ is a circulant matrix with $g_{\ell-m}=g_{\ell-m+N}$. Third, the sum of entries per row in any indefinite conductance matrix has to vanish, because no currents flow if all contacts are at identical non-zero potential \cite{Haykin,Balabanian,Simonyi}.

If we map a regular symmetric Hall-plate to the unit disk, its contact $C_m$ ranges from azimuthal angles 
\begin{equation}\label{eq:RMFoCD100b}
\alpha_m=2\pi \frac{m-1}{N} \;\text{ to }\; \beta_m=\alpha_m+\chi \frac{\pi}{N} , 
\end{equation}
for $\chi\in\left( 0,2\right)$ and $m=1,2,\ldots,N$. Contacts are small for $\chi\to 0$, whereas contact spacings are small for $\chi\to 2$. The case $\chi = 1$ gives \emph{strictly regular symmetric} Hall-plates, whose contacts are equally large as the contacts spacings (see Fig. \ref{fig:regular-vs-strictly-regular-Hall-plates}). 

We can write down the eigenvalues and eigenvectors of a real-valued circulant matrix $\bm{{^i}\!G}$, see Ref.~\onlinecite{GrayCirculant} 
\begin{equation}\label{eq:RMFoCD52}\begin{split} 
& \bm{{^i}\!G} = \bm{Q} \;\bm{{^i}\!\Gamma} \bm{Q}^C \quad\text{with } \bm{{^i}\!\Gamma} = \mathrm{diag}({^i}\!\gamma_1,{^i}\!\gamma_2,\ldots ,{^i}\!\gamma_N) ,\\
%& 2\times\bm{1}-\bm{\hat{1}}-\bm{\hat{1}}^T = \bm{Q} \bm{\Delta} \bm{Q}^C \quad\text{with} \\ 
& (\bm{Q})_{k,\ell} = \frac{1}{\sqrt{N}} \exp\left(\frac{2\pi \mathbbm{i} k \ell }{N}\right)  \;\forall k,\ell\in\{1,2,\ldots,N\} .
\end{split}\end{equation} 
$\bm{Q}^C$ is the complex conjugate of $\bm{Q}$, and $\mathbbm{i}=\sqrt{-1}$. $\bm{{^i}\!\Gamma}$ is a diagonal matrix, which comprises all eigenvalues ${^i}\!\gamma_\ell$ of $\bm{{^i}\!G}$. It holds $\bm{Q}=\bm{Q}^T$ and $\bm{Q}\bm{Q}^C=\bm{Q}^C\bm{Q}=\bm{1}$. The eigenvectors of $\bm{{^i}\!G}$ are the columns of $\bm{Q}$, which are coefficient vectors of the inverse discrete Fourier transform (IDFT). It generally holds 
\begin{equation}\label{eq:RMFoCD69a}
{^i}\!\gamma_{N-\ell}={^i}\!\gamma_\ell^C ,
\end{equation} 
for the eigenvalues of any real circulant matrix, where $z^C$ is the complex conjugate of $z$.

\section{A Closed Form Solution for the Eigenvalues of $\bm{{^i}\!G}$} 
\label{sec:iG}

\noindent We use Ref.~ \onlinecite{Homentcovschi2019}. Their equation (17) reads in our notation
\begin{equation}\label{eq:RMFoCD100}
\sum_{m=1}^N B_{k,m} \phi_m = \sum_{m=1}^N  \frac{-A_{k,m} \psi_m}{\cos(\theta_H)} \text{ for }k\in\{1,\ldots,N-1\} ,
\end{equation} 
whereby $\phi_m$ is the potential on $C_m$ (= the $m$-th peripheral contact), and $\psi_m$ is the stream function on the insulating peripheral segment between $C_m$ and $C_{m+1}$, with $C_N=C_0$. 
%(For the definition of the stream function $\psi$ see Ref.~\onlinecite{Ausserlechner2019b}.)
%The Hall-plate is mapped to the unit disk, with $C_m$ ranging from azimuthal angle 
%begin{equation}\label{eq:RMFoCD100b}
%\alpha_m=2\pi \frac{m-1}{N} \;\text{ to }\; \beta_m=\alpha_m+\frac{\pi}{N} \text{, for } m=1,2,\ldots,N . 
%\end{equation} 
In Ref.~\onlinecite{Homentcovschi2019} the entries of the matrices $\bm{A},\bm{B}$ are defined as
\begin{equation}\label{eq:RMFoCD101}\begin{split}
& A_{k,m} = \int_{\beta_m}^{\alpha_{m+1}}\frac{h(\tau)\,\mathrm{d}\tau}{\sin((\tau-\beta_k)/2)\sin((\tau-\beta_N)/2)} , \\
& B_{k,m} = \int_{\alpha_m}^{\beta_m}\frac{h(\tau)\,\mathrm{d}\tau}{\sin((\tau-\beta_k)/2)\sin((\tau-\beta_N)/2)}, \\
&\text{with } h(\tau) = \prod_{\ell=1}^N \left| \frac{\sin((\tau-\beta_\ell)/2)}{\sin((\tau-\alpha_\ell)/2)} \right|^{(1/2-\theta_H/\pi)} .
\end{split}\end{equation} 
Note that in (\ref{eq:RMFoCD100}) the sums extend over $N$ terms, whereas in Ref.~\onlinecite{Homentcovschi2019} they had only $N-1$ terms. Hence, in contrast to Ref.~\onlinecite{Homentcovschi2019}, we do not require $\phi_N=0$ and $\psi_N=0$. This enables us to work with the indefinite conductance matrix, which is circulant, and therefore it lends for a simpler mathematical treatment than the definite conductance matrix. Therefore, our matrices $\bm{A},\bm{B}$ have $N$ columns and $N-1$ rows. Equation (\ref{eq:RMFoCD100}) must hold, if we add an arbitrary constant to all $\phi_k$ or to all $\psi_k$. Consequently it must hold 
\begin{equation}\label{eq:RMFoCD102}
\sum_{m=1}^N A_{k,m} = 0 \text{ and } \sum_{m=1}^N B_{k,m} = 0 \text{ for }k\in\{1,\ldots,N-1\} .
\end{equation}
Multiplying the left equation of (\ref{eq:RMFoCD102}) with $\psi_N/\cos(\theta_H)$ and adding it to (\ref{eq:RMFoCD100}) gives 
\begin{equation}\label{eq:RMFoCD103}
\sum_{m=1}^N B_{k,m} \phi_m = \sum_{m=1}^N -A_{k,m} \frac{\psi_m-\psi_N}{\cos(\theta_H)} \text{ for }k\in\{1,\ldots,N-1\} .
\end{equation} 
From (\ref{eq:RMFoCD1d}) it follows 
\begin{equation}\label{eq:RMFoCD104}
I_k = \frac{\psi_k-\psi_{k-1}}{R_\mathrm{sheet}} .
\end{equation} 
This means 
\begin{equation}\label{eq:RMFoCD105}\begin{split}
& \left.\begin{array}{rcl} I_1&=&(\psi_1-\psi_N)/R_\mathrm{sheet}\\ I_1+I_2&=&(\psi_2-\psi_N)/R_\mathrm{sheet}\\ & \vdots & \\ I_1+\ldots I_{N-1}&=&(\psi_{N-1}-\psi_N)/R_\mathrm{sheet}\\ I_1+\ldots I_N&=&0 \end{array} \right\} \\
&\qquad\quad \Rightarrow \bm{L_1}\bm{I} = \frac{1}{R_\mathrm{sheet}}\left(\begin{array}{c}\psi_1-\psi_N\\ \psi_2-\psi_N\\ \vdots\\ \psi_{N-1}-\psi_N\\0 \end{array}\right) , \\
& \qquad\qquad\text{with } \bm{L_1} = 
 \left(\! {\begin{array}{ccccc} 1 & 0 & 0 & \cdots & 0 \\ 1 & 1 & 0 & \cdots & 0 \\ 1 & 1 & 1 & \cdots & 0 \\ \vdots & \vdots & \vdots &\ddots & \vdots \\ 1 & 1 & 1 & \cdots & 1 \\ \end{array} }\!\right) .
\end{split}\end{equation} 
Inserting (\ref{eq:RMFoCD105}) into (\ref{eq:RMFoCD103}) gives (in matrix form) 
\begin{equation}\label{eq:RMFoCD106}
\bm{B} \bm{\phi} = -R_\mathrm{sq}\bm{A}\bm{L_1}\bm{I} = -R_\mathrm{sq}\bm{A}\bm{L_1}\bm{{^i}\!G}\bm{\phi} .
\end{equation} 
This equation is valid for arbitrary voltage vectors $\bm{\phi}$, therefore we can skip $\bm{\phi}$ on both sides of (\ref{eq:RMFoCD106}). 
Next let us introduce the matrix 
\begin{equation}\label{eq:RMFoCD8}\begin{split}
& \bm{\hat{1}} = 
 \left(\! {\begin{array}{ccccc} 0 & 1 & 0 & \cdots & 0 \\ 0 & 0 & 1 & \cdots & 0 \\ \vdots & \vdots & \vdots &\ddots & \vdots \\ 0 & 0 & 0 & \cdots & 1 \\ 1 & 0 & 0 & \cdots & 0 \\ \end{array} }\!\right) .
\end{split}\end{equation}
$\bm{\hat{1}}$ is a circulant $N\times N$ matrix, which is obtained from the identity matrix $\bm{1}$ by shifting all entries up once, in a rolling way (= by cyclic permutation). It holds $\bm{\hat{1}}^{-1}=\bm{\hat{1}}^T$, and $\mathrm{det}(\bm{1}-\bm{\hat{1}})=0$.
Multiplying (\ref{eq:RMFoCD106}) from right with $(\bm{1}\!-\!\bm{\hat{1}}^T\!)$ gives 
\begin{equation}\label{eq:RMFoCD107}\begin{split}
\bm{B} (\bm{1}\!-\!\bm{\hat{1}}^T\!) & = -R_\mathrm{sq}\bm{A}\bm{L_1}\bm{{^i}\!G}(\bm{1}\!-\!\bm{\hat{1}}^T\!) \\ 
& = -R_\mathrm{sq}\bm{A}\bm{L_1}(\bm{1}\!-\!\bm{\hat{1}}^T\!)\bm{{^i}\!G} ,
\end{split}\end{equation} 
whereby I used the fact that $\bm{{^i}\!G}$ and $(\bm{1}\!-\!\bm{\hat{1}}^T\!)$ are circulant matrices, and therefore they commute. Writing out the product in detail gives $\bm{A}\bm{L_1}(\bm{1}-\bm{\hat{1}}^T) = $ 
\begin{equation}\label{eq:RMFoCD108}\begin{split}
& \bm{A} \left(\! {\begin{array}{ccccc} 1 & 0 & \cdots & 0 \\ 1 & 1 & \ddots & 0 \\ \vdots & \vdots &\ddots & \ddots \\ 1 & 1 & \cdots & 1 \\ \end{array} }\!\right)
\left(\!\begin{array}{ccccc} 1&0&\cdots &0&-1\\ -1&1&\ddots &0&0\\0&-1&\ddots &0 &0\\ \vdots & \ddots & \ddots & 1 & 0 \\ 0&0&\ddots &-1&1 \end{array}\!\right) \\
& = \bm{A}\left(\!\begin{array}{ccccc} 1&0&\cdots &0&-1\\ 0&1&\ddots &0&-1\\0&\ddots &\ddots &\ddots &\vdots \\ 0 & \ddots & \ddots & 1 & -1 \\ 0&0&\cdots &0&0 \end{array}\!\right) = \bm{A} ,
\end{split}\end{equation} 
whereby I used (\ref{eq:RMFoCD102}) in the last equality of (\ref{eq:RMFoCD108}). %Equation (\ref{eq:RMFoCD108}) is remarkable, because $\bm{A}\bm{L_1}(\bm{1}-\bm{\hat{1}}^T) = \bm{A}$ even though $\bm{L_1}(\bm{1}-\bm{\hat{1}}^T)\ne\bm{1}$. 
Inserting (\ref{eq:RMFoCD108}) into (\ref{eq:RMFoCD107}) gives 
\begin{equation}\label{eq:RMFoCD109}\begin{split}
\bm{B} (\bm{1}\!-\!\bm{\hat{1}}^T\!) & = -R_\mathrm{sq}\bm{A}\bm{{^i}\!G} \\ 
\text{with (\ref{eq:RMFoCD52}) }\;\Rightarrow \bm{B} (\bm{1}\!-\!\bm{\hat{1}}^T\!)\bm{Q} & = -R_\mathrm{sq}\bm{A}\bm{Q} \,\bm{{^i}\!\Gamma} . 
\end{split}\end{equation} 
It holds 
\begin{equation}\label{eq:RMFoCD110}\begin{split}
& (\bm{B} \bm{\hat{1}}^T \bm{Q})_{j,m} = \sum_{k=1}^N \sum_{\ell=1}^N B_{j,k} \delta_{k,1+(\ell\,\mathrm{mod}\,N)} Q_{\ell ,m} \\ 
& = \sum_{\ell=1}^N B_{j,1+(\ell\,\mathrm{mod}\,N)} Q_{\ell ,m} =  \sum_{k=0}^{N-1} B_{j,k+1} Q_{k,m} . 
\end{split}\end{equation} 
%\vspace{20mm}
%\noindent 
Inserting this into (\ref{eq:RMFoCD109}) gives 
\begin{equation}\label{eq:RMFoCD111}\begin{split}
{^i}\!\gamma_m R_\mathrm{sq} \sum_{k=1}^N A_{j,k} Q_{k,m} = -\sum_{k=1}^N B_{j,k} ( Q_{k,m}-Q_{k-1,m} ) .
\end{split}\end{equation} 
Inserting (\ref{eq:RMFoCD100b}) into (\ref{eq:RMFoCD101}) and using the product formula \cite{RyshikGradstein} gives 
\begin{equation}\label{eq:RMFoCD112} 
h(\tau) = \sin\left(\frac{\chi\pi}{2}\right) \left| \cot\left(\frac{\chi\pi}{2}\right) - \cot\left(\frac{N\tau}{2}\right) \right|^{(1/2-\theta_H/\pi)} .
\end{equation} 
Back-inserting this into (\ref{eq:RMFoCD101}) gives after some manipulation 
\begin{equation}\label{eq:RMFoCD113}\begin{split}
& A_{k,m} = \frac{-2}{N} (2-\chi) \sin\left(\frac{\chi\pi}{2}\right) \\
&\quad\quad\quad\times\! \int_0^{\pi/2} \!\frac{\left| \cot\left(\frac{\chi\pi}{2}\right) - \cot\left(\frac{\chi\pi}{2}+z(2-\chi)\right) \right|^{(1/2-\theta_H/\pi)}}{\sin((2\!-\!\chi)z/N\!+\!\pi m/N)} \\ 
&\qquad\qquad\qquad\times \frac{\,\mathrm{d}z}{\sin((2\!-\!\chi)z/N\!+\!\pi (m\!-\!k)/N)}, 
\end{split}\end{equation} 
and 
\begin{equation}\label{eq:RMFoCD113b}\begin{split}
&\quad B_{k,m} = \frac{-2}{N}\chi \sin\left(\frac{\chi\pi}{2}\right) \\
&\quad\quad\quad\times\! \int_0^{\pi/2} \!\frac{\left| \cot\left(\frac{\chi\pi}{2}\right) - \cot\left(\chi z\right) \right|^{(1/2-\theta_H/\pi)}}{\sin(\chi (2z\!-\!\pi)/(2N)\!+\!\pi m/N)} \\
&\qquad\qquad\qquad\times \frac{\,\mathrm{d}z}{\sin(\chi (2z\!-\!\pi)/(2N)\!+\!\pi (m\!-\!k)/N)} .
\end{split}\end{equation} 
Let us insert (\ref{eq:RMFoCD113}), (\ref{eq:RMFoCD113b}) into (\ref{eq:RMFoCD111}) and use the following identity (see Appendix \ref{sec:sum}) 
\begin{equation}\label{eq:RMFoCD125}\begin{split}
& \sum_{k=1}^N \frac{\exp\left(2\pi \mathbbm{i} k m / N\right)}{\sin(x/N\!+\!\pi k/N)\sin(x/N\!+\!\pi (k\!-\!j)/N)} \\
& = 2N\frac{\sin\left(\pi j m / N\right)}{\sin\left(\pi j / N\right)} \left(\mathbbm{i}\cot(x)\!-\!1\right)\exp\left(\mathbbm{i}m \frac{j\pi\!-\!2x}{N}\right) .
\end{split}\end{equation} 
Finally, this gives the complex eigenvalues ${^i}\!\gamma_m$ for $m\in\{1,2,\ldots, N\}$ of the indefinite conductance matrix $\bm{{^i}\!G}$ of a regular symmetric Hall-plate with $N$ peripheral contacts of size $\chi$ as a function of the Hall-angle $\theta_H$, 
\begin{widetext}
\begin{equation}\label{eq:RMFoCD1500}\begin{split}
{^i}\!\gamma_m = & \frac{-2\chi\mathbbm{i}}{R_\mathrm{sq}(2-\chi)}\exp\left(-\mathbbm{i}\pi\frac{m}{N}\right) \sin\left(\pi\frac{m}{N}\right) \\ 
&\times\frac{\int_0^1 \left(1+\mathbbm{i}\cot\left(\frac{\pi}{2}\chi\tau\right)\right) \exp\left(\mathbbm{i}\pi\chi\frac{m}{N}\tau\right) \left[ \cot\left(\frac{\pi}{2}\chi (1-\tau)\right) -\cot\left(\frac{\pi}{2}\chi\right) \right]^{(1/2-\theta_H/\pi)}\,\mathrm{d}\tau}{\int_0^1 \left(1-\mathbbm{i}\cot\left(\frac{\pi}{2}(2-\chi)\tau\right)\right) \exp\left(-\mathbbm{i}\pi(2-\chi)\frac{m}{N}\tau\right) \left[ \cot\left(\frac{\pi}{2}\chi\right) -\cot\left(\frac{\pi}{2}\left(\chi+\tau(2-\chi)\right)\right) \right]^{(1/2-\theta_H/\pi)}\,\mathrm{d}\tau} .
\end{split}\end{equation}
\end{widetext}
Let us replace $\chi=1+\epsilon$ with $\epsilon\in(-1,1)$, and $\tau\pi/2=y$ with $y\in[0,\pi/2]$. We use 
\begin{equation}\label{eq:RMFoCD1501}
\cot(a)-\cot(a+b) = \frac{\sin(b)}{\sin(a)\sin(a+b)} ,
\end{equation}
and
\begin{equation}\label{eq:RMFoCD1501b}
\cot(a-c)-\cot(a) = \frac{\sin(c)}{\sin(a)\sin(a-c)} 
\end{equation}
for $a=(1+\epsilon)\pi/2$, $b=(1-\epsilon)y$, $c=(1+\epsilon)y$. After a few re-arrangements we get 
\begin{widetext}
\begin{equation}\label{eq:RMFoCD1502}\begin{split}
{^i}\!\gamma_m = & \frac{2}{R_\mathrm{sq}} \sin\left(\pi\frac{m}{N}\right) \frac{1+\epsilon}{1-\epsilon} \exp\left(\mathbbm{i}\frac{\pi}{2}\epsilon\left(1-\frac{2m}{N}\right)\right) %\\ &\times
\frac{\int_0^{\pi/2}\frac{\exp\left(-\mathbbm{i}y(1+\epsilon)(1-2m/N)\right)}{\sqrt{\sin(y(1+\epsilon))\sin((\pi/2-y)(1+\epsilon))}} \left(\frac{\sin(y(1+\epsilon))}{\sin((\pi/2-y)(1+\epsilon))}\right)^{-\theta_H/\pi}\;\mathrm{d}y}{\int_0^{\pi/2}\frac{\exp\left(-\mathbbm{i}(\pi/2-y)(1-\epsilon)(1-2m/N)\right)}{\sqrt{\sin(y(1-\epsilon))\sin((\pi/2-y)(1-\epsilon))}} \left(\frac{\sin(y(1-\epsilon))}{\sin((\pi/2-y)(1-\epsilon))}\right)^{-\theta_H/\pi}\;\mathrm{d}y} .
\end{split}\end{equation}
\end{widetext}
Splitting up the integration range $[0,\pi/2]$ into $[0,\pi/4]$ and $[\pi/4,\pi/2]$ and replacing $y=x+\pi/4$ finally gives 
\begin{widetext}
\begin{equation}\label{eq:RMFoCD1503}\begin{split}
%{^i}\!\gamma_m = & \frac{2}{R_\mathrm{sq}} \sin\left(\pi\frac{m}{N}\right) \frac{\int_0^{\pi/4} f(\epsilon) \left(g(\epsilon,\theta_H) +g^C(\epsilon,-\theta_H) \right)\;\mathrm{d}x }{\int_0^{\pi/4} f(-\epsilon) \left(g^C(-\epsilon,\theta_H) +g(-\epsilon,-\theta_H) \right)\;\mathrm{d}x } \\
%& f\left(\epsilon\right) = \frac{(1+\epsilon)}{\sqrt{\sin\left((\pi/4+x)(1+\epsilon)\right)\sin\left((\pi/4-x)(1+\epsilon)\right)}} \\
%& g\left(\epsilon,\theta_H\right) = \exp\left(-\mathbbm{i}x(1+\epsilon)\left(1-\frac{2m}{N}\right)\right) \left(\frac{\sin\left((\pi/4-x)(1+\epsilon)\right)}{\sin\left((\pi/4+x)(1+\epsilon)\right)}\right)^{\theta_H/\pi} .
& {^i}\!\gamma_m = \frac{2}{R_\mathrm{sq}}\;\frac{1+\epsilon}{1-\epsilon} \,\sin\left(\pi\frac{m}{N}\right)\frac{h(\epsilon,\theta_H,\frac{m}{N}) + h^C(\epsilon,-\theta_H,\frac{m}{N})}{h(-\epsilon,-\theta_H,\frac{m}{N}) + h^C(-\epsilon,\theta_H,\frac{m}{N})} \\ 
& h\left(\epsilon,\theta_H,\frac{m}{N}\right) = \int_0^{\pi/4} \!\!\exp\!\left(\!\!-\mathbbm{i}x(1\!+\!\epsilon)\left(\!1\!-\!\frac{2m}{N}\right)\!\!\right) \left[\sin\left(\!\!\left(\frac{\pi}{4}\!-\!x\right)(1\!+\!\epsilon)\right)\!\!\right]^{\!\!\frac{\theta_H}{\pi}\!-\!\frac{1}{2}} \left[\sin\left(\!\!\left(\frac{\pi}{4}\!+\!x\right)(1\!+\!\epsilon)\right)\!\!\right]^{\!\!\frac{-\theta_H}{\pi}\!-\!\frac{1}{2}} \mathrm{d}x  .
\end{split}\end{equation}
\end{widetext}

\begin{figure*}
\vspace{1mm}
  \centering
        \begin{subfigure}[t]{0.49\textwidth}
                \centering
                \includegraphics[width=1.0\textwidth]{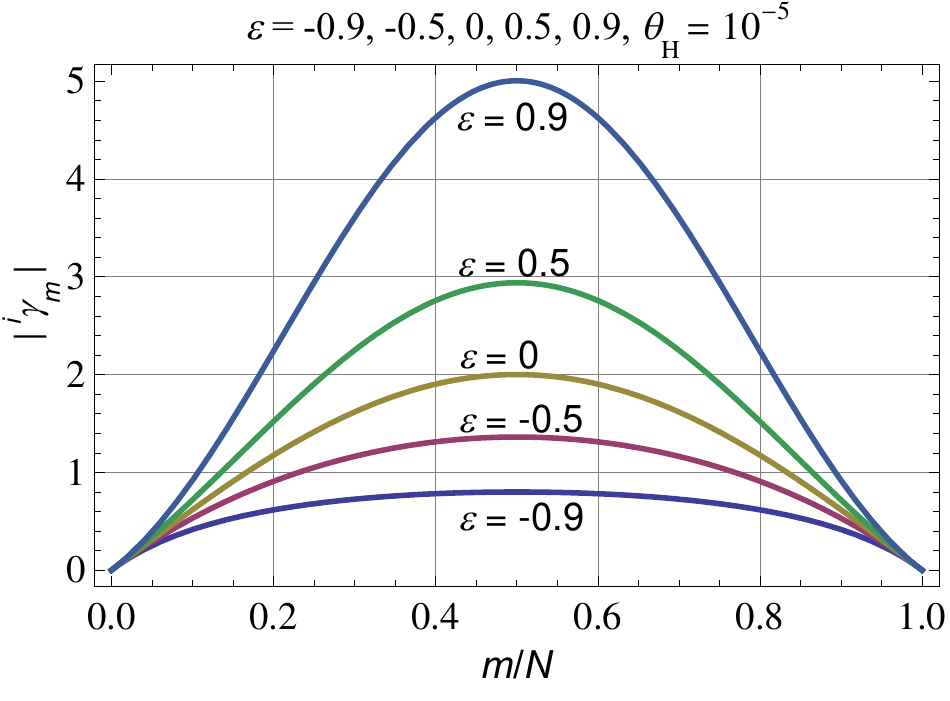}
                \caption{For $\theta_H\to 0$ (weak magnetic field limit). }
                \label{fig:absigamma-vs-mdN-vs-eps}
        \end{subfigure}
        \hfill  % An dieser Stelle kann ein zusätzlicher Zwischenraum eingebunden werden: ~, \quad, \qquad, \hfill usw.
          % Eine leere Zeile erzwingt, dass die zweite Grafik darunter erscheint.
        \begin{subfigure}[t]{0.5\textwidth}
                \centering
                \includegraphics[width=1.0\textwidth]{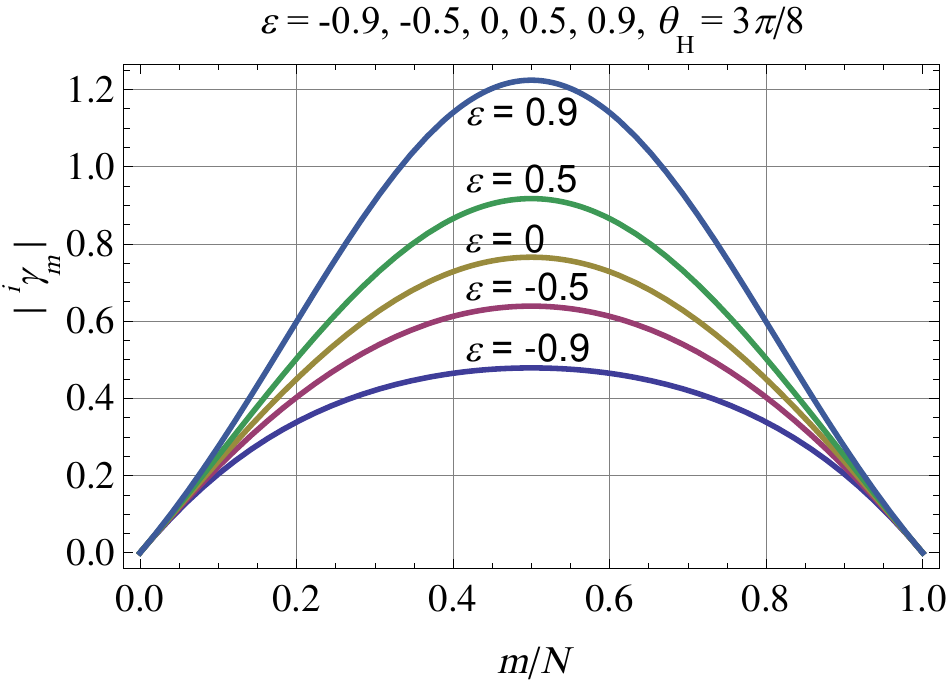}
                \caption{For $\theta_H=3\pi/8$ (strong magnetic field). }
                \label{fig:absigamma-vs-mdN-vs-eps_largeB}
        \end{subfigure}
    \caption{Magnitudes of the eigenvalues ${^i}\!\gamma_m$ of regular symmetric Hall-plates versus $m/N$ for various contacts sizes, $\epsilon$, after (\ref{eq:RMFoCD1503}). }
   \label{fig:absigamma}
\end{figure*}
% original file: HomentcovschiMurray_regular_HHalls_eigenvalues5.nb
%
%
\begin{figure*}
\vspace{1mm}
  \centering
        \begin{subfigure}[t]{0.496\textwidth}
                \centering
                \includegraphics[width=1.0\textwidth]{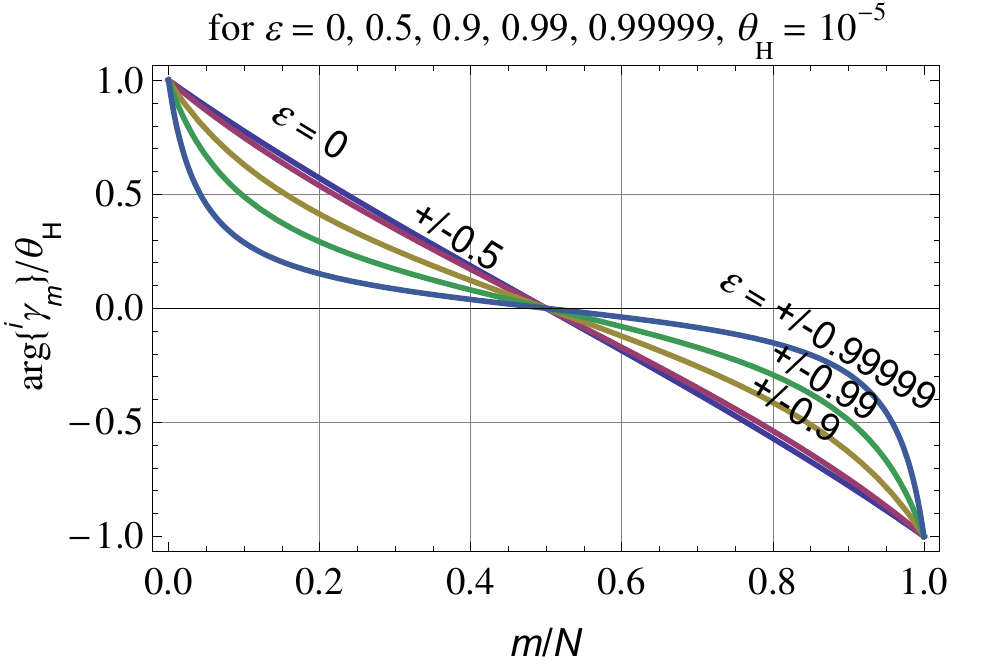}
                \caption{For $\theta_H\to 0$ (weak magnetic field limit). }
                \label{fig:argigamma-vs-mdN-various-eps_smallB}
        \end{subfigure}
        \hfill  % An dieser Stelle kann ein zusätzlicher Zwischenraum eingebunden werden: ~, \quad, \qquad, \hfill usw.
          % Eine leere Zeile erzwingt, dass die zweite Grafik darunter erscheint.
        \begin{subfigure}[t]{0.499\textwidth}
                \centering
                \includegraphics[width=1.0\textwidth]{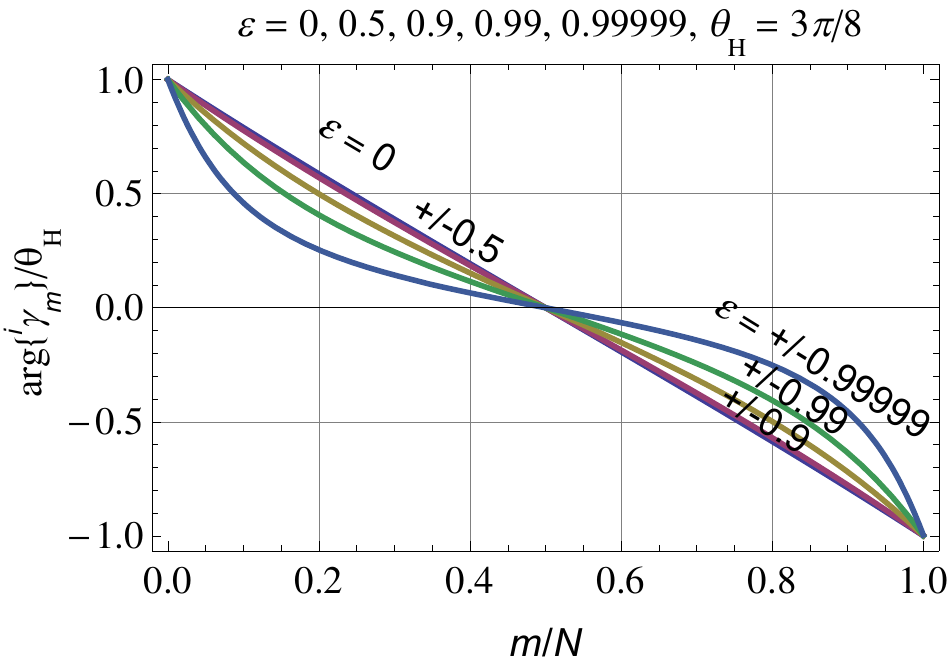}
                \caption{For $\theta_H=3\pi/8$ (strong magnetic field). }
                \label{fig:argigamma-vs-mdN-various-eps_largeB}
        \end{subfigure}
    \caption{Arguments of the eigenvalues ${^i}\!\gamma_m$ of regular symmetric Hall-plates versus $m/N$ for various contacts sizes, $\epsilon$, after (\ref{eq:RMFoCD1503}). }
   \label{fig:argigamma}
\end{figure*}
% original file: HomentcovschiMurray_regular_HHalls_eigenvalues5.nb

\section{Some Properties of the Eigenvalues ${^i}\!\gamma_m$} 
\label{sec:igamma}

\subsection{General Symmetry Properties of ${^i}\!\gamma_m$} 
\label{sec:igamma-sym}

\noindent If we reverse the magnetic field in (\ref{eq:RMFoCD1503}) it holds 
\begin{equation}\label{eq:RMFoCD1504}
{^i}\!\gamma_m (-\theta_H) = {^i}\!\gamma_m^C (\theta_H) = {^i}\!\gamma_{N-m} (\theta_H) ,
\end{equation}
where I used (\ref{eq:RMFoCD69a}) for the rightmost equality. Thus the argument of each eigenvalue is an odd function of the Hall-angle, 
\begin{equation}\label{eq:RMFoCD1504a}
\mathrm{arg}\{{^i}\!\gamma_m (-\theta_H)\} = -\mathrm{arg}\{{^i}\!\gamma_m (\theta_H)\} .
\end{equation}
Consequently, at zero magnetic field the eigenvalues become real numbers. Moreover, the magnitudes of the eigenvalues are independent of the polarity of the magnetic field 
\begin{equation}\label{eq:RMFoCD1504aa}
\left|{^i}\!\gamma_m (-\theta_H)\right| = \left|{^i}\!\gamma_m (\theta_H)\right| .
\end{equation}
At large positive Hall-angle, $\theta_H\to\pi/2$ the integrals in the numerator and in the denominator of (\ref{eq:RMFoCD1503}) grow unboundedly. The $h$-functions at $(-\theta_H)$ are responsible for this. Therefore, we can discard the $h$-functions at $\theta_H$ in this asymtotic limit. Let us first study the two limits 
\begin{equation}\label{eq:RMFoCD1504b}\begin{split}
& \lim_{\alpha\to 1-0} \int_0^{\pi/4} (1\pm\epsilon)\frac{\exp\left(\mp\mathbbm{i}y(1\pm\epsilon)(2m/N-1)\right)}{\left(\sin(y(1\pm\epsilon))\right)^\alpha} \,\mathrm{d}y \\ 
= & \lim_{\alpha\to 1-0} \int_0^{\pi/4} \left[ (1\pm\epsilon)\frac{\exp\left(\mp\mathbbm{i}y(1\pm\epsilon)(2m/N-1)\right)}{\left(\sin(y(1\pm\epsilon))\right)^\alpha} \right. \\
& \qquad\qquad\quad \left. -\frac{1}{y^\alpha} \right] \,\mathrm{d}y + \int_0^{\pi/4}\frac{\mathrm{d}y}{y^\alpha} %= \lim_{\alpha\to 1-0} \int_0^{\pi/4} \frac{\mathrm{d}\,y}{y^\alpha}
 \\ 
= & \lim_{\alpha\to 1-0} \left. \frac{y^{1-\alpha}}{1-\alpha} \right|_0^{\pi/4} = \lim_{\alpha\to 1-0} \frac{1}{1-\alpha} = +\infty .
\end{split}\end{equation}
In the second line of (\ref{eq:RMFoCD1504b}) the first integral is finite---we can neglect it against the infinite second integral. We use (\ref{eq:RMFoCD1504b}) in (\ref{eq:RMFoCD1503}). After a few rearrangements this gives 
\begin{equation}\label{eq:RMFoCD1504c}
\lim_{\theta_H\to\pi/2} {^i}\!\gamma_m = 2\frac{\cos(\theta_H)}{R_\mathrm{sheet}} \sin\!\left(\!\pi\frac{m}{N}\!\right) \exp\!\left(\!\mathbbm{i}\frac{\pi}{2}\!\left(\!1\!-\!\frac{2m}{N}\!\right)\!\right) .
\end{equation}
It means that the eigenvalues of the indefinite conductance matrices of regular symmetric Hall-plates at strong magnetic field are proportional to $\cos(\theta_H)$. They tend to zero, while their arguments are linear functions of $m/N$, aequidistantly  distributed between $-90$° and $90$°. Thereby, the size of the contacts is irrelevant; $\epsilon$ does not show up in (\ref{eq:RMFoCD1504c}).

From (\ref{eq:RMFoCD1503}) it also follows 
\begin{equation}\label{eq:RMFoCD1505}
{^i}\!\gamma_m (-\epsilon,-\theta_H) {^i}\!\gamma_m (\epsilon,\theta_H) = \left(\frac{2}{R_\mathrm{sq}}\sin\left(\pi \frac{m}{N}\right)\right)^{\!\!2} ,
\end{equation}
which is yet another form of the \emph{Reverse Magnetic Field on Complementary Device}-principle (RMFoCD)\cite{rspa2,Ausserlechner2019}, relating the eigenvalues of a regular symmetric Hall-plate, ${^i}\!\gamma_m (\epsilon,\theta_H)$, to the eigenvalues of its complementary Hall-plate at reverse magnetic field, $\overline{{^i}\!\gamma_m (\epsilon,-\theta_H)}={^i}\!\gamma_m (-\epsilon,-\theta_H)$. A complementary Hall-plate is obtained from an original Hall-plate by swapping all insulating and conducting segments on the boundary. All properties of complementary Hall-plates are denoted by an overbar. In matrix form (\ref{eq:RMFoCD1505}) reads 
\begin{equation}\label{eq:RMFoCD1505b}\begin{split}
& \overline{\bm{{^i}}\!\bm{\Gamma}(-\theta_H)} \bm{{^i}\!\Gamma}(\theta_H) = \frac{\bm{\Delta}}{R_\mathrm{sq}^2} \Rightarrow \overline{\bm{{^i}\!\Gamma}(-\theta_H)} = \frac{\bm{\Delta}}{R_\mathrm{sq}^2} \bm{{^i}\!\Gamma}^+(\theta_H) , \\ %\text{ with } \\
& \overline{\bm{{^i}}\!\bm{\Gamma}} = \mathrm{diag}(\overline{{^i}\!\gamma}_1,\overline{{^i}\!\gamma}_2,\ldots ,\overline{{^i}\!\gamma}_N) , \text{ with } \overline{{^i}\!\gamma}_\ell(\epsilon) = {^i}\!\gamma_\ell(-\epsilon) , \\
& (\bm{\Delta})_{k,\ell} = \left(2\sin\left(\frac{\pi\ell}{N}\right)\right)^{\!\!2} \delta_{k,\ell}, \forall k,\ell\in\{1,2,\ldots,N\} .
\end{split}\end{equation}
$\bm{\Delta}$ is a diagonal matrix \cite{ ArXiv2023}. 
%with the eigenvalues of the matrix $2*\bm{1}-\bm{\hat{1}}-\bm{\hat{1}}^T$ on its main diagonal \cite{ ArXiv2023} (here, $*$ means a multiplication of a scalar with a matrix). 
$\delta_{k,\ell}$ is Kronecker's delta. $\bm{{^i}\!\Gamma}$ is a diagonal matrix. $\bm{{^i}\!\Gamma}^+$ is the pseudo-inverse (= Moore-Penrose inverse) of $\bm{{^i}\!\Gamma}$. In this simple case of a diagonal matrix it means that all zeros remain zeros and all non-zeros are replaced by their reciprocals. From (\ref{eq:RMFoCD1505}) and (\ref{eq:RMFoCD1504aa}) we get 
\begin{equation}\label{eq:RMFoCD1505b2}
\left|\overline{{^i}\!\gamma_m}\right| \left|{^i}\!\gamma_m\right| = \left(\frac{2}{R_\mathrm{sq}}\sin\left(\pi \frac{m}{N}\right)\right)^{\!\!2} ,
\end{equation}
where we simply wrote $\overline{{^i}\!\gamma_m}$ instead of $\overline{{^i}\!\gamma_m (\epsilon,\theta_H)}$ and ${^i}\!\gamma_m$ instead of ${^i}\!\gamma_m (\epsilon,\theta_H)$. From (\ref{eq:RMFoCD1505b2}) it follows 
\begin{equation}\label{eq:RMFoCD1505b3}
\frac{\left|{^i}\!\gamma_m\right|}{\left|\overline{{^i}\!\gamma_m}\right|} =  \left(\frac{R_\mathrm{sq}\left|{^i}\!\gamma_m\right|}{2\sin\left(\pi \frac{m}{N}\right)}\right)^{\!\!2} = \left(\frac{\left|{^i}\!\gamma_m\right|}{\left|{^i}\!\gamma_m (\epsilon=0)\right|}\right)^{\!\!2} .
\end{equation}
From (\ref{eq:RMFoCD1505}) and (\ref{eq:RMFoCD1504a}) it follows that the arguments of the original and the complementary Hall-plates are the same, 
\begin{equation}\label{eq:RMFoCD1505c}
\mathrm{arg}\left\{{^i}\!\gamma_m (-\epsilon,\theta_H)\right\} = \mathrm{arg}\left\{{^i}\!\gamma_m (\epsilon,\theta_H)\right\} .
\end{equation}

\noindent Figures \ref{fig:absigamma} and \ref{fig:argigamma} show plots of the magnitude and the argument of ${^i}\!\gamma_m$ versus $m/N$, for various sizes of contacts, $\epsilon$, for weak and strong magnetic field. The magnitude becomes smaller with larger Hall-angle, which is plausible, because the conductances get smaller due to the magneto-resistive effect. Moreover, it is also obvious that the magnitudes of the eigenvalues become smaller for smaller contacts, $\epsilon<0$, because smaller contacts naturally give smaller conductances, too. Third, the magnitudes tend to zero for $m\to 0$ and $m\to N$, whereas they show a maximum at $m=N/2$. These properties are reflected in (\ref{eq:RMFoCD1505}). The arguments of the eigenvalues are close to the straight line $(1-2m/N)\theta_H$ for very large Hall-angles or for \emph{almost} strictly regular symmetry, $|\epsilon|<0.5$. Conversely, at moderate or weak magnetic field, very large or very small contacts pull $\mathrm{arg}\{{^i}\!\gamma_m\}/\theta_H$ towards zero, wherey the sign of $\epsilon$ does not change the arguments of the eigenvalues -- small and large contacts with $\pm\epsilon$ have the same $\mathrm{arg}\{{^i}\!\gamma_m\}$, see (\ref{eq:RMFoCD1505c}).

\subsection{Eigenvalues ${^i}\!\gamma_m$ at Zero Magnetic Field} 
\label{sec:igamma-zero-field}

\noindent For even $N$ and $\theta_H=0$ equation (\ref{eq:RMFoCD1503}) gives (see Appendix \ref{sec:igamma-zero}) 
\begin{equation}\label{eq:RMFoCD1550}\begin{split}
& {^i}\!\gamma_{N/2}(\theta_H=0) = \frac{2}{R_\mathrm{sheet}} \\ 
& \;\;*\frac{\!F\!\!\left(\!\!\sqrt{\!\frac{\!\cos\!(\!(\!1\!-\!\epsilon\!)\!\pi/4\!)}{\!\sin\!(\!(\!1\!+\!\epsilon\!)\!\pi/4\!)}},\!\sqrt{\!\cos\!(\!(\!1\!-\!\epsilon\!)\!\pi/4\!)\!\sin\!(\!(\!1\!+\!\epsilon\!)\!\pi/4\!)}\right)}{\!F\!\!\left(\!\!\sqrt{\!\frac{\!\cos\!(\!(\!1\!+\!\epsilon\!)\!\pi/4\!)}{\!\sin\!(\!(\!1\!-\!\epsilon\!)\!\pi/4\!)}},\!\sqrt{\!\cos\!(\!(\!1\!+\!\epsilon\!)\!\pi/4\!)\!\sin\!(\!(\!1\!-\!\epsilon\!)\!\pi/4\!)}\right)} .
\end{split}\end{equation}
Here, $*$ means a multiplication of two scalars. In (\ref{eq:RMFoCD1550}) we used the incomplete elliptic integral of the first kind, which is available in in Mathematica, $F\!(\sin(\phi),k)=\mathrm{EllipticF}[\phi,k^2]$. It is defined by  
\begin{equation}\label{eq:RMFoCD1553}
F\!(w,k) = \int_0^w \left(1-x^2\right)^{-1/2} \left(1-k^2 x^2\right)^{-1/2}\,\mathrm{d}x . 
\end{equation}

\noindent If $N$ is an integer multiple of three, equation (\ref{eq:RMFoCD1503}) can also be integrated in closed form for $m=N/3$. However, this gives a long expression for ${^i}\!\gamma_{N/3}(\theta_H=0)$, which I do not report here. Later we will get a simpler formula anyhow, see (\ref{eq:RMFoCD1573}). \\
%\begin{equation}\label{eq:RMFoCD1551}\begin{split}
%& {^i}\!\gamma_{N/3}(\theta_H=0) = \frac{2}{R_\mathrm{sheet}} \\ 
%& XXXXXXX .
%\end{split}\end{equation}

\noindent If $N$ is an integer multiple of four, equation (\ref{eq:RMFoCD1503}) gives (see Appendix \ref{sec:igamma-zero}) 
\begin{equation}\label{eq:RMFoCD1552}\begin{split}
{^i}\!\gamma_{N/4}(\theta_H=0) & = \frac{\sqrt{2}}{R_\mathrm{sheet}} \sqrt{\frac{\sqrt{2}+\sqrt{1+\sin(\pi\epsilon/2)}}{\sqrt{2}+\sqrt{1-\sin(\pi\epsilon/2)}}} \\ 
& *\frac{K\!\!\left(\sqrt{\frac{\sqrt{2}-\sqrt{1-\sin(\pi\epsilon/2)}}{\sqrt{2}+\sqrt{1-\sin(\pi\epsilon/2)}}}\right)}{K\!\!\left(\sqrt{\frac{\sqrt{2}-\sqrt{1+\sin(\pi\epsilon/2)}}{\sqrt{2}+\sqrt{1+\sin(\pi\epsilon/2)}}}\right)} ,
\end{split}\end{equation}
wherein $K(k)=F\!(1,k)$ is the complete elliptic integral of the first kind, also available in Mathematica, $K\!(k)=\mathrm{EllipticK}[k^2]$. 

\begin{figure}
%\vspace{1mm}
  \centering
        \begin{subfigure}[b]{0.245\textwidth}
                \centering
                \includegraphics[width=1.0\textwidth]{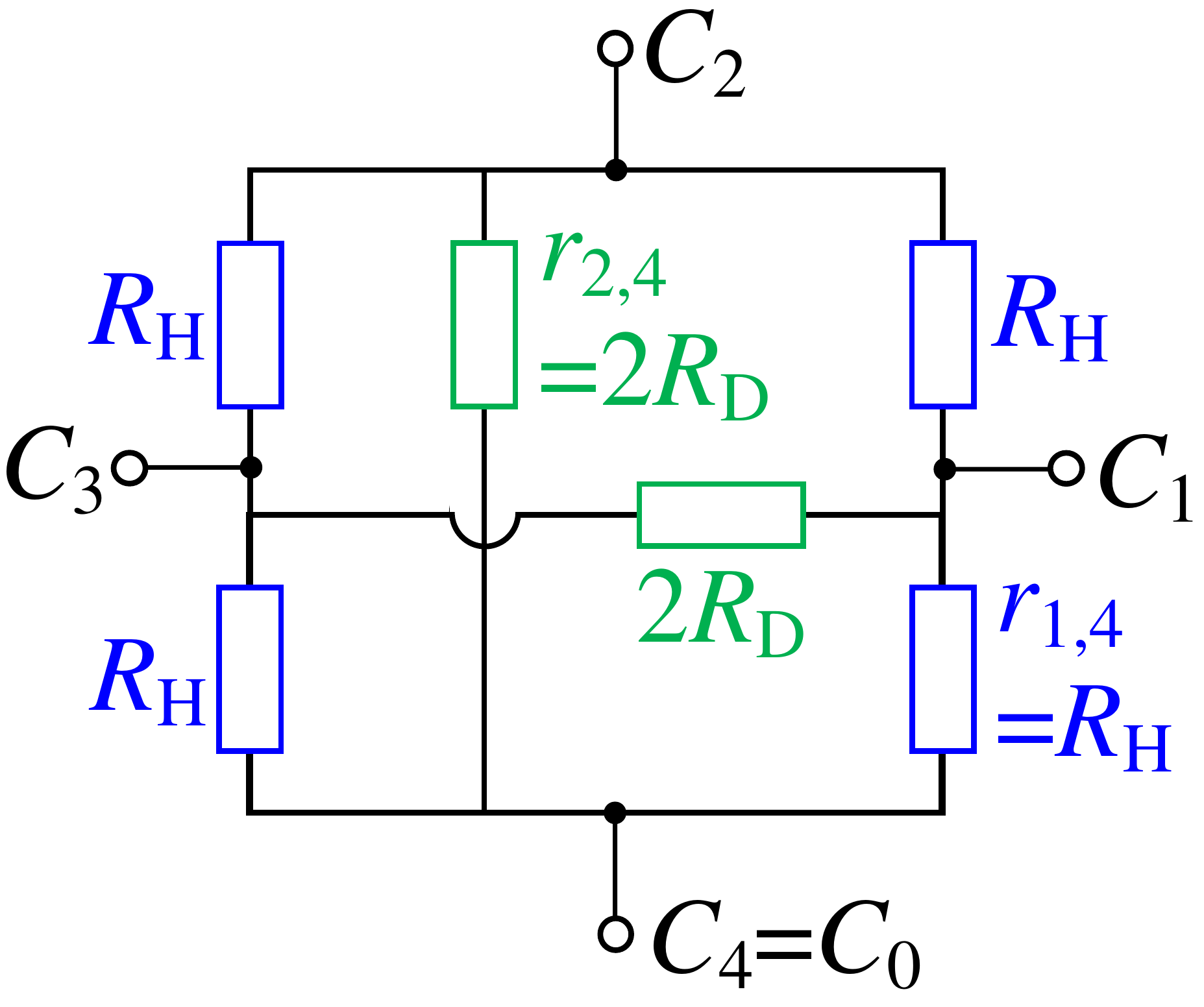}
                \caption{ERC. }
                \label{fig:ERC-of-regular-4C-Hall}
        \end{subfigure}
	\hfill
        \begin{subfigure}[b]{0.23\textwidth}
                \centering
                \includegraphics[width=1.0\textwidth]{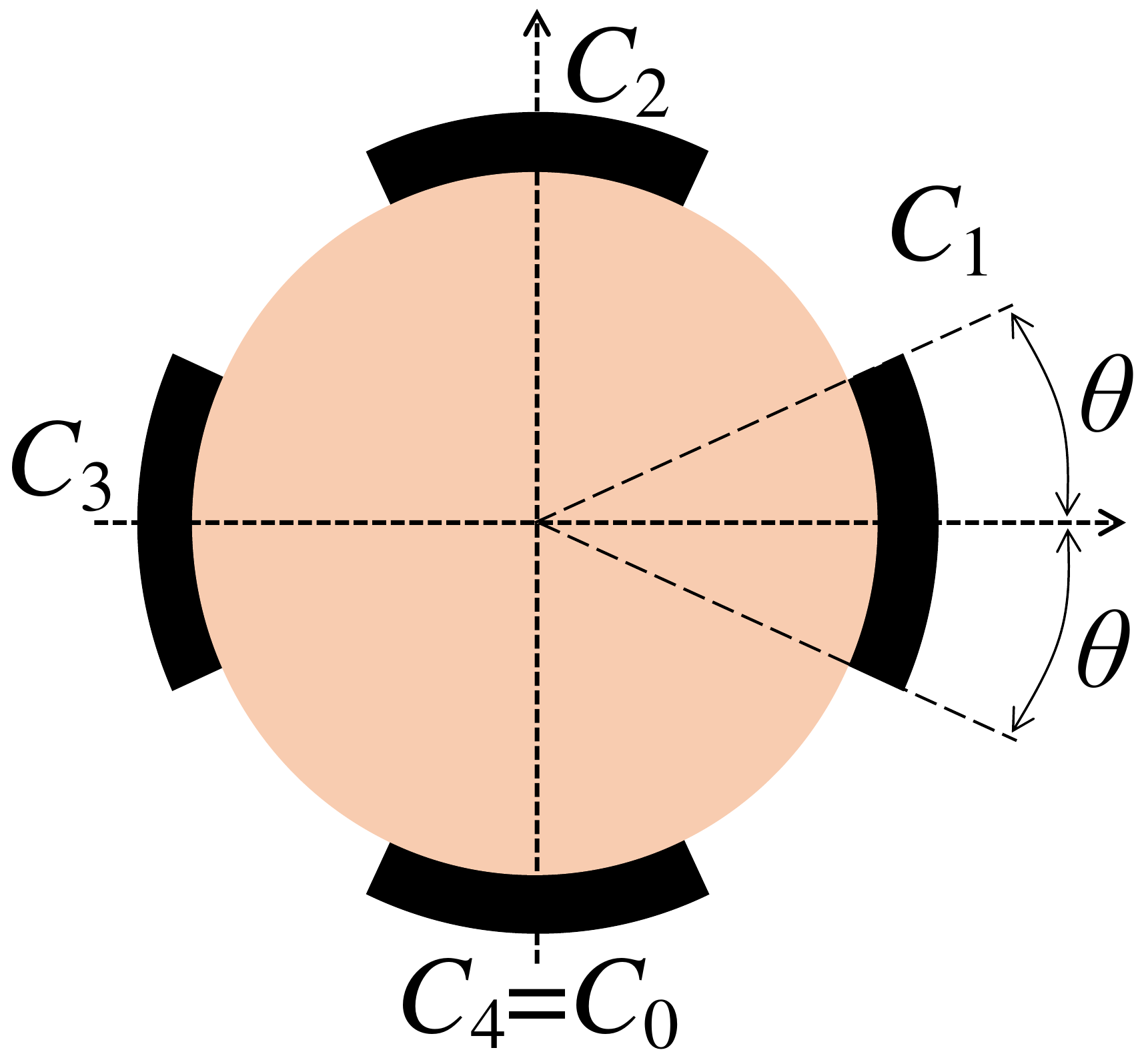}
                \caption{Disk shaped Hall-plate. }
                \label{fig:regular-4C-Hall-disk}
        \end{subfigure}
    \caption{Equivalent resistor circuit (ERC) and disk-shape representation of a regular Hall-plate with four contacts. }
   \label{fig:regular-4C-Hall}
\end{figure}

\noindent We can get simpler expressions than (\ref{eq:RMFoCD1550}), (\ref{eq:RMFoCD1552}) if we use prior literature on symmetric Hall-plates with three and four contacts, because eigenvalues do not depend on the absolute values of $m$ and $N$, but only on the ratio $m/N$, see (\ref{eq:RMFoCD1503}) (this holds with or without applied magnetic field). Therefore, ${^i}\!\gamma_{N/2}$ for even $N$ is identical to ${^i}\!\gamma_2$ for $N=4$. Likewise, for $N$ being an integer multiple of four, the eigenvalues ${^i}\!\gamma_{N/4}$ are the same as ${^i}\!\gamma_1$ for $N=4$. We use (\ref{eq:RMFoCD52}) to link the eigenvalues with the conductances for $N=4$, 
\begin{equation}\label{eq:RMFoCD1555}\begin{split}
G_{1,2}(\theta_H=0) & = \frac{-{^i}\!\gamma_2(\theta_H=0)}{4} , \text{ and }\\ 
G_{1,3}(\theta_H=0) & = \frac{-{^i}\!\gamma_1(\theta_H=0)}{2}+\frac{{^i}\!\gamma_2(\theta_H=0)}{4} .
\end{split}\end{equation}
Next, we relate the conductances to the resistances in the equivalent resistor circuit (ERC) according to Ref.~\onlinecite{Ausserlechner2020a}, 
\begin{equation}\label{eq:RMFoCD1556}\begin{split}
G_{1,2}(\theta_H=0) & = \frac{-1}{r_{1,4}(\theta_H=0)}, \text{ and }\\
G_{1,3}(\theta_H=0) & = \frac{-1}{r_{2,4}(\theta_H=0)} ,
\end{split}\end{equation}
whereby $r_{\ell,m}$ are the lumped resistors between contacts $C_\ell$ and $C_m$ as in Fig. \ref{fig:ERC-of-regular-4C-Hall}. 
Due to the symmetry of the Hall-plate the ERC has only two types of lumped resistors: $R_H$ is between neighboring contacts and $2R_D$ is between non-neighboring contacts, see Fig. \ref{fig:ERC-of-regular-4C-Hall}.

\noindent A key parameter of Hall-plates with four contacts is the \emph{effective number of squares}, $\lambda$. It is defined as the ratio of the zero magnetic field resistance measured between two non-neighoring contacts over the sheet resistance, 
\begin{equation}\label{eq:4C-Hall6a}
\lambda = \frac{R_{2,2}(\theta_H=0)}{R_\mathrm{sheet}} \text{ for } N=4 .
\end{equation}
A rectangle with end contacts and aspect ratio equal to $\lambda$ has the same resistance as the four contacts Hall-plate at zero magnetic field. 
Note that in the ERC of the Hall-plate the resistor $r_{2,4}=2R_D$ lies between the contacts $C_2$ and $C_4$, whereas the resistance measured between these contacts is $R_{2,2}$, with $R_{2,2}\ne 2R_D$.

\noindent Moreover we define the \emph{ cross-resistance}, $R_x$, measured between two terminals, where the contacts with even numbers are connected to the first terminal and the contacs with odd numbers are connected to the second terminal. At zero magnetic field it holds 
\begin{equation}\label{eq:4C-Hall6ab}
\lambda_x = \frac{R_x(\theta_H=0)}{R_\mathrm{sheet}}, 
\end{equation}
with $\lambda_x$ being the number of cross squares.
It is straightforward to show how the resistors in the ERC are related to $\lambda,\lambda_x$ (see also  Ref.~\onlinecite{AusserlechnerVdP2}), 
\begin{equation}\label{eq:RMFoCD1557}\begin{split}
r_{1,4}(\theta_H=0) & = R_H(\theta_H=0) = 4 R_\mathrm{sheet} \lambda_x ,\text{ and }\\ 
r_{2,4}(\theta_H=0) & = 2R_D(\theta_H=0) = \frac{4 R_\mathrm{sheet} \lambda_x \lambda}{4\lambda_x-\lambda} .
\end{split}\end{equation}
From Ref.~\onlinecite{AusserlechnerVdP2} we also know that $\lambda$ and $\lambda_x$ are not independent, 
\begin{equation}\label{eq:RMFoCD1557b}
4\lambda_x=\frac{K'\!\left(L(\lambda)\right)}{K\!\left(L(\lambda)\right)} \;\Leftrightarrow\; L(4\lambda_x)=\left(L(\lambda)\right)^{\!\!2} ,
\end{equation}
with $K'(k)=K(\sqrt{1-k^2})$. $L(y)$ is the modular lambda function\cite{Nehari1952} defined by $L(K'(k)/K(k))=k^2 \;\forall k\in [0,1]$. It is defined in Mathematica  $L(y)=\mathrm{ModularLambda}[\mathbbm{i}y]$. 

\noindent Finally, we express the effective number of squares, $\lambda$, by the azimuthal angle, $2\theta$, subtended by a contact of the Hall-plate when the plate is mapped conformally to the unit disk \cite{GH0approx} (see Fig. \ref{fig:regular-4C-Hall-disk}) 
\begin{equation}\label{eq:RMFoCD1559}
\lambda = 2\frac{K\!(k_1)}{K'\!(k_1)} \text{ with } k_1=\frac{1-\tan{(\theta)}}{1+\tan{(\theta)}} .
\end{equation}
(Do not mix up the geometrical parameter $\theta$ with the Hall-angle $\theta_H$.) According to (\ref{eq:RMFoCD100b}) and with $\chi=1+\epsilon$ it holds 
\begin{equation}\label{eq:RMFoCD1560}
\theta=\pi\frac{1+\epsilon}{8} . 
\end{equation}
Appendix \ref{sec:igamma-zero} shows how to combine (\ref{eq:RMFoCD1555})---(\ref{eq:RMFoCD1560}) to get the eigenvalues of a four-contacts regular symmetric Hall-plate at zero magnetic field, 
\begin{equation}\label{eq:RMFoCD1565a}\begin{split}
{^i}\!\gamma_{N/2}(\theta_H=0) & = \left.{^i}\!\gamma_2(\theta_H=0)\right|_{N=4} \\
& = \frac{4}{R_\mathrm{sheet}} \underbrace{\frac{K}{K'}\!\!\left(\left[\tan\left(\pi\frac{1+\epsilon}{8}\right)\right]^2\right)}_{=1/(4\lambda_x) \text{ see (\ref{apx-popularHalls12})}} , 
\end{split}\end{equation} 
\begin{equation}\label{eq:RMFoCD1565b}\begin{split}
{^i}\!\gamma_{N/4}(\theta_H=0) & = \left.{^i}\!\gamma_1(\theta_H=0)\right|_{N=4} \\
& = \frac{2}{R_\mathrm{sheet}} \underbrace{\frac{K}{K'}\!\!\left(\tan\left(\pi\frac{1+\epsilon}{8}\right)\right)}_{=1/\lambda \text{ see (\ref{apx-popularHalls12c})}} ,
\end{split}\end{equation}
with the abbreviation 
\begin{equation}\label{eq:RMFoCD1565b2}
\frac{K}{K'}\!\!\left(k\right) = \frac{K(k)}{K'(k)} = \frac{K(k)}{K\left(\sqrt{1-k^2}\right)} .
\end{equation}
Equations (\ref{eq:RMFoCD1565a}) and (\ref{eq:RMFoCD1565b}) are simpler than (\ref{eq:RMFoCD1550}) and (\ref{eq:RMFoCD1552}). 

Now we apply the same method to a regular symmetric Hall-plate with three contacts, $N=3$. From (\ref{eq:RMFoCD52}) we get $G_{1,2}(\theta_H=0) = -{^i}\!\gamma_1(\theta_H=0)/3$. The ERC has only three identical resistors between each pair of contacts. We use Ref.~\onlinecite{3C-Hall}, where this resistor is called $R_d(\theta_H=0) = R_\mathrm{sheet}\lambda_d$ with 
\begin{equation}\label{eq:RMFoCD1570}\begin{split}
& \lambda_d = \frac{K'\!(k_d)}{2K\!(k_d)} , \text{ with } k_d=\frac{\left(\!\sqrt{W_1\!-\!W_3}-\sqrt{W_2\!-\!W_3}\right)^{2}}{W_1-W_2} , \\ 
& \text{with } W_1=-\left(\!\tan\!\left(\!\pi\frac{1\!-\!\epsilon}{12}\right)\!\!\right)^{\!\!2} , \\
& W_2=-\left(\!\tan\!\left(\!\pi\frac{3\!+\!\epsilon}{12}\right)\!\!\right)^{\!\!2} , W_3=-\left(\!\tan\!\left(\!\pi\frac{5\!-\!\epsilon}{12}\right)\!\!\right)^{\!\!2} .
\end{split}\end{equation}
It holds $G_{1,2}(\theta_H=0)=-1/R_d(\theta_H=0)$. Combining all these facts gives the eigenvalue of a three-contacts regular symmetric Hall-plate at zero magnetic field, 
\begin{equation}\label{eq:RMFoCD1573}\begin{split}
& {^i}\!\gamma_{N/3}(\theta_H=0) = \left.{^i}\!\gamma_1(\theta_H=0)\right|_{N=3} \\ 
& = \frac{3}{R_\mathrm{sheet}} \frac{K}{K'}\!\!\left(\frac{\cos\left(\pi\frac{5-\epsilon}{12}\right)}{\sqrt{2}\cos\left(\pi\frac{3+\epsilon}{12}\right)} \sqrt{1+\sqrt{3}\tan\left(\frac{\pi\epsilon}{6}\right)}\right)
%= \frac{6}{R_\mathrm{sheet}} \frac{K\!(k_d)}{K'\!(k_d)} , \\ & \text{with } k_d=\frac{\sqrt{3}(2z^2+2z-1)-(1+2z)\sqrt{4z^2-1}}{2(1-z)\sqrt{1-z^2}} , \\ & \text{with } z = \cos\left(\pi\frac{1+\epsilon}{6}\right) .
\end{split}\end{equation}
In the limit of vanishing spacings between the contacts, $\epsilon\to 1$, the eigenvalues of (\ref{eq:RMFoCD1565a}), (\ref{eq:RMFoCD1565b}) and (\ref{eq:RMFoCD1573}) diverge logarithmically like $K(k\to 1)=\ln\left(4/\sqrt{1-k^2}\right)$. Conversely, in the limit of vanishing contacts sizes, $\epsilon\to -1$, these eigenvalues go to zero like $1/K'(k\to 0)=1/\ln\left(4/k\right)$.

\subsection{Eigenvalues for Small Contacts or Contacts Spacings} 
\label{sec:small-contacts}

\noindent Fig. \ref{fig:argigamma} shows that the arguments of the eigenvalues vanish in the limit $\epsilon\to\pm 1$, except for $m/N\to 0$ or $m/N\to 1$ or $\theta_H\to\pm\pi/2$, however, they diminish only slowly. Thus, in practical cases $|\epsilon|$ will hardly every be larger than $0.99$, and therefore we should not expect to see the arguments vanish in real life. 
%In this case, the slope of the curves, $\mathrm{d}\left(\mathrm{arg}\{{^i}\gamma_m\}/\theta_H\right)/\mathrm{d}m$ for $m\to 0$ and $m\to N/2$ gives a good impression on how the arguments vary versus contacts sizes.

The strongest deviation of $\mathrm{arg}\{{^i}\gamma_m\}/\theta_H$ from the straight line in Fig. \ref{fig:argigamma} occurs in the weak magnetic field limit. There, we may replace $z^{\theta_H/\pi}\to 1+(\theta_H/\pi)\ln(z)$ in (\ref{eq:RMFoCD1503}), analogous to (63) in Ref.~\onlinecite{ArXiv2023}. This formula is lengthy and seems to give no new insight, therefore I do not report it here.

Regular symmetric Hall-plates with very small contacts spacings, $\epsilon\to 1$, or with very small contacts $\epsilon\to -1$, are linked by (\ref{eq:RMFoCD1505}). 
Let us study the case of small contacts explicitly---with (\ref{eq:RMFoCD1505}) one can deduce the converse case. 
%The interesting point about these two extremes is that for zero contacts spacings the Hall-effect vanishes \cite{Haeusler1967,Ausserlechner2019c} (i.e., the Hall-geometry factor of a four-contacts Hall-plate is zero, $G_H^{(4C)}=0$, see Section \ref{sec:4C-Hall}), whereas for point-sized contacts the Hall-effect is maximized (i.e., the Hall-geometry factor equals one, $G_H^{(4C)}=1$). However, in both cases the argument of the eigenvalue tends to zero, $\mathrm{arg}\{{^i}\!\gamma_1\}/\theta_H\to 0$, see Fig. \ref{fig:argigamma}. XXXXXXXXXXXXXXX

In Section \ref{sec:4C-Hall} we compute the output voltage of a regular Hall-plate with four contacts, see (\ref{eq:4C-Hall1}). Alternatively, we may compute the output voltage via the resistance matrix, 
\begin{equation}\label{eq:smallC-5}\begin{split}
\frac{V_3 - V_1}{I_\mathrm{supply}} = \left(R_{3,2}-R_{1,2}\right) = 2\frac{\sin\left(\mathrm{arg}\{{^i}\gamma_1\}\right)}{|{^i}\gamma_1|} ,
\end{split}\end{equation}
where we used (\ref{eq:RMFoCD1515}). In the limit of weak magnetic field we can insert (\ref{eq:RMFoCD1565b}) into (\ref{eq:smallC-5}) and set the result equal to (\ref{eq:4C-Hall1}). This gives 
\begin{equation}\label{eq:smallC-6}
G_{H0}^{(4C)} = \underbrace{\frac{K'}{K}\!\!\left(\!\tan\!\left(\pi\frac{1+\epsilon}{8}\right)\!\right)}_{=\lambda \text{ see (\ref{apx-popularHalls12c})}} \lim_{\theta_H\to 0} \frac{\sin\left(\mathrm{arg}\{{^i}\!\gamma_1\}\right)}{\sin\left(\theta_H\right)} .
\end{equation}
If we use (\ref{eq:4C-Hall4}) - (\ref{eq:4C-Hall5}) for $G_{H0}^{(4C)}$ in (\ref{eq:smallC-6}) we can compute the argument of the eigenvalue as a function of the size of the contacts, $\epsilon$. Thereby, we link the integral in $G_{H0}^{(4C)}$ (see (\ref{eq:4C-Hall5})) with the integral in ${^i}\!\gamma_1$ (see (\ref{eq:RMFoCD1503}))---both integrals are not solvable in closed form for general $\epsilon$. However, in the rare cases where we know an exact value for $G_{H0}^{(4C)}$, (\ref{eq:smallC-6}) returns an exact value for $\mathrm{arg}\{{^i}\!\gamma_1\}$. One such case is small contacts, $\epsilon= -1+\mathrm{d}\epsilon$ with $\mathrm{d}\epsilon\to 0$, where it holds $G_{H0}^{(4C)}\to 1$. This gives 
\begin{equation}\label{eq:smallC-8}\begin{split}
& \lim_{\theta_H\to 0}\lim_{\mathrm{d}\epsilon\to 0} \frac{\mathrm{arg}\{{^i}\!\gamma_{N/4}\}}{\theta_H} = \lim_{\mathrm{d}\epsilon\to 0} \frac{K}{K'}\!\!\left(\tan\!\left(\pi\frac{\mathrm{d}\epsilon}{8}\right)\!\right) \\ 
& \qquad = \frac{\pi}{2}\;\frac{1}{5\ln(2)-\ln(\pi\,\mathrm{d}\epsilon)} .
\end{split}\end{equation}
For $\epsilon=0.99999$ and $m/N=0.25$ (\ref{eq:smallC-8}) gives a value of $0.113547$, which differs from the exact value in Fig. \ref{fig:argigamma-vs-mdN-various-eps_smallB} only by $3.2$ppm. For $\epsilon=0.99$ the relative error of (\ref{eq:smallC-8}) increases to $0.32\%$, and even for $\epsilon=0.9$ the relative error of (\ref{eq:smallC-8}) is still only $3.3\%$.

\begin{figure*}
\vspace{1mm}
  \centering
        \begin{subfigure}[t]{0.48\textwidth}
                \centering
                \includegraphics[width=1.0\textwidth]{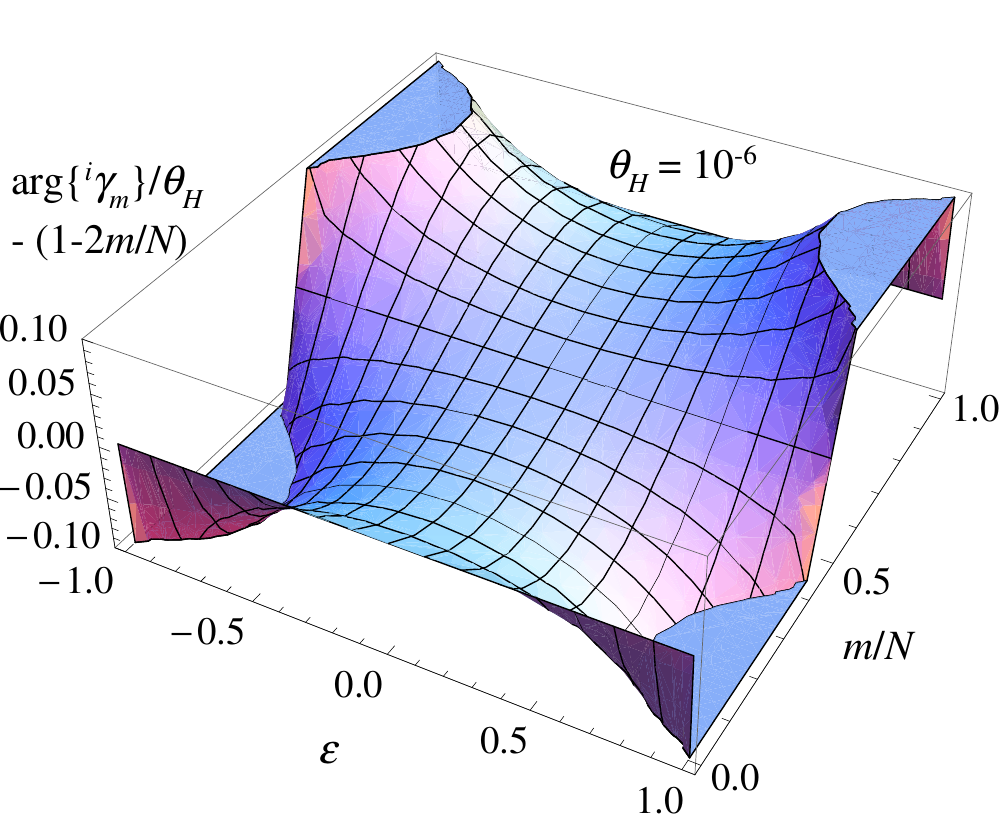}
                \caption{For $\theta_H\to 0$ (weak magnetic field limit). }
                \label{fig:argigamma-vs-mdN-vs-eps_smallB}
        \end{subfigure}
        \qquad  % An dieser Stelle kann ein zusätzlicher Zwischenraum eingebunden werden: ~, \quad, \qquad, \hfill usw.
          % Eine leere Zeile erzwingt, dass die zweite Grafik darunter erscheint.
        \begin{subfigure}[t]{0.48\textwidth}
                \centering
                \includegraphics[width=1.0\textwidth]{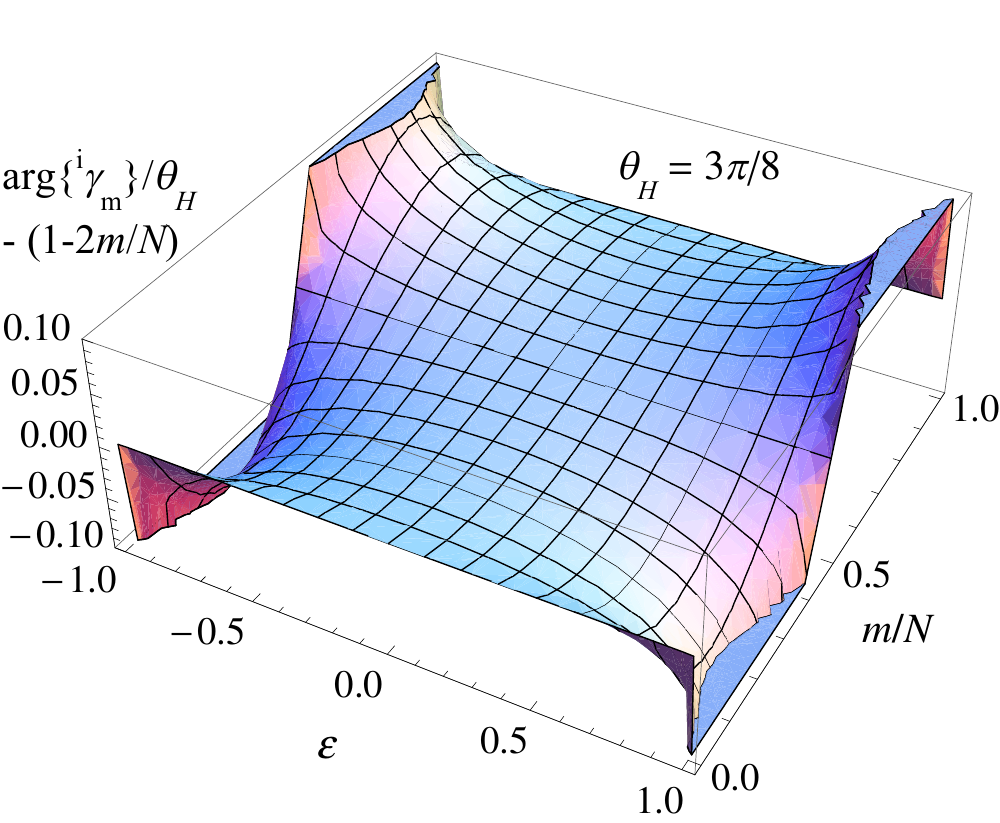}
                \caption{For $\theta_H=3\pi/8$ (strong magnetic field). }
                \label{fig:argigamma-vs-mdN-vs-eps_largeB}
        \end{subfigure}
    \caption{Absolute error of the approximation (\ref{eq:almost1})  versus $\epsilon, \theta_H$. }
   \label{fig:argigamma-over-thetaH-minus1-plus2mdN_vs_eps_mdN}
\end{figure*}
% original file: HomentcovschiMurray_regular_HHalls_eigenvalues5.nb

\begin{figure}[t]
%\vspace{1mm}
  \centering
                \includegraphics[width=0.49\textwidth]{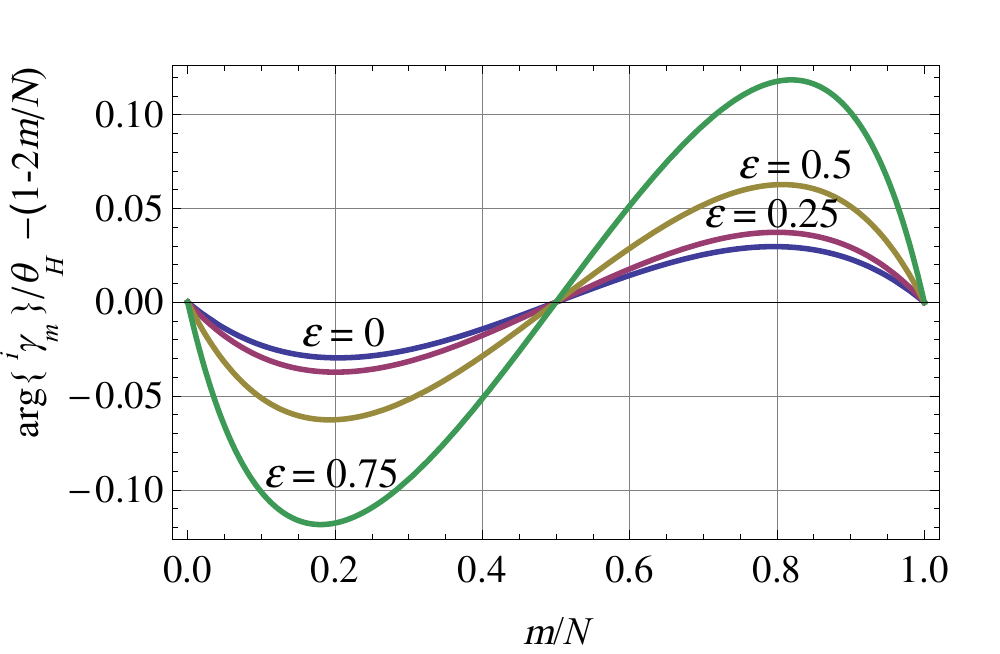}
    \caption{Absolute error of the approximation (\ref{eq:almost1}) for $\theta_H\to 0$. For larger Hall-angles the error becomes smaller. For $|\epsilon|<0.5$ the worst case error is below $\pm 0.063$. }
   \label{fig:argigamma-over-thetaH-minus1-plus2mdN_thetaH-eq_1u-vs_mdN-various_eps}
\end{figure}
% original file: HomentcovschiMurray_regular_HHalls_eigenvalues5.nb

With the same method we can compute $\mathrm{arg}\{{^i}\!\gamma_{N/3}\}$, based on $G_{H0}^{(3C)}$ from Ref.~\onlinecite{3C-Hall} and (\ref{eq:RMFoCD1573}). In Ref.~\onlinecite{3C-Hall} the Hall-geometry factor for Hall-plates with three contacts was defined by 
\begin{equation}\label{eq:smallC-10}
V_2(\theta_H)-V_2(-\theta_H) = R_\mathrm{sheet} \tan(\theta_H) G_H^{(3C)} I_1 ,
\end{equation}
whereby contact $C_0$ was grounded, current $I_1$ was supplied by a current source into contact $C_1$ and the potential at $C_2$ was sampled at positive and negative magnetic field with a volt-meter. Alternatively, we may write $V_2=R_{2,1} I_1$, which gives 
\begin{equation}\label{eq:smallC-11}
V_2(\theta_H)-V_2(-\theta_H) = 2\sqrt{3}\frac{\sin\left(\mathrm{arg}\{{^i}\!\gamma_1\}\right)}{\left|{^i}\!\gamma_1\right|} I_1 ,
\end{equation}
whereby we used (\ref{eq:RMFoCD1514}). Setting (\ref{eq:smallC-10}) equal to (\ref{eq:smallC-11}) at weak magnetic field gives 
\begin{equation}\label{eq:smallC-12}\begin{split}
& \lim_{\theta_H\to 0} \frac{\mathrm{arg}\{{^i}\!\gamma_{N/3}\}}{\theta_H} = G_{H0}^{(3C)}\,\frac{\sqrt{3}}{2}\, \\ 
& \quad * \frac{K}{K'}\!\!\left(\!\!\frac{\cos\left(\pi\frac{5-\epsilon}{12}\right)}{\sqrt{2}\cos\left(\pi\frac{3+\epsilon}{12}\right)} \sqrt{1\!+\!\sqrt{3}\tan\left(\frac{\pi\epsilon}{6}\right)}\right) , 
\end{split}\end{equation}
where we used (\ref{eq:RMFoCD1573}). For small contacts, $\epsilon= -1+\mathrm{d}\epsilon$ with $\mathrm{d}\epsilon\to 0$, it holds $G_{H0}^{(3C)}\to 1$. This gives 
\begin{equation}\label{eq:smallC-18}\begin{split}
& \lim_{\theta_H\to 0}\,\lim_{\mathrm{d}\epsilon\to 0} \frac{\mathrm{arg}\{{^i}\!\gamma_{N/3}\}}{\theta_H} = \lim_{\mathrm{d}\epsilon\to 0} \frac{\sqrt{3}}{2} \, \frac{K}{K'}\!\!\left(\!\frac{\pi^{3/2}}{2}\; \frac{\mathrm{d}\epsilon^{3/2}}{3^{9/4}}\!\right) \\ 
& \qquad = \frac{\pi}{\sqrt{3}}\;\frac{1}{4\ln(2)+3\ln(3)-2\ln(\pi\,\mathrm{d}\epsilon)} .
\end{split}\end{equation}
For $\epsilon=0.99999$ and $m/N=1/3$ (\ref{eq:smallC-18}) gives a value of $0.0676669$, which differs from the exact value in Fig. \ref{fig:argigamma-vs-mdN-various-eps_smallB} only by $3.7$ppm. For $\epsilon=0.99$ the relative error of (\ref{eq:smallC-18}) increases to $0.37\%$, and for $\epsilon=0.9$ the relative error of (\ref{eq:smallC-18}) is $3.8\%$.

\subsection{Eigenvalues for Strictly Regular Symmetry} 
\label{sec:strictly}

\begin{figure*}
\vspace{1mm}
  \centering
        \begin{subfigure}[t]{0.48\textwidth}
                \centering
                \includegraphics[width=1.0\textwidth]{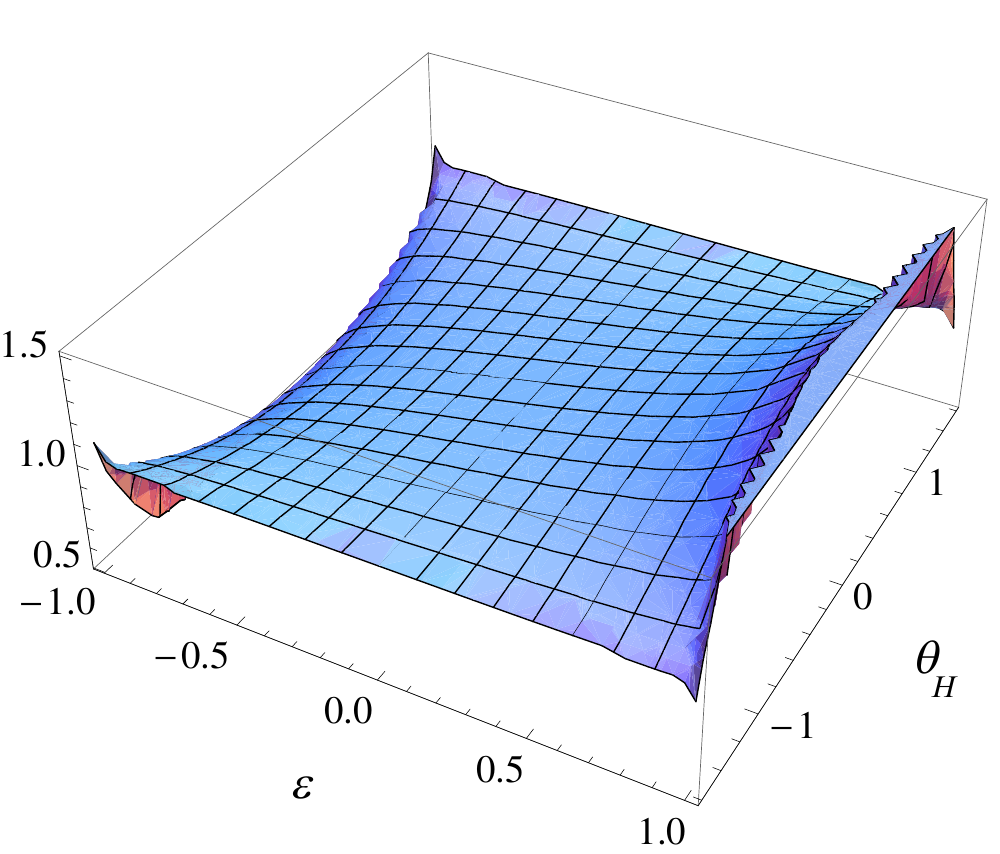}
                \caption{For $m/N=1/10$. }
                \label{fig:absigamma-Rsq-over-2sinPimdN_vs_eps_thetaH-mdN_eq_100m}
        \end{subfigure}
        \qquad  % An dieser Stelle kann ein zusätzlicher Zwischenraum eingebunden werden: ~, \quad, \qquad, \hfill usw.
          % Eine leere Zeile erzwingt, dass die zweite Grafik darunter erscheint.
        \begin{subfigure}[t]{0.48\textwidth}
                \centering
                \includegraphics[width=1.0\textwidth]{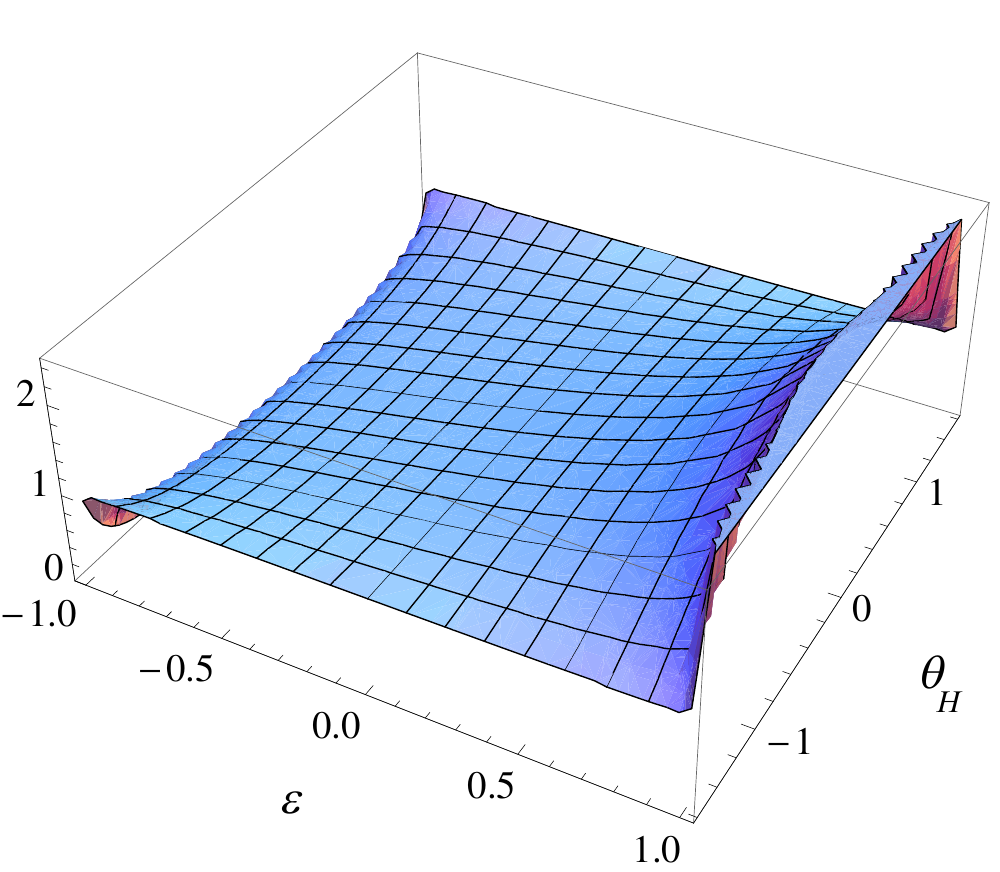}
                \caption{For $m/N=1/2$. }
                \label{fig:absigamma-Rsq-over-2sinPimdN_vs_eps_thetaH-mdN_eq_500m}
        \end{subfigure}
    \caption{Ratio $\left|{^i}\!\gamma_m \right| R_\mathrm{sq}/\left(2\sin\left(\pi m/N\right)\right)$ versus $\epsilon, \theta_H$, compare (\ref{eq:almost2}). }
   \label{fig:absigamma-Rsq-over-2sinPimdN_vs_eps_thetaH}
\end{figure*}
% original file: HomentcovschiMurray_regular_HHalls_eigenvalues5.nb

\begin{figure}[t]
%\vspace{1mm}
  \centering
                \includegraphics[width=0.49\textwidth]{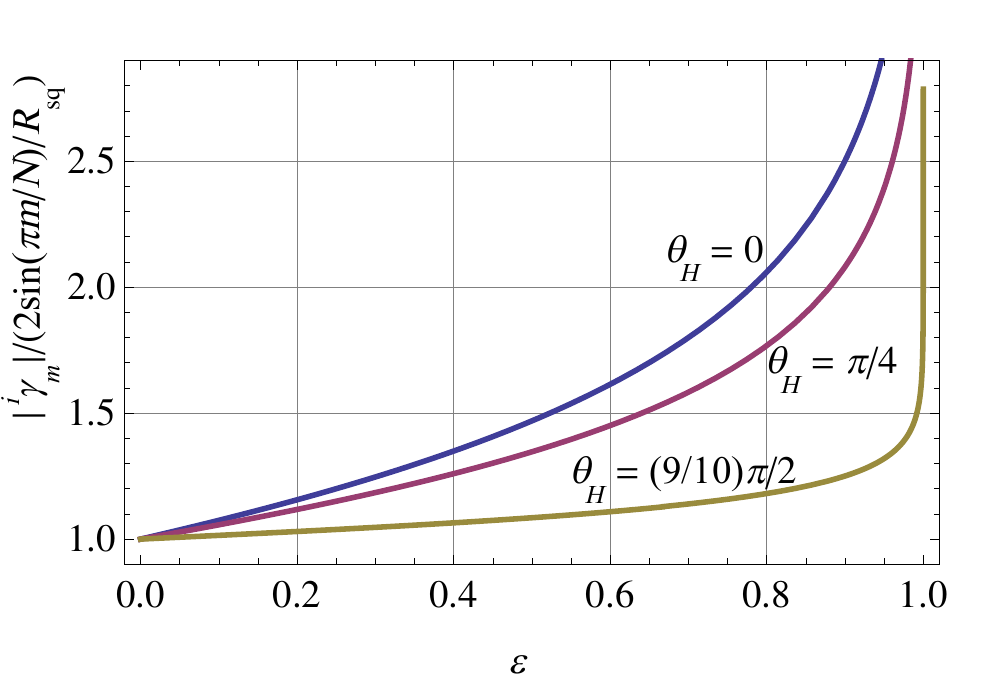}
    \caption{$\left|{^i}\!\gamma_m \right|R_\mathrm{sq}/\left(2\sin\left(\pi m/N\right)\right)$ versus size of contacts, for $\theta_H = 0, \pi/4, (9/10)*\pi/2$ and $m/N=1/2$. At zero magnetic field, $\theta_H=0$, the curve is given in closed form by (\ref{eq:RMFoCD1565a}). }
   \label{fig:absigamma-Rsq-over-2sinPimdN_vs_eps_various-thetaH-mdN_eq_500m}
\end{figure}
% original file: HomentcovschiMurray_regular_HHalls_eigenvalues5.nb

\noindent The special case of stricly regular symmetric Hall-plates ($\chi=1 \leftrightarrow \epsilon=0$) was thoroughly discussed in Ref.~\onlinecite{ArXiv2023}. Here I only make a small addendum concerning the arguments of the eigenvalues at weak magnetic field. For $N=4$ and $\epsilon=0$ it holds $G_{H0}^{(4C)}\to 2/3$, see Ref.~\onlinecite{Haeusler1966}. Inserting this into (\ref{eq:smallC-6}) gives 
\begin{equation}\label{eq:strictly-1}
\lim_{\theta_H\to 0} \frac{\mathrm{arg}\{{^i}\!\gamma_{N/4}\}}{\theta_H} = \frac{2}{3} \frac{K}{K'}\!\!\left(\!\tan\!\left(\frac{\pi}{8}\right)\right) = \frac{\sqrt{2}}{3} .
\end{equation}
Here we used $\tan(\pi/8)=\sqrt{2}-1$ and $K/K'(\sqrt{2}-1)=1/\sqrt{2}$. The right hand side of (\ref{eq:strictly-1}) is the same as (66) in Ref.~\onlinecite{ArXiv2023}. \\
For $N=3$ and $\epsilon=0$ it holds $G_{H0}^{(3C)}\to 0.622157$, see Ref.~\onlinecite{3C-Hall}. Inserting this into (\ref{eq:smallC-12}) gives 
\begin{equation}\label{eq:strictly-2}\begin{split}
\lim_{\theta_H\to 0} \frac{\mathrm{arg}\{{^i}\!\gamma_{N/3}\}}{\theta_H} & = 0.622157 \frac{\sqrt{3}}{2} \frac{K}{K'}\!\!\left(\!\cos\!\left(\frac{5\pi}{12}\right)\right) \\ 
& = 0.311078705 .
\end{split}\end{equation}
Here we used $\cos(5\pi/12)=\sqrt{2-\sqrt{3}}/2$ and $K/K'(\sqrt{2-\sqrt{3}}/2)=1/\sqrt{3}$. The right hand side of (\ref{eq:strictly-2}) is the same as (66) in Ref.~\onlinecite{ArXiv2023}.

\subsection{Eigenvalues for Almost Strictly Regular Symmetry} 
\label{sec:almost}

\noindent By \emph{almost strictly regular symmetric} Hall-plates, we mean regular symmetric Hall-plates with small $|\epsilon|$, say $|\epsilon|<0.5$. Then, according to Fig. \ref{fig:argigamma} the arguments of the eigenvalues are close to a straight line in the plot, which means 
\begin{equation}\label{eq:almost1}
\frac{\mathrm{arg}\left\{{^i}\!\gamma_m \right\}}{\theta_H} \approx 1-2\frac{m}{N} ,
\end{equation}
which has an absolute accuracy of $\pm 0.063$, whereby the error is smaller for large magnetic fields than for weak ones, see Fig. \ref{fig:argigamma-over-thetaH-minus1-plus2mdN_vs_eps_mdN}. The largest error occurs near $m/N\approx 0.2$ and $m/N\approx 0.8$, see Fig. \ref{fig:argigamma-over-thetaH-minus1-plus2mdN_thetaH-eq_1u-vs_mdN-various_eps}. 
A numerical inspection of the magnitudes of the eigenvalues gives the following approximation (see Figs. \ref{fig:absigamma-Rsq-over-2sinPimdN_vs_eps_thetaH} and \ref{fig:absigamma-Rsq-over-2sinPimdN_vs_eps_various-thetaH-mdN_eq_500m}; see also (\ref{eq:RMFoCD1505b3}))
\begin{equation}\label{eq:almost2}\begin{split}
& \frac{\left|{^i}\!\gamma_m \right|}{(2/R_\mathrm{sq})\sin\left(\pi m/N\right)} = \frac{\left|{^i}\!\gamma_m\right|}{\left|{^i}\!\gamma_m (\epsilon=0)\right|} \\ 
& \quad \approx 1+0.71777*\epsilon\left(\cos(\theta_H)\sin(\pi m/N)\right)^{0.8} .
\end{split}\end{equation}
The accuracy of (\ref{eq:almost2}) is worst for large $|\epsilon|$ and $m/N=1/2$. For $|\epsilon|\le 0.5$ the relative error is bounded by $+0/-0.0751$, for $|\epsilon|\le 0.25$ it is bounded by $+0/-0.01714$. %ueberprueft in HomentcovschiMurray_regular_HHalls_eigenvalues5.nb
The right hand side of (\ref{eq:almost2}) gives the tangents on the curves in Fig. \ref{fig:absigamma-Rsq-over-2sinPimdN_vs_eps_various-thetaH-mdN_eq_500m} in $\epsilon=0$ for arbitrary $\theta_H$ and $m/N$.

\subsection{Eigenvalues of Regular Symmetric Hall-Plates with a Hole} 
\label{sec:hole}

\noindent Inversion of (\ref{eq:RMFoCD52}) gives the eigenvalues in terms of the indefinite conductance matrix, see (\ref{eq:RMFoCD1517}). This works also for regular symmetric Hall-plates \emph{with holes}, because it only relies on the fact that the indefinite conductance matrix is a circulant matrix, regardless if the Hall-plate is a singly- or multiply-connected domain. The conductance matrix of Hall-plates with four contacts and a single hole with insulating hole boundary at weak magnetic field is given in Ref.~\onlinecite{Ausserlechner2022rspa}. For regular symmetry this gives the eigenvalues 
\begin{equation}\label{eq:hole1} 
{^i}\!\gamma_1 = \frac{2}{R_\mathrm{sheet}\lambda}\left(\!1\!+\!\mathbbm{i} G_{H0}^{(4C1H)} \frac{\tan(\theta_H)}{\lambda} \right) ,  
\;{^i}\!\gamma_2 = \frac{1}{\lambda_x R_\mathrm{sheet}} ,
\end{equation}
with $\lambda,\lambda_x$ as defined in (\ref{eq:4C-Hall6a}),(\ref{eq:4C-Hall6ab}). Thereby, $G_{H0}^{(4C1H)}$ is the Hall-geometry factor for regular symmetric Hall-plates with a hole with insulated hole boundary. For singly-connected Hall-plates $\lambda_x$ is a function of $\lambda$ only (see (\ref{eq:RMFoCD1557b})), however, for doubly connected Hall-plates with regular symmetry it becomes a function of $\lambda$ \emph{and} of the size of the hole \cite{eq2.34rspa}. Then it holds $\lambda/4<\lambda_x<\lambda/2$. 

According to Ref.~\onlinecite{Ausserlechner2022rspa} the complementary Hall-plate has the hole boundary clad with a contact, that is not connected (no current flows in or out of this 'floating' contact, and therefore this contact does not correspond to any row or column in the conductance matrix).  For regular symmetry this gives the eigenvalues 
\begin{equation}\label{eq:hole4}
\overline{{^i}\!\gamma_1} = \frac{1}{R_\mathrm{sheet}}\left(\!\lambda\!+\!\mathbbm{i} G_{H0}^{(4C1H)} \tan(\theta_H) \right) ,  
\; \overline{{^i}\!\gamma_2} = \frac{4 \lambda_x}{R_\mathrm{sheet}} .
\end{equation}
With (\ref{apx-popularHalls15}) and (\ref{apx-popularHalls16}) and with (5.1) from Ref.~\onlinecite {Ausserlechner2022rspa} we can also write 
\begin{equation}\label{eq:hole5}
\overline{{^i}\!\gamma_1} = \frac{2}{R_\mathrm{sheet}\overline{\lambda}}\left(\!1\!+\!\mathbbm{i} \overline{G}_{H0}^{(4C1H)} \frac{\tan(\theta_H)}{\overline{\lambda}} \right) ,  
\; \overline{{^i}\!\gamma_2} = \frac{1}{\overline{\lambda}_x R_\mathrm{sheet}} ,
\end{equation}
with $\overline{G}_{H0}^{(4C1H)}$ being the Hall-geometry factor of the complementary Hall-plate, whose hole has a floating contact. 
According to Ref.~\onlinecite{Ausserlechner2022rspa} it holds 
\begin{equation}\label{eq:hole5b}
\overline{\lambda} = \frac{2}{\lambda}, \;\; \overline{\lambda}_x = \frac{1}{4\lambda_x}, \;\; \overline{G}_{H0}^{(4C1H)} = \frac{2}{\lambda^2} G_{H0}^{(4C1H)} .
\end{equation}
Equations (\ref{eq:hole1}) and (\ref{eq:hole5}) are symmetric, if we swap all overbars. 
Both equations (\ref{eq:hole1}) and (\ref{eq:hole4}) use the same Hall geometry factor $G_{H0}^{(4C1H)}$, which is a complicated function of $\lambda$ and $\lambda_x$, see Ref.~\onlinecite{eq3.28rspa}. 

Multiplying the eigenvalues of the original and the complementary doubly-connected Hall-plates gives 
\begin{equation}\label{eq:hole6}
\lim_{\theta_H\to 0} \overline{{^i}\!\gamma_m (-\theta_H)}\;{^i}\!\gamma_m (\theta_H) = \left(\frac{2}{R_\mathrm{sheet}}\sin\left(\pi \frac{m}{N}\right)\right)^{\!\!2} ,
\end{equation}
for $m\in\{1,2\},\, N=4$.
Thus, at weak magnetic field the eigenvalues of complementary regular symmetric Hall-plates with four contacts are related according to (\ref{eq:RMFoCD1505}), \emph{regardless if they have a hole or not}. 
%\newpage

\section{The Resistance Matrix $\bm{R}$} 
\label{sec:R}

\noindent We can make good use of (\ref{eq:RMFoCD1505}), (\ref{eq:RMFoCD1505b}) if we want to compute the resistance matrix in closed form. Thereby, we start with the RMFoCD-principle in matrix form (see equation (8) in Ref.~\onlinecite{ ArXiv2023}). Solving it for $\bm{R}=\bm{G}^{-1}$ gives 
\begin{equation}\label{eq:RMFoCD1510}
\bm{R} =R_ \mathrm{sq}^2 \bm{L_1}^T \overline{\bm{G}^T} \bm{L_1} .
\end{equation}
The RMFR-principle \cite{Sample1987,Cornils2008} says $\overline{\bm{G}^T}= \overline{\bm{G}(-\theta_H)}$ and inserting (\ref{eq:RMFoCD1505b}) into (\ref{eq:RMFoCD52}) gives  
\begin{equation}\label{eq:RMFoCD1511}
\overline{\bm{{^i}\!G}(-\theta_H)} = \frac{1}{R_\mathrm{sq}^2} \bm{Q}\,\bm{{^i}\!\Gamma}^+ \bm{\Delta} \bm{Q}^C .
\end{equation}
In (\ref{eq:RMFoCD1510}) the matrices have $N-1$ rows and columns, whereas in (\ref{eq:RMFoCD1511}) they have $N$ rows and columns. Next, we combine (\ref{eq:RMFoCD1510}) and (\ref{eq:RMFoCD1511}), and therefore we use index notation. Inserting (\ref{eq:RMFoCD1511}) into (\ref{eq:RMFoCD1510}) gives 
\begin{equation}\label{eq:RMFoCD1512}\begin{split}
%(\bm{R})_{j,p} & 
R_{j,p} \!=\! \frac{4}{N} \!\sum_{k=j}^{N-1}\sum_{m=p}^{N-1}\sum_{\ell=1}^{N-1} % \\ & \qquad
\frac{1}{{^i}\!\gamma_\ell}\left(\!\!\sin\!\left(\!\!\frac{\pi\ell}{N}\!\right)\!\!\right)^{\!\!2} \!\exp\!\left(\!\!2\pi\mathbbm{i}\frac{k\!-\!m}{N}\ell\!\right) 
\end{split}\end{equation}
With the identity 
\begin{equation}\label{eq:RMFoCD1513}\begin{split}
& \sum_{k=j}^{N-1}\sum_{m=p}^{N-1} \exp\!\left(\!\!2\pi\mathbbm{i}\frac{k\!-\!m}{N}\ell\!\right) = 
\exp\left(\mathbbm{i}\pi\frac{j\!-\!p}{N}\ell\right) \\ 
& \qquad\times\sin\!\left(\!\!\frac{\pi j\ell}{N}\!\right)\sin\!\left(\!\!\frac{\pi p\ell}{N}\!\right) \left(\sin\!\left(\!\!\frac{\pi\ell}{N}\!\right)\right)^{\!\!-2} ,
\end{split}\end{equation}

\noindent valid for $k,m,\ell\in\{1,2,\ldots,N-1\}$ we get
\begin{equation}\label{eq:RMFoCD1514}\begin{split}
R_{j,p} & \!=\! \frac{4}{N} \!\sum_{\ell=1}^{N-1} \frac{(-1)^{(j-p)\ell/N}}{{^i}\!\gamma_\ell} \sin\!\left(\!\frac{\pi j\ell}{N}\!\right) \sin\!\left(\!\frac{\pi p\ell}{N}\!\right) \\
& = \sum_{\ell=1}^{N-1} \frac{1}{{^i}\!\gamma_\ell} \left(Q_{j,\ell}-\frac{1}{\sqrt{N}}\right) \left(Q_{p,\ell}^C-\frac{1}{\sqrt{N}}\right) .
\end{split}\end{equation}
In matrix form this reads 
\begin{equation}\label{eq:RMFoCD1514b}\begin{split}
\bm{R}_{N\times N} & = \left(\bm{Q}-\frac{1}{\sqrt{N}}\bm{F}\right) \bm{{^i}\!\Gamma}^+ \left(\bm{Q}^C-\frac{1}{\sqrt{N}}\bm{F}\right) , \\ 
\bm{R}_{N\times N} & = \left(\!\!\begin{array}{cc} (\bm{R})&0\\ 0&0 \end{array} \!\!\right), \quad 
\bm{F} = \left(\!\!\begin{array}{ccc} 1&\cdots &1\\ \vdots &\ddots &\vdots\\ 1&\cdots&1 \end{array} \!\!\right)
\end{split}\end{equation}
where $\bm{R}$ has $N-1$ rows and columns and all other matrices have $N$ rows and columns. Decomposition into even (magneto-resistive) and odd (Hall) terms gives 
\begin{widetext}
\begin{equation}\label{eq:RMFoCD1515}\begin{split}
R_{\mathrm{ev}j,p} & = \frac{R_{j,p}+R_{p,j}}{2} = 
\frac{4}{N} \frac{1}{{^i}\!\gamma_{\lfloor N/2\rfloor}} (-1)^{(j-p)/2} \sin\!\left(\!\frac{\pi j}{2}\!\right) \sin\!\left(\!\frac{\pi p}{2}\!\right) \left(\left\lfloor\frac{N}{2}\right\rfloor -\left\lfloor\frac{N-1}{2}\right\rfloor\right) \\
&\qquad\qquad\qquad + \frac{8}{N} \sum_{\ell=1}^{\lfloor(N-1)/2\rfloor} \frac{1}{|{^i}\!\gamma_\ell|} \cos\left(\mathrm{arg}\{{^i}\!\gamma_\ell\}\right) \cos\left(\pi\ell\frac{j-p}{N}\right) \sin\left(\frac{\pi j\ell}{N}\right) \sin\left(\frac{\pi p\ell}{N}\right) , \\ 
R_{\mathrm{odd}j,p} & = \frac{R_{j,p}-R_{p,j}}{2} = \frac{8}{N} \sum_{\ell=1}^{\lfloor(N-1)/2\rfloor} \frac{1}{|{^i}\!\gamma_\ell|} \sin\left(\mathrm{arg}\{{^i}\!\gamma_\ell\}\right) \sin\left(\pi\ell\frac{j-p}{N}\right) \sin\left(\frac{\pi j\ell}{N}\right) \sin\left(\frac{\pi p\ell}{N}\right) ,
\end{split}\end{equation}
\end{widetext}
whereby $\lfloor x\rfloor$ is the integer part of a positive real number $x$. If the resistance matrix or the conductance matrix is given, we can compute the eigenvalues by inverting (\ref{eq:RMFoCD1514b}). This gives 
\begin{widetext}
\begin{equation}\label{eq:RMFoCD1516}\begin{split}
\bm{{^i}\!\Gamma}_{(N-1)\times (N-1)} = \left(\bm{Q}_{(N-1)\times (N-1)}^C-\frac{1}{\sqrt{N}}\bm{F}_{(N-1)\times (N-1)}\right) \bm{G}  \left(\bm{Q}_{(N-1)\times (N-1)}-\frac{1}{\sqrt{N}}\bm{F}_{(N-1)\times (N-1)}\right)
\end{split}\end{equation}
\end{widetext}
where all $(N-1)\times (N-1)$ matrices are obtained from the corresponding $N\times N$ matrices by deleting the last row and column. However, it is simpler to use only $N\times N$ matrices, 
\begin{equation}\label{eq:RMFoCD1517}
\bm{{^i}\!\Gamma} = \bm{Q}^C \;\bm{{^i}\!G} \bm{Q} ,
\end{equation}
which follows from (\ref{eq:RMFoCD52}). This equation also works for regular symmetric Hall-plates \emph{with holes}, because it only relies on the fact that the indefinite conductance matrix is a circulant matrix.

\section{Regular Symmetric Hall-Plates with Four Contacts} 
\label{sec:4C-Hall}

\begin{figure}
%\vspace{1mm}
  \centering
                \includegraphics[width=0.45\textwidth]{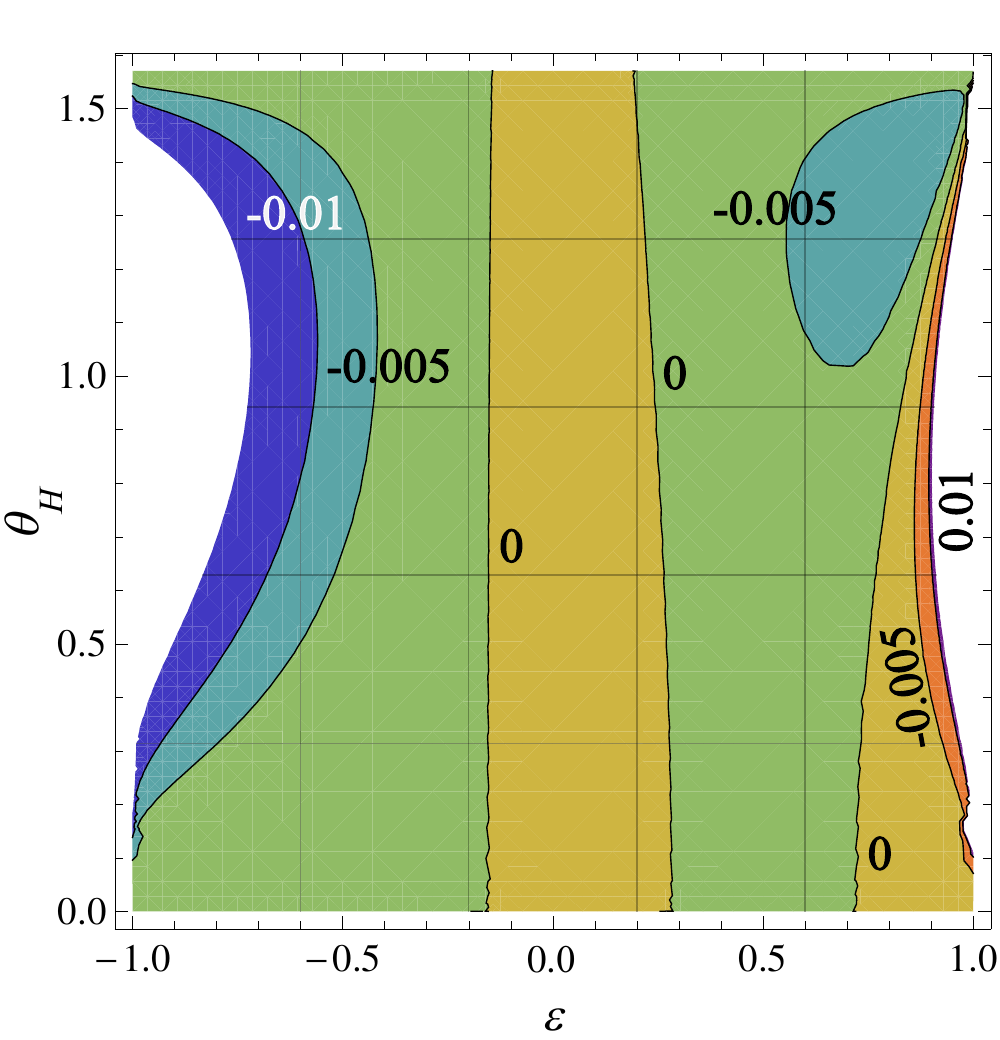}
    \caption{Contour plot of the relative (=percental) accuracy of the $G_H^{(4C)}$-approximation (\ref{eq:4C-Hall2approx}) for regular symmetric Hall-plates vs. contacts size $\epsilon$ and Hall-angle $\theta_H$. }
   \label{fig:GH4C-approximation-contour2}
\end{figure}
% original file: HomentcovschiMurray_regular_HHalls_eigenvalues5.nb

\noindent The vast majority of modern Hall-plates uses contact commutation schemes \cite{Munter,AusserlechnerOffset2004,IFX-current-sensor2012,Mosser} to cancel out  offset errors (= zero-point errors). These schemes rely solely on the RMFR-principle \cite{Sample1987,Cornils2008}, and therefore they work for all electrically linear Hall-plates regardless of their symmetry. However, for the electronic circuits which supply the Hall-plate with power and which amplify its output signal, it is advantageous to use Hall-plates with $90$° symmetry (= regular symmetry). Therefore a formula for the output signal of regular symmetric Hall-plates with four contacts is of great practical relevance. The output voltage of an ordinary Hall-plate with four contacts can be expressed as \cite{ArXiv2023} 
\begin{equation}\label{eq:4C-Hall1}\begin{split}
V_3 - V_1 = R_\mathrm{sheet}\tan(\theta_H) I_\mathrm{supply} G_H^{(4C)} ,
\end{split}\end{equation}
whereby $I_\mathrm{supply}$ flows into contact $C_2$ and through the Hall-plate towards the opposite grounded contact $C_0=C_4$, and an ideal voltmeter (i.e., with infinite impedance) samples the voltage across the output contacts $C_3-C_1$. 
Combining (\ref{eq:RMFoCD1515}) and (\ref{eq:4C-Hall1}) gives the Hall-geometry factor,   
\begin{equation}\label{eq:4C-Hall2}
G_H^{(4C)} = \frac{2}{R_\mathrm{sq}|{^i}\!\gamma_1|}\;\frac{\sin\left(\mathrm{arg}\{{^i}\!\gamma_1\}\right)}{\sin(\theta_H)} ,
\end{equation}
which holds for regular symmetric Hall-plates. In the special case of strictly regular symmetry the prefactor simplifies, $2/(R_\mathrm{sq}|{^i}\!\gamma_1|)\to\sqrt{2}$. At weak and strong magnetic fields closed formulae for the eigenvalue ${^i}\!\gamma_1$ are given in (\ref{eq:RMFoCD1565b}), (\ref{eq:smallC-6}), and (\ref{eq:RMFoCD1504c}). In Ref.~\onlinecite{ArXiv2023} we saw that for strictly regular symmetric Hall-plates $\mathrm{arg}\{{^i}\!\gamma_1\}$ varies in good approximation quadratically versus the Hall-angle. Now we try the same approximation for regular symmetric Hall-plates. 
\begin{equation}\label{eq:4C-Hall2-b}\begin{split}
\text{(\ref{eq:smallC-6}) } \Rightarrow & \lim_{\theta_H\to 0}\mathrm{arg}\{{^i}\!\gamma_1\} = \theta_H \frac{G_{H0}^{(4C)}}{\lambda} , \\
\text{(\ref{eq:RMFoCD1504c}) } \Rightarrow & \lim_{\theta_H\to\pi/2}\mathrm{arg}\{{^i}\!\gamma_1\} = \theta_H \frac{1}{2} , 
\end{split}\end{equation}
And we try the same $\theta_H^2$-law for the reciprocal of the pre-factor, $(R_\mathrm{sq}|{^i}\!\gamma_1|)/2$. 
\begin{equation}\label{eq:4C-Hall2-c}\begin{split}
\text{(\ref{eq:RMFoCD1565b}) } \Rightarrow & \lim_{\theta_H\to 0} R_\mathrm{sq}\frac{|{^i}\!\gamma_1|}{2} = \frac{1}{\lambda} , \\
\text{(\ref{eq:RMFoCD1504c}) } \Rightarrow & \lim_{\theta_H\to\pi/2} R_\mathrm{sq}\frac{|{^i}\!\gamma_1|}{2} = \frac{1}{\sqrt{2}} . 
\end{split}\end{equation}
This gives  
\begin{equation}\label{eq:4C-Hall2approx}
G_H^{(4C)} \approx \frac{\sin\!\!\left(\theta_H\left[\frac{G_{H0}^{(4C)}}{\lambda}+\left(\frac{1}{2}-\frac{G_{H0}^{(4C)}}{\lambda}\right)\left(\!\frac{\theta_H}{\pi/2}\!\right)^{\!\!2}\right]\right)}{\left[\frac{1}{\lambda}+\left(\frac{1}{\sqrt{2}}-\frac{1}{\lambda}\right)\left(\!\frac{\theta_H}{\pi/2}\!\right)^{\!\!2}\right]\sin(\theta_H)} , 
\end{equation}
with $\lambda$ from (\ref{apx-popularHalls12c}), and with the weak magnetic field limit of the Hall-geometry factor, 
\begin{equation}\label{eq:4C-Hall2approx-b}
 G_{H0}^{(4C)}=\lim_{\theta_H\to 0} G_H^{(4C)} .
\end{equation}
The accuracy of (\ref{eq:4C-Hall2approx}) is $\pm 1\%$ for $-0.55<\epsilon<0.89$, see Fig. \ref{fig:GH4C-approximation-contour2}. Equation (\ref{eq:4C-Hall2approx}) expresses $G_H^{(4C)}$ as a function of  the two parameters $\lambda$ and $G_{H0}^{(4C)}$, which are valid in the limit of weak magnetic field. Thereby $G_{H0}^{(4C)}$ is again a function of $\lambda$ only, see (\ref{eq:4C-Hall4}) - (\ref{eq:4C-Hall4})). Thus, the effective number of squares at zero magnetic field, $\lambda$, is the essential parameter, which fully determines $G_{H0}^{(4C)}$ and $G_{H}^{(4C)}(\theta_H)$.

For regular symmetric Hall-plates with four contacts we know that their complementary counterparts have the same Hall-output voltages, if they are supplied by the same input voltage \cite{Ausserlechner2019}. (Another short way to show this, is to combine (\ref{eq:4C-Hall20}) with (\ref{eq:RMFoCD1505c}).) Hence, for regular symmetric Hall-plates with four contacts the ratio of the Hall-geometry factor over the input resistance is identical for the original and for the complementary plate at arbitrary magnetic field, 
\begin{equation}\label{eq:4C-Hall2b}
\frac{\overline{G}_H^{(4C)}}{\overline{R}_{2,2}} = \frac{G_H^{(4C)}}{R_{2,2}} .
\end{equation}
Inserting (\ref{eq:RMFoCD1505c}) and (\ref{eq:RMFoCD1515}) into (\ref{eq:4C-Hall2b}) gives 
\begin{equation}\label{eq:4C-Hall2c}
\frac{\overline{G}_H^{(4C)}}{G_H^{(4C)}} = \frac{\left|{^i}\!\gamma_1\right|}{\left|\overline{{^i}\!\gamma_1}\right|} = \left( \frac{\left|{^i}\!\gamma_1\right|}{\left|{^i}\!\gamma_1(\epsilon=0)\right|}\right)^{\!\!2} = \frac{1}{2} R_\mathrm{sq}^{2} \left|{^i}\!\gamma_1\right|^{\!\!2} .
\end{equation}
It also holds 
\begin{equation}\label{eq:4C-Hall2d}
\overline{G}_H^{(4C)} G_H^{(4C)} = \left(\sqrt{2}\;\frac{\sin\left(\mathrm{arg}\{{^i}\!\gamma_1\}\right)}{\sin(\theta_H)} \right)^{\!\!2} .
\end{equation}

%\begin{figure}
%\vspace{1mm}
%                \centering
%                \includegraphics[width=0.45\textwidth]{figures/fig_GH4C-approximation.pdf}
%                \caption{3D-plot of the relative accuracy of (\ref{eq:4C-Hall3}) versus $\epsilon,\theta_H$. A contour plot of the same function is given in Fig. \ref{fig:GH4C-approximation-contour}. }
%   \label{fig:GH4C-approximation-3D-plot}
%\end{figure}
% original file: HomentcovschiMurray_regular_HHalls_eigenvalues5.nb

\begin{figure}
%\vspace{1mm}
  \centering
%        \begin{subfigure}[c]{0.45\textwidth}
%                \centering
%                \includegraphics[width=1.0\textwidth]{figures/fig_GH4C-approximation-contourb.pdf}
%                \caption{$G_H^{(4C)}$-approximation of this work: (\ref{eq:4C-Hall3}). }
%                \label{fig:GH4C-approximation-contour}
%        \end{subfigure}
%	\hfill
%        \begin{subfigure}[c]{0.45\textwidth}
%                \centering
                \includegraphics[width=0.45\textwidth]{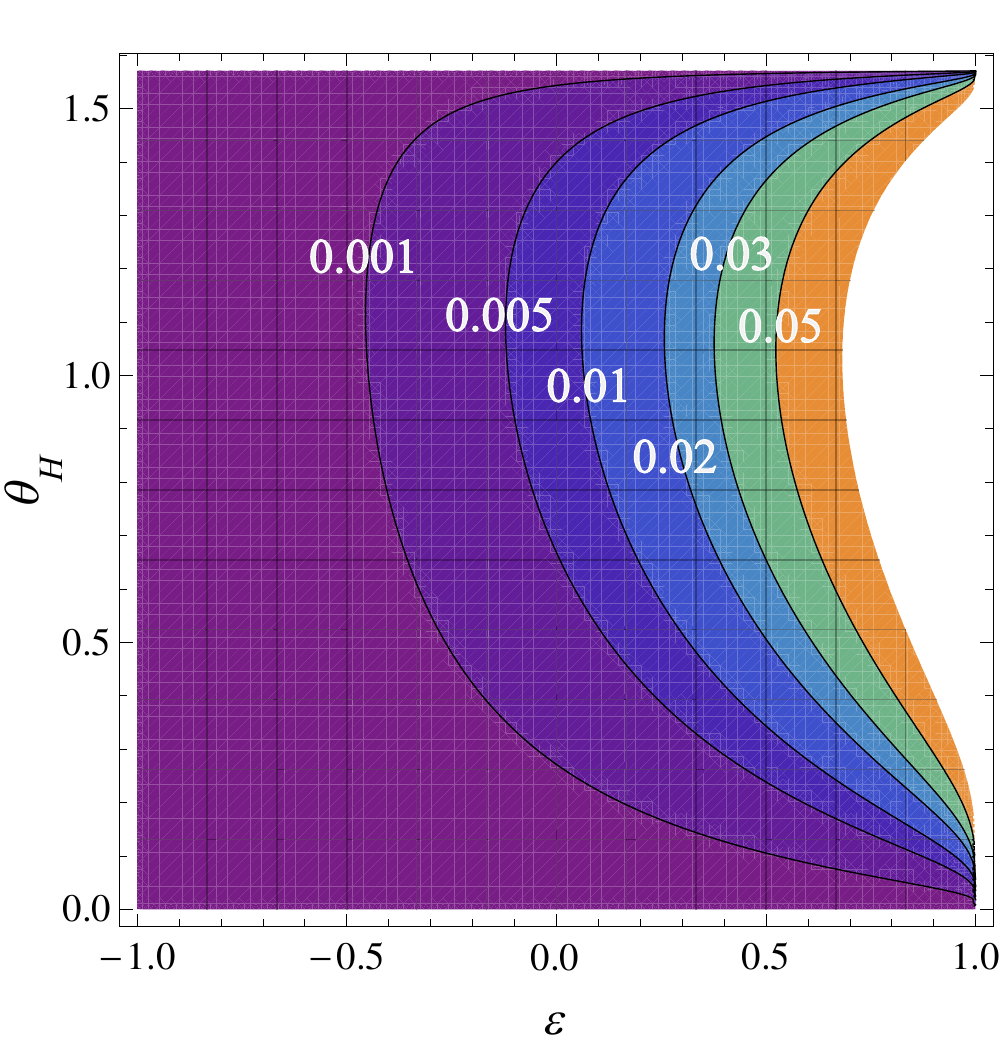}
%                \caption{$G_H^{(4C)}$-approximation of Versnel \cite{Versnel}: (\ref{eq:4C-Hall9}). }
%                \label{fig:fig_GH4C-approximation-contour_Versnel}
%        \end{subfigure}
    \caption{Contour plot of the relative (=percental) accuracy of Versnel's $G_H^{(4C)}$-approximation (\ref{eq:4C-Hall9}) for regular symmetric Hall-plates vs. contacts size $\epsilon$ and Hall-angle $\theta_H$. 
%$G_H^{(4C)}$ is an even function of $\theta_H$. 
}
   \label{fig:GH4C-approximation-contour}
\end{figure}
% original file: HomentcovschiMurray_regular_HHalls_eigenvalues5.nb

In Ref.~\onlinecite{Versnel} Versnel gave a simple approximation for the Hall-geometry factor of regular symmetric Hall-plates with four extended contacts at arbitrary magnetic field. In our notation it reads 
\begin{equation}\label{eq:4C-Hall9}
G_H^{(4C)} \approx 1 - \left(1-G_{H0}^{(4C)}\right)\frac{\theta_H}{\tan\left(\theta_H\right)} .
\end{equation}
Fig. \ref{fig:GH4C-approximation-contour} shows the relative accuracy of (\ref{eq:4C-Hall9}) for all contacts sizes and Hall-angles: The error is below $1$\% as long as $\epsilon\le 0.06$, and even for larger contacts, $\epsilon > 0.06$, it is similarly small if the Hall-angle is small, $\theta_H<0.1$. 
Versnel expressed $G_{H0}^{(4C)}$ as a function of geometrical parameters for several common geometries like disks, squares, octagons, and Greek crosses. Alternatively, we can give $G_{H0}^{(4C)}$ as a single function of a single parameter $\lambda$, and only in a second step we relate $\lambda$ to the geometrical parameters of the various shapes of Hall-plates (see Appendix \ref{sec:popularHalls}). The parameter $\lambda$ has a physical meaning: it is the effective number of squares of the input resistance of the Hall-plate at zero magnetic field in (\ref{eq:4C-Hall6a}), wherein $R_{2,2}(\theta_H=0)$ follows from the first eigenvalue,    
\begin{equation}\label{eq:4C-Hall6b}
R_{2,2}(\theta_H=0) = \frac{2}{{^i}\!\gamma_1(\theta_H=0)} 
\end{equation}
according to (\ref{eq:RMFoCD1515}). Large $\lambda$ means small contacts and small $\lambda$ means large contacts. However, the largest signal-to-thermal-noise-ratio-per-power is obtained for $\lambda=\sqrt{2}$ which corresponds to medium-sized contacts, for which it holds $\epsilon=0$ (see Section \ref{sec:noise} or more detailed in Ref.~\onlinecite{AusserlechnerSNR2017}).
There exist several formulae for the weak magnetic field limit of the Hall-geometry factor. One simple approximation with an accuracy of $+0/-2\%$ is   
\begin{equation}\label{eq:4C-Hall4}
G_{H0}^{(4C)} \approx \frac{\lambda^2}{\sqrt{\lambda^4+\lambda^2/2+4}} ,
\end{equation}
with $\lambda$ from (\ref{apx-popularHalls12c}). A more accurate approximation with an error of $-0.006\%\ldots +0.02\%$ is 
\begin{equation}\label{eq:4C-Hall4b}\begin{split}
& G_{H0}^{(4C)} \approx \frac{\lambda^2}{\sqrt{\lambda^4+\lambda^2/2+4}}   \\ 
&*\left[1+\Lambda^2\exp\left(-c_0-(c_2\Lambda)^2+(c_4\Lambda)^4-(c_6\Lambda)^6 \right)\right] ,
\end{split}\end{equation}
with $\Lambda = \ln(\lambda/\sqrt{2})$, and $c_0=2.279$, $c_2=1.394$, $c_4=0.6699$, and $c_6=0.4543$. Equations (\ref{eq:4C-Hall4}) and (\ref{eq:4C-Hall4b}) are explained in Ref.~\onlinecite{GH0approx}. An exact formula for $G_{H0}^{(4C)}$ is also known \cite{AusserlechnerSNR2017},
\begin{equation}\label{eq:4C-Hall5}\begin{split}
& G_{H0}^{(4C)} = \left(\!K\!\left(\!\!\sqrt{L}\right)\!\!\right)^{\!\!\!-2}\!\!\int_0^{\pi/2}\!\!\!\!\!\frac{F\!\left(\!\sin(\alpha),\sqrt{1\!-\!L}\right)\,\mathrm{d}\alpha}{\sqrt{(\sin(\alpha)\!)^2\!+\!(\cos(\alpha)\!)^2 L}} \\ 
&\text{with } L=L(\lambda)=\left(\!\tan\!\left(\pi\frac{1+\epsilon}{8}\right)\!\!\right)^{\!\!2} . 
\end{split}\end{equation}

%With the help of (\ref{eq:4C-Hall2c}) or (\ref{eq:4C-Hall2d}) we can improve the accuracy of Versnel's $G_H^{(4C)}$-approximation for Hall-plates with large contacts, $\epsilon>0$. Ausarbeiten, welche der beiden Optionen die bessere Genauigkeit liefert!xxxxxxxxxxxxxx

According to (\ref{eq:RMFoCD52}) and (\ref{eq:RMFoCD69a}) we can express the conductance matrix in terms of the eigenvalues  
\begin{equation}\label{eq:4C-Hall8}\begin{split}
G_{1,1} & = \frac{1}{2}\Re\{{\^i}\!\gamma_1\} + \frac{1}{4}{\^i}\!\gamma_2=G_{2,2}=G_{3,3}, \\ 
G_{1,2} & = \frac{1}{2}\Im\{{\^i}\!\gamma_1\} - \frac{1}{4}{\^i}\!\gamma_2=G_{2,3}, \\ 
G_{1,3} & = \frac{-1}{2}\Re\{{\^i}\!\gamma_1\} + \frac{1}{4}{\^i}\!\gamma_2=G_{3,1}, \\ 
G_{2,1} & = \frac{-1}{2}\Im\{{\^i}\!\gamma_1\} - \frac{1}{4}{\^i}\!\gamma_2=G_{3,2},
\end{split}\end{equation}
with $\Re\{z\}$ and $\Im\{z\}$ being the real and imaginary parts of a complex number $z$. Check: the sum of these four entries vanishes, since they are the first line of the \emph{indefinite} conductance matrix \cite{Haykin,Balabanian,Simonyi} (see also (\ref{G-symmetry})).

\begin{figure}
%\vspace{1mm} 
  \centering
        \begin{subfigure}[c]{0.24\textwidth}
                \centering
                \includegraphics[width=1.0\textwidth]{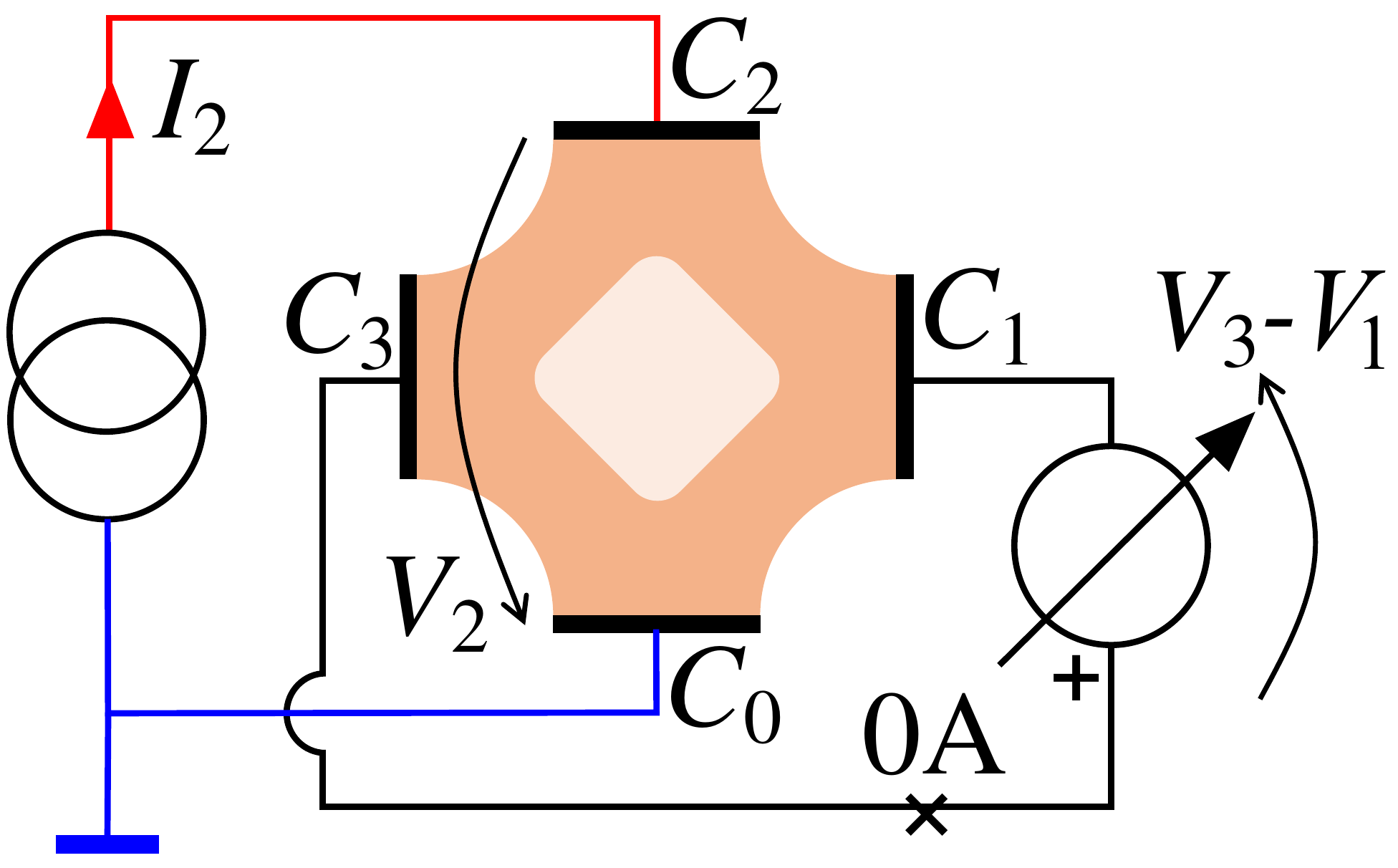}
                \caption{Circuit for ${^i}\!\gamma_1$. }
                \label{fig:igamma1-measurement}
        \end{subfigure}
	\hfill
        \begin{subfigure}[c]{0.23\textwidth}
                \centering
                \includegraphics[width=1.0\textwidth]{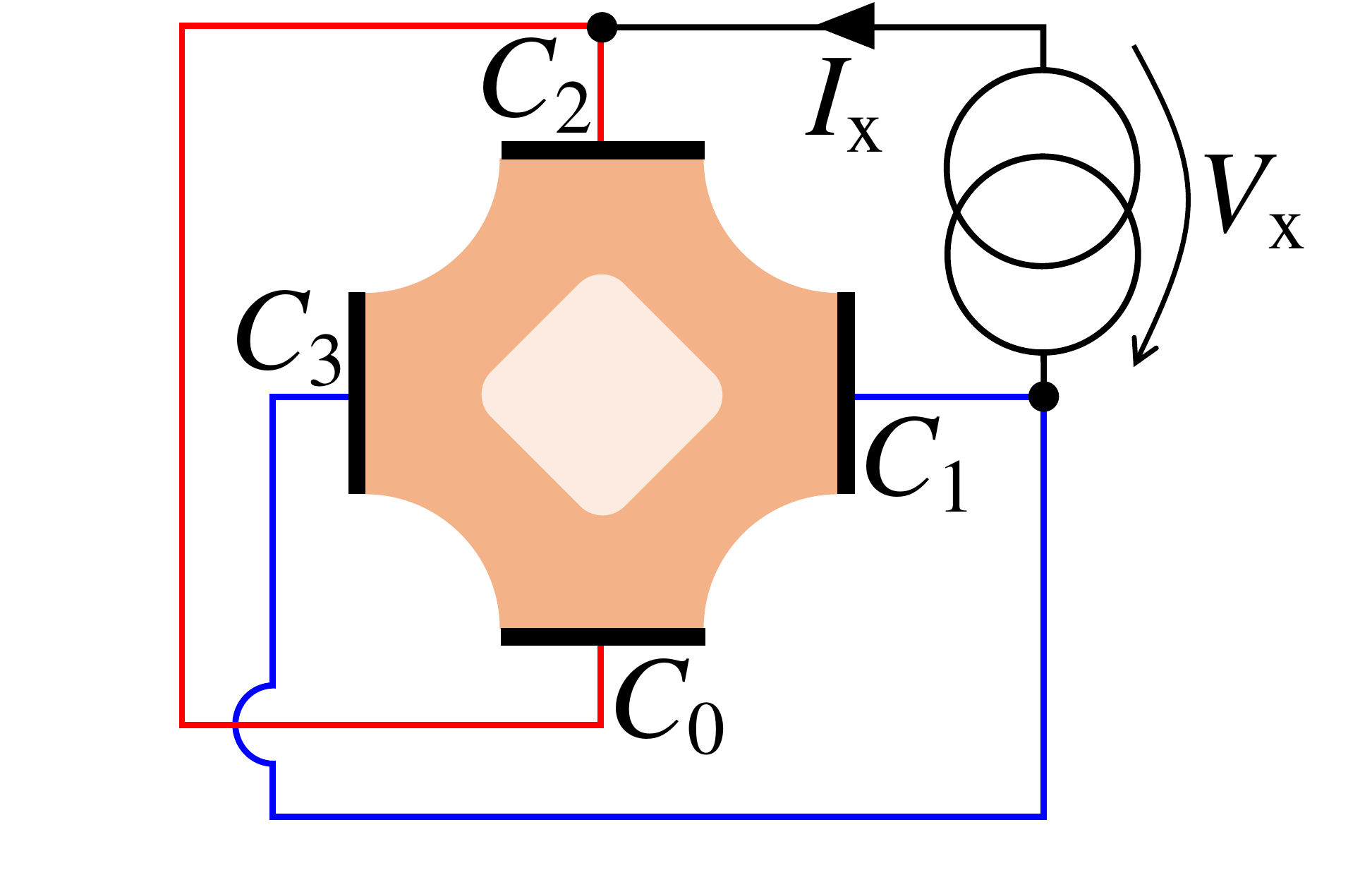}
                \caption{Circuit for ${^i}\!\gamma_2$. }
                \label{fig:igamma2-measurement}
        \end{subfigure}
    \caption{Circuits for the measurements of the eigenvalues of regular symmetric Hall-plates with four contacts (with or without a hole) at arbitrary magnetic field. }
   \label{fig:igamma-measurement}
\end{figure}

Inverting the $\bm{G}$-matrix gives the $\bm{R}$-matrix, 
\begin{equation}\label{eq:4C-Hall8b}\begin{split}
& \bm{R} = \frac{1}{|{\^i}\!\gamma_1|} \!\!\left(\!\!\begin{array}{ccc} |{\^i}\!\gamma_1|/{\^i}\!\gamma_2\!+\!\cos_1 & -\!\cos_1\!-\!\sin_1 & |{\^i}\!\gamma_1|/{\^i}\!\gamma_2\!+\!\sin_1 \\ -\!\cos_1\!+\!\sin_1 & 2\cos_1 & -\!\cos_1\!-\!\sin_1\\ |{\^i}\!\gamma_1|/{\^i}\!\gamma_2\!-\!\sin_1 & -\!\cos_1\!+\!\sin_1 & |{\^i}\!\gamma_1|/{\^i}\!\gamma_2\!+\!\cos_1 \end{array} \!\!\right) \\ 
&\text{with } \cos_1=\cos\left(\mathrm{arg}\{{\^i}\!\gamma_1\}\right) \;\;\text{and } \sin_1=\sin\left(\mathrm{arg}\{{\^i}\!\gamma_1\}\right).
\end{split}\end{equation}

We can measure the eigenvalues with the circuits of Fig. \ref{fig:igamma-measurement}. Let us start with the argument of the first eigenvalue. Its tangent equals the open-loop Hall-output voltage over the supply voltage in Fig.\ref{fig:igamma1-measurement} , 
\begin{equation}\label{eq:4C-Hall20}\begin{split}
& \frac{V_3-V_1}{V_2} = \frac{R_{3,2}-R_{1,2}}{R_{2,2}} \\ 
& \underbrace{\Rightarrow}_{\text{with }(\ref{eq:4C-Hall8b})}\;\; \frac{V_3-V_1}{V_2} = \tan\left(\mathrm{arg}\{{^i}\!\gamma_1\}\right) .
\end{split}\end{equation}
For the magnitude of the first eigenvalue we can use the same circuit of Fig. \ref{fig:igamma1-measurement}, 
\begin{equation}\label{eq:4C-Hall21}
\frac{I_2}{V_2} = \frac{1}{R_{2,2}} \;\;\underbrace{\Rightarrow}_{\text{with }(\ref{eq:4C-Hall8b})}\;\;
\left|{^i}\!\gamma_1\right| = 2\,\frac{I_2}{V_2}\,\cos\left(\mathrm{arg}\{{^i}\!\gamma_1\}\right) .
\end{equation}
For the second eigenvalue we need the circuit of Fig. \ref{fig:igamma2-measurement}, which measures the cross-resistance
\begin{equation}\label{eq:4C-Hall22}\begin{split}
& I_x = -I_1 - I_3 = -V_x (G_{1,1}+G_{1,3}+G_{3,1}+G_{3,3}) \\
& \underbrace{\Rightarrow}_{\text{with }(\ref{eq:4C-Hall8})} {^i}\!\gamma_2 = \frac{1}{R_x} = \frac{I_x}{V_x} .
\end{split}\end{equation}

\section{The Maximum Noise Efficiency of Regular Symmetric Hall-Plates } 
\label{sec:noise}

\begin{figure}[t]
%\vspace{1mm}
  \centering
                \includegraphics[width=0.49\textwidth]{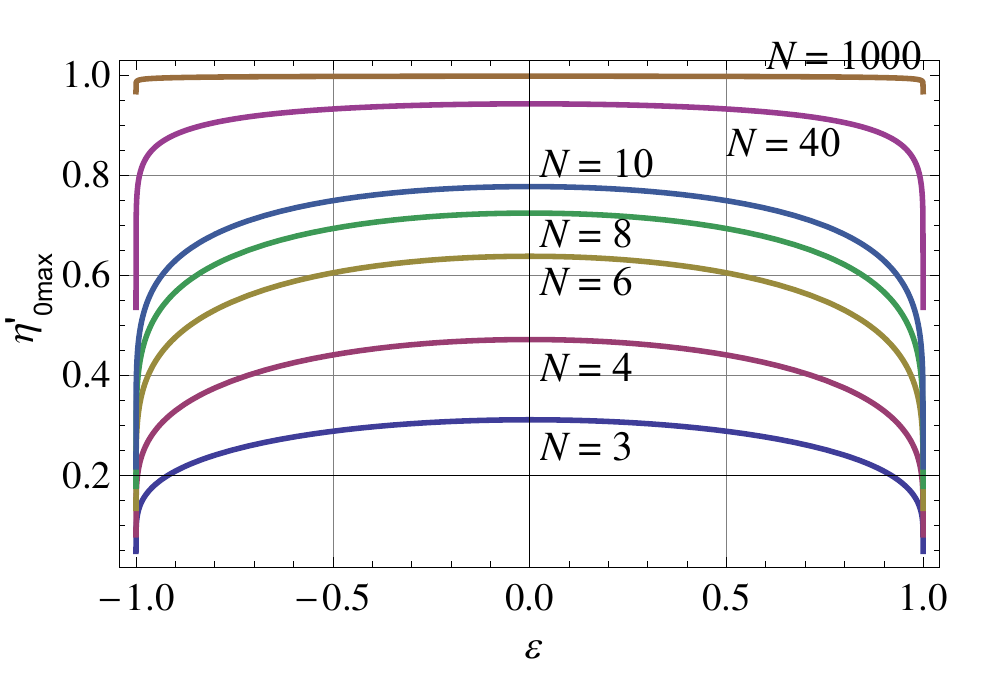}
    \caption{Maximum noise efficiency of regular symmetric Hall-plates with $N$ contacts, versus contacts size $\epsilon$ in the weak magnetic field limit, from (\ref{eq:noise-efficiency6b}). }
   \label{fig:eta0_vs_eps_and_N_regular_Halls}
\end{figure}

\noindent In Ref.~\onlinecite{Ausserlechner2020hybrid} I defined the noise efficiency of Hall-plates by 
\begin{equation}\label{eq:noise-efficiency6}\begin{split}
\eta' & = \frac{\mathrm{SNR}}{\left|\tan(\theta_H)\right|}\;\sqrt{\frac{4 k_b T \Delta_f}{P_d}} . 
\end{split}\end{equation}
SNR is the ratio of the output signal over the thermal noise, $k_b$ is Boltzmann's constant, $T$ is the absolute temperature, $\Delta_f$ is the effective noise bandwidth in which one observes the output signal, and $P_d$ is the power dissipated in the Hall-plate. For optimized patterns of supply voltages or currents and optimized weighting coefficients in the linear combinations of the output signals the maximum noise efficiency at weak magnetic field was found~\cite{ ArXiv2023}  
\begin{equation}\label{eq:noise-efficiency6b}\begin{split}
\eta'_{0\mathrm{max}} & = \lim_{\theta_H\to 0} \eta'_\mathrm{max} \\ 
& = \lim_{\theta_H\to 0} \frac{\tan\left( \mathrm{arg}\{{^i}\!\gamma_1\} \right)}{\tan(\theta_H)} = \lim_{\theta_H\to 0} \frac{\mathrm{arg}\{{^i}\!\gamma_1\}}{\theta_H} . 
\end{split}\end{equation}
(\ref{eq:noise-efficiency6b}) holds for all Hall-plates, regardless of their symmetry, provided that the eigenvalues are sorted in such a way that ${^i}\!\gamma_1$ has the largest argument. Fig. \ref{fig:eta0_vs_eps_and_N_regular_Halls} shows $\eta'_{0\mathrm{max}}$  versus the contacts size of the regular symmetric Hall-plates, $\epsilon$, for various numbers of contacts. Clearly, the maximum noise efficiency of strictly regular symmetric Hall-plates (in $\epsilon = 0$) is larger than of regular symmetric Hall-plates with $\epsilon \ne 0$. However, the maximum is flat, and it is all the flatter the more contacts the Hall-plate has. Hence, the noise efficiency of disk-shaped Hall-plates is not much reduced if the size of the contacts deviates slightly from the size of the contacts spacings. The symmetry of the curves for $\pm\epsilon$ is a consequence of (\ref{eq:RMFoCD1505c}). Note that in the weak magnetic field limit $\eta'_{0\mathrm{max}}$ is the same, regardless if output voltages or currents (or combinations of both) are measured \cite{Ausserlechner2020hybrid}. This is in striking contrast to Refs. \onlinecite{Heidari,XuWangHuJiang}, which claim that current mode operation of Hall-plates might have superior noise performance \cite{meinComment,Ausserlechner2020a}.

\section{Van der Pauw's Method for Regular Symmetric Hall-Plates with Four Contacts} 
\label{sec:VdP}

\noindent Van der Pauw's original method of measuring the sheet resistance $R_\mathrm{sheet}$ works for plane, uniform, singly connected domains with four peripheral point-sized contacts \cite{VdP}. His main achievement is to give a single simple equation which relates the sheet resistance to measureable quantities, the so-called trans-resistances. These are voltages between two contacts divided by currents through the other contacts, and they can also be expressed as differences of entries of the resistance matrix, 
\begin{equation}\label{VdP0}
\exp\!\left(\pi\frac{R_{3,1}-R_{2,1}}{R_\mathrm{sheet}}\right) + \exp\!\left(\pi\frac{R_{3,1}-R_{3,2}}{R_\mathrm{sheet}}\right) = 1 .
\end{equation}
Note that for point-sized contacts all entries $R_{\ell,m}$ of the resistance matrix tend to infinity, yet the trans-resistances $R_{3,1}-R_{2,1}$ and $R_{3,1}-R_{3,2}$ remain finite. Van der Pauw's equation (\ref{VdP0}) is nonlinear and needs to be solved numerically for $R_\mathrm{sheet}$, whereby $R_{3,1}-R_{2,1}$ and $R_{3,1}-R_{3,2}$ are known from two  measurements, 
\begin{equation}\label{VdP0b}\begin{split}
R_{3,1}-R_{2,1} & = \left.\frac{V_3-V_2}{I_1}\right|_{I_2=I_3=0 \;\land\; V_4=0} , \\ 
R_{3,1}-R_{3,2} & = \left.\frac{V_3}{I_1}\right|_{I_3=0 \;\land\; I_2=-I_1 \;\land\; V_4=0} .
\end{split}\end{equation}
%In (\ref{VdP0c}) we used the RMFR-principle in the first equation, and in the last equation we used the fact that in Hall-plates with point-sized contacts all points on the boundary between current carrying contacts have identical Hall output voltage \cite{Ausserlechner2019b} (i.e., the voltage between them does not depend on the sign of the magnetic field).
The restriction to point-sized contacts brings the big advantages that van der Pauw's equation is independent of the shape of the Hall-plate (as long as it remains singly connected and the contacts are at the circumference) and it is also independent of the applied magnetic field (the latter is a consequence of (16c) in Ref.~\onlinecite{Ausserlechner2019b}). Moreover, the resistances of the leads and of the physical contacts at the interface between the probe tips and the Hall-plate do not cause measurement errors---this follows from the simple fact that they are in series to the point-sized contacts, which have infinite resistance, anyhow. In practice, the point-sized contacts are also a drawback, because they cause infinite electric field and current density. This leads to velocity saturation, self heating, inhomogeneous temperature, and mechanical stress, which may cause again errors in the measurement of the sheet resistance.

\begin{figure}[t]
%\vspace{1mm}
  \centering
                \includegraphics[width=0.30\textwidth]{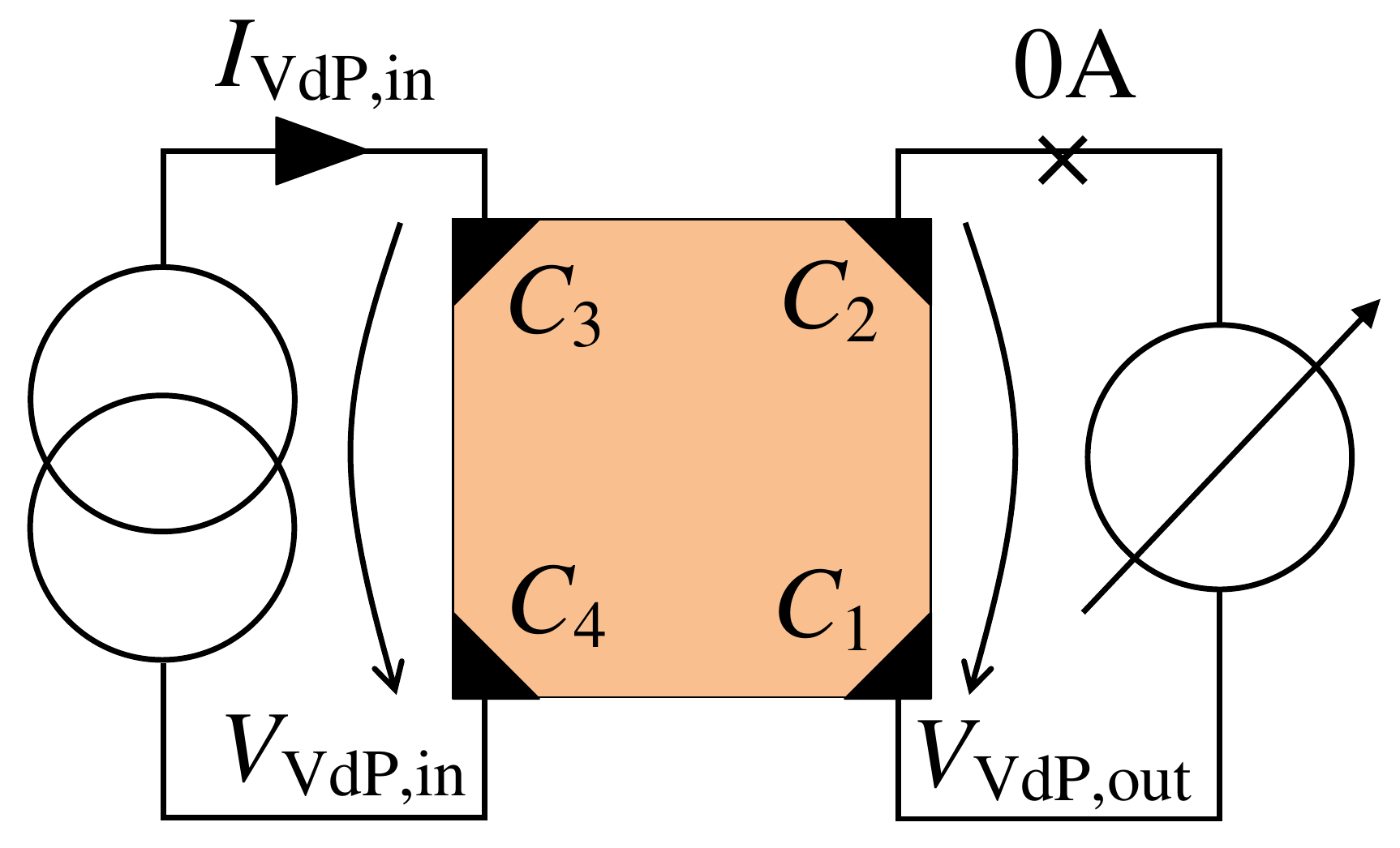}
    \caption{Circuit for a generalized Van-der-Pauw measurement \cite{AusserlechnerVdP2}. }
   \label{fig:VdP}
\end{figure}

\begin{figure*}
%\vspace{1mm}
  \centering
        \begin{subfigure}[t]{0.48\textwidth}
                \centering
                \includegraphics[width=1.0\textwidth]{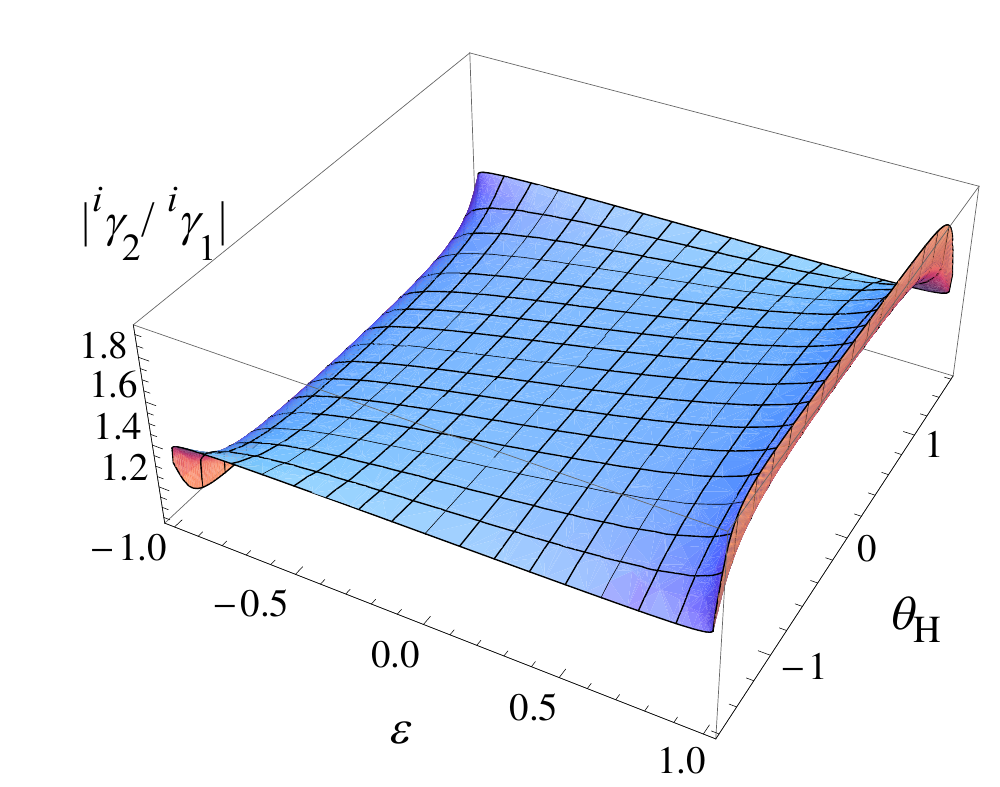}
                \caption{$\left|{^i}\!\gamma_2/{^i}\!\gamma_1 \right|$. }
                \label{fig:Plot3D_igamma2durch1_vs_eps_thetaH}
        \end{subfigure}
        \hfill  % An dieser Stelle kann ein zusätzlicher Zwischenraum eingebunden werden: ~, \quad, \qquad, \hfill usw.
          % Eine leere Zeile erzwingt, dass die zweite Grafik darunter erscheint.
        \begin{subfigure}[t]{0.48\textwidth}
                \centering
                \includegraphics[width=1.0\textwidth]{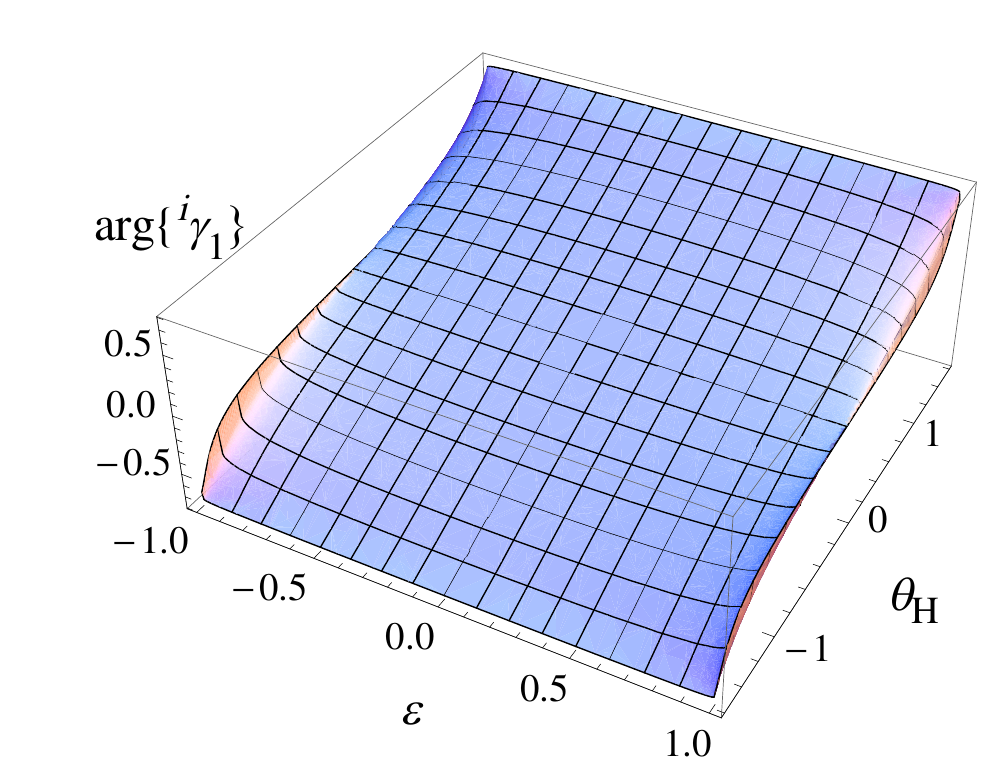}
                \caption{$\mathrm{arg}\left\{{^i}\!\gamma_1\right\}$. }
                \label{fig:Plot3D_arg_igamma1_vs_eps_thetaH}
        \end{subfigure}
	 \\
        \begin{subfigure}[t]{0.47\textwidth}
                \centering
                \includegraphics[width=1.0\textwidth]{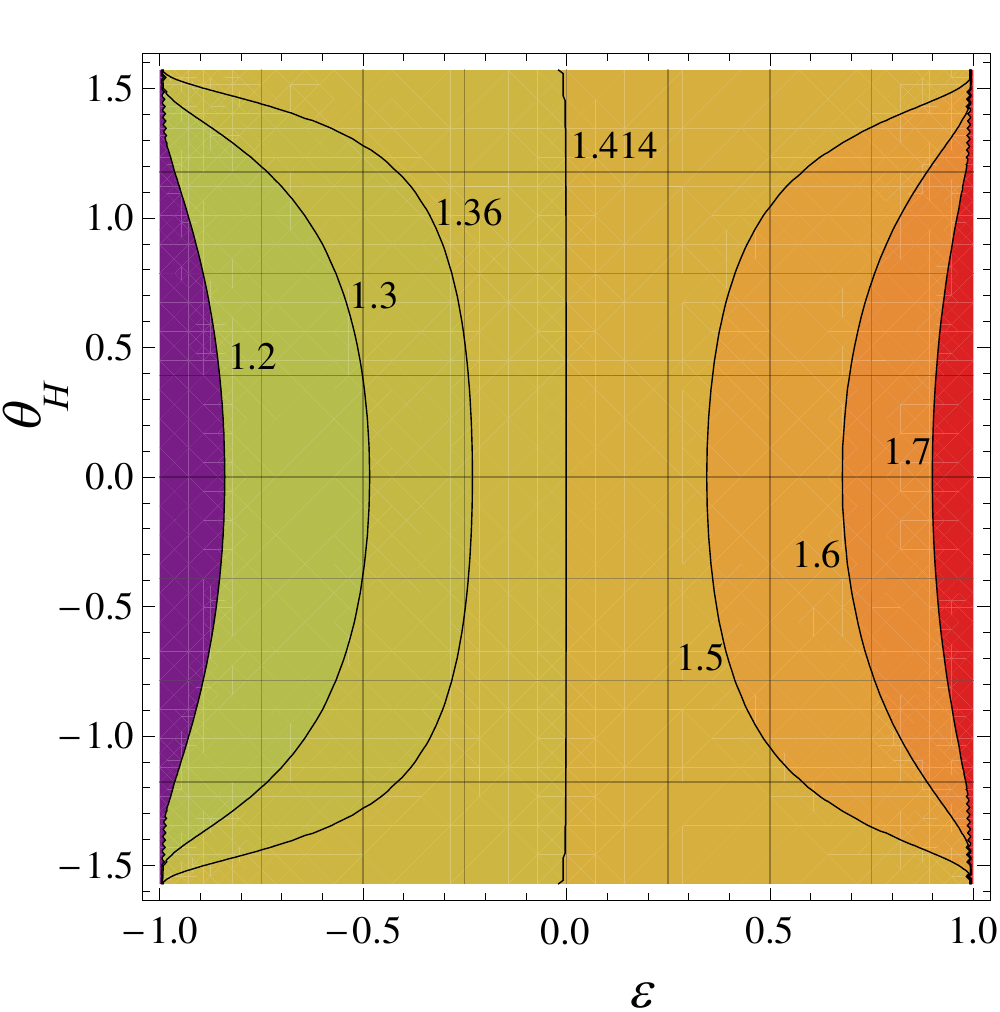}
                \caption{$\left|{^i}\!\gamma_2/{^i}\!\gamma_1 \right|=\mathrm{constant}$. }
                \label{fig:contour_igamma2durch1_vs_eps_thetaH}
        \end{subfigure}
        \hfill  
        \begin{subfigure}[t]{0.47\textwidth}
                \centering
                \includegraphics[width=1.0\textwidth]{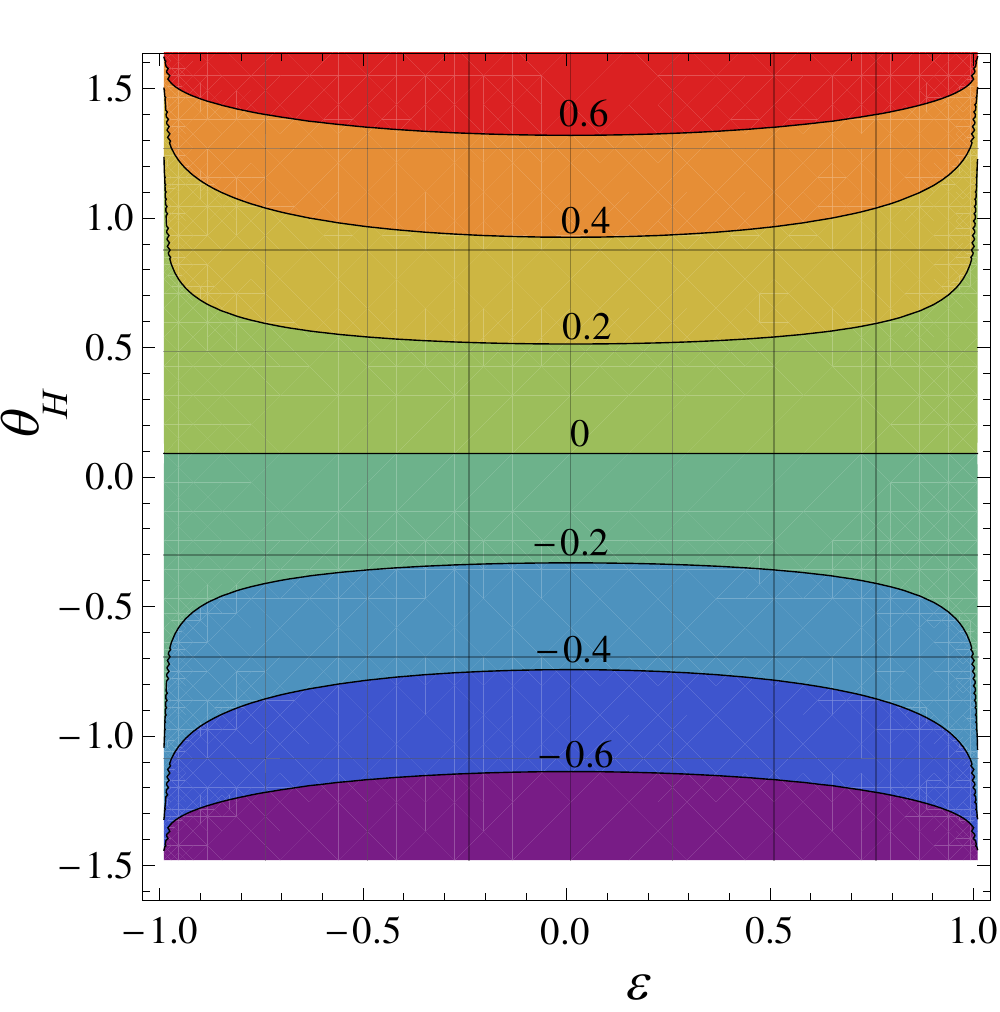}
                \caption{$\mathrm{arg}\left\{{^i}\!\gamma_1\right\}=\mathrm{constant}$. }
                \label{fig:contour_abs_igamma2_vs_eps_thetaH}
        \end{subfigure}
    \caption{$\left|{^i}\!\gamma_2/{^i}\!\gamma_1\right|$ and $\mathrm{arg}\left\{{^i}\!\gamma_1\right\}$ for regular symmetric Hall-plates with $N=4$ contacts versus contacts size $\epsilon$ and Hall-angle $\theta_H$ (both functions are independent of $R_\mathrm{sheet}$). The upper row shows 3D-plots, and the lower row shows contour plots. A generalized van-der-Pauw method determines the unique point, where a contour line in the left lower plot crosses a contour line in the right lower plot. Such a crossing point always exists for $1\le\left|{^i}\!\gamma_2/{^i}\!\gamma_1 \right|\le 2$, see (\ref{eq:VdP10}). }
   \label{fig:igamma_vs_eps_thetaH}
\end{figure*}
% original file: HomentcovschiMurray_regular_HHalls_eigenvalues5.nb

If the contacts are extended, the situation becomes much more complicated, yet one can still generalize van der Pauw's method. The basic idea is to measure the full resistance matrix. If we can express the resistance matrix as a function of the contact geometry, the Hall-angle and the sheet resistance, we can solve this system of equations (at least numerically), provided that this set of nonlinear equations has a unique solution and that there are enough equations. 

For regular symmetric geometries with extended contacts at zero Hall angle, $\theta_H=0$, one can use a simple generalized van der Pauw method \cite{AusserlechnerVdP2}. Equation (5b) in Ref.~\onlinecite{AusserlechnerVdP2} becomes 
\begin{equation}\label{VdP1}
\left(\!L\!\left(\!\frac{V_{\mathrm{VdP,in}}\!+\!V_{\mathrm{VdP,out}}}{R_\mathrm{sheet}\,I_{\mathrm{VdP,in}}}\!\right)\!\right)^{\!\!2} = L\!\left(2\frac{V_{\mathrm{VdP,in}}\!-\!V_{\mathrm{VdP,out}}}{R_\mathrm{sheet}\, I_{\mathrm{VdP,in}}}\!\right) 
\end{equation}
whereby the voltages and currents are explained in Fig. \ref{fig:VdP}. Equation (\ref{VdP1}) is a nonlinear implicit equation for $R_\mathrm{sheet}$, if we insert numbers for all measured voltages and currents. 
\noindent Let us express the voltages in terms of the resistance matrix $\bm{R}$ and the input current. This gives 
\begin{equation}\label{VdP2}\begin{split}
V_{\mathrm{VdP,in}}+V_{\mathrm{VdP,out}} & = R_{2,2} \;I_{\mathrm{VdP,in}} , \\
V_{\mathrm{VdP,in}}-V_{\mathrm{VdP,out}} & = \frac{R_{1,1}-R_{1,2}+R_{1,3}}{2}\; I_{\mathrm{VdP,in}} ,
\end{split}\end{equation}
with all resistances evaluated at $\theta_H=0$. Using (\ref{eq:4C-Hall8b}) in (\ref{VdP2}) and inserting it into (\ref{VdP1}) gives 
\begin{equation}\label{VdP3}
\left(\!L\!\left(\!\frac{2}{R_\mathrm{sheet}\,{^i}\!\gamma_1(\theta_H=0)}\!\right)\!\right)^{\!\!2} = L\!\left(\frac{4}{R_\mathrm{sheet}\,{^i}\!\gamma_2(\theta_H=0)}\!\right) .
\end{equation}
% numerically checked in HomentcovschiMurray_regular_HHalls_eigenvalues5.nb
If we insert (\ref{eq:RMFoCD1550}) and (\ref{eq:RMFoCD1552}) into (\ref{VdP3}) it seems astonishing that (\ref{VdP3}) is valid for all $\epsilon\in [-1,1]$. Yet, if we insert (\ref{eq:RMFoCD1565a}) and (\ref{eq:RMFoCD1565b}) into (\ref{VdP3}) the identity is obvious in view of the definition of the modular lambda function $L(y)$ from \ref{eq:RMFoCD1557b}. Equation (\ref{VdP3}) means that at zero magnetic field there is a relation between both eigenvalues and the sheet resistance regardless of the size of the contacts. 
%We can write (\ref{VdP3}) also like this 
%\begin{widetext}
%\begin{equation}\label{eq:VdP4}\begin{split}
%& \text{If } \frac{K'(k)}{K(k)} = \sqrt{2} \frac{\int_0^{\pi/4} f(-\epsilon)\cos((1-\epsilon)x/2)\,\mathrm{d}x}{\int_0^{\pi/4} f(\epsilon)\cos((1+\epsilon)x/2)\,\mathrm{d}x} \\ 
%& \text{then } \frac{K'(k^2)}{K(k^2)} = 2 \frac{\int_0^{\pi/4} f(-\epsilon)\,\mathrm{d}x}{\int_0^{\pi/4} f(\epsilon)\,\mathrm{d}x} \quad\forall\epsilon\in[-1,1] ,
%\end{split}\end{equation}
%\end{widetext}
%with $f(\epsilon)$ from (\ref{eq:RMFoCD1503}).

Now let us turn to the case of regular symmetric Hall-plates with four extended contacts at \emph{non-vanishing} magnetic field. The electrical properties at the terminals of these plates are defined by three parameters, $R_\mathrm{sheet},\epsilon,\theta_H$. If we measure the full condutance or resistance matrix, $\bm{G}, \bm{R}$, we can derive all eigenvalues ${^i}\!\gamma_1, {^i}\!\gamma_2$. Thus, we can say that we 'measure' the eigenvalues. ${^i}\!\gamma_1$ is complex, whereas ${^i}\!\gamma_2$ is real. This gives three equations for the three unknown real numbers. The only question left is, if this set of nonlinear equations has a unique solution, which we show in the sequel by eliminating $R_\mathrm{sheet}$ from these equations: consider the expressions $\left|{^i}\!\gamma_2/{^i}\!\gamma_1\right|$ and $\mathrm{arg}\left\{{^i}\!\gamma_1\right\}$. Both are known from the measurements in Fig. \ref{fig:igamma-measurement}. They give two equations for two unknowns, $\epsilon,\theta_H$---$R_\mathrm{sheet}$ has dropped out. Fig. \ref{fig:igamma_vs_eps_thetaH} shows the 3D-plots and the contour lines of these expressions, from where it is obvious that a unique solution exists for arbitrary values of $\left|{^i}\!\gamma_2/{^i}\!\gamma_1\right|$ and $\mathrm{arg}\left\{{^i}\!\gamma_1\right\}$. This solution determines $\epsilon$ and $\theta_H$, which we insert into (\ref{eq:RMFoCD1503}) to compute ${^i}\!\gamma_2 R_\mathrm{sheet}$. Finally, comparison with the measured value for ${^i}\!\gamma_2$ gives $R_\mathrm{sheet}$. This shows that a generalized van-der-Pauw method for regular symmetric Hall-plates can determine the sheet resistance, regardless of the applied magnetic field. 

With (\ref{eq:RMFoCD1505}) it follows a symmetry relation of the surface in Fig. \ref{fig:Plot3D_igamma2durch1_vs_eps_thetaH}, 
\begin{equation}\label{VdP7}
\left|\frac{{^i}\!\gamma_2(-\epsilon,\theta_H)}{{^i}\!\gamma_1(-\epsilon,\theta_H)}\right| \; \left|\frac{{^i}\!\gamma_2(\epsilon,\theta_H)}{{^i}\!\gamma_1(\epsilon,\theta_H)}\right| = 2 .
\end{equation}
Thus, for strictly regular symmetric Hall-plates at arbitrary magnetic field it holds 
\begin{equation}\label{VdP8}
\left|\frac{{^i}\!\gamma_2(\epsilon=0,\theta_H)}{{^i}\!\gamma_1(\epsilon=0,\theta_H)}\right| = \sqrt{2} .
\end{equation}
On the other hand, $|{^i}\!\gamma_2/{^i}\!\gamma_1|$ depends stronger on the contacts size $\epsilon$ at zero magnetic field than at non-zero magnetic field. From (\ref{eq:RMFoCD1565a}) and (\ref{eq:RMFoCD1565b}) it follows   
\begin{equation}\label{VdP9}\begin{split}
\lim_{\epsilon\to -1}\left|\frac{{^i}\!\gamma_2(\theta_H=0)}{{^i}\!\gamma_1(\theta_H=0)}\right| & = 1 , \\ 
\lim_{\epsilon\to 1}\left|\frac{{^i}\!\gamma_2(\theta_H=0)}{{^i}\!\gamma_1(\theta_H=0)}\right| & = 2 .
\end{split}\end{equation}
We used the asymptotic limit $K(x\to 1)=\ln(4/\sqrt{1-x^2})$ and de l'Hopital's rule. Consequently, the ratio of the magnitudes of both eigenvalues of a four-contacts regular symmetric Hall-plate at arbitrary magnetic field is limited 
\begin{equation}\label{eq:VdP10}
1\le \left|\frac{{^i}\!\gamma_2}{{^i}\!\gamma_1}\right|\le 2 \quad\forall\theta_H.
\end{equation}
Simultaneously, the argument of ${^i}\!\gamma_1$ is also limited to
\begin{equation}\label{VdP13}\begin{split}
\frac{-\theta_H}{2} \le \mathrm{arg}\{{^i}\!\gamma_1\} \le \frac{\theta_H}{2} \quad\forall\theta_H ,
\end{split}\end{equation}
according to Fig. \ref{fig:Plot3D_arg_igamma1_vs_eps_thetaH} and (\ref{eq:RMFoCD1504c}).

\section{Conclusion} 
\label{sec:conclusion}

\noindent This work discussed plane, uniform, isotropic, singly-connected Hall-plates with an arbitrary number of $N$ extended contacts and with $N$-fold symmetry (i.e., transforming the Hall-plate conformally into the unit disk and rotating the disk by integer multiples of $360$°$/N$ gives again the same disk). A special case of great practical relevance is Hall-plates with four contacts and $90$° symmetry. The starting point for this paper was a matrix equation from Homentcovschi, Bercia, and Murray \cite{HomentcovschiBercia,Homentcovschi2019}, which links the potentials to the currents at the contacts of a general uniform singly connected Hall-plate with peripheral contacts. There, I introduced the indefinite conductance matrix, which is a  circulant matrix, due to the assumed regular symmetry of the Hall-plates. This admits a closed analytical solution for the complex eigenvalues of the indefinite conductance matrix. The resistance and conductance matrices are sums of weighted eigenvalues. It was shown how the eigenvalues change with contacts size and Hall-angle, and simple approximations were given. The Hall-geometry factor describes how the shape of the Hall-plate affects the magnitude of the output voltage and the nonlinearity of the output voltage versus magnetic field. Exact and approximate formulae for the Hall-geometry factor of plates with four contacts were given and compared to prio art literature. Based on the eigenvalues it was also shown that strictly regular symmetric Hall-plates have better noise efficiency than regular symmetric ones---highest possible degree of symmetry is favorable for low thermal noise. Finally, a generalized van-der-Pauw method for the measurement of the sheet resistance was explained, and it was proven that a unique solution for the sheet resistance exists, regardless of the applied magnetic field.

Up to my knowledge this work gives the most comprehensive, simple, and read-to-use body of equations to quantify \emph{all electrical parameters} of regular symmetric electrically linear Hall-plates at arbitrary magnetic field. This paper is also published in Ref.~\onlinecite{ResearchGate}.

\appendix

\section{How to compute the sum in (\ref{eq:RMFoCD125})}
\label{sec:sum}

\begin{figure}[t]
%\vspace{1mm}
  \centering
                \includegraphics[width=0.20\textwidth]{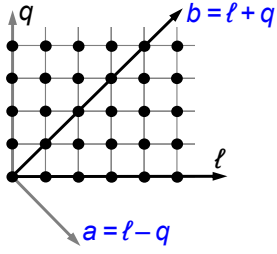}
    \caption{Transformation of indices for a double sum. }
   \label{fig:sum-index-transformation}
\end{figure}

\noindent We compute the sum 
\begin{equation}\label{eq:sum1}\begin{split}
\sum_{k=1}^N \frac{\exp\left(2\pi \mathbbm{i} k m / N\right)}{\sin(x/N\!+\!\pi k/N)\sin(x/N\!+\!\pi (k\!-\!j)/N)} ,
\end{split}\end{equation} 
for $N\ge 2$, $j\in\{1,2,\ldots,N-1\}$, and integer $m$.
First we replace $k=N$ by $k=0$, then we express the sines by complex exponentials, 
\begin{equation}\label{eq:sum2}\begin{split}
& \sum_{k=0}^{N-1} \frac{(2\mathbbm{i})^2 \exp\left(2\pi \mathbbm{i} k m / N\right)}{\exp\left(\mathbbm{i}(x/N\!+\!\pi k/N)\right) \exp\left(\mathbbm{i}(x/N\!+\!\pi (k\!-\!j)/N)\right)} \\
&\qquad \times \left[1\!-\!\exp\left(-2\mathbbm{i}\frac{x\!+\!\pi k}{N}\right)\right]^{-1} \\
&\qquad \times \left[1\!-\!\exp\left(-2\mathbbm{i}\frac{x\!+\!\pi (k\!-\!j)}{N}\right)\right]^{-1} .
\end{split}\end{equation} 
Then we make twice a Taylor series expansion of the type $[1-z]^{-1}=1+z+z^2+z^3+\ldots$. This gives 
\begin{equation}\label{eq:sum3}\begin{split}
& \sum_{\ell=0}^\infty\sum_{q=0}^\infty -4\exp\left(-2\mathbbm{i}x\frac{1+\ell+q}{N}\right) \exp\left(\mathbbm{i}\pi j\frac{1+2q}{N}\right) \\
& \qquad\quad \times \sum_{k=0}^{N-1} \exp\left(2\pi \mathbbm{i}k\frac{m-1-\ell-q}{N}\right) .
\end{split}\end{equation} 
The sum over index $k$ vanishes, except for $m-1-\ell-q=-p N$ with $p=0,1,2\ldots$, hence it equals $N\delta_{q,m-1-\ell+pN}$. We re-arrange the double sum over indices $\ell, q$ as a double sum over indices $a,b$ with $a=\ell-q, b=\ell+q$ (see Fig. \ref{fig:sum-index-transformation}). With $b=m-1+p N$ it holds 
\begin{equation}\label{eq:sum4}\begin{split}
& \sum_{p=0}^\infty -4 N \exp\left[\mathbbm{i}\left(\pi j\!-\!2x\right)\frac{m\!+\!pN}{N}\right] \\
& \qquad \times \sum_{a=-(m-1+pN),\Delta a=2}^{m-1+pN} \!\exp\left(\frac{-\mathbbm{i}\pi j a}{N}\right) ,
\end{split}\end{equation} 
whereby the index $a$ is incremented by $\Delta a=2$. We replace the index $a$ by $c$ via $a=2c-m+1-pN$. This gives 
\begin{equation}\label{eq:sum5}\begin{split}
& \sum_{a=-(m-1+pN),\Delta a=2}^{m-1+pN} \exp\left(\frac{-\mathbbm{i}\pi j a}{N}\right) \\ 
& = \exp\left(\mathbbm{i}\pi j\frac{m-1+pN}{N}\right) \sum_{c=0}^{m-1+pN} \exp\left(-2\pi\mathbbm{i}\frac{j}{N}c\right) \\ 
& = (-1)^{j p}\frac{\sin\left(\pi j m/N\right)} {\sin\left(\pi j/N\right)} .
\end{split}\end{equation} 
Inserting(\ref{eq:sum5}) into (\ref{eq:sum4}) and computing the infinite sum over index $p$ finally gives
\begin{equation}\label{eq:sum6}\begin{split}
2N\frac{\sin\left(\pi j m / N\right)}{\sin\left(\pi j / N\right)} \left(\mathbbm{i}\cot(x)-1\right)\exp\left(\mathbbm{i} m \frac{j\pi-2x}{N}\right) .
\end{split}\end{equation}
%which proves (\ref{eq:RMFoCD125}).

\section{How to compute (\ref{eq:RMFoCD1550}), (\ref{eq:RMFoCD1552}), (\ref{eq:RMFoCD1565a}), (\ref{eq:RMFoCD1565b})}
\label{sec:igamma-zero}

\noindent First we compute (\ref{eq:RMFoCD1550}). We start with (\ref{eq:RMFoCD1503}) for $m=N/2$ and $\theta_H=0$. Then, the integral in the numerator of (\ref{eq:RMFoCD1503}) is 
\begin{equation}\label{apx1}\begin{split}
& \int_0^{\pi/4} \frac{(1+\epsilon)\,\mathrm{d}x}{\sqrt{\sin\!\left((\pi/4\!+\!x)(1\!+\!\epsilon)\!\right)\sin\!\left((\pi/4\!-\!x)(1\!+\!\epsilon)\!\right)}} = \\ 
& \int_0^{\pi/4} \frac{2\sqrt{2}(1+\epsilon)\,\mathrm{d}x}{\sqrt{\cos\left(2(1+\epsilon)x\right)+\sin(\pi\epsilon/2)}} .
\end{split}\end{equation}
We replace $(1+\epsilon)x=\pi/2-y$ and $\sin(\pi\epsilon/2)=1-2 \kappa^2$ with $0\le\kappa\le 1$. This gives 
\begin{equation}\label{apx2}\begin{split}
& \frac{1}{\kappa}\int_{\pi(1-\epsilon)/4}^{\pi/2} \frac{2\,\mathrm{d}y}{\sqrt{\left(\sin(y)/\kappa\right)^2-1}} \\ 
& = 2 F\!\left(\!\frac{\cos\left(\pi(1-\epsilon)/4\right)}{\sqrt{1-\kappa^2}},\sqrt{1-\kappa^2}\!\right) \\ 
&\text{with } 1-\kappa^2 = \sin\left(\pi(1+\epsilon)/4\right)\cos\left(\pi(1-\epsilon)/4\right) .
\end{split}\end{equation}
The integration in (\ref{apx2}) is tabulated in Ref.~\onlinecite{Prudnikov1}. Equation (\ref{eq:RMFoCD1550}) follows immediately from (\ref{apx2}), q.e.d. 

Next we compute (\ref{eq:RMFoCD1552}). We start again with (\ref{eq:RMFoCD1503}) for $m=N/4$ and $\theta_H=0$. Then, the integral in the numerator of (\ref{eq:RMFoCD1503}) is 
\begin{equation}\label{apx4}\begin{split}
& \int_0^{\pi/4} \frac{2(1+\epsilon)\cos\left((1+\epsilon)x/2\right)\,\mathrm{d}x}{\sqrt{\sin\!\left((\pi/4\!+\!x)(1\!+\!\epsilon)\!\right)\sin\!\left((\pi/4\!-\!x)(1\!+\!\epsilon)\!\right)}} = \\ 
& \int_0^{\pi(1+\epsilon)/8} \frac{4\cos(\alpha)\,\mathrm{d}\alpha}{\sqrt{\left(\cos\left(2\alpha\right)\right)^2-\kappa^2}} ,
\end{split}\end{equation}
whereby we made the substitution $\alpha=(1+\epsilon)x/2$, and we used he same $\kappa$ as above. Now we replace $x=(\cos(\alpha))^2$ and get 
\begin{equation}\label{apx5}\begin{split}
& \int_{(1+\kappa)/2}^1 \frac{\mathrm{d}x}{\sqrt{1\!-\!y}\sqrt{y\!-\!(1\!+\!\kappa)/2}\sqrt{y\!-\!(1\!-\!\kappa)/2}} \\ 
& = \frac{2\sqrt{2}}{1+\kappa} K\left(\sqrt{\frac{1-\kappa}{1+\kappa}}\right) .
\end{split}\end{equation}
This integral is tabulated in Ref.~\onlinecite{Prudnikov2}. Equation (\ref{eq:RMFoCD1552}) follows immediately from (\ref{apx5}), q.e.d. 

Next we compute (\ref{eq:RMFoCD1565a}) based on (\ref{eq:RMFoCD1555})---(\ref{eq:RMFoCD1560}). 
\begin{equation}\label{apx10}\begin{split}
{^i}\!\gamma_2(\theta_H=0) & = -4 G_{1,2}(\theta_H=0) = \frac{4}{r_{1,4}(\theta_H=0)} \\ 
& = \frac{4}{4R_\mathrm{sheet}\lambda_x} = \frac{4}{R_\mathrm{sheet}}\; \frac{K\left(L\left(2\frac{K(k_1)}{K'(k_1)}\right)\right)}{K'\left(L\left(2\frac{K(k_1)}{K'(k_1)}\right)\right)} .
\end{split}\end{equation}
With the modular equation \cite{ELEN} it holds 
\begin{equation}\label{apx11}
2\frac{K(k_1)}{K'(k_1)} = \frac{K\left(\frac{2\sqrt{k_1}}{1+k_1}\right)}{K'\left(\frac{2\sqrt{k_1}}{1+k_1}\right)} = \frac{K'\left(\frac{1-k_1}{1+k_1}\right)}{K\left(\frac{1-k_1}{1+k_1}\right)} .
\end{equation}
Inserting (\ref{eq:RMFoCD1559}) into (\ref{apx11}) gives 
\begin{equation}\label{apx12}
L\left(2\frac{K(k_1)}{K'(k_1)}\right) = \left(\frac{1-k_1}{1+k_1}\right)^2 = \left(\tan(\theta)\right)^2 .
\end{equation}
Inserting (\ref{apx12}) into (\ref{apx10}) readily gives ${^i}\!\gamma_2(\theta_H=0)$ in (\ref{eq:RMFoCD1565a}), q.e.d. 

For ${^i}\!\gamma_1$ in (\ref{eq:RMFoCD1565b}) we combine (\ref{eq:RMFoCD1555})---(\ref{eq:RMFoCD1560}) like this 
\begin{equation}\label{apx15}\begin{split}
& {^i}\!\gamma_1(\theta_H=0) = \frac{{^i}\!\gamma_2(\theta_H=0)}{2}-2 G_{1,3}(\theta_H=0) \\ 
& \qquad = \frac{2}{r_{1,4}(\theta_H=0)}+\frac{2}{r_{2,4}(\theta_H=0)} = \frac{2}{R_\mathrm{sheet}\lambda} \\
& \qquad = \frac{1}{R_\mathrm{sheet}}\; \frac{K'(k_1)}{K(k_1)} = \frac{1}{R_\mathrm{sheet}}\; \frac{K(\sqrt{1-k_1^2})}{K'(\sqrt{1-k_1^2})} .
\end{split}\end{equation}
Inserting (\ref{eq:RMFoCD1559}) and (\ref{eq:RMFoCD1560}) into (\ref{apx15}) gives ${^i}\!\gamma_1(\theta_H=0)$ in (\ref{eq:RMFoCD1565b}), q.e.d.

%\clearpage
\section{Popular shapes of Hall-plates with four contacts}
\label{sec:popularHalls}

\begin{figure}
%\vspace{1mm}
  \centering
        \begin{subfigure}[c]{0.22\textwidth}
                \centering
                \includegraphics[width=1.0\textwidth]{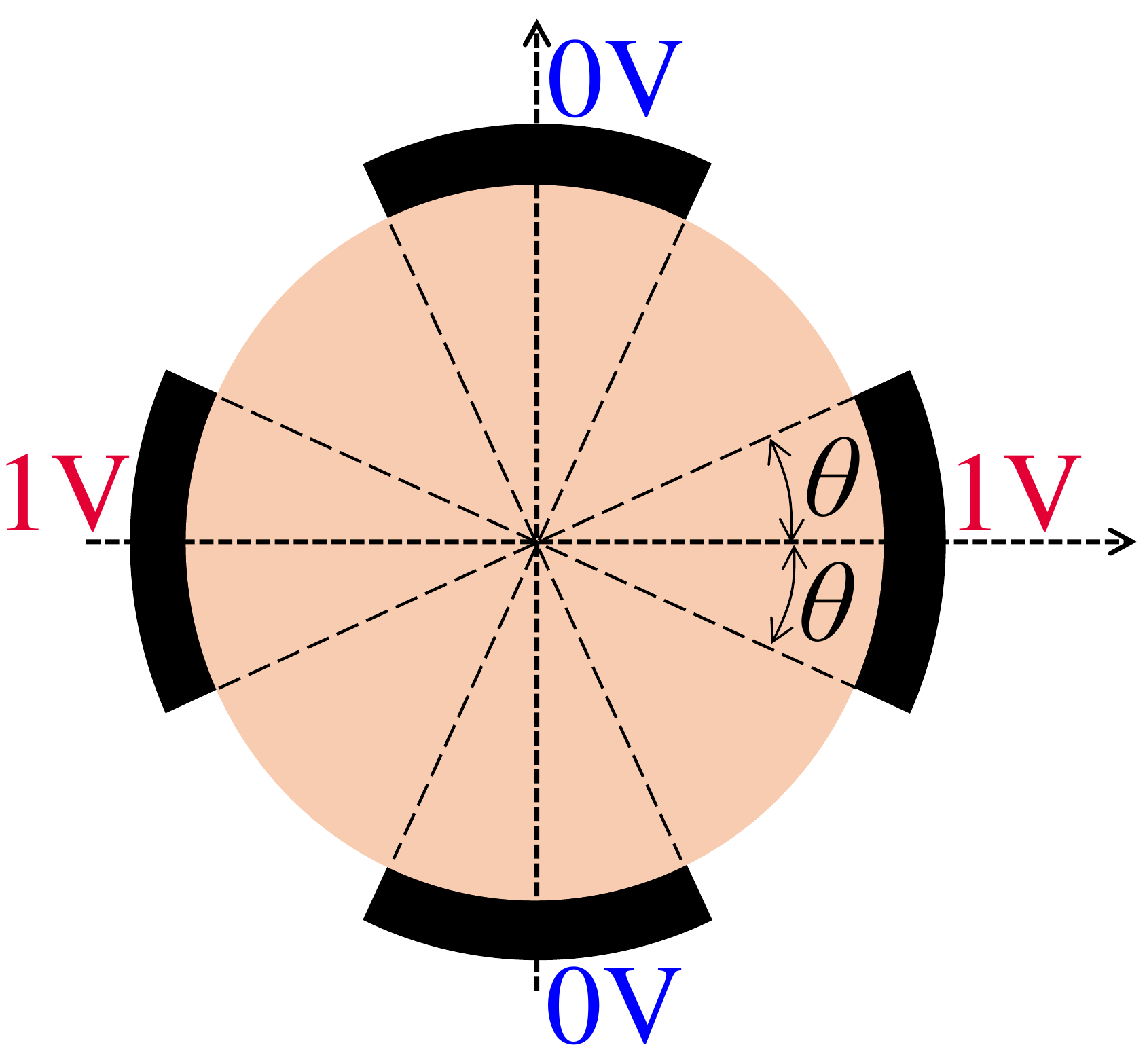}
                \caption{Complete Hall-plate in the $z$-plane. Resistance = $R_x$. }
                \label{fig:regular-4C-Hall-disk_Rx_z-plane1}
        \end{subfigure}
	\hfill
        \begin{subfigure}[c]{0.22\textwidth}
                \centering
                \includegraphics[width=1.0\textwidth]{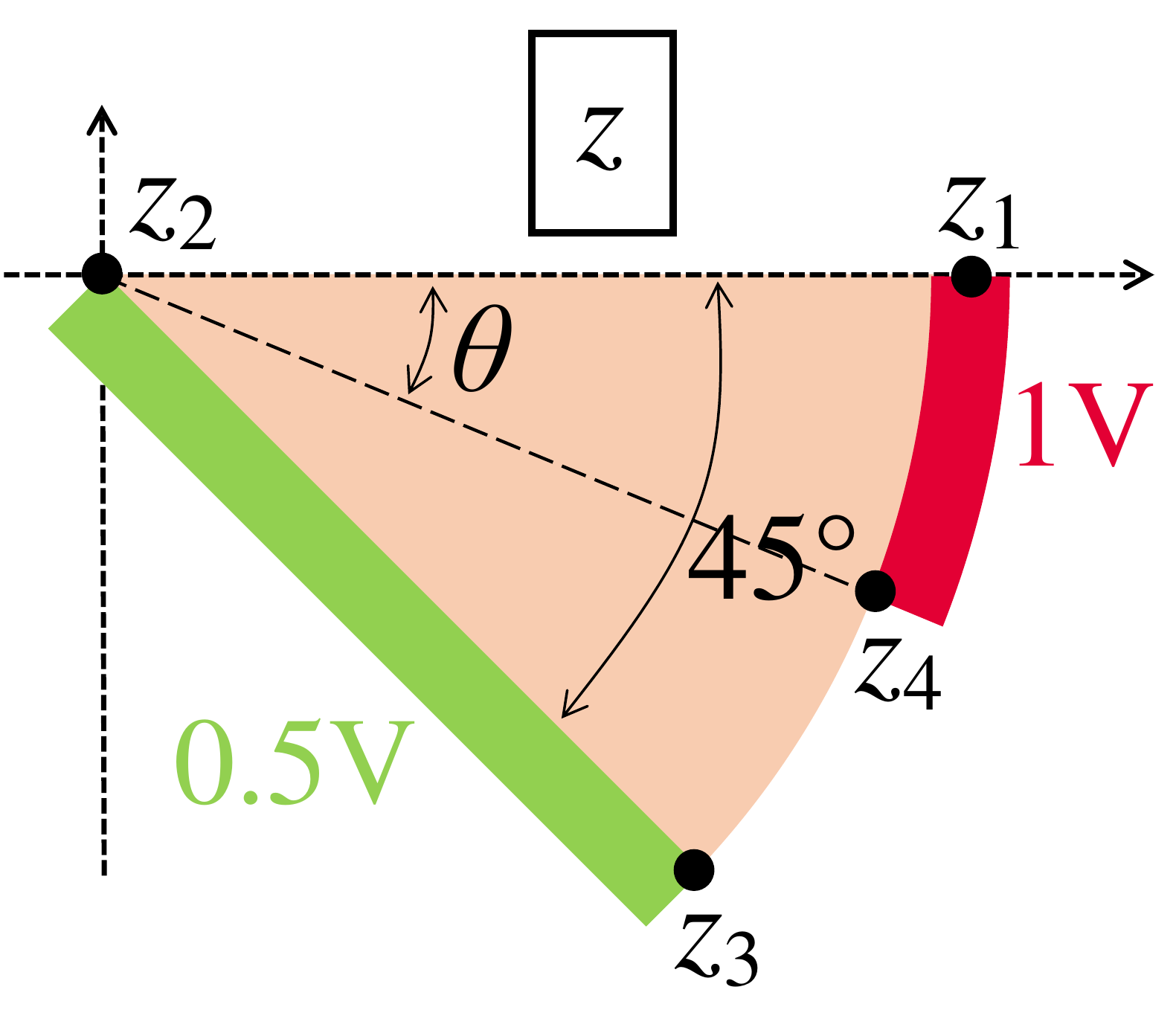}
                \caption{Unit cell of Fig. \ref{fig:regular-4C-Hall-disk_Rx_z-plane1}. Resistance = $2R_x$. }
                \label{fig:regular-4C-Hall-disk_Rx_z-plane2}
        \end{subfigure}
	\\
        \begin{subfigure}[c]{0.42\textwidth}
                \centering
                \includegraphics[width=1.0\textwidth]{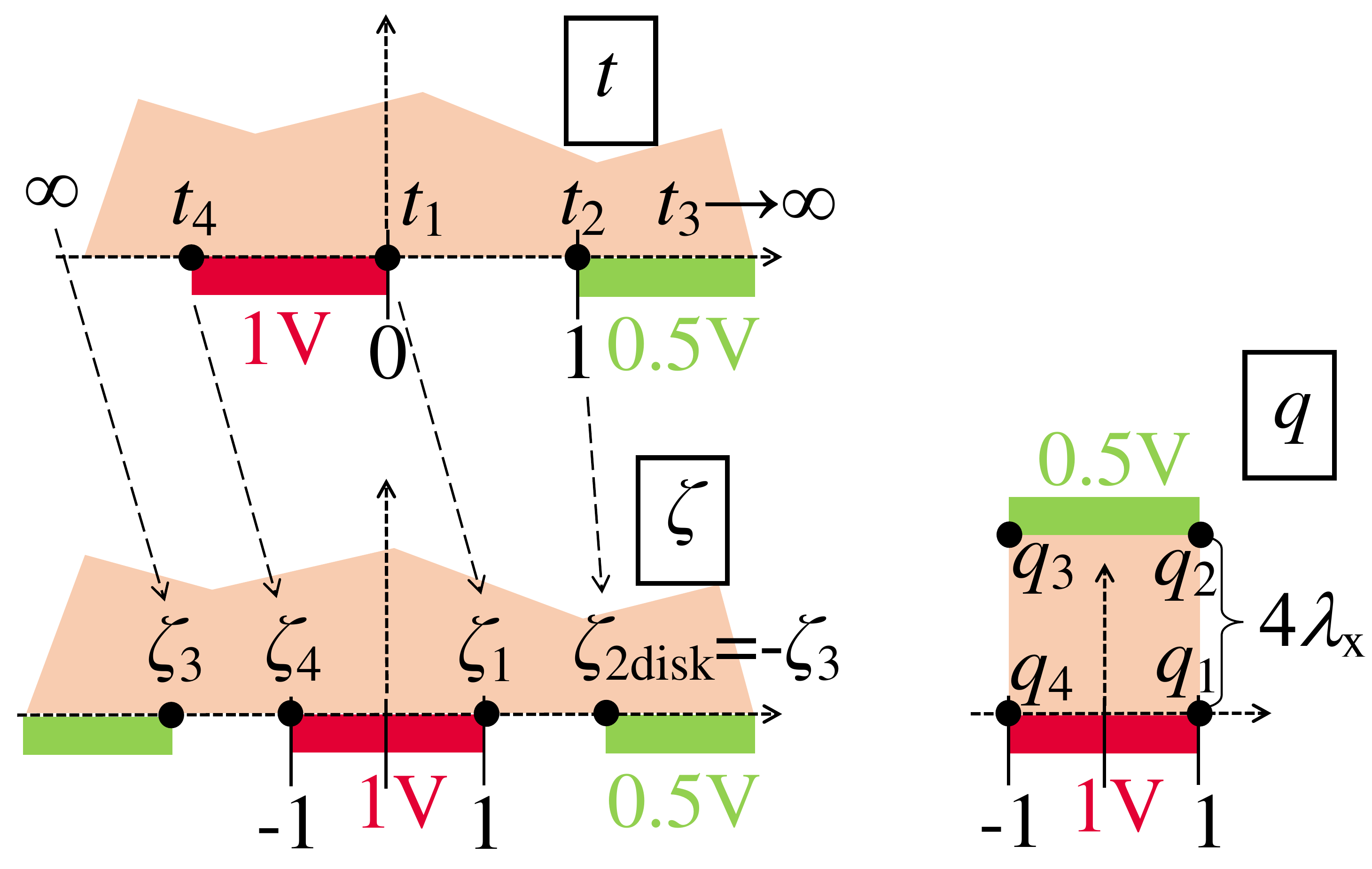}
                \caption{Unit cell mapped to $t$-, $\zeta$-, $q$-planes. Resistance = $2R_x$. }
                \label{fig:regular-4C-Hall-disk_Rx_t-and-zeta-plane}
        \end{subfigure}
    \caption{Conformal mapping of a disk-shaped Hall-plate with $90$° symmetry to compute the cross resistance $R_x$. }
   \label{fig:regular-4C-Hall-disk_Rx}
\end{figure}

\begin{figure*}[t]
%\vspace{1mm}
  \centering
                \includegraphics[width=0.78\textwidth]{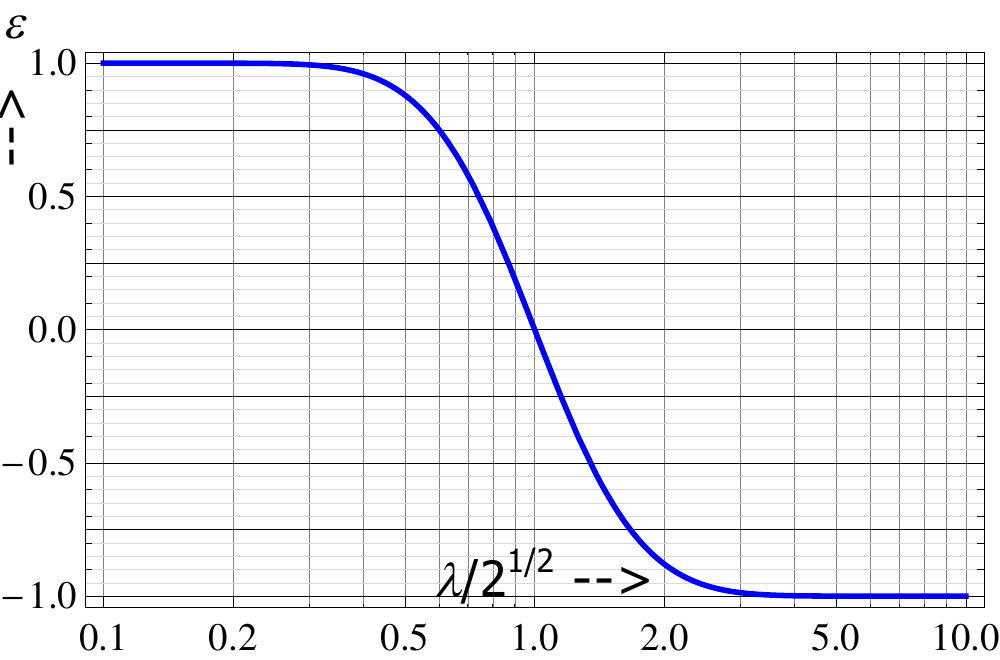}
    \caption{Size of contacts of disk-shaped regular symmetric Hall-plate, $\epsilon$, versus the effective number of squares (normalized by $\sqrt{2}$), $\lambda/\sqrt{2}$, according to (\ref{apx-popularHalls12c}) and (\ref{eq:RMFoCD1560}). }
   \label{fig:eps_vs_lamda}
\end{figure*}

\begin{figure*}[t]
%\vspace{1mm}
  \centering
                \includegraphics[width=0.78\textwidth]{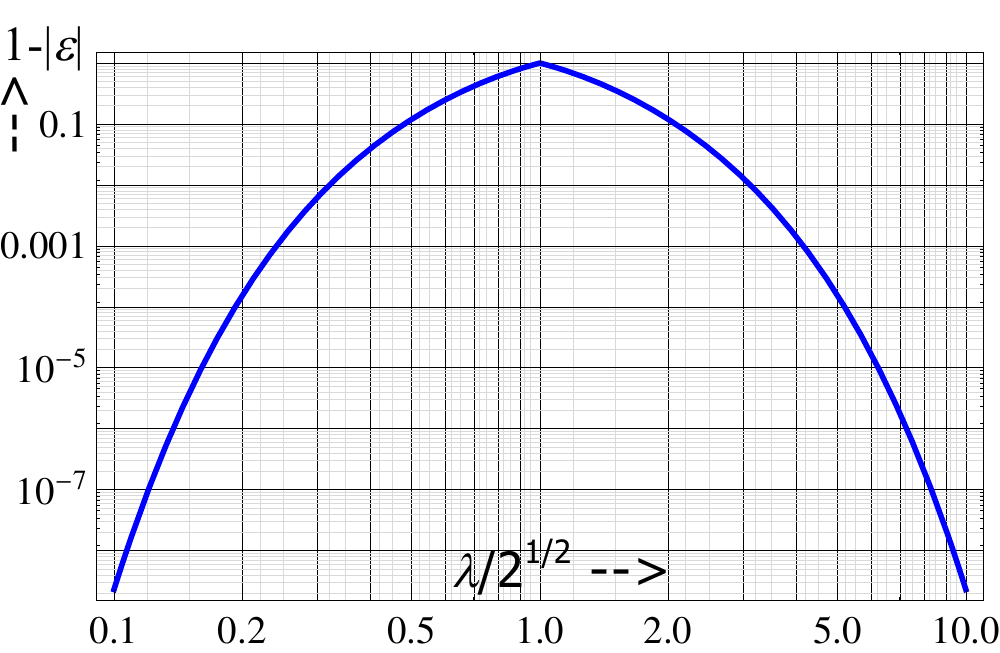}
    \caption{Details of Fig. \ref{fig:eps_vs_lamda} for small and large effective number of squares: $1-|\epsilon|$ versus $\lambda/\sqrt{2}$. }
   \label{fig:1minus-abs_esp_vs_lamda_logarithmic}
\end{figure*}

\noindent The main text of this paper discussed the electrical properties (= the eigenvalues) of circular disk-shaped regular symmetric Hall-plates. This appendix shows how to find other well-known singly connected geometries with peripheral contacts (Greek crosses, octagons, rectangles, squares), which have identical electrical properties (= identical eigenvalues) as the disks. The fundamental idea is, to figure out how large the contacts for these shapes must be, in order to have the same resistance like the disk \emph{at zero magnetic field}. According to (\ref{eq:4C-Hall4}) and (\ref{eq:4C-Hall5}) the weak field magnetic sensitivity depends only on the effective number of squares, $\lambda=R_{2,2}(\theta_H)/R_\mathrm{sheet}$, and all properties at larger magnetic field relate uniquely to the zero-field properties via conformal transformation---consequently, they are also identical to the properties of the disk. Thereby, we can choose $\lambda$ or $\lambda_x$, because both are related by (\ref{eq:RMFoCD1557}). In contrast to Versnel \cite{Versnel} I use $\lambda_x$, because in some cases it leads to simpler conformal mappings: the higher degree of symmetry in the current flowlines allows for splitting up the entire geometry into smaller unit cells with less vertices, and this gives fewer terms in the denominator of the Schwartz-Christoffel transformation. Morever, if two Hall-plate geometries have identical $\lambda_x$ at zero magnetic field, they will also have identical $R_x/R_\mathrm{sheet}$ at any other magnetic field. Hence, we are free to choose the magnetic field (except for $\theta_H=\pm\pi/2$ where all finite contact sizes collaps to point contacts and all resistances in the ERC become infinite). Of course, the simplest way is, to use zero magnetic field. Fig. \ref{fig:regular-4C-Hall-disk_Rx} shows how to map the unit cell of the electric potential during the measurement of $R_x$ of a disk-shaped Hall-plate from the $z$-plane to the upper half of the $t$-plane via the transformation \cite{Koppenfels}
\begin{equation}\label{apx-popularHalls10}
z = \left(\frac{1-\sqrt{t}}{1+\sqrt{t}}\right)^{\!\!1/4} .
\end{equation}
It maps the interior of the circular segment in the $z$-plane onto the upper half of the $t$-plane, with $z_\ell\to t_\ell, (\ell=1,\ldots,4)$. The point $z_4=\exp(-\mathbbm{i}\theta)$ corresponds to $t_4=-(\tan(2\theta))^2$. The bilinear mapping 
\begin{equation}\label{apx-popularHalls11}\begin{split}
& \zeta = \frac{at+b}{ct+1},\quad\text{with } b=1, \\ 
&  a=\frac{\left(\tan(\theta)\right)^{-2}-1}{2}, \;\; c=\frac{\left(\tan(\theta)\right)^{2}-1}{2}, 
\end{split}\end{equation}
maps the upper half of the $t$-plane onto the upper half of the $\zeta$-plane whereby the contacts are made symmetric with respect to the origin $\zeta=0$. Finally, the point $t=t_2=1$ is mapped to $\zeta=\zeta_2>1$ with 
\begin{equation}\label{apx-popularHalls11b}
\zeta_{2\mathrm{disk}} =\left(\tan(\theta)\right)^{-2} = \left(\tan(\pi\frac{1+\epsilon}{8})\right)^{\!\!-2} , 
\end{equation}
where we used (\ref{eq:RMFoCD1560}) for the equation on the right hand side. A Schwartz-Christoffel transformation maps the upper half of the $\zeta$-plane onto the interior of a rectangle in the $q$-plane,
\begin{equation}\label{apx-popularHalls11b}
q = C \int_{\alpha=0}^\zeta \frac{\mathrm{d}\,\alpha}{\sqrt{1-\alpha^2}\sqrt{1-\alpha^2/\zeta_{2\mathrm{disk}}^2}} .
\end{equation}
It gives the resistance 
\begin{equation}\label{apx-popularHalls12}
4 \frac{R_x(\theta_H=0)}{R_\mathrm{sheet}} = 4\lambda_x =\frac{K'}{K}\!\!\left(\left(\tan(\theta)\right)^2\right).
\end{equation}
With 
\begin{equation}\label{apx-popularHalls12b}
L(4\lambda_x) = \left(\tan(\theta)\right)^4 = \left(L(\lambda)\right)^2 
\end{equation}
from (\ref{eq:RMFoCD1557b}), it follows (see Figs. \ref{fig:eps_vs_lamda} and \ref{fig:1minus-abs_esp_vs_lamda_logarithmic})
\begin{equation}\label{apx-popularHalls12c}
\frac{R_{2,2}(\theta_H=0)}{R_\mathrm{sheet}} = \lambda =\frac{K'}{K}\!\!\left(\tan(\theta)\right) .
\end{equation}
In the subsequent paragraphs we will map the unit cells of the electric potential during the measurement of $R_x$ of other geometries of Hall-plates to the upper half of the $\zeta$-plane, whereby we require the same $\zeta_{2\mathrm{disk}}$ as in (\ref{apx-popularHalls11b}).

Occasionally, we will make reference to complementary Hall-plates, where contacts and insulating boundary segments are swapped. If a Hall-plate has normalized resistances $\lambda$ and $\lambda_x$, the complementary Hall-plate has normalized resistances $\overline{\lambda}$ and $\overline{\lambda_x}$. The unit cell in Fig. \ref{fig:regular-4C-Hall-disk_Rx_z-plane2} has a normalized resistance $2\lambda_x$. If we swap conducting and insulating segments, this unit cell has the normalized resistance $2\overline{\lambda_x}$. If we map the unit cell to the rectangle in the $q$-plane, it is obvious that the first resistance ($2\lambda_x$) is measured between two opposite edges, while the second resistance ($2\overline{\lambda_x}$) is measured between the other two opposite edges. Hence it follows 
\begin{equation}\label{apx-popularHalls15}
2\overline{\lambda_x}=\frac{1}{2\lambda_x} \quad\Rightarrow\quad \overline{\lambda_x}=\frac{1}{4\lambda_x} .
\end{equation} 
(This was also shown in (13a) of Ref.~\onlinecite{AusserlechnerVdP2}.) On the other hand, we can combine (\ref{eq:RMFoCD1505}) for $m=1$, $N=4$, (\ref{eq:4C-Hall6a}), and (\ref{eq:4C-Hall6b}) to get
\begin{equation}\label{apx-popularHalls16}
\overline{\lambda}=\frac{2}{\lambda} .
\end{equation} 
(This was also shown in (17b) of Ref.~\onlinecite{AusserlechnerSNR2017}.)

\clearpage
\subsection{Greek cross shaped Hall-plates}
\label{sec:Greek-crosses}

\begin{figure}[t]
%\vspace{1mm}
  \centering
                \includegraphics[width=0.22\textwidth]{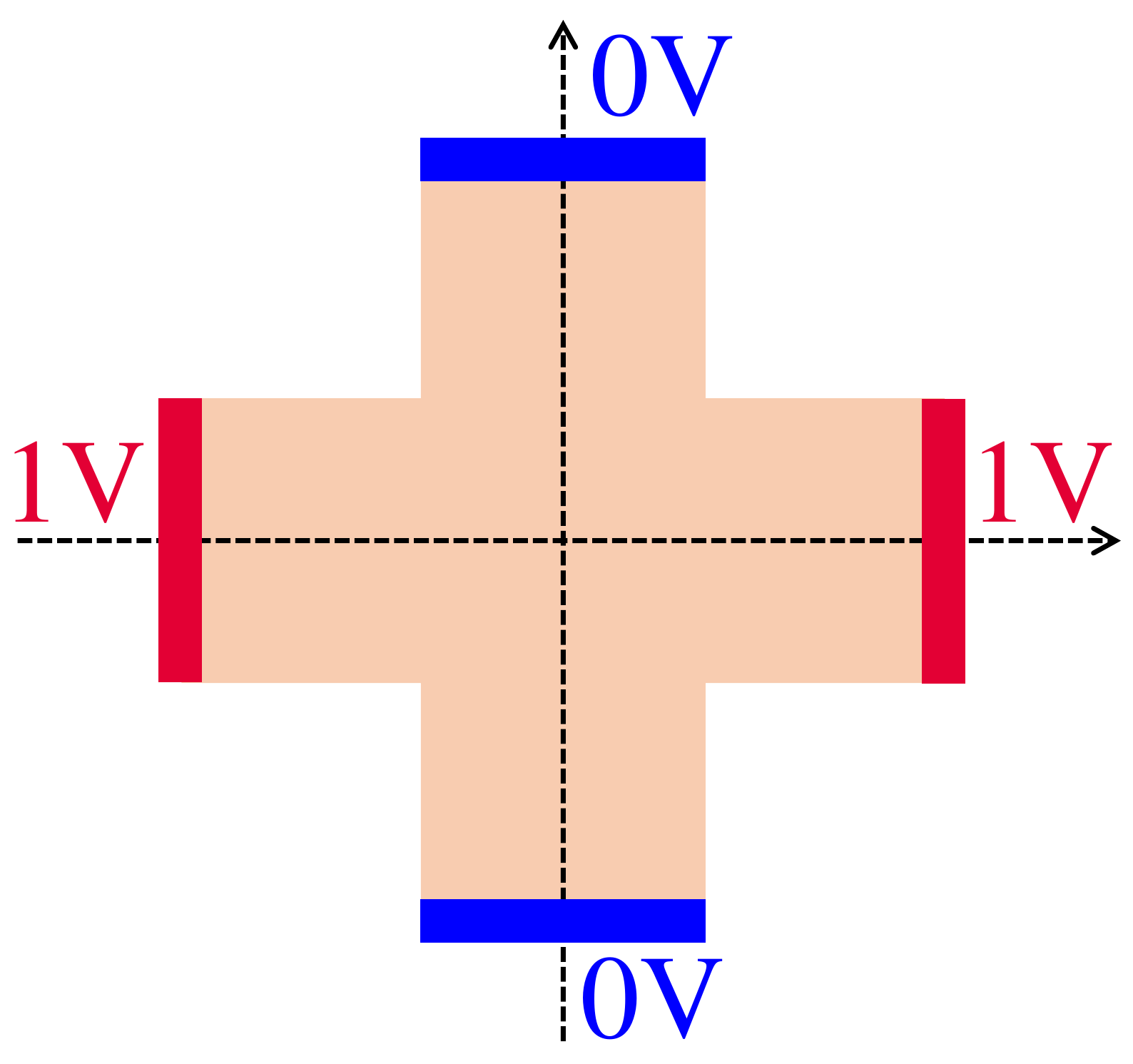}
    \caption{Greek cross with end contacts. Resistance = $R_x$. }
   \label{fig:Greek-cross_Rx_z-plane1}
\end{figure}

\begin{figure}
%\vspace{1mm}
  \centering
        \begin{subfigure}[c]{0.19\textwidth}
                \centering
                \includegraphics[width=1.0\textwidth]{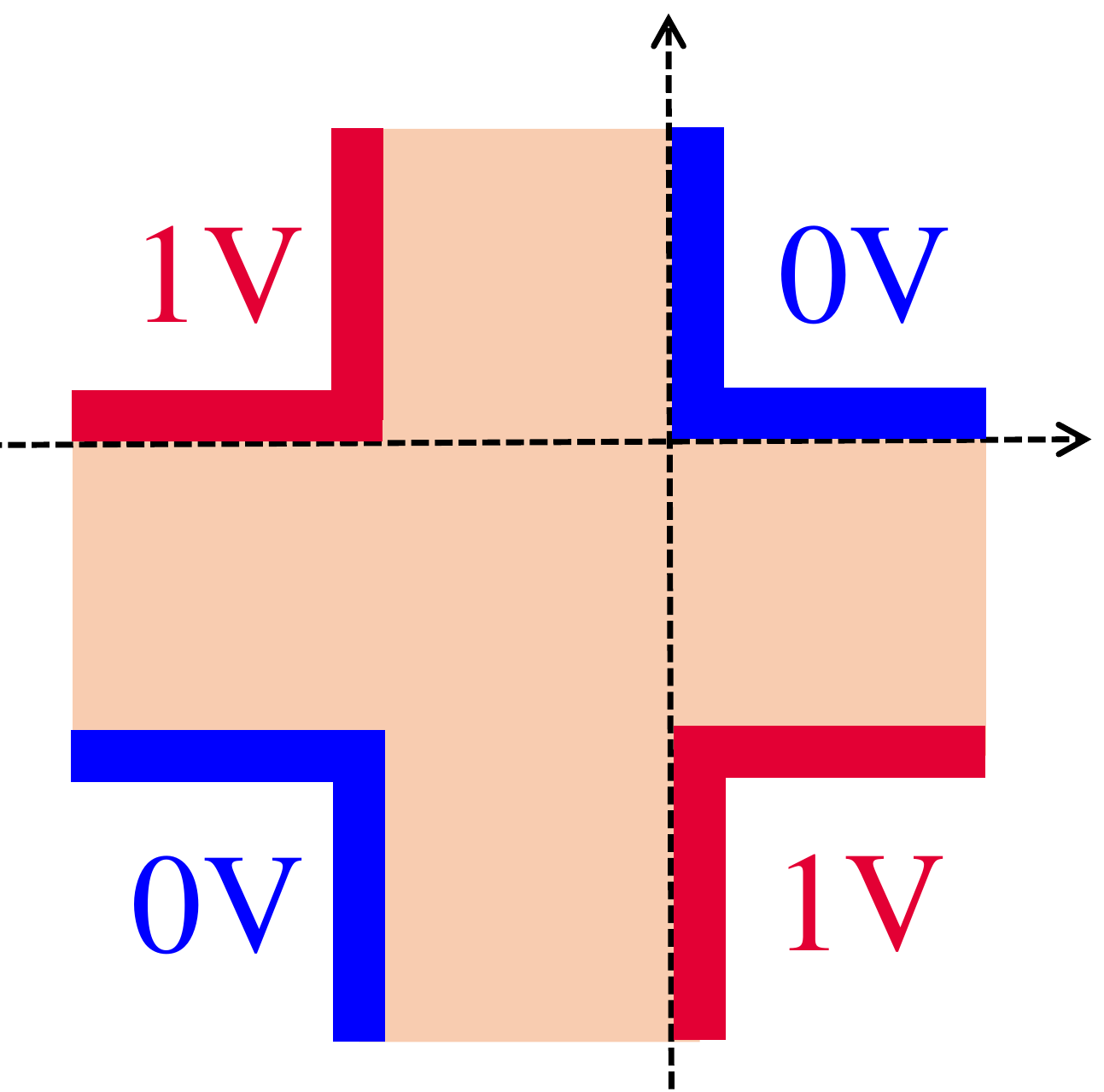}
                \caption{Greek cross with corner contacts during the measurement of $\overline{R}_x$. }
                \label{fig:inv-Greek-cross_Rxq_z-plane1}
        \end{subfigure}
	\hfill
        \begin{subfigure}[c]{0.26\textwidth}
                \centering
                \includegraphics[width=1.0\textwidth]{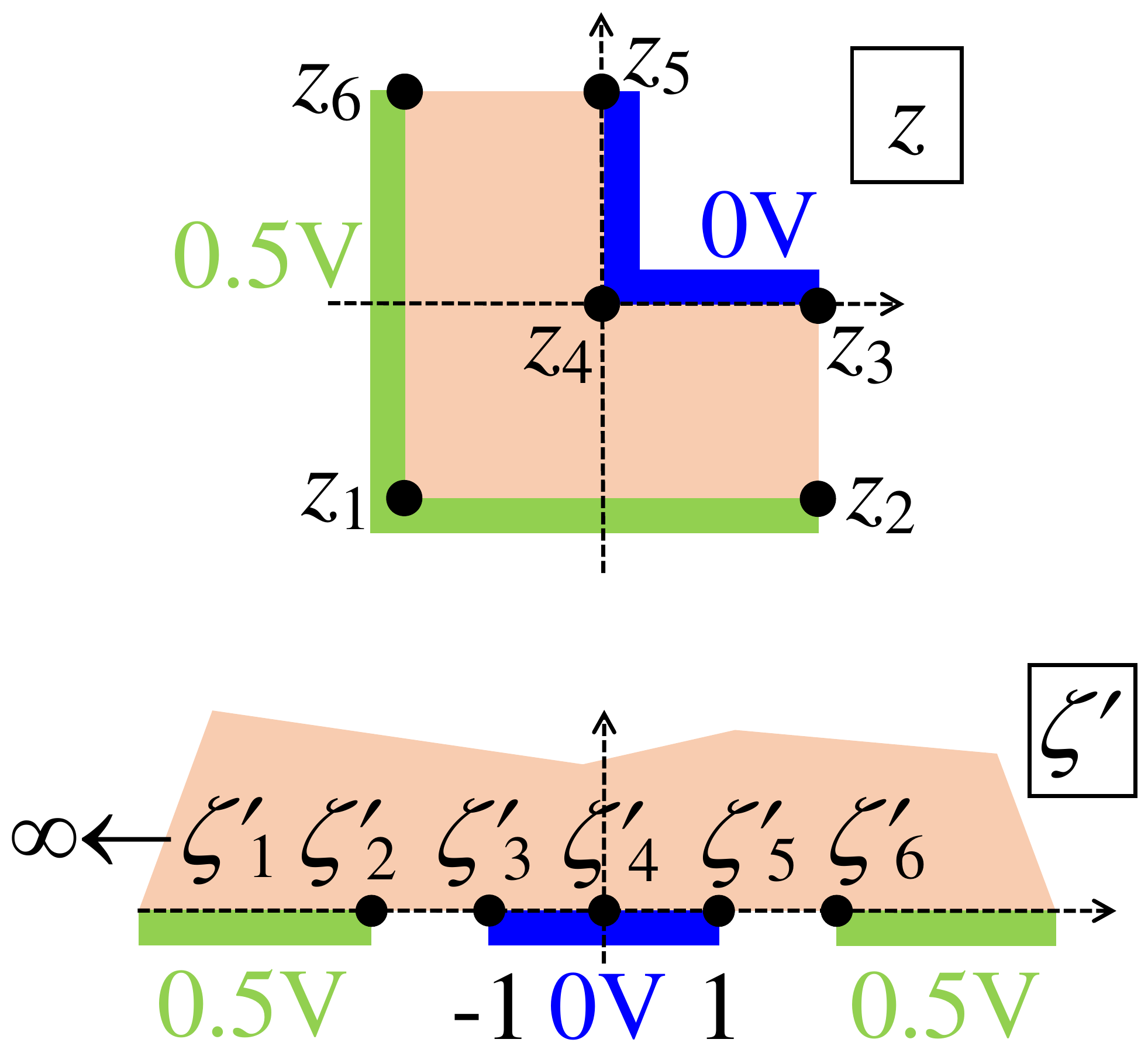}
                \caption{One quarter of Fig. \ref{fig:inv-Greek-cross_Rxq_z-plane1} and mapping to the $\zeta'$-plane. $\overline{R}_x$. }
                \label{fig:inv-Greek-cross_Rxq_z-plane2}
        \end{subfigure}
    \caption{Conformal mapping of a Greek cross with corner contacts to compute the cross resistance $\overline{R}_x$. }
   \label{fig:Greek-cross_with_corner-contacts_Rxq}
\end{figure}

A Greek cross can have end contacts as in Fig. \ref{fig:Greek-cross_Rx_z-plane1} or corner contacts as in Fig. \ref{fig:inv-Greek-cross_Rxq_z-plane1}. We call the Hall-plate with end contacts 'original device', and the Hall-plate with corner contacts 'complementary device'. Therefore, the potentials indicated in Fig. \ref{fig:Greek-cross_Rx_z-plane1} occur during the measurement of $R_x$, whereas Fig. \ref{fig:inv-Greek-cross_Rxq_z-plane1} shows the potentials at the contacts during a measurement of $\overline{R}_x$. A quarter of the complementary device in the $z$-plane is shown in Fig. \ref{fig:inv-Greek-cross_Rxq_z-plane2}. It is mapped to the upper half of the $\zeta'$-plane in the lower part of Fig. \ref{fig:inv-Greek-cross_Rxq_z-plane2}. The resistance between the two contacts in Fig. \ref{fig:inv-Greek-cross_Rxq_z-plane2} is also $\overline{R}_x$. The mapping is  
\begin{equation}\label{apx-Greek-cross1}
z = C' \!\!\int_0^{\zeta'}\!\! \frac{\sqrt{\zeta'}\mathrm{d}\,\zeta'}{\sqrt{{\zeta'}^2-1}\sqrt{{\zeta'}^2-{\zeta'_6}^2}} .
\end{equation}
The shape in the upper part of Fig. \ref{fig:inv-Greek-cross_Rxq_z-plane2} is made up of two unit cells. I chose this shape, because its symmetry leads to $\zeta'_1\to\infty$, which simplifies the transformation formula in (\ref{apx-Greek-cross1}). A comparison of the lower part of Fig. \ref{fig:inv-Greek-cross_Rxq_z-plane2} with Fig. \ref{fig:regular-4C-Hall-disk_Rx_t-and-zeta-plane} gives 
\begin{equation}\label{apx-Greek-cross2}
\overline{\lambda}_x = \frac{1}{2}\,\frac{K'}{K}\!\left(\frac{1}{\zeta'_6}\right) .
\end{equation}
With (\ref{apx-popularHalls15}) we get 
\begin{equation}\label{apx-Greek-cross3}
4\lambda_x = 2\,\frac{K}{K'}\!\!\left(\!\frac{1}{\zeta'_6}\right) = \frac{K}{K'}\!\!\left(\!\frac{\frac{2}{\sqrt{\zeta'_6}}}{1\!+\!\frac{1}{\zeta'_6}}\right) = \frac{K'}{K}\!\!\left(\!\frac{\zeta'_6\!-\!1}{\zeta'_6\!+\!1}\right) ,
\end{equation}
wherein we used (27) from Ref.~\onlinecite{ELEN}. Comparing (\ref{apx-Greek-cross3}) with (\ref{apx-popularHalls12}) specifies the Greek cross with corner contacts in terms of an equivalent disk-shaped Hall-plate 
\begin{equation}\label{apx-Greek-cross4}
\frac{\zeta'_6-1}{\zeta'_6+1} = \left(\!\tan\!\left(\frac{\pi}{4}\!-\!\theta\right)\!\right)^{\!\!2} \;\;\Rightarrow\;\; \frac{1}{\zeta'_6} = \cos\!\left(\frac{\pi}{2}\!-\!2\theta\right) .
\end{equation}
Thereby we took into account that the complementary device (with $4\overline{\lambda}_x$) corresponds to a disk with contacts size $2\theta$, and therefore the original device (with $4\lambda_x$) corresponds to a disk with contacts size $\pi/2-2\theta$. We define the filling factor $f_{CC}$ as the sum of the lengths of all contacts in Fig. \ref{fig:inv-Greek-cross_Rxq_z-plane1} over the entire perimeter. Then it holds 
\begin{equation}\label{apx-Greek-cross5}\begin{split}
f_{CC} & = \frac{2|z_3-z_4|}{2|z_3-z_4|+2|z_2-z_3|} \\
\Rightarrow \frac{1}{f_{CC}} & = 1+\left|\frac{\int_1^{\zeta'_6} \frac{\sqrt{\zeta'}\mathrm{d}\zeta'}{\sqrt{{\zeta'}^2-1}\sqrt{{\zeta'}^2-{\zeta'_6}^2}}}{\int_0^1 \frac{\sqrt{\zeta'}\mathrm{d}\zeta'}{\sqrt{{\zeta'}^2-1}\sqrt{{\zeta'}^2-{\zeta'_6}^2}}} \right| .
\end{split}\end{equation}
The integrals are solvable with Mathematica in terms of Gauss' hyper-geometric function, and we use the identity \cite{hypergeo} 
\begin{equation}\label{apx-Greek-cross6}\begin{split}
& {_2}\!F_1\!\left(\frac{-1}{2},\frac{3}{4},\frac{1}{4},z\right) +(z-1)\;{_2}\!F_1\!\left(\frac{1}{2},\frac{3}{4},\frac{1}{4},z\right) \\
& = 2z\;{_2}\!F_1\!\left(\frac{1}{2},\frac{3}{4},\frac{5}{4},z\right) .
\end{split}\end{equation}
The result is 
\begin{equation}\label{apx-Greek-cross7}\begin{split}
f_{CC} & = \frac{4\pi\Gamma(3/4)}{\left(\Gamma(1/4)\right)^3} \sqrt{2\cos\left(\pi\frac{1-\epsilon}{4}\right)} \\ 
& \quad *\frac{{_2}\!F_1\!\left(\frac{1}{2},\frac{3}{4},\frac{5}{4},\left(\cos\left(\pi\frac{1-\epsilon}{4}\right)\right)^2\right)}{{_2}\!F_1\!\left(\frac{1}{4},\frac{1}{2},\frac{3}{4},\left(\cos\left(\pi\frac{1-\epsilon}{4}\right)\right)^2\right)} .
\end{split}\end{equation}
In (\ref{apx-Greek-cross7}) a disk with contacts size $\epsilon$ is equivalent to a Greek cross with corner contacts having the filling factor $f_{CC}(\epsilon)$. The complementary device is a disk with contacts size $-\epsilon$, which is equivalent to a Greek cross with end contacts having a filling factor $1-f_{CC}(\epsilon)$. Hence, a disk with contacts size $\epsilon$ corresponds to a Greek cross with end contacts with a filling factor $f_{CE}(\epsilon)=1-f_{CC}(-\epsilon)$, which means 
\begin{equation}\label{apx-Greek-cross8}\begin{split}
f_{CE} & = 1-\frac{4\pi\Gamma(3/4)}{\left(\Gamma(1/4)\right)^3} \sqrt{2\cos\left(\pi\frac{1+\epsilon}{4}\right)} \\ 
& \qquad\;\; *\frac{{_2}\!F_1\!\left(\frac{1}{2},\frac{3}{4},\frac{5}{4},\left(\cos\left(\pi\frac{1+\epsilon}{4}\right)\right)^2\right)}{{_2}\!F_1\!\left(\frac{1}{4},\frac{1}{2},\frac{3}{4},\left(\cos\left(\pi\frac{1+\epsilon}{4}\right)\right)^2\right)} .
\end{split}\end{equation}

%At the same time, the complementary disk has contacts size $-\epsilon$, and it corresponds to the complementary Greek-cross, which is the very same Greek-cross, yet with corner contacts instead of end contacts. Hence, if we define $f_\mathrm{CC}$ as the ratio of the length of all four contacts of a Greek-cross with corner contacts from  Fig. \ref{fig:inv-Greek-cross_Rx_z-plane1} over the length of the entire boundary, it holds 
%\begin{equation}\label{apx-Greek-cross14}\begin{split}
%f_\mathrm{CC} & = 1+\frac{1}{\sqrt{\overline{\omega}}}\;\frac{\Im\{{_2}F_1\left(1/4,1/2,3/4,\overline{\omega}^{\;-2}\right)\}}{{_2}F_1\left(1/4,1/2,3/4,\overline{\omega}^{\;2}\right)} \\
%& \text{with } \overline{\omega} = \omega(-\epsilon) = \cos\left(\pi\frac{1-\epsilon}{4}\right) .
%\end{split}\end{equation}

\begin{figure*}[t]
%\vspace{1mm}
  \centering
                \includegraphics[width=0.98\textwidth]{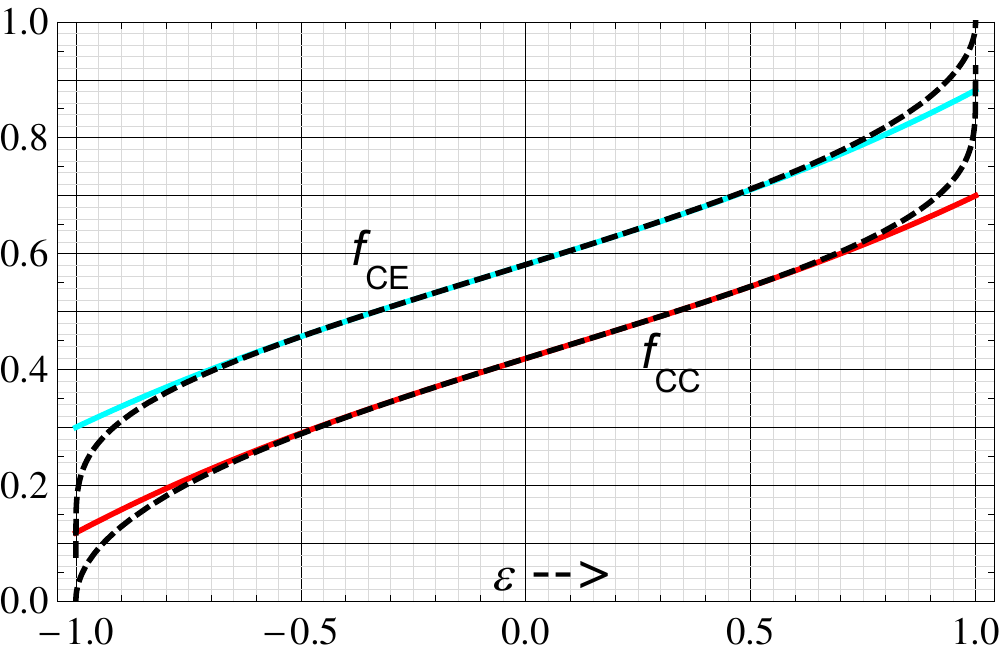}
    \caption{Plot of filling factors of $90$° symmetric Greek cross Hall-plates with sharp corners ($f_{CE}$ for end contacts, $f_{CC}$ for angled corner contacts), versus the size of contacts of equivalent disk-shaped Hall-plates, $\epsilon$. The black dashed curves are the exact filling factors from (\ref{apx-Greek-cross7}), (\ref{apx-Greek-cross8}). The solid colored curves are the approximations from (\ref{apx-Greek-cross15}). }
   \label{fig:fCC_fCE_vs_eps}
\end{figure*}

For medium-sized contacts we can develop (\ref{apx-Greek-cross7}) and (\ref{apx-Greek-cross8}) into Taylor series in $\epsilon$, 
\begin{equation}\label{apx-Greek-cross15}\begin{split}
f_\mathrm{CE} & \approx 0.581+0.241\epsilon+0.01\epsilon^2+0.05\epsilon^3 , \\
f_\mathrm{CC} & \approx 0.419+0.241\epsilon-0.01\epsilon^2+0.05\epsilon^3 .
\end{split}\end{equation}
The relative errors of both approximations in (\ref{apx-Greek-cross15}) are less than $1\%$ for $-0.6<\epsilon<0.7$. Fig. \ref{fig:fCC_fCE_vs_eps} shows a plot of (\ref{apx-Greek-cross7}), (\ref{apx-Greek-cross8}), and (\ref{apx-Greek-cross15}).

\begin{figure*}
%\vspace{1mm}
  \centering
        \begin{subfigure}[t]{0.63\textwidth}
                \centering
                \includegraphics[width=1.0\textwidth]{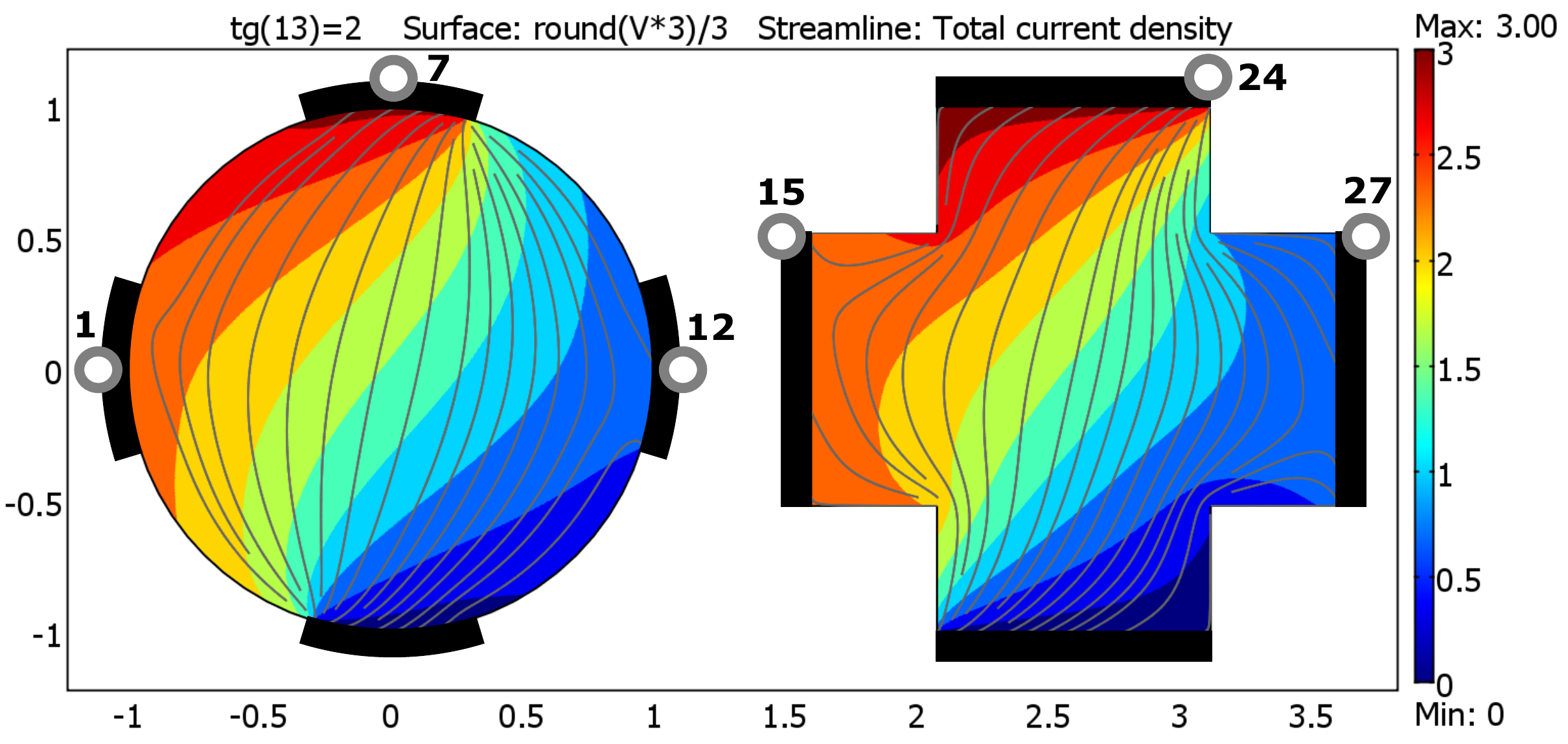}
                \caption{Current streamlines and potential at $\tan(\theta_H)=2$ in a disk-shaped Hall-plate with contacts size of $\epsilon=-1/4$ and an equivalent Greek cross with end contacts with the filling factor $f_{CE}=0.520614$. }
                \label{fig:streamlines_tg_eq_2_eps_eq_m250m_Greek-crosses1}
        \end{subfigure}
        \hfill  % An dieser Stelle kann ein zusätzlicher Zwischenraum eingebunden werden: ~, \quad, \qquad, \hfill usw.
          % Eine leere Zeile erzwingt, dass die zweite Grafik darunter erscheint.
        \begin{subfigure}[t]{0.36\textwidth}
                \centering
                \includegraphics[width=1.0\textwidth]{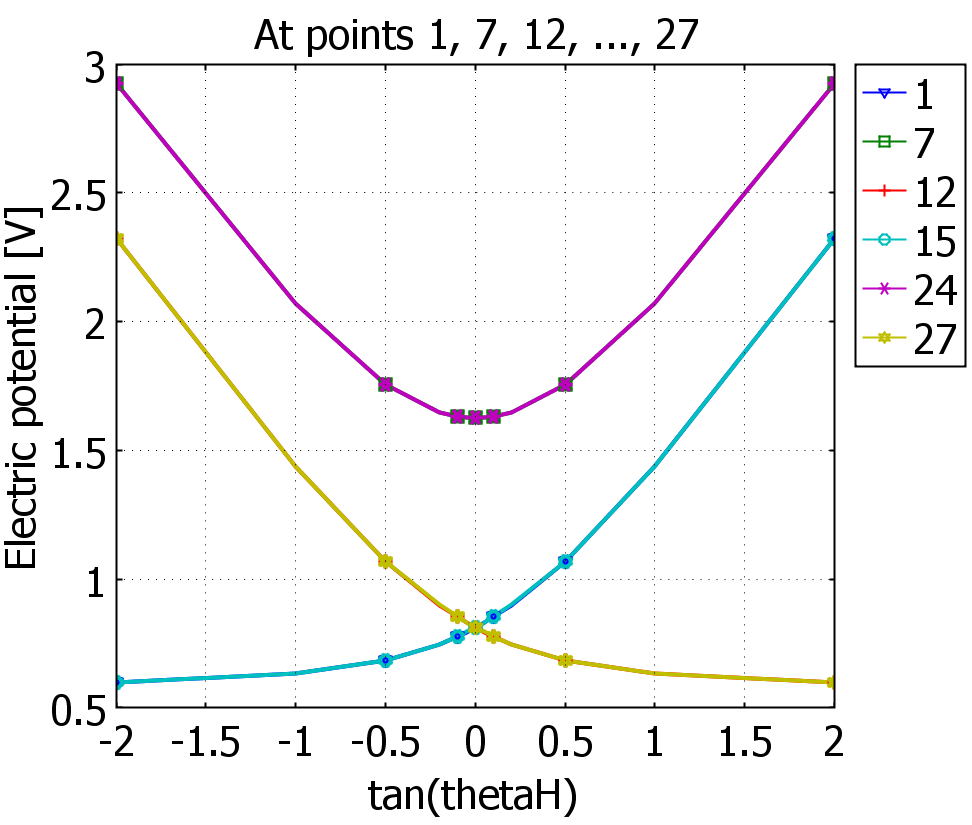}
                \caption{Potentials at the contacts versus $\tan(\theta_H)$ for the Hall-plates of Fig. \ref{fig:streamlines_tg_eq_2_eps_eq_m250m_Greek-crosses1}. These pairs of curves are nearly identical: 1 - 15, 7 -24, and 12 - 27. }
                \label{fig:eps_eq_m250m_Greek-crosses1}
        \end{subfigure}
	 \\
        \begin{subfigure}[t]{0.63\textwidth}
                \centering
                \includegraphics[width=1.0\textwidth]{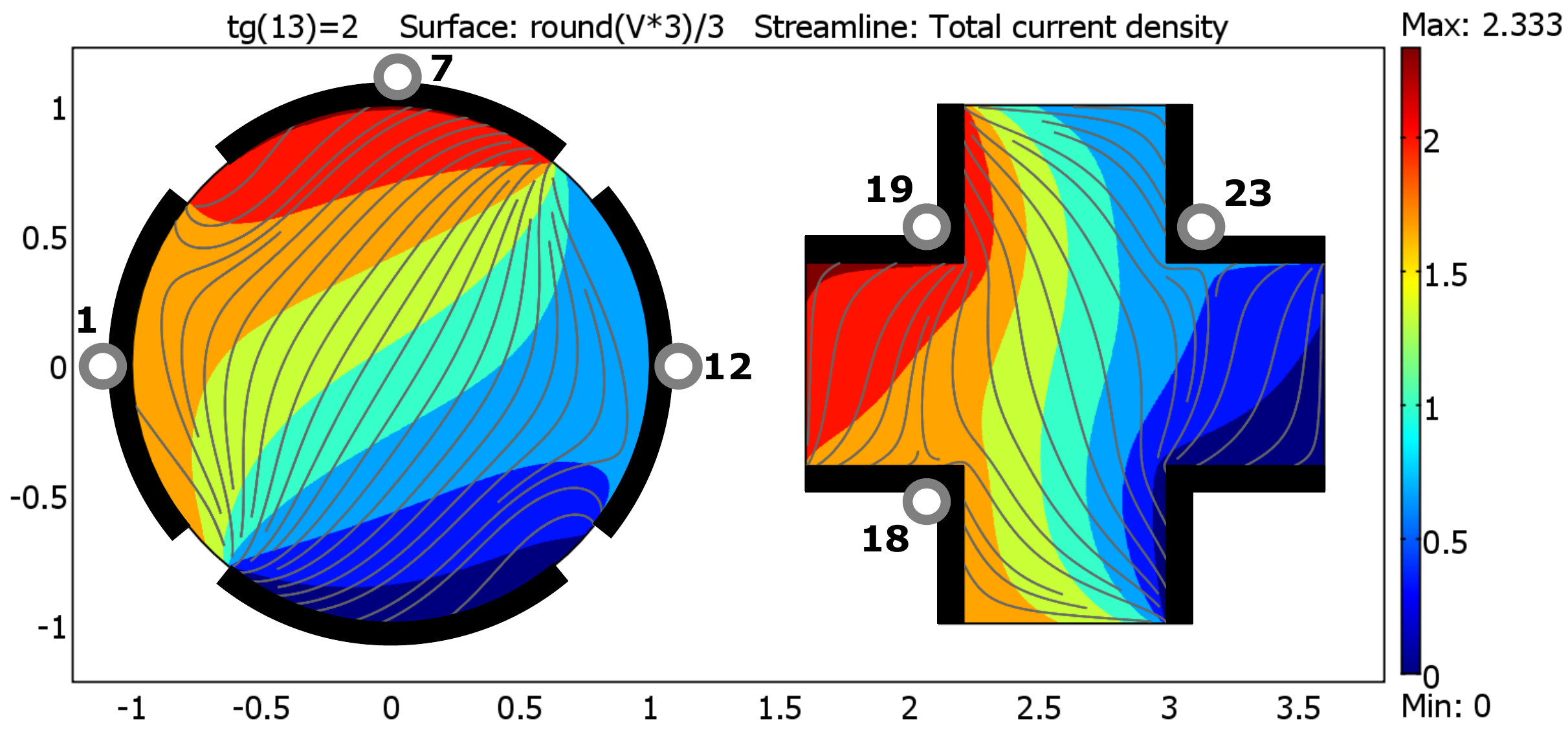}
                \caption{Current streamlines and potential at $\tan(\theta_H)=2$ in a disk-shaped Hall-plate with contacts size of $\epsilon=5/7$ and an equivalent Greek cross with corner contacts with the filling factor $f_{CC}=0.608121$. }
                \label{fig:streamlines_tg_eq_2_eps_eq_5durch7_inv-Greek-crosses6}
        \end{subfigure}
        \hfill  
        \begin{subfigure}[t]{0.36\textwidth}
                \centering
                \includegraphics[width=1.0\textwidth]{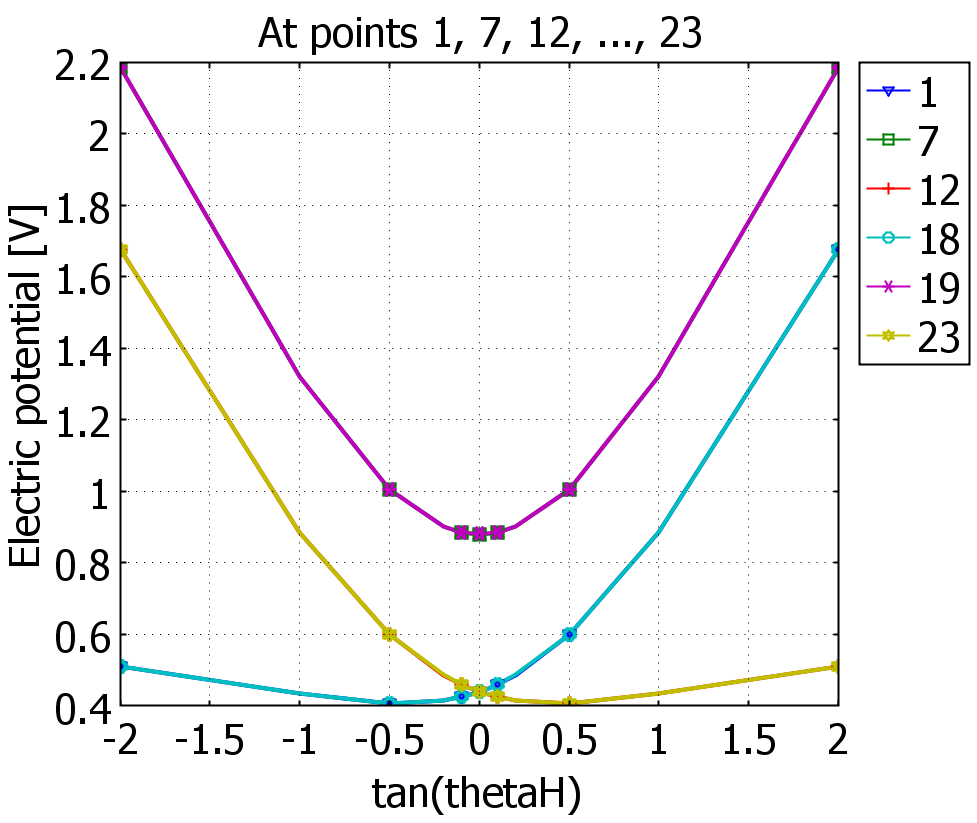}
                \caption{Potentials at the contacts versus $\tan(\theta_H)$ for the Hall-plates of Fig. \ref{fig:streamlines_tg_eq_2_eps_eq_5durch7_inv-Greek-crosses6}. These pairs of curves are nearly identical: 1 - 18, 7 -19, and 12 - 23. }
                \label{fig:eps_eq_5durch7_inv-Greek-crosses6}
        \end{subfigure}
    \caption{Results of FEM-simulations of Hall-plates with the shapes of disks and Greek crosses. Contacts without any labelled circle are grounded. A current of $1$ A enters contacts comprising circles 7, 24, and 19. All of them are diametrically opposite to the grounded contacts. $R_\mathrm{sheet}=1\;\Omega$. The potentials at corresponding points differ by up to  500 ppm due to numerical inaccuracies. }
   \label{fig:Greek-crosses-checks}
\end{figure*}

I made a numerical check with finite element simulation by the commercial code from COMSOL Multiphysics. There I used the application mode emdc (= conductive media dc) in two dimensions with Lagrange multipliers (weak constraints: on, constraint type: non-ideal). The Lagrange multipliers are necessary to increase the accuracy of current integration near the singular points at the vertices of the contacts. The mesh had approximately $7\times 10^5$ elements per Hall-plate. The sheet resistance was $R_\mathrm{sheet}=1\;\Omega$, and the conductivity tensor was 
\begin{equation}\label{apx-Greek-cross17}
\bm{\kappa} =  \frac{1}{1+(\tan(\theta_H))^2}\;\left(\begin{array}{cc} 1&\tan(\theta_H) \\ -\tan(\theta_H)&1 \end{array} \right) .
\end{equation}
Disk-shaped Hall-plates and Greek crosses with end contacts and with corner contacts were modelled for three cases, 
\begin{equation}\label{apx-Greek-cross19}
\begin{array}{c|cccc} \mathrm{case} & \epsilon & \theta & f_\mathrm{CE} & f_\mathrm{CC} \\ \hline
1 & -1/4 & 3\pi/32 & 0.520614 & 0.357187 \\
2 & 0 & \pi/8 & 0.580980 & 0.419020 \\
3 & 5/7 & 3\pi/14 & 0.782991 & 0.608121 \end{array}
%\begin{split}
%\text{1st case: } & \epsilon = \frac{-1}{4} \;\;\leftrightarrow\;\; \theta = \frac{3\pi}{32} \;\;\leftrightarrow\;\; \\ & f_\mathrm{CE} = 0.520614 \;\;\leftrightarrow\;\; f_\mathrm{CC} = 0.357187, \\
%\text{2nd case: } & \epsilon = 0 \;\;\leftrightarrow\;\; \theta = \frac{\pi}{8} \;\;\leftrightarrow\;\; \\ & f_\mathrm{CE} = 0.58098 \;\;\leftrightarrow\;\; f_\mathrm{CC} = 0.41902, \\
%\text{3rd case: } & \epsilon = \frac{5}{7} \;\;\leftrightarrow\;\; \theta = \frac{3\pi}{14} \;\;\leftrightarrow\;\; \\ & f_\mathrm{CE} = 0.782991 \;\;\leftrightarrow\;\; f_\mathrm{CC} = 0.608121.
%\end{split}
\end{equation}
A current of $1$ A was injected into contact $C_2$, while $C_0=C_4$ was grounded, and the other contacts $C_1, C_3$ were not connected. The potentials at $C_1,C_2,C_3$ were computed for $-2\le\tan(\theta_H)\le 2$. The potentials of the disks and the Greek crosses differed by less than $500$ ppm (see the plots in Figs. \ref{fig:eps_eq_m250m_Greek-crosses1} and \ref{fig:eps_eq_5durch7_inv-Greek-crosses6} and the potentials and the current streamlines in Figs. \ref{fig:streamlines_tg_eq_2_eps_eq_m250m_Greek-crosses1} and \ref{fig:streamlines_tg_eq_2_eps_eq_5durch7_inv-Greek-crosses6} for cases 1 and 3). The small differences seem to be caused by insufficiently fine meshing, because coarser meshes give larger differences. Nevertheless, this check is a strong indication for the correctness of (\ref{apx-Greek-cross7}) and (\ref{apx-Greek-cross8}).

\clearpage
\subsection{Greek cross with rounded corners}
\label{sec:Greek-rounded}

\begin{figure}
%\vspace{1mm}
  \centering
        \begin{subfigure}[c]{0.21\textwidth}
                \centering
                \includegraphics[width=1.0\textwidth]{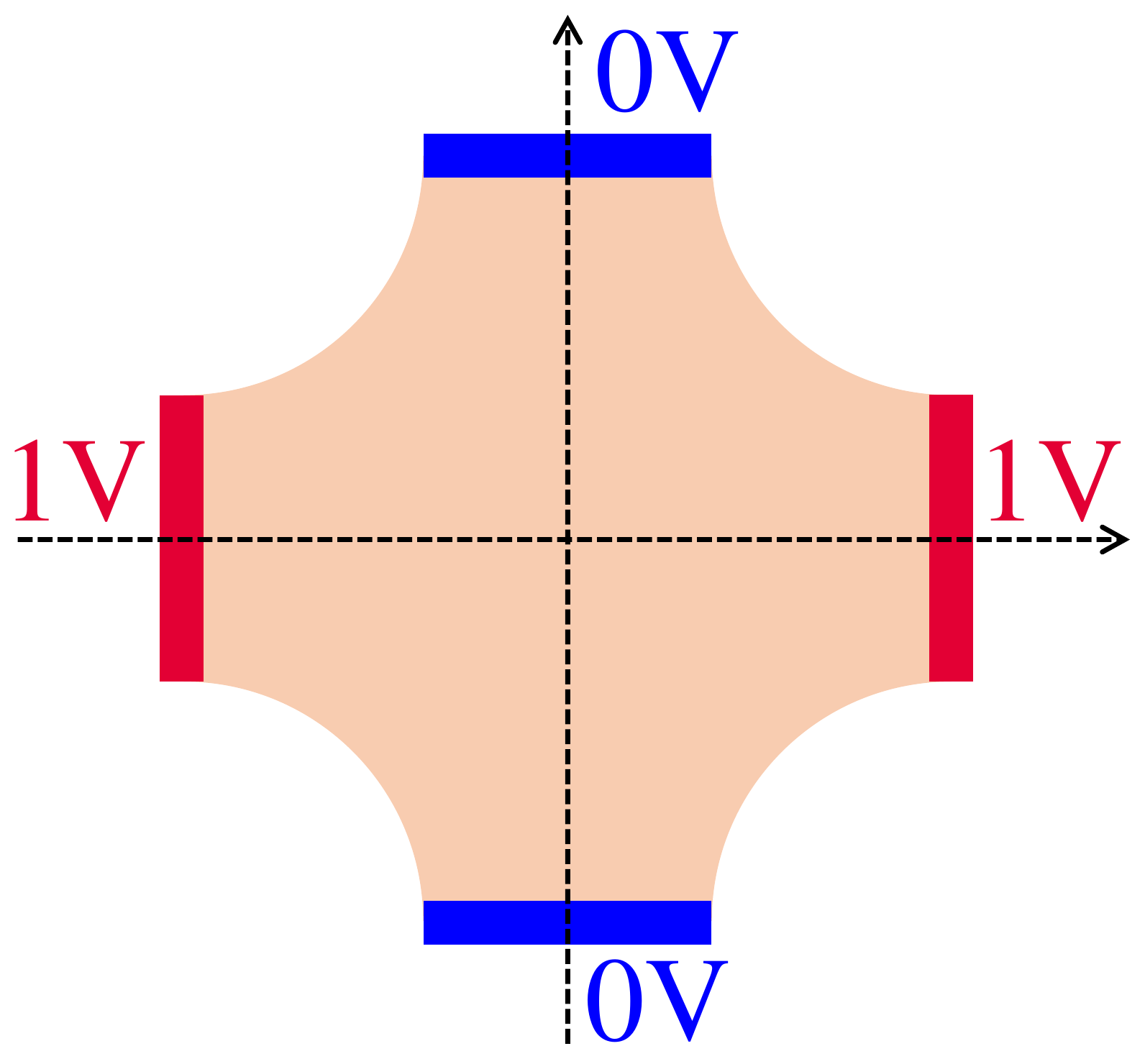}
                \caption{Complete Hall-plate in the $z$-plane. Resistance $=R_x$. }
                \label{fig:rounded-Greek-cross_Rx_z-plane1}
        \end{subfigure}
	\hfill
        \begin{subfigure}[c]{0.19\textwidth}
                \centering
                \includegraphics[width=1.0\textwidth]{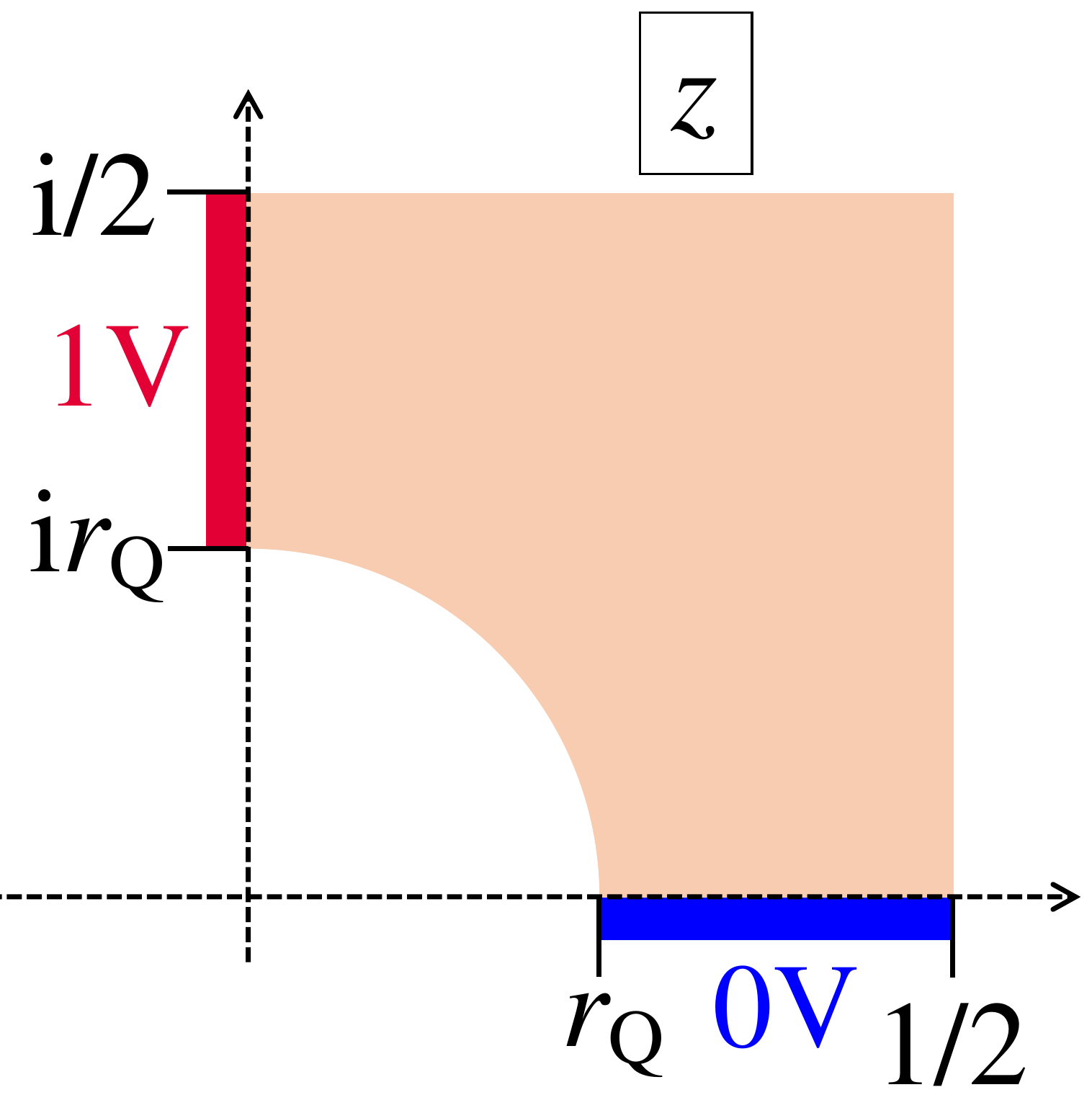}
                \caption{A quarter of Fig. \ref{fig:rounded-Greek-cross_Rx_z-plane1}. Resistance $=4R_x$. }
                \label{fig:rounded-Greek-cross_Rx_z-plane2}
        \end{subfigure}

        \begin{subfigure}[c]{0.21\textwidth}
                \centering
                \includegraphics[width=1.0\textwidth]{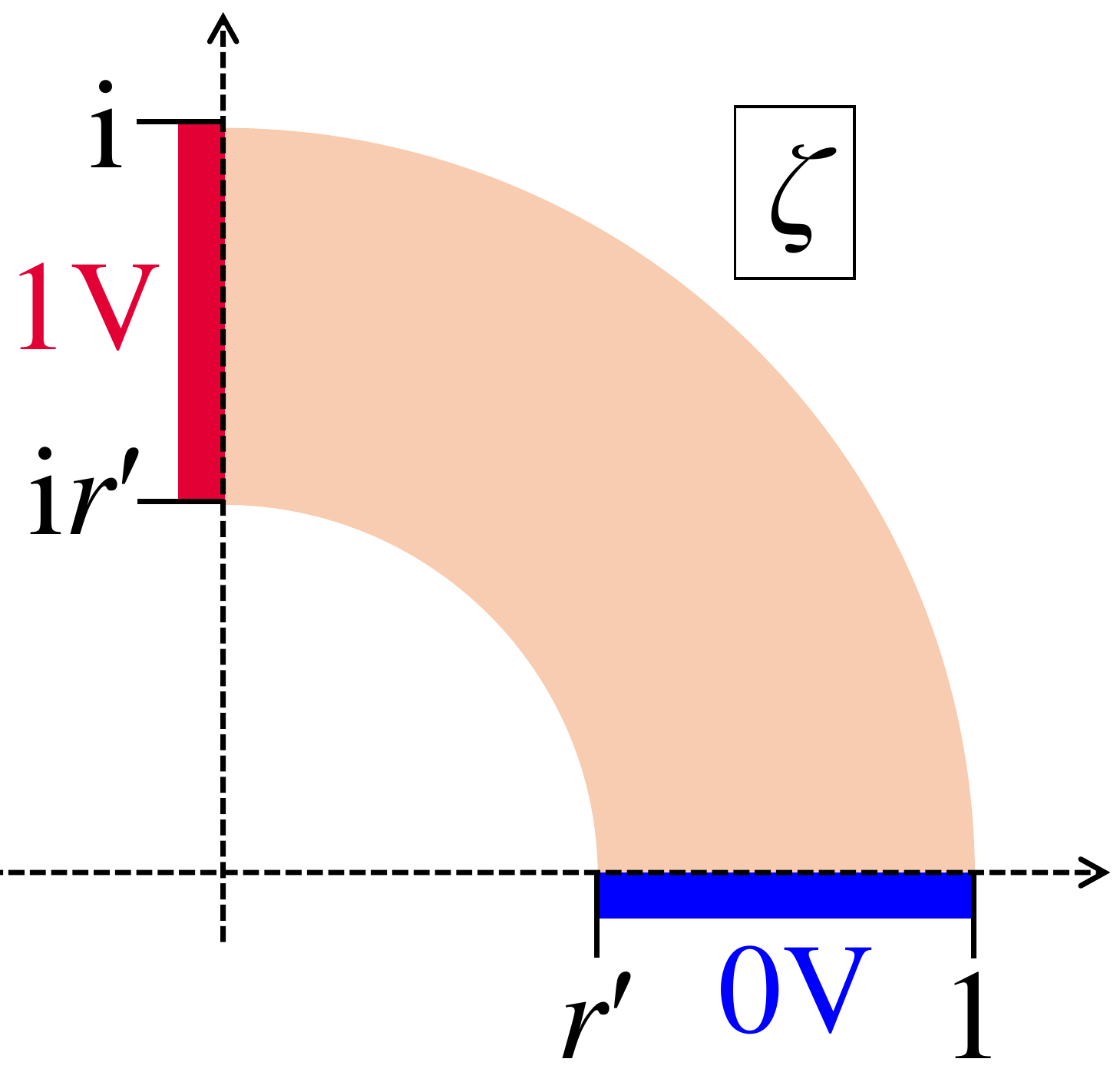}
                \caption{Fig. \ref{fig:rounded-Greek-cross_Rx_z-plane2} mapped to the $\zeta$-plane. Resistance $=4R_x$. }
                \label{fig:rounded-Greek-cross_Rx_zeta-plane}
        \end{subfigure}
	\hfill
        \begin{subfigure}[c]{0.15\textwidth}
                \centering
                \includegraphics[width=1.0\textwidth]{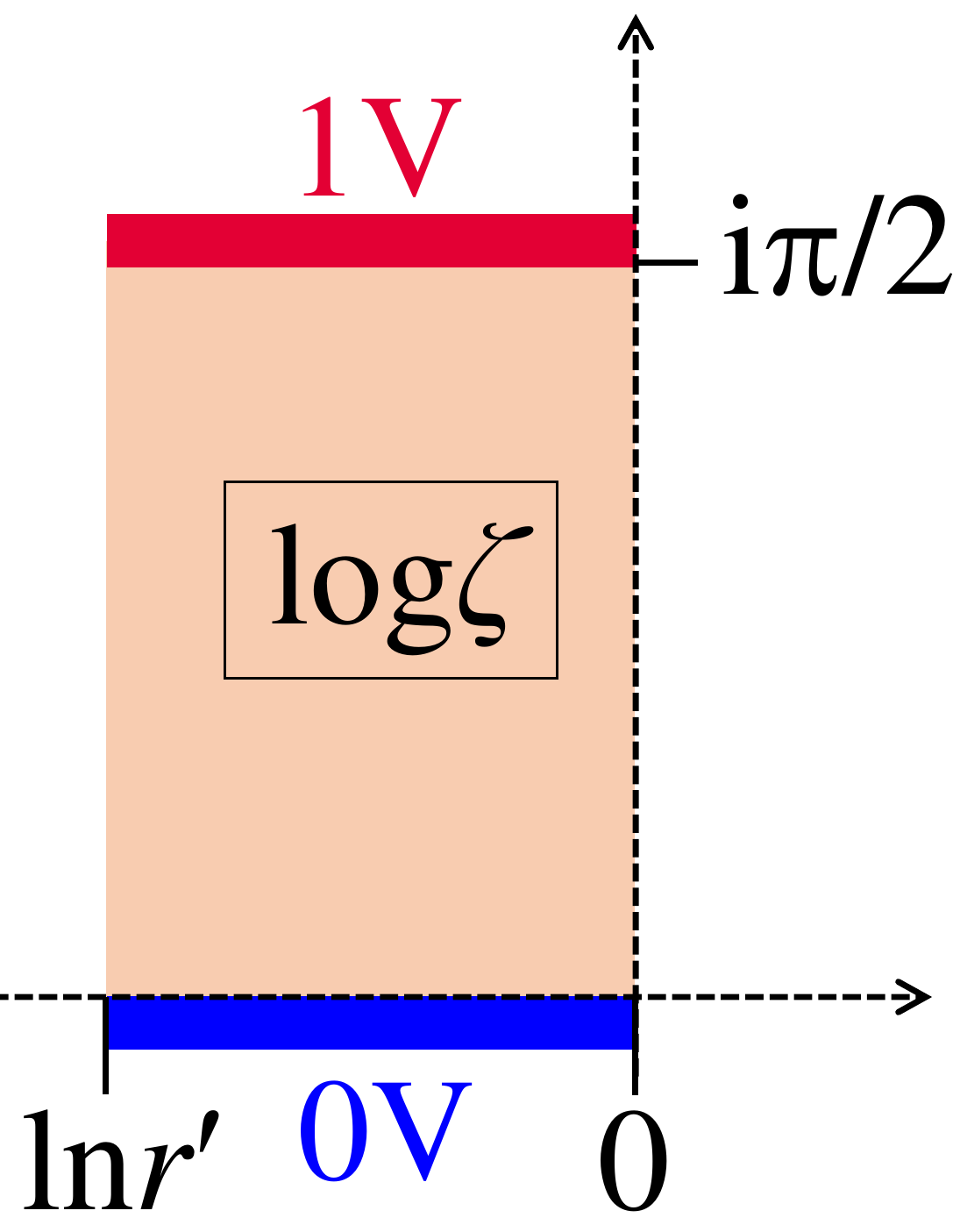}
                \caption{Fig. \ref{fig:rounded-Greek-cross_Rx_zeta-plane} mapped to the $\log(\zeta)$-plane. Resistance  $=4R_x$. }
                \label{fig:rounded-Greek-cross_Rx_log-zeta-plane}
        \end{subfigure}
    \caption{Conformal mapping of a rounded Greek cross with end contacts to compute the approximate cross resistance $R_x$. }
   \label{fig:Greek-cross_with_end-contacts_Rx}
\end{figure}

Fig. \ref{fig:rounded-Greek-cross_Rx_z-plane1} shows a Greek cross with rounded corners and end contacts. Compared to the Greek crosses from Appendix \ref{sec:Greek-crosses} the inner corners are rounded off by quarter circles. The shape is obtained by taking a square with edge length $1$ and cutting off quarter disks of radius $r_Q$ which are centered at the corners. I call $r_Q$ the \emph{normalized} quarter circle radius, because the edge length is $1$. This geometry is one of the best for Hall-plates, because it avoids singularities of the electric field near the corners. Unfortunately, there is no simple conformal transformation of this shape to the upper half of the $\zeta$-plane. Note that the general  transformation from these so-called circular arc polygonal domains (CAPDs) or curvilinear polygons is known \cite{NehariCAPD,Bauer}, but it is difficult to find the involved parameters. Alternatively, see Refs.~\onlinecite{Pergamentseva1,Pergamentseva2}. %We define $c_{CSE}$ as the ratio of the sum of lengths of all straight segments (= the contacts) over the entire boundary ('CSE' stands for Cross with Smooth corners and End contacts).
Yet, for small and moderate $r_Q$ one can use the following approximation. 

\begin{figure}
%\vspace{1mm}
  \centering
        \begin{subfigure}[c]{0.19\textwidth}
                \centering
                \includegraphics[width=1.0\textwidth]{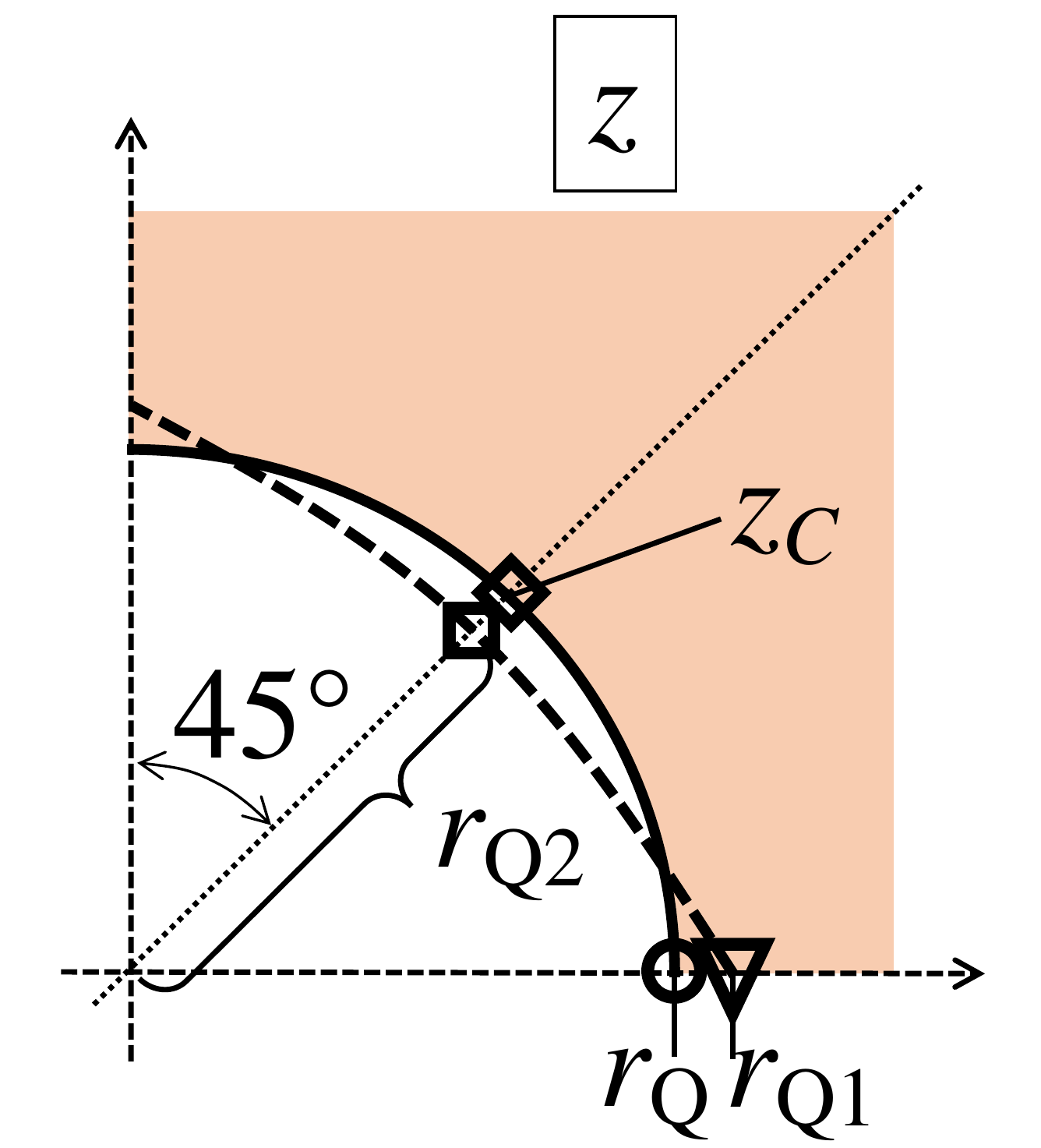}
                \caption{In the $z$-plane. }
                \label{fig:rounded-Greek-cross_approx_z-plane}
        \end{subfigure}
	\hfill
        \begin{subfigure}[c]{0.19\textwidth}
                \centering
                \includegraphics[width=1.0\textwidth]{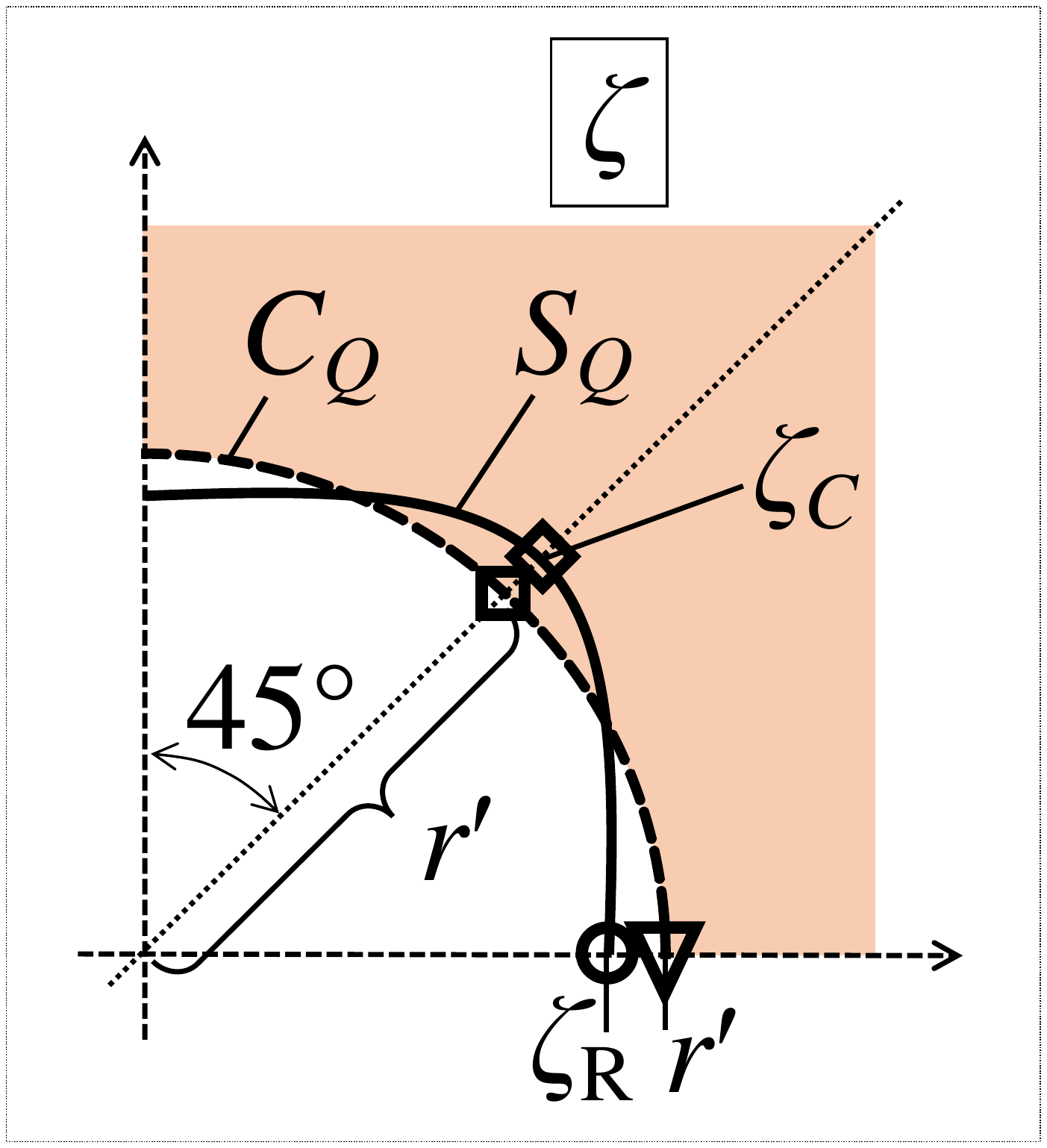}
                \caption{In the $\zeta$-plane. }
                \label{fig:rounded-Greek-cross_approx_zeta-plane}
        \end{subfigure}
    \caption{Approximation of the quarter circles in (\ref{apx-Greek-rounded3}). Corresponding points are labelled with identical symbols ($\square,\bigcirc,\Diamond,\bigtriangledown$). }
   \label{fig:rounded-Greek-cross_approx}
\end{figure}

During the measurement of $R_x$ a quarter of the Hall-plate is shown in Fig. \ref{fig:rounded-Greek-cross_Rx_z-plane2}. It has the resistance $4R_x$ between the contacts on the real and imaginary axes. We map the circumscribing square from the $z$-plane to the unit disk in the $\zeta$-plane in Fig. \ref{fig:rounded-Greek-cross_Rx_zeta-plane} via the conformal transformation 
\begin{equation}\label{apx-Greek-rounded1}
z = A' \int_0^\zeta \frac{\mathrm{d}t}{\sqrt{1+t^4}} . 
\end{equation}
This formula is given in Refs.~\onlinecite{Durand,Betz} (it is a simple combination of two mappings, one from the square to the upper half plane, and the second from the disk to the upper half plane). The constant $A'$ follows from $z=1/2\mapsto\zeta=1$, 
\begin{equation}\label{apx-Greek-rounded2}
\frac{1}{2} = A' \!\int_0^1 \!\!\frac{\mathrm{d}t}{\sqrt{1\!+\!t^4}} \;\Rightarrow\;  A'=\frac{4\sqrt{\pi}}{(\Gamma(1/4))^2} \approx \frac{1.0787}{2} .
\end{equation}
This transformation mapps the circle $|z|=r_Q$ onto a segment $S_Q$ in the $\zeta$-plane, which deviates slightly from a quarter circle (see Fig. \ref{fig:rounded-Greek-cross_approx}): the distance from a point on this segment to the origin is smallest, if the point is on the real or imaginary $\zeta$-axis, $\zeta=\zeta_R$, and it increases a bit, when the point moves on the segment $S_Q$ towards its center $\zeta_C=|\zeta_C|\exp(\mathbbm{i}\pi/4)$. Thus, the segment is obtained by streching a quarter circle slightly in the direction $\zeta=1+\mathbbm{i}$. Our approximation replaces the segment $S_Q$ by a quarter circle $C_Q$ with a radius $r'$, which is in-between these two extremes, $\zeta_R<r'<|\zeta_C|$. If we map $\zeta_=r'$ to the $z$-plane, it will give $z=r_{Q1}>r_Q$, and if we map $\zeta=r'\exp(\mathbbm{i}\pi/4)$ to the $z$-plane, it will give $z=r_{Q2}\exp(\mathbbm{i}\pi/4)$ with $r_{Q2}<r_Q$. We choose $r'$ such that the average of $r_{Q1}$ and $r_{Q2}$ equals $r_Q$,  
\begin{equation}\label{apx-Greek-rounded3}
r_Q\approx\frac{A'}{2}\!\left(\!\int_0^{r'}\!\!\!\!\!\frac{\mathrm{d}t}{\sqrt{1\!+\!t^4}} \!+\! \exp\!\left(\!\!\frac{\!-\!\mathbbm{i}\pi}{4}\!\!\right)\!\!\!\int_0^{r'\sqrt{\mathbbm{i}}}\!\!\!\!\!\frac{\mathrm{d}t}{\sqrt{1\!+\!t^4}}\!\!\right) .
\end{equation}
For small $r'$ we may approximate the integrands by $1-t^4/2$, which gives 
\begin{equation}\label{apx-Greek-rounded4}
r_Q \approx A' r' .
\end{equation}
The final transformation $z\mapsto\log(z)$ maps the quarter of the annular region in Fig. \ref{fig:rounded-Greek-cross_Rx_zeta-plane} onto the rectangle in Fig. \ref{fig:rounded-Greek-cross_Rx_log-zeta-plane}, which has the resistance $4R_x=4\lambda_x R_\mathrm{sheet}$ with 
\begin{equation}\label{apx-Greek-rounded5}
4\lambda_x = \frac{\pi/2}{|\ln(r')|} \quad\Rightarrow\quad r'=\exp\left(\frac{-\pi}{8\lambda_x}\right) .
\end{equation}
Finally, (\ref{apx-Greek-rounded5}) and (\ref{apx-Greek-rounded4}) give the cross-resistance $R_x$ as a function of the normalized quarter-circle radius $r_Q$
\begin{equation}\label{apx-Greek-rounded6}
\lambda_x \approx \frac{\pi}{8\ln\!\left(\frac{4\sqrt{\pi} r_Q}{(\Gamma(1/4))^2}\right)} \;\Rightarrow\; r_Q \approx \frac{4\sqrt{\pi}\exp\!\left(-\pi/(8\lambda_x)\right)}{(\Gamma(1/4))^2} .
\end{equation}

\begin{figure*}[t]
%\vspace{1mm}
  \centering
                \includegraphics[width=0.98\textwidth]{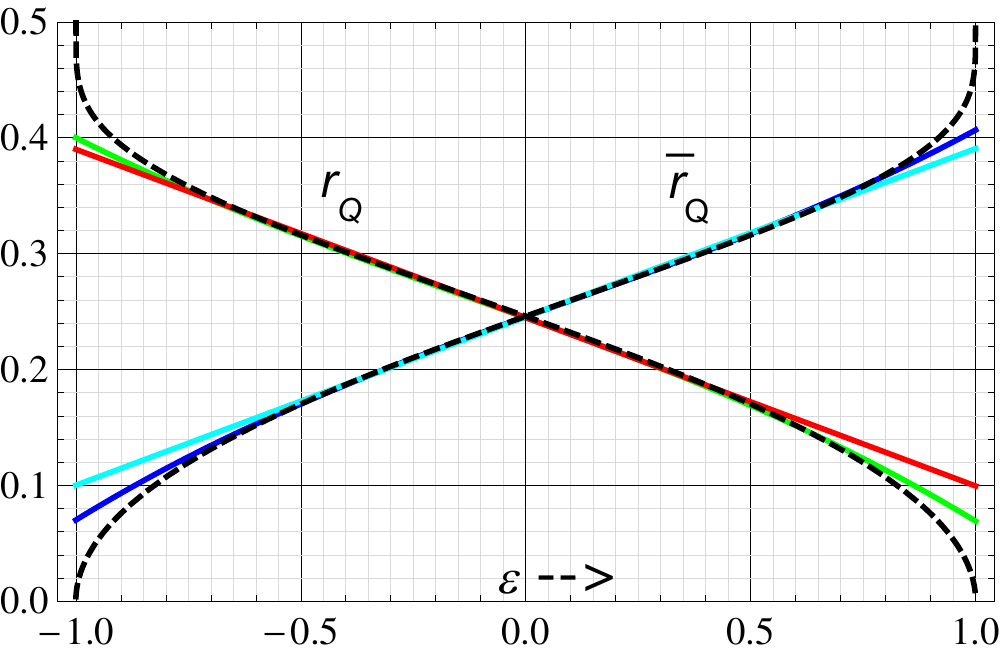}
    \caption{Plot of the normalized quarter circle radius of $90$° symmetric Greek cross Hall-plates with rounded corners ($r_Q$ for straight end contacts, $\overline{r}_Q$ for quarter circle contacts), versus the size of contacts of equivalent disk-shaped Hall-plates, $\epsilon$. The black dashed curves are the filling factors from (\ref{apx-Greek-rounded7}), (\ref{apx-Greek-rounded12}). The solid colored curves are the approximations from (\ref{apx-Greek-rounded1000}), (\ref{apx-Greek-rounded1001}). }
   \label{fig:rQ_rQq_vs_eps}
\end{figure*}

\noindent This approximation for $r_Q$ is accurate up to $0.4\%$ at $r_Q=0.4$ (which means $\lambda_x=1.33164$ and $\lambda=3.10440$), and even more accurate at smaller $r_Q$ (which means smaller $\lambda_x$ and $\lambda$). 
A Greek cross with rounded corners and end contacts is equivalent to a disk-shaped Hall-plate with contacts size $2\theta$, if its normalized quarter-circle radius equals 
\begin{equation}\label{apx-Greek-rounded7}
r_Q \approx \frac{4\sqrt{\pi}}{(\Gamma(1/4))^2} \exp\left(\frac{-\pi}{2}\frac{K}{K'}\!\!\left(\left(\tan(\theta)\right)^2\right)\right) .
\end{equation}
A numerical fit of (\ref{apx-Greek-rounded7}) gives 
\begin{equation}\label{apx-Greek-rounded1000}\begin{split}
r_Q & \approx 0.4 - 0.5* \theta + 0.43* \theta^2 - 0.42* \theta^3 \\ 
& \approx 0.39 - 0.37* \theta ,
\end{split}\end{equation}
with $\theta$ in units of radians in (\ref{apx-Greek-rounded1000}). The cubic fit deviates from (\ref{apx-Greek-rounded7}) less than $\pm 1$\% for $11$° $<2\theta<76$°, and the linear fit has the same accuracy for $12$° $<2\theta<67$°.

\begin{figure*}
%\vspace{1mm}
        \begin{subfigure}[t]{0.62\textwidth}
                \centering
                \includegraphics[width=1.0\textwidth]{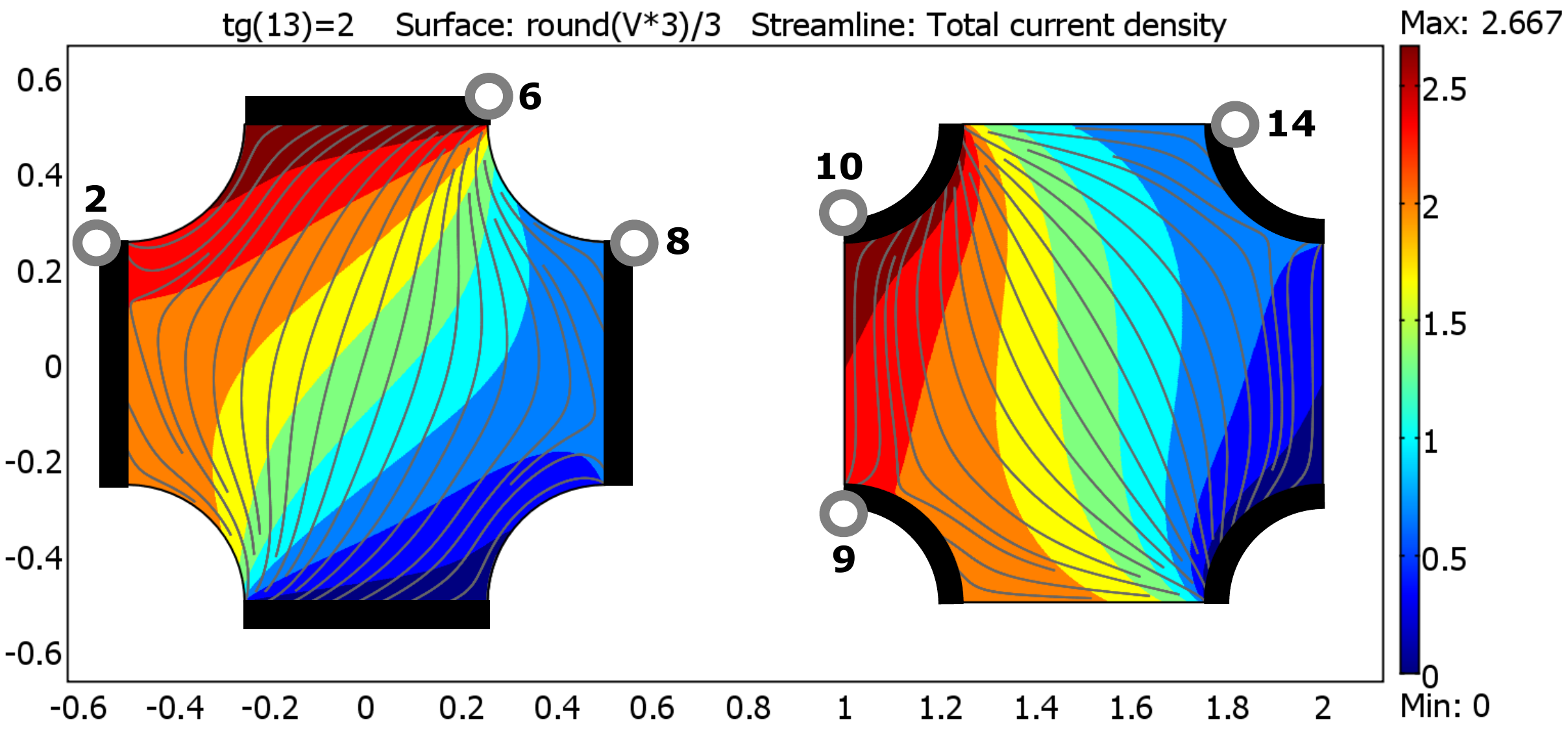}
                \caption{Current streamlines and potential at $\tan(\theta_H)=2$ in Greek crosses with rounded corners. The left Hall-plate has end contacts, the right one has quarter-circle contacts. Both Hall-plates have $r_Q=\overline{r}_Q=0.245911$, which corresponds to a disk-shaped Hall-plate with $\epsilon=0$ (= strictly regular symmetry). }
                \label{fig:streamlines_tg_eq_2_eps_eq_0_rounded-Greek-crosses}
        \end{subfigure}
        \hfill  
        \begin{subfigure}[t]{0.37\textwidth}
                \centering
                \includegraphics[width=1.0\textwidth]{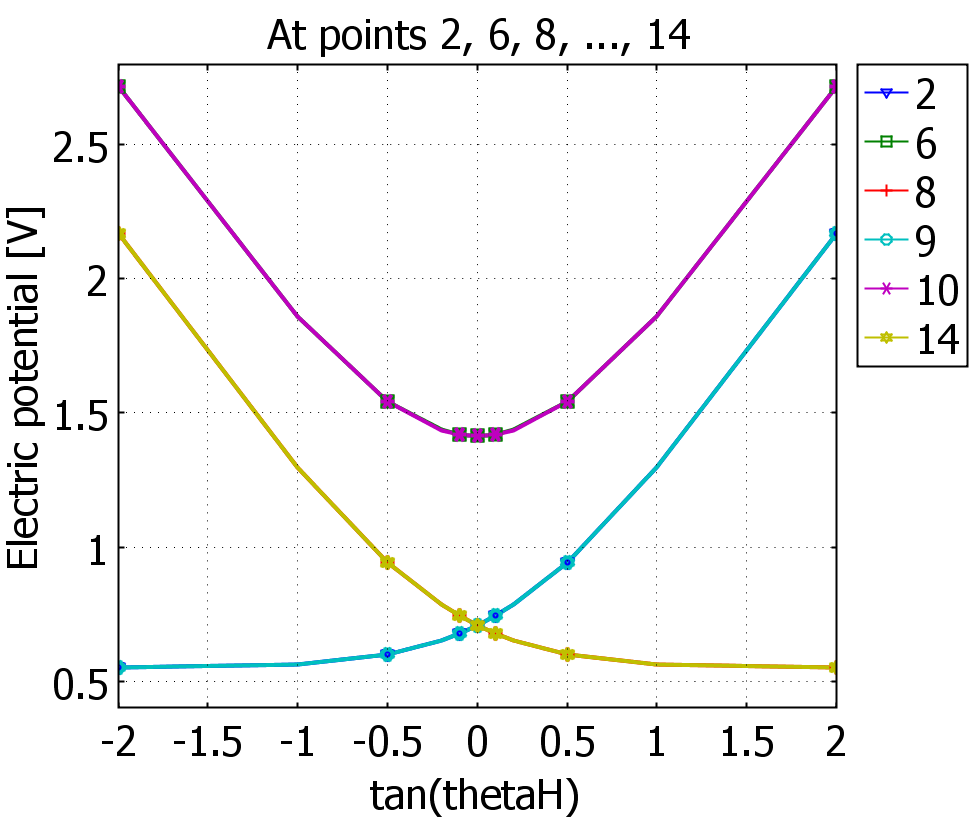}
                \caption{Potentials at the contacts versus $\tan(\theta_H)$ for the Hall-plates of Fig. \ref{fig:streamlines_tg_eq_2_eps_eq_0_rounded-Greek-crosses}. These pairs of curves are nearly identical: 2 - 9, 6 -10, and 8 - 14. }
                \label{fig:eps_eq_0_rounded-Greek-crosses}
        \end{subfigure}
	\\
  \centering
        \begin{subfigure}[t]{0.62\textwidth}
                \centering
                \includegraphics[width=1.0\textwidth]{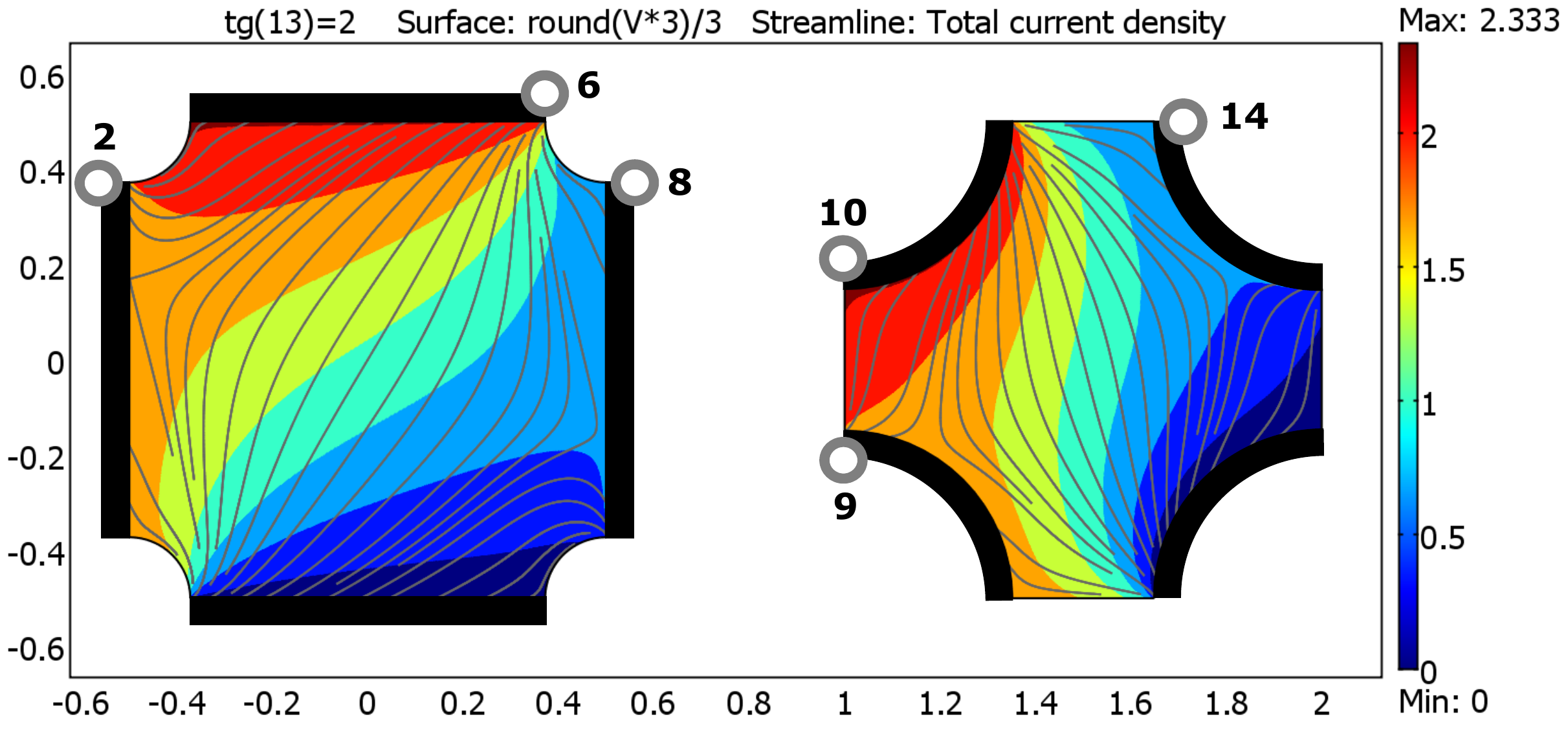}
                \caption{Current streamlines and potential at $\tan(\theta_H)=2$ in Greek crosses with rounded corners. The left Hall-plate has end contacts with $r_Q=0.128018$, the right Hall-plate has quarter-circle contacts with $\overline{r}_Q=0.351238$. Both are equivalent to a disk-shaped Hall-plate with contacts size of $\epsilon=5/7$ from Fig. \ref{fig:streamlines_tg_eq_2_eps_eq_5durch7_inv-Greek-crosses6}. }
                \label{fig:streamlines_tg_eq_2_eps_eq_5durch7_rounded-Greek-crosses}
        \end{subfigure}
        \hfill  % An dieser Stelle kann ein zusätzlicher Zwischenraum eingebunden werden: ~, \quad, \qquad, \hfill usw.
          % Eine leere Zeile erzwingt, dass die zweite Grafik darunter erscheint.
        \begin{subfigure}[t]{0.37\textwidth}
                \centering
                \includegraphics[width=1.0\textwidth]{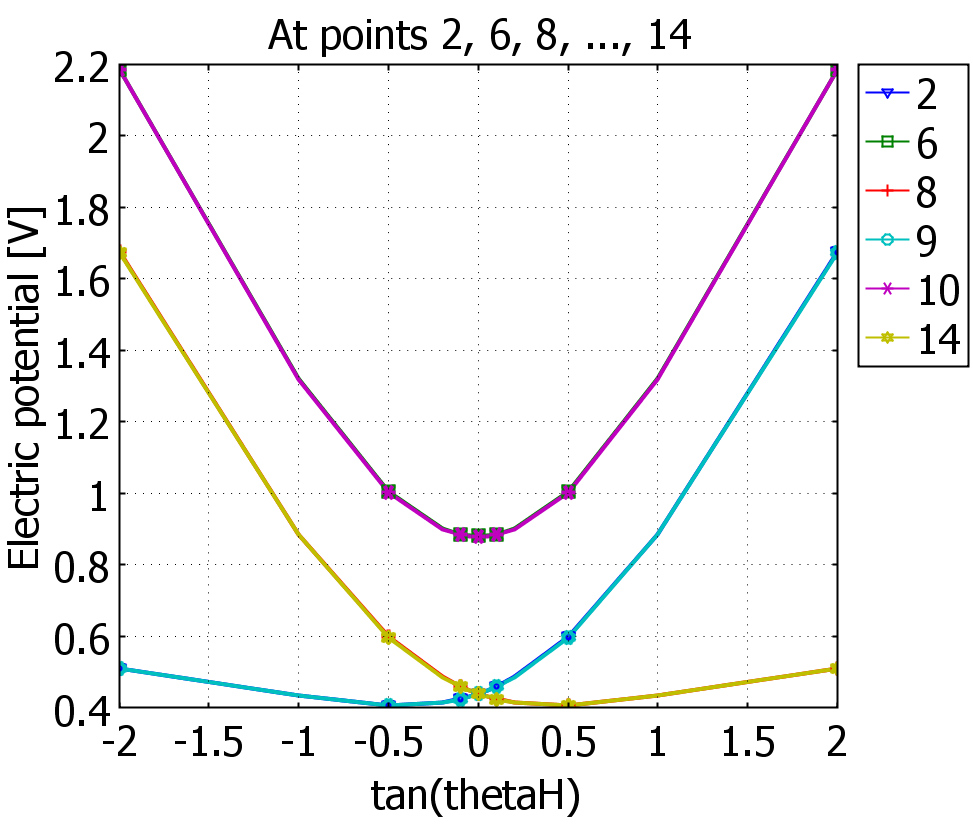}
                \caption{Potentials at the contacts versus $\tan(\theta_H)$ for the Hall-plates of Fig. \ref{fig:streamlines_tg_eq_2_eps_eq_5durch7_rounded-Greek-crosses}. These pairs of curves are nearly identical: 2 - 9, 6 -10, and 8 - 14. They are also identical to the ones in Fig. \ref{fig:eps_eq_5durch7_inv-Greek-crosses6}. }
                \label{fig:eps_eq_5durch7_rounded-Greek-crosses}
        \end{subfigure}
    \caption{Results of FEM-simulations of Greek cross shaped Hall-plates with rounded corners. Contacts without any labelled circle are grounded. A current of $1$ A enters contacts comprising circles 6 and 10. They are diametrically opposite to the grounded contacts. $R_\mathrm{sheet}=1\;\Omega$. The differences in the potentials of corresponding points $2-9,6-10,8-14$ are negligible. They come from the approximation explained in Fig. \ref{fig:rounded-Greek-cross_approx}. }
   \label{fig:rounded-Greek-crosses-checks}
\end{figure*}

The complementary Hall-plate is a Greek-cross with quarter circle contacts (instead of end contacts). If the original disk plate with contacts size $2\theta$ corresponds to a Greek-cross with rounded corners with $r_Q$ having end contacts, the complementary disk plate has contacts size $2\overline{\theta}=\pi/2-2\theta$, and it corresponds to the very same Greek-cross shape, i.e. the same $r_Q$, yet with quarter circle contacts instead of end contacts. Hence, for the Greek cross with quarter circle contacts we simply have to replace $\theta\to\pi/4-\theta$ in (\ref{apx-Greek-rounded7}). It holds 
\begin{equation}\label{apx-Greek-rounded10}\begin{split}
& \tan\!\left(\!\frac{\pi}{4}\!-\!\theta\!\right) = \frac{1-\tan(\theta)}{1+\tan(\theta)} \;\;\text{and} \\ 
& \frac{K}{K'}\!\left(\!\left(\!\frac{1\!-\!\tan(\theta)}{1\!+\!\tan(\theta)}\!\right)^{\!\!2}\right) = \frac{1}{4}\frac{K'}{K}\!\left(\left(\tan(\theta)\right)^{\!2}\right) ,
\end{split}\end{equation}
whereby the last identity comes from (28) in Ref.~\onlinecite{ELEN}. (Note that $K$ and $K'$ in the fraction are swapped.) Consequently, a Greek cross with quarter circle contacts of normalized radius $\overline{r}_Q$ is equivalent to a disk-shaped Hall-plate with contacts size $2\theta$, if it holds 
\begin{equation}\label{apx-Greek-rounded12}
\overline{r}_Q \approx \frac{4\sqrt{\pi}}{(\Gamma(1/4))^2} \exp\left(\frac{-\pi}{8}\frac{K'}{K}\!\!\left(\left(\tan(\theta)\right)^2\right)\right) .
\end{equation}
In (\ref{apx-Greek-rounded12}) we used the overbar to distinguish it from (\ref{apx-Greek-rounded7}): $r_Q$ is the radius of the quarter circle segment in the smooth Greek cross with end contacts, whereas $\overline{r}_Q$ is the radius of the quarter circle segment in the smooth Greek cross with quarter circle contacts.
A numerical fit of (\ref{apx-Greek-rounded12}) gives 
\begin{equation}\label{apx-Greek-rounded1001}\begin{split}
\overline{r}_Q & \approx 0.07 + 0.61* \theta - 0.6* \theta^2 + 0.47* \theta^3 \\ 
& \approx 0.1 + 0.37* \theta ,
\end{split}\end{equation}
with $\theta$ in units of radians in (\ref{apx-Greek-rounded1001}). The cubic fit deviates from (\ref{apx-Greek-rounded12}) less than $\pm 1$\% for $17$° $<2\theta<83$°, and the linear fit has the same accuracy for $24$° $<2\theta<78$°. Fig. \ref{fig:rQ_rQq_vs_eps} shows a plot of (\ref{apx-Greek-rounded7}), (\ref{apx-Greek-rounded1000}), (\ref{apx-Greek-rounded12}), and (\ref{apx-Greek-rounded1001}).

Equations (\ref{apx-Greek-rounded7}) and (\ref{apx-Greek-rounded12}) give the following normalized quarter circle radii for three cases, 
\begin{equation}\label{apx-Greek-rounded1002}
\begin{array}{c|cccc} \mathrm{case} & \epsilon & \theta & r_Q & \overline{r}_Q \\ \hline
1 & -1/4 & 3\pi/32 & 0.280304 & 0.210164 \\
2 & 0 & \pi/8 & 0.245911 & 0.245911 \\
3 & 5/7 & 3\pi/14 & 0.128018 & 0.351238 \end{array}
\end{equation}

Numerical checks of these equations, analogous to the ones in the preceding section, are shown in Fig. \ref{fig:rounded-Greek-crosses-checks}. The mesh had about 1.2 million elements per Hall-plate. The corresponding supply potentials of both Hall-plates in Fig. \ref{fig:streamlines_tg_eq_2_eps_eq_5durch7_rounded-Greek-crosses} differ $0.25\%$ at zero magnetic field and $0.11\%$ at $\tan(\theta_H)=2$, and the Hall-output voltages differ $0.42\%$ at $\tan(\theta_H)=0.05$ and $0.22\%$ at $\tan(\theta_H)=2$. The differences in Fig. \ref{fig:streamlines_tg_eq_2_eps_eq_0_rounded-Greek-crosses} were smaller: the supply potentials of left and right Hall-plates differ only $143$ ppm at zero magnetic field and $72$ ppm at $\tan(\theta_H)=2$, and the Hall-output voltages differ $143$ ppm at $\tan(\theta_H)=0.05$ and $72$ ppm at $\tan(\theta_H)=2$. The reason for the larger discrepancy in the first two Hall-plates is very likely \emph{not} a numerical inaccuracy of the finite element simulation---instead, it seems to come from the approximations made in (\ref{apx-Greek-rounded7}) and (\ref{apx-Greek-rounded12}). Nevertheless, the accuracy is sufficient for practical use, and in view of the approximations, these numerical results corroborate the analytical theory.

\clearpage
\subsection{Octagonal Hall-plates}
\label{sec:Octagons}

\begin{figure}
%\vspace{1mm}
  \centering
        \begin{subfigure}[c]{0.21\textwidth}
                \centering
                \includegraphics[width=1.0\textwidth]{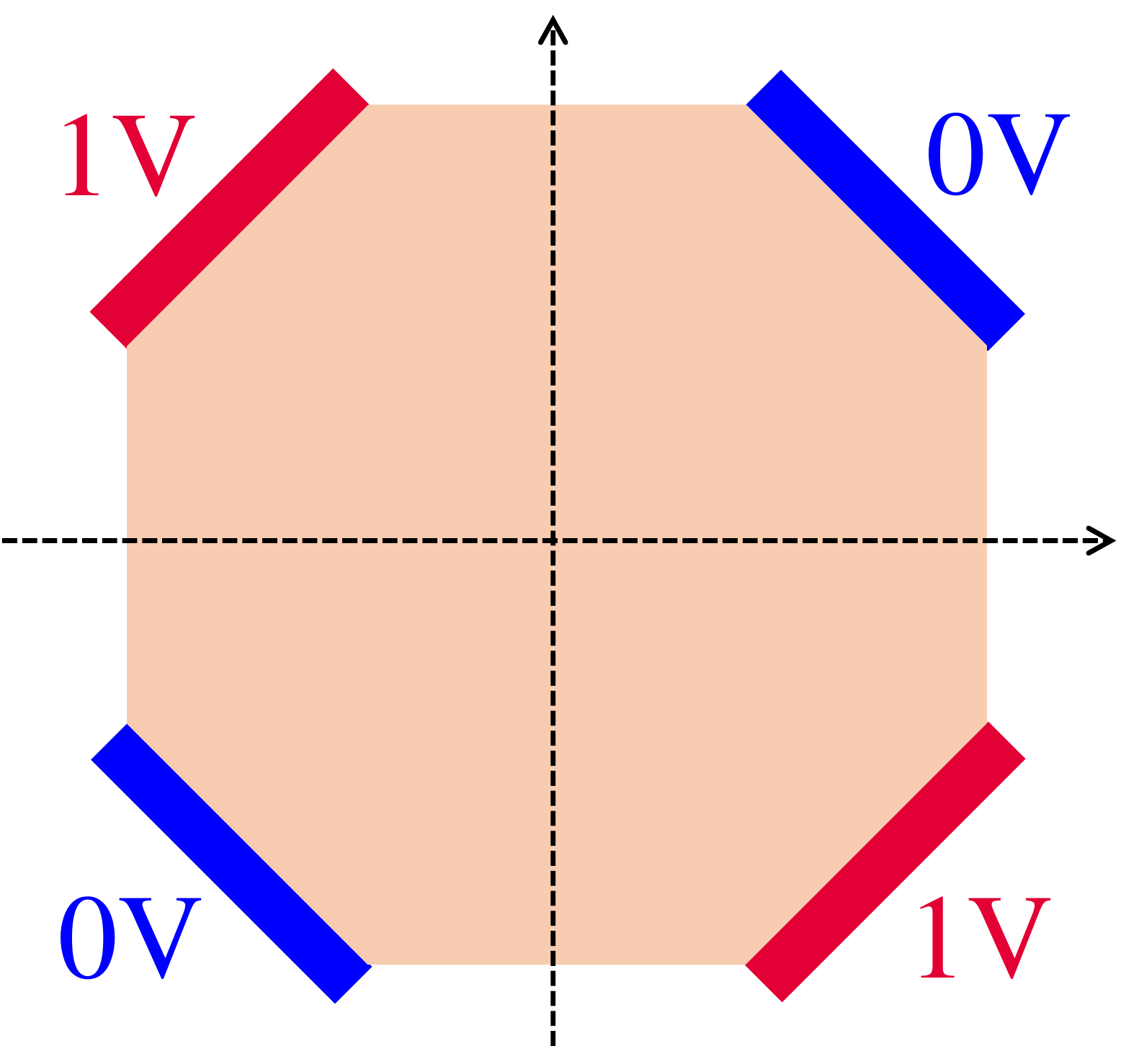}
                \caption{Complete Hall-plate in the $z$-plane. Resistance = $\overline{R}_x$. }
                \label{fig:compl_octagon_Rxq_z-plane1}
        \end{subfigure}
	\hfill
        \begin{subfigure}[c]{0.2\textwidth}
                \centering
                \includegraphics[width=1.0\textwidth]{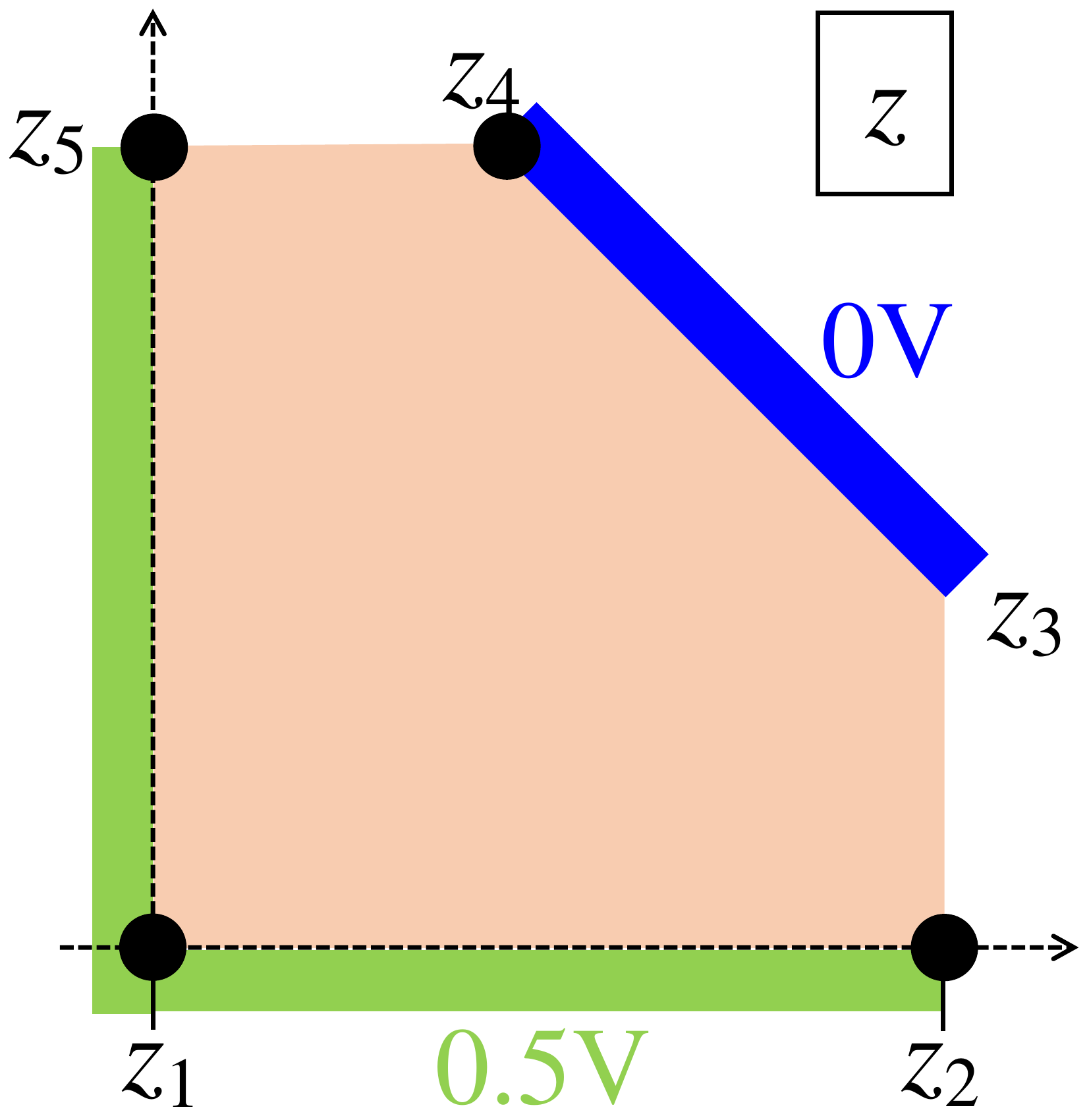}
                \caption{A quarter of Fig. \ref{fig:compl_octagon_Rxq_z-plane1}. Resistance = $\overline{R}_x$. }
                \label{fig:compl_octagon_Rxq_z-plane2}
        \end{subfigure}
	\\
        \begin{subfigure}[c]{0.29\textwidth}
                \centering
                \includegraphics[width=1.0\textwidth]{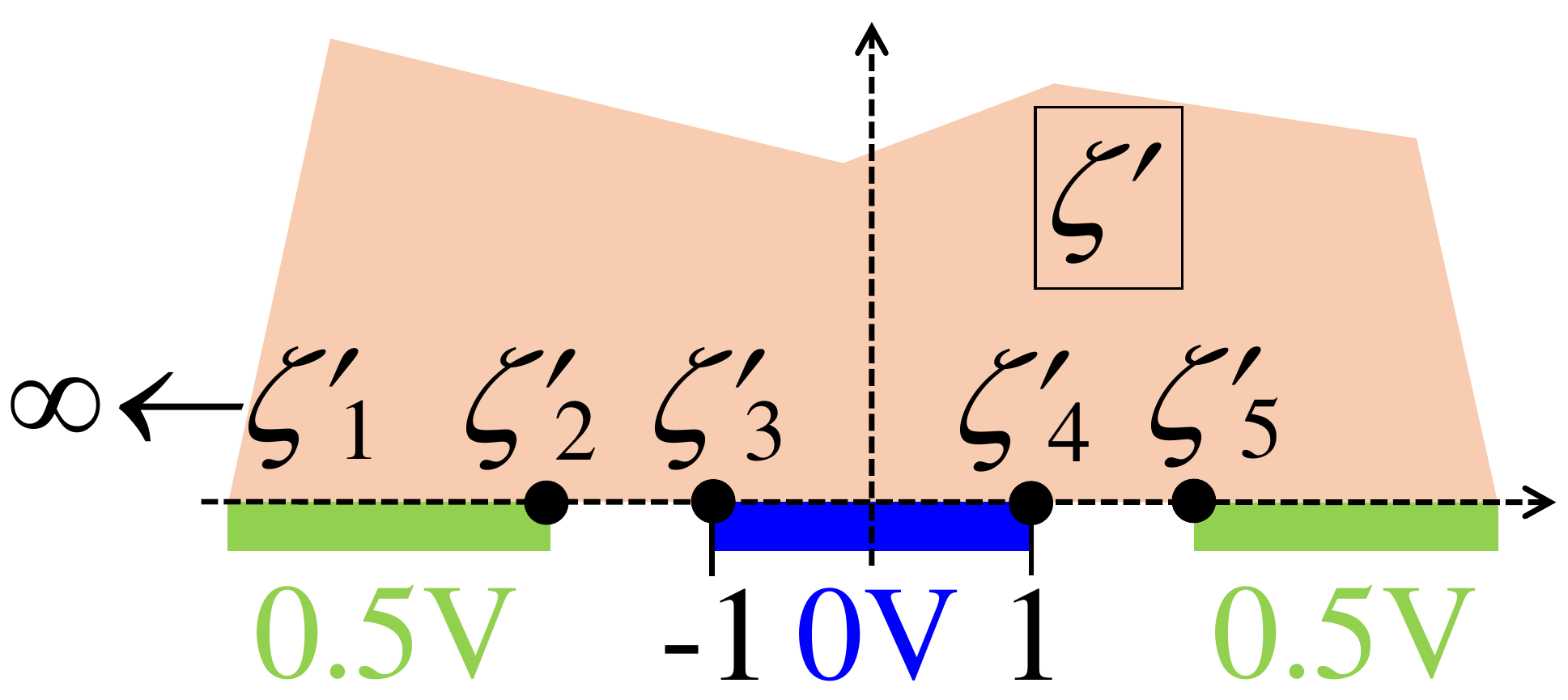}
                \caption{Mapped to $\zeta'$-plane. Resistance = $\overline{R}_x$. }
                \label{fig:compl_octagon_Rxq_zetas-plane}
        \end{subfigure}
    \caption{Conformal mapping of a complementary octagonal Hall-plate to compute the cross resistance $\overline{R}_x$. }
   \label{fig:compl_octagon_Rxq}
\end{figure}

If a Hall-plate is octagonal, its complementary plate is also octagonal. Let us call the plate in Fig. \ref{fig:compl_octagon_Rxq_z-plane1} complementary. The potentials in this figure occur during a measurement of $\overline{R}_x$. Fig. \ref{fig:compl_octagon_Rxq_z-plane2} shows one quarter of the octagonal plate. It has again the resistance $\overline{R}_x$ between its contacts. In Fig. \ref{fig:compl_octagon_Rxq_zetas-plane} we map it to the upper half of the $\zeta'$-plane according to 
\begin{equation}\label{apx-oct10}
z = C' \int_0^{\zeta'} \frac{\mathrm{d}\zeta}{\left({\zeta'}^2-1\right)^{1/4}\sqrt{{\zeta'}^2-{\zeta'}_5^2}} , \;\;\text{with }{\zeta'}_5>1 .
\end{equation}
The shape in Fig. \ref{fig:compl_octagon_Rxq_z-plane2} is made up of two unit cells. I chose this shape, because its symmetry leads to $\zeta'_1\to\infty$, which simplifies the transformation formula in (\ref{apx-oct10}). A comparison of Fig. \ref{fig:compl_octagon_Rxq_zetas-plane} and Fig. \ref{fig:regular-4C-Hall-disk_Rx_t-and-zeta-plane} gives 
\begin{equation}\label{apx-oct12}
\overline{\lambda}_x = \frac{1}{2}\,\frac{K'}{K}\!\left(\frac{1}{\zeta_5}\right) .
\end{equation}
With (\ref{apx-popularHalls15}) we get 
\begin{equation}\label{apx-oct14}
4\lambda_x = 2\,\frac{K}{K'}\!\!\left(\!\frac{1}{\zeta_5}\right) = \frac{K}{K'}\!\!\left(\!\frac{\frac{2}{\sqrt{\zeta_5}}}{1\!+\!\frac{1}{\zeta_5}}\right) = \frac{K'}{K}\!\!\left(\!\frac{\zeta_5\!-\!1}{\zeta_5\!+\!1}\right) ,
\end{equation}
wherein we used (27) from Ref.~\onlinecite{ELEN}. Comparing (\ref{apx-oct14}) with (\ref{apx-popularHalls12}) specifies the octagonal Hall-plate in terms of an equivalent disk-shaped Hall-plate 
\begin{equation}\label{apx-oct16}
\frac{\zeta_5-1}{\zeta_5+1} = \left(\tan\!\left(\frac{\pi}{4}-\theta\right)\right)^{\!\!2} \;\;\Rightarrow\;\; \frac{1}{\zeta_5} = \cos\!\left(\frac{\pi}{2}-2\theta\right) .
\end{equation}
Thereby we took into account that the complementary device (with $4\overline{\lambda}_x$) corresponds to a disk with contacts size $2\theta$, and therefore the original device (with $4\lambda_x$) corresponds to a disk with contacts size $\pi/2-2\theta$. We define the filling factor $\overline{f_{OC}}$ as the sum of the lengths of all contacts in Fig. \ref{fig:compl_octagon_Rxq_z-plane1} over the entire perimeter. It holds 
\begin{equation}\label{apx-oct18}\begin{split}
\overline{f_{OC}} & = \frac{|z_3-z_4|}{|z_3-z_4|+2|z_4-z_5|} \\
\Rightarrow \frac{1}{\overline{f_{OC}}} & = 1+\left|\frac{\int_1^{\zeta_5} \frac{\mathrm{d}\zeta}{\left(\zeta^2-1\right)^{1/4}\sqrt{\zeta^2-\zeta_5^2}}}{\int_0^1 \frac{\mathrm{d}\zeta}{\left(\zeta^2-1\right)^{1/4}\sqrt{\zeta^2-\zeta_5^2}}} \right| \\ 
& = 1-\sqrt{2}+\frac{1}{4\sqrt{\cos\left(\pi\frac{1-\epsilon}{4}\right)}}\left(\frac{\Gamma(1/4)}{\Gamma(3/4)}\right)^{\!\!2} \\
& \quad * \frac{{_2}F_1\left(1/4,1/4,3/4,\left(\cos(\pi\frac{1-\epsilon}{4})\right)^{\!2}\right)}{{_2}F_1\left(1/2,1/2,5/4,\left(\cos(\pi\frac{1-\epsilon}{4})\right)^{\!2}\right)} .
\end{split}\end{equation}
A numerical fit of (\ref{apx-oct18}) gives 
\begin{equation}\label{apx-oct19}\begin{split}
\overline{f_{OC}} & \approx 0.502 + 0.34 *\epsilon + 0.06 *\epsilon^3 \\ 
& \approx 0.501 + 0.35 *\epsilon .
\end{split}\end{equation}
In the (\ref{apx-oct19}) the linear approximation is accurate up to $\pm 1\%$ for $-0.59<\epsilon<0.68$ and the cubic one for $-0.78<\epsilon<0.87$. Fig. \ref{fig:fSC_fSM_vs_eps} shows a plot of (\ref{apx-Square6}), (\ref{apx-Square7}), and (\ref{apx-Square15}).

\begin{figure*}[t]
%\vspace{1mm}
  \centering
                \includegraphics[width=0.98\textwidth]{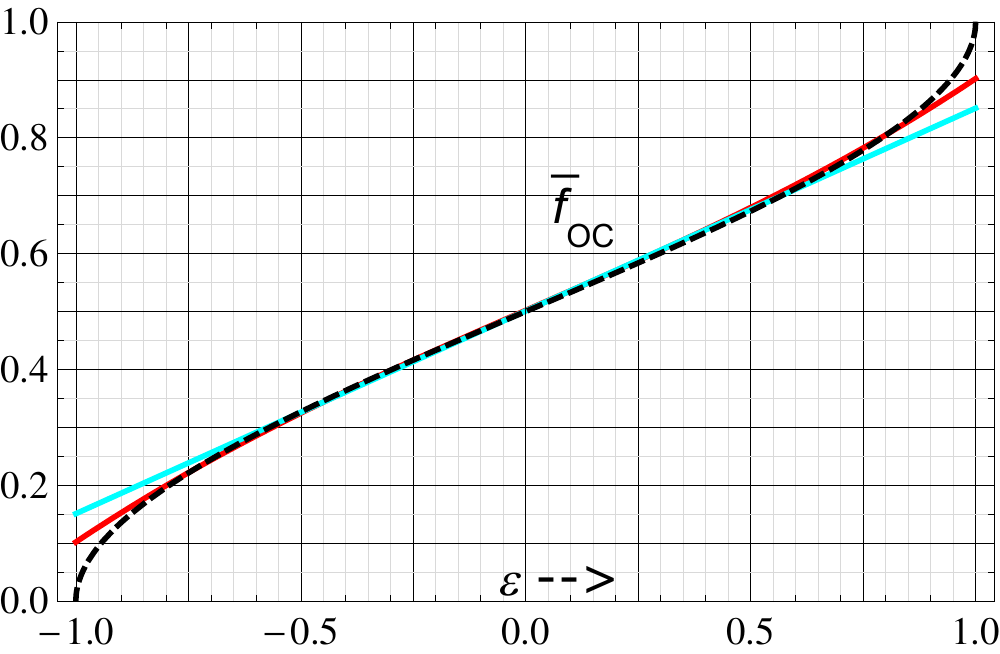}
    \caption{Plot of the filling factor of $90$° symmetric octagonal Hall-plates, $\overline{f_{OC}}$, versus the size of contacts of equivalent disk-shaped Hall-plates, $\epsilon$. The black dashed curve is the exact filling factor from (\ref{apx-oct18}). The solid colored curves are the approximations (\ref{apx-oct19}). }
   \label{fig:fqOC_vs_eps}
\end{figure*}

Equation (\ref{apx-oct18}) gives the following filling factors for three cases, 
\begin{equation}\label{apx-oct19b}
\begin{array}{c|ccc} \mathrm{case} & \epsilon & \theta & \overline{f_{OC}} \\ \hline
1 & -1/4 & 3\pi/32 & 0.416197 \\
2 & 0 & \pi/8 & 0.5 \\
3 & 5/7 & 3\pi/14 & 0.761779 \end{array}
\end{equation}

\begin{figure*}
%\vspace{1mm}
  \centering
        \begin{subfigure}[t]{0.63\textwidth}
                \centering
                \includegraphics[width=1.0\textwidth]{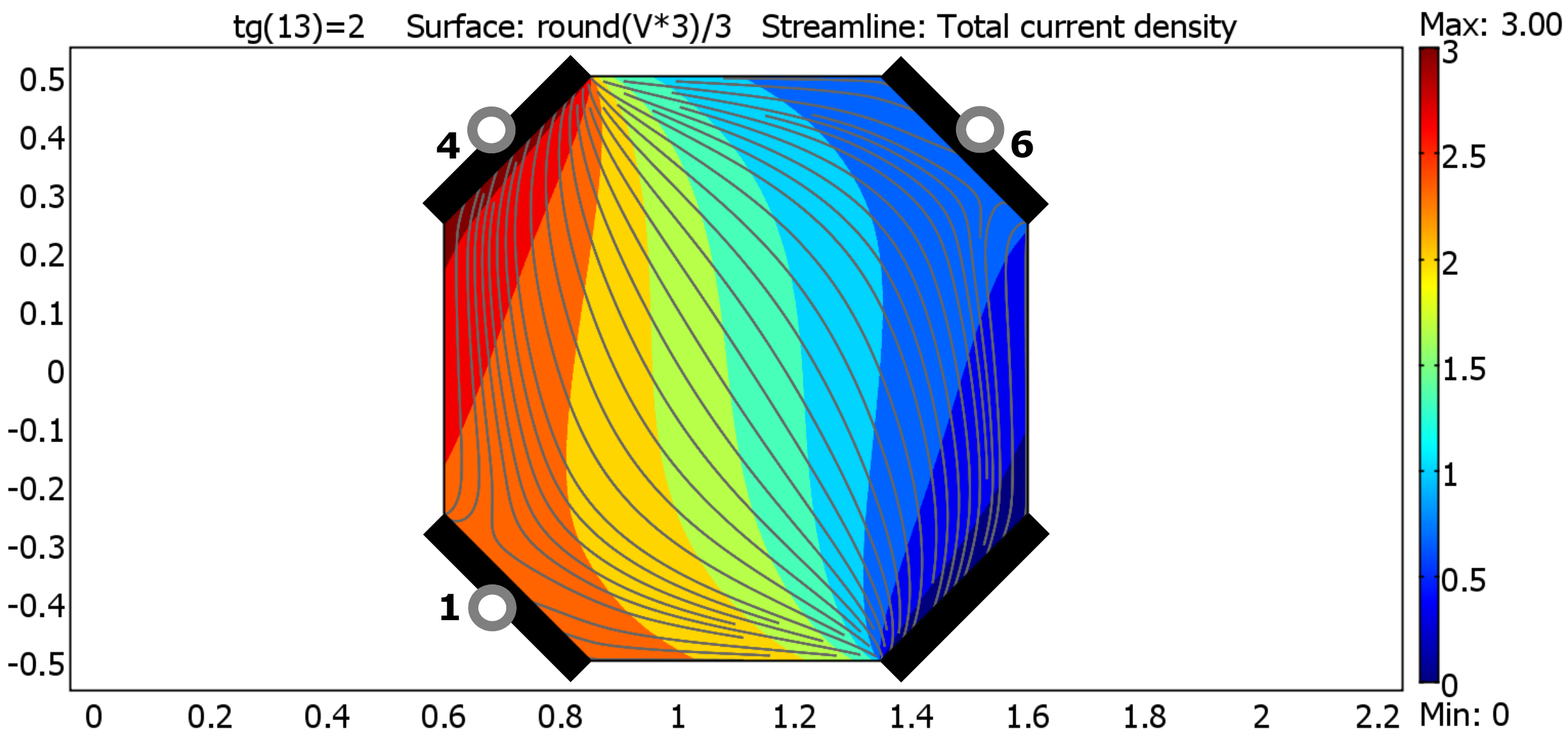}
                \caption{Current streamlines and potential at $\tan(\theta_H)=2$ in an octagonal Hall-plate with the filling factor $\overline{f_{OC}}=0.416197$. It is equivalent to a disk-shaped Hall-plate with contacts size of $\epsilon=-1/4$ from Fig. \ref{fig:streamlines_tg_eq_2_eps_eq_m250m_Greek-crosses1}. }
                \label{fig:streamlines_tg_eq_2_eps_eq_m250m_Octagon}
        \end{subfigure}
        \hfill  % An dieser Stelle kann ein zusätzlicher Zwischenraum eingebunden werden: ~, \quad, \qquad, \hfill usw.
          % Eine leere Zeile erzwingt, dass die zweite Grafik darunter erscheint.
        \begin{subfigure}[t]{0.35\textwidth}
                \centering
                \includegraphics[width=1.0\textwidth]{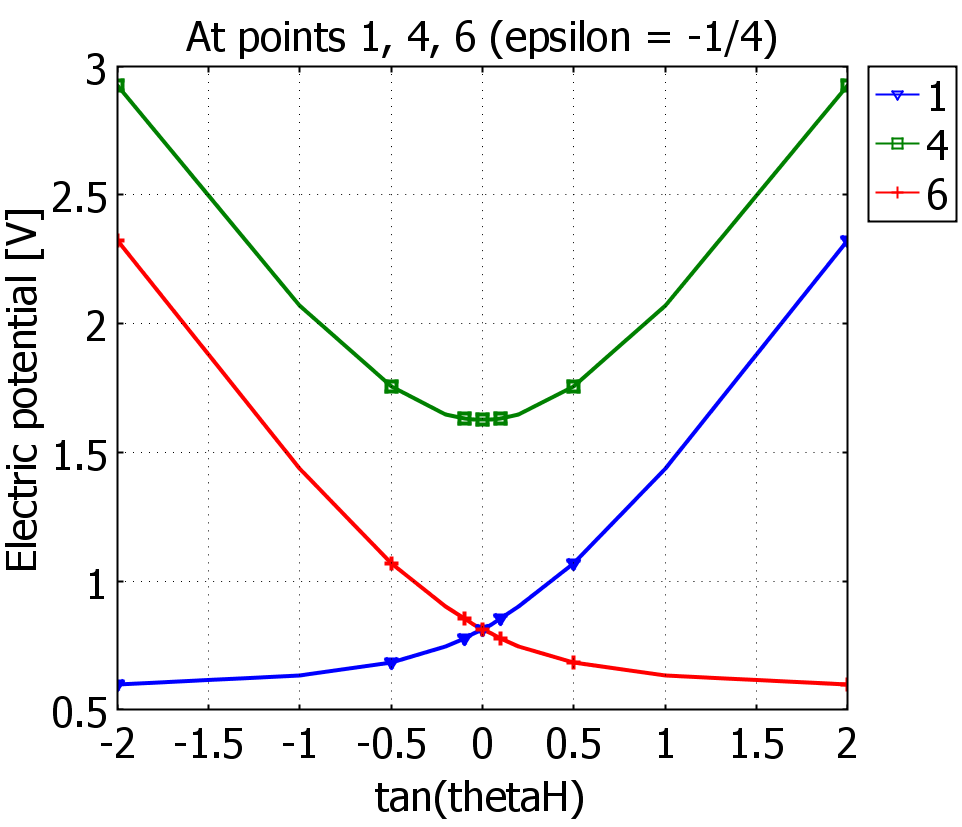}
                \caption{Potentials at the contacts versus $\tan(\theta_H)$ for the Hall-plates of Fig. \ref{fig:streamlines_tg_eq_2_eps_eq_m250m_Octagon}. These curves are identical to the ones in Fig. \ref{fig:eps_eq_m250m_Greek-crosses1}. }
                \label{fig:eps_m250m_Octagon}
        \end{subfigure}
	 \\
        \begin{subfigure}[t]{0.63\textwidth}
                \centering
                \includegraphics[width=1.0\textwidth]{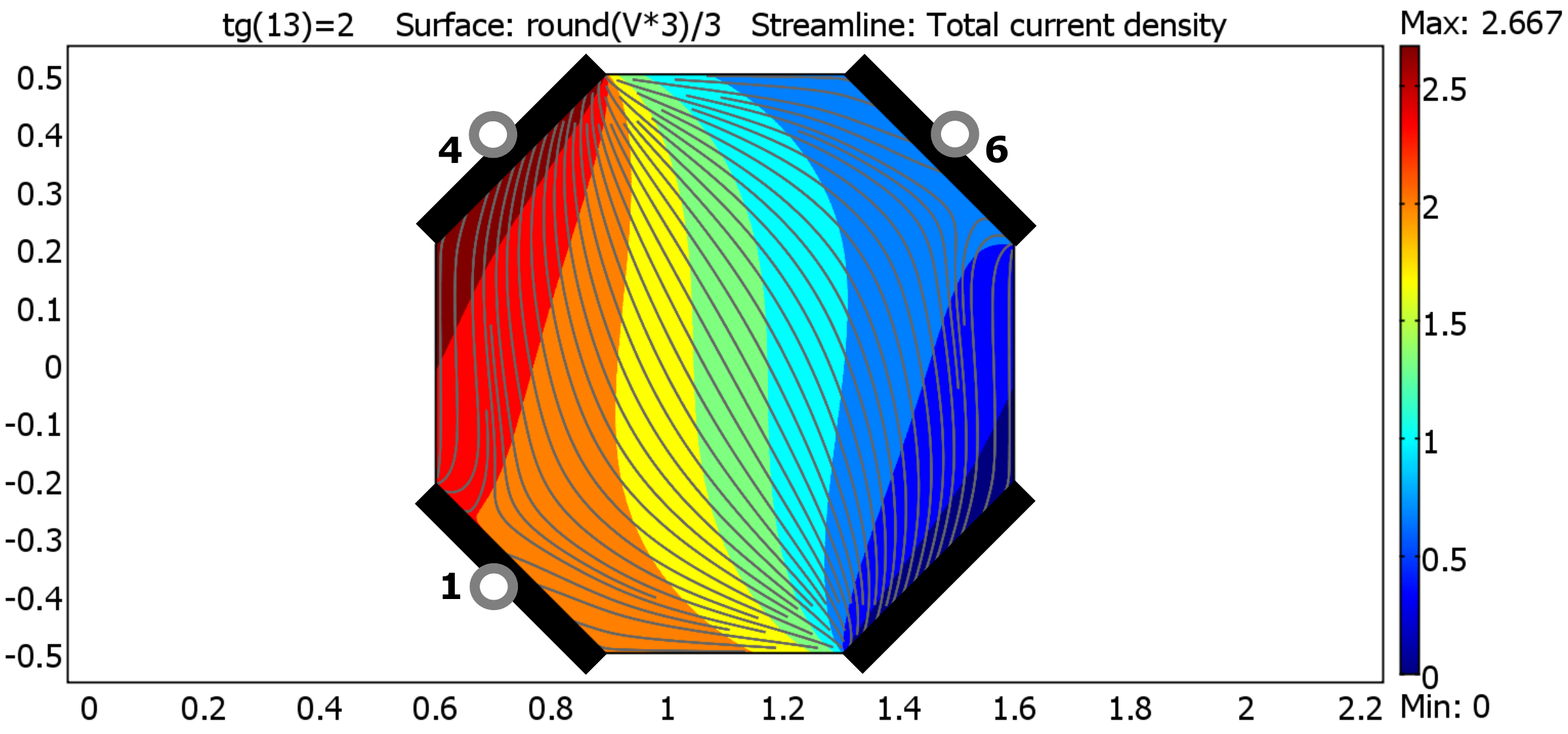}
                \caption{Current streamlines and potential at $\tan(\theta_H)=2$ in an octagonal Hall-plate with the filling factor $\overline{f_{OC}}=0.5$. It is equivalent to a disk-shaped Hall-plate with contacts size of $\epsilon=0$ (= strictly regular symmetry) or to smooth Greek crosses from Fig. \ref{fig:streamlines_tg_eq_2_eps_eq_0_rounded-Greek-crosses}. }
                \label{fig:streamlines_tg_eq_2_eps_eq_0_Octagon}
        \end{subfigure}
        \hfill  
        \begin{subfigure}[t]{0.35\textwidth}
                \centering
                \includegraphics[width=1.0\textwidth]{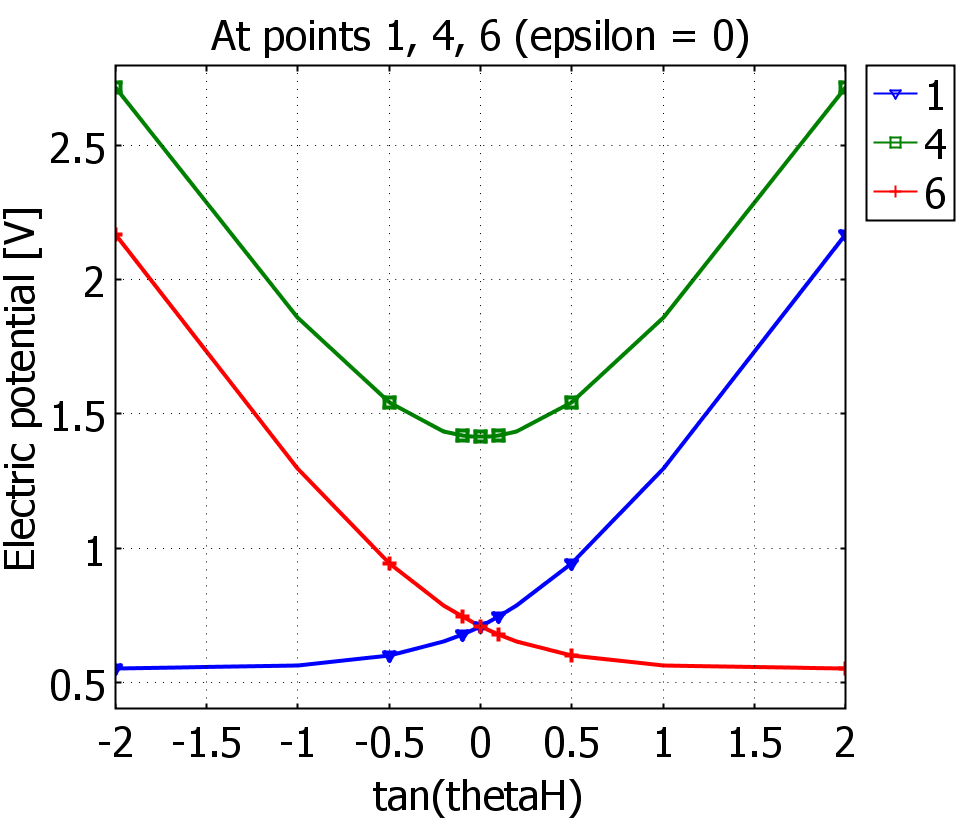}
                \caption{Potentials at the contacts versus $\tan(\theta_H)$ for the Hall-plates of Fig. \ref{fig:streamlines_tg_eq_2_eps_eq_0_Octagon}. These curves are identical to the ones in Fig. \ref{fig:eps_eq_0_rounded-Greek-crosses}. }
                \label{fig:eps_eq_0_Octagon}
        \end{subfigure}
	 \\
        \begin{subfigure}[t]{0.63\textwidth}
                \centering
                \includegraphics[width=1.0\textwidth]{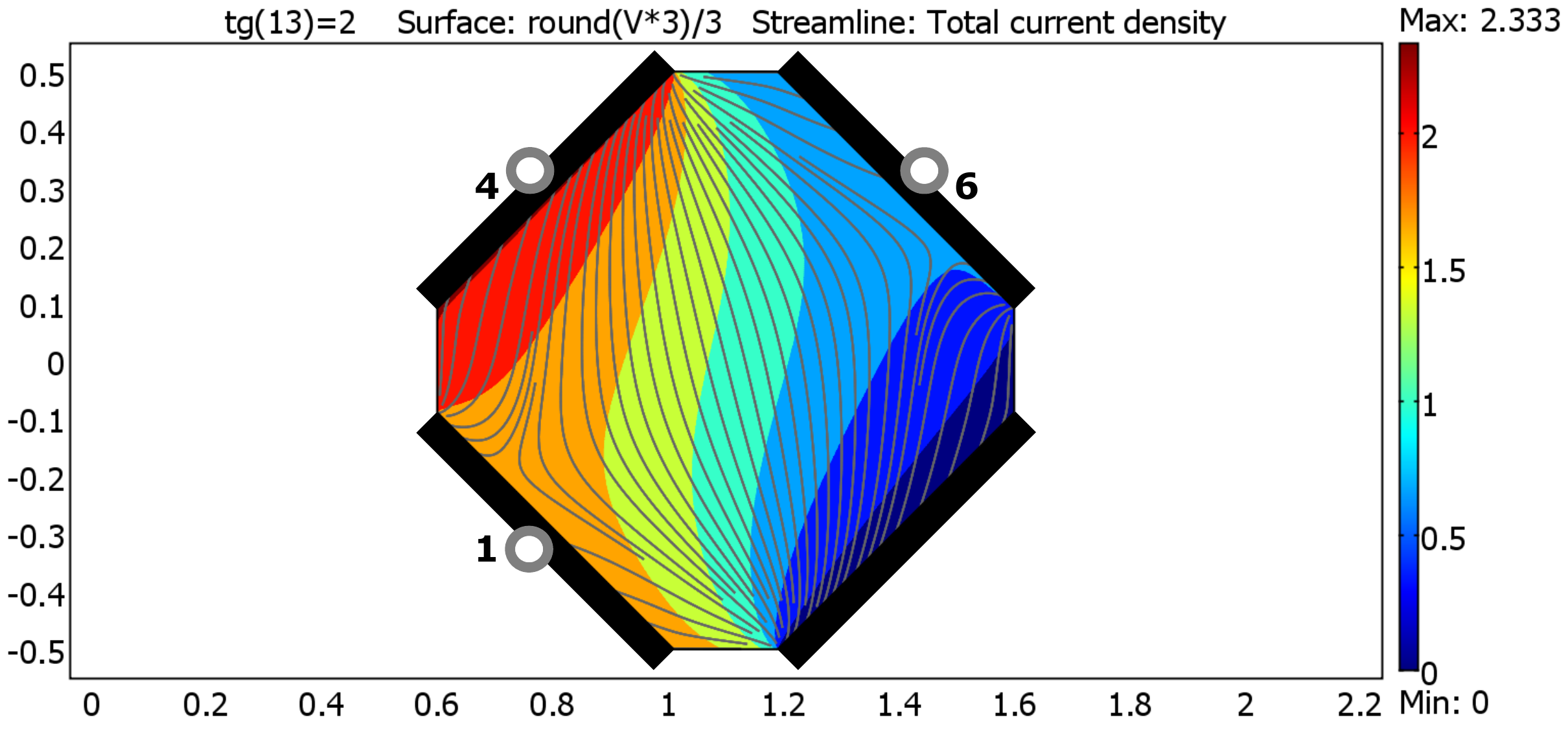}
                \caption{Current streamlines and potential at $\tan(\theta_H)=2$ in an octagonal Hall-plate with the filling factor $\overline{f_{OC}}=0.761779$. It is equivalent to a disk-shaped Hall-plate with contacts size of $\epsilon=5/7$ from Fig. \ref{fig:streamlines_tg_eq_2_eps_eq_5durch7_inv-Greek-crosses6}. }
                \label{fig:streamlines_tg_eq_2_eps_eq_5durch7_Octagon}
        \end{subfigure}
        \hfill  
        \begin{subfigure}[t]{0.35\textwidth}
                \centering
                \includegraphics[width=1.0\textwidth]{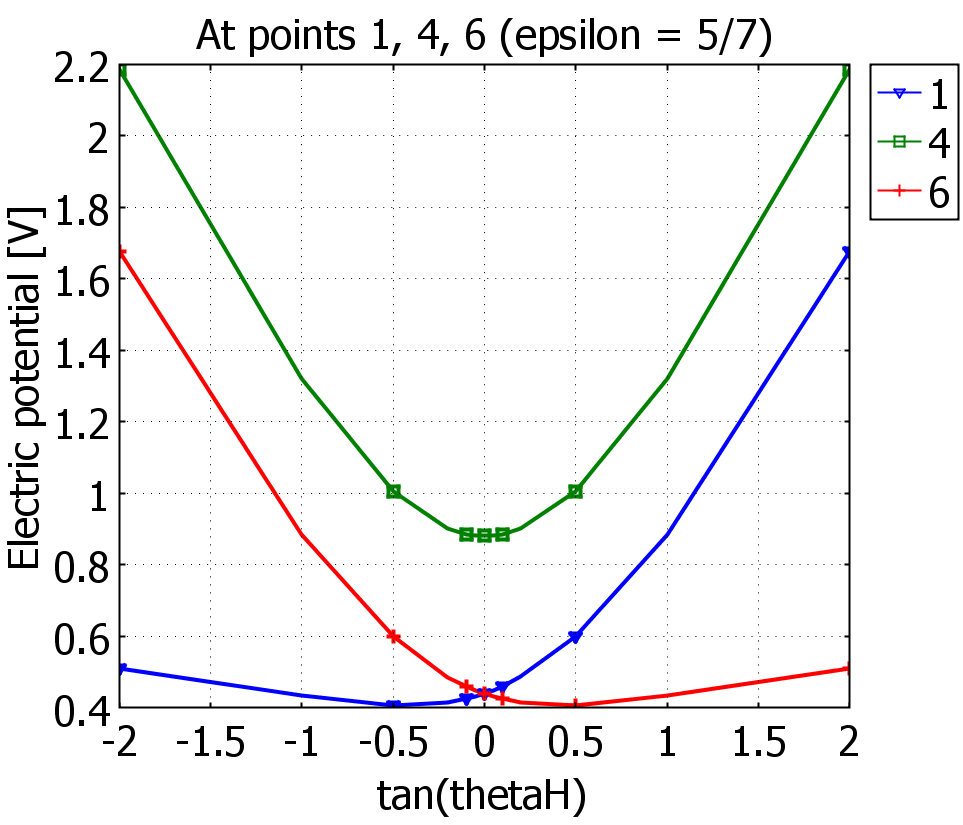}
                \caption{Potentials at the contacts versus $\tan(\theta_H)$ for the Hall-plates of Fig. \ref{fig:streamlines_tg_eq_2_eps_eq_5durch7_Octagon}. These curves are identical to the ones in Fig. \ref{fig:eps_eq_5durch7_inv-Greek-crosses6}. }
                \label{fig:eps_eq_5durch7_Octagon}
        \end{subfigure}
    \caption{Results of FEM-simulations of octagonal Hall-plates. The contacts without any labelled circles are grounded. A current of $1$ A enters the contact with circle 4. It is diametrically opposite to the grounded contact. $R_\mathrm{sheet}=1\;\Omega$. }
   \label{fig:octagons-checks}
\end{figure*}

Numerical checks of (\ref{apx-oct18}), analogous to the ones in the preceding sections, are shown in Fig. \ref{fig:octagons-checks}, where the plots of the potentials versus magnetic field are identical to the ones in the preceding sections. 
% The mesh had roughly 1.8 to 2.4 million elements per Hall-plate.

In practice, ocatgonal Hall-plates are the second best choice (after Greek crosses with rounded corners), because the electric field and the current density have less pronounced singularities at the vertices of the contacts \cite{vanBladel}. This leads to smaller electrical non-linearity caused by velocity saturation, and finally, this gives smaller residual offset errors of spinning current schemes.

\clearpage
\subsection{Square Hall-plates}
\label{sec:Squares}

\begin{figure}[t]
%\vspace{1mm}
  \centering
                \includegraphics[width=0.22\textwidth]{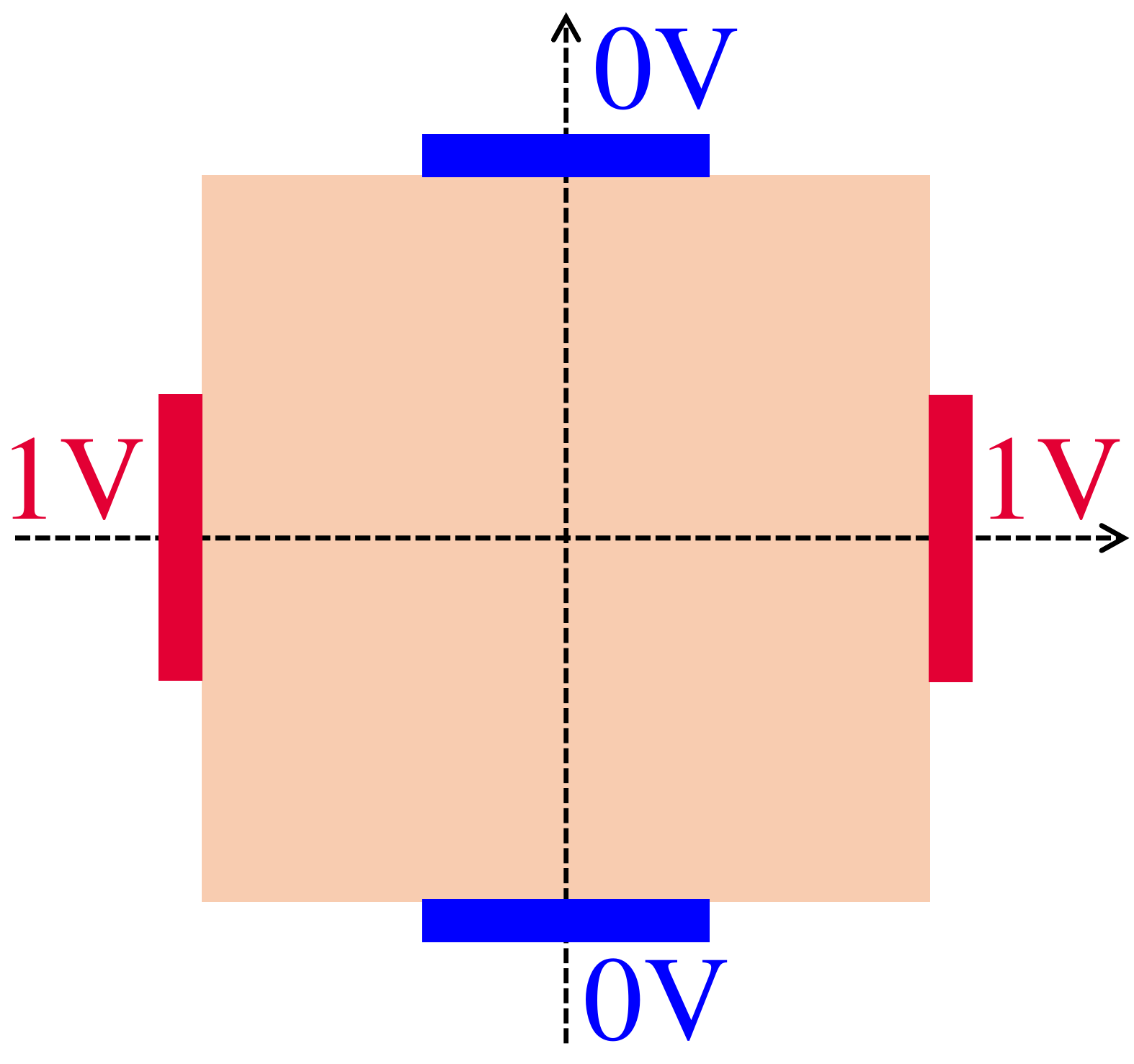}
    \caption{Square with mid-contacts. Resistance = $R_x$. }
   \label{fig:Square_Rx_z-plane1}
\end{figure}

\begin{figure}
%\vspace{1mm}
  \centering
        \begin{subfigure}[c]{0.19\textwidth}
                \centering
                \includegraphics[width=1.0\textwidth]{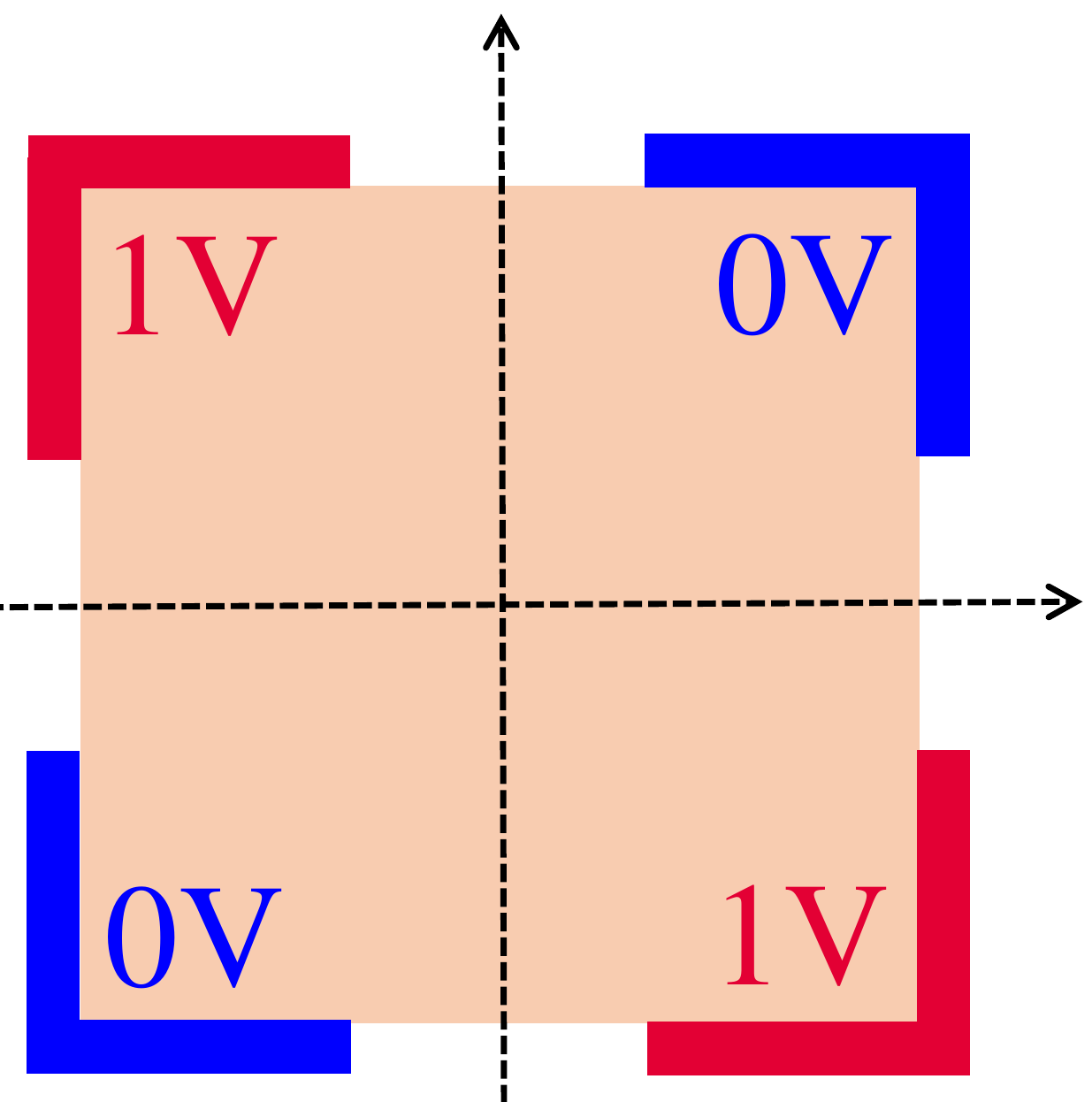}
                \caption{Square with corner contacts during the measurement of $\overline{R}_x$. }
                \label{fig:inv-Square_Rxq_z-plane1}
        \end{subfigure}
	\hfill
        \begin{subfigure}[c]{0.26\textwidth}
                \centering
                \includegraphics[width=1.0\textwidth]{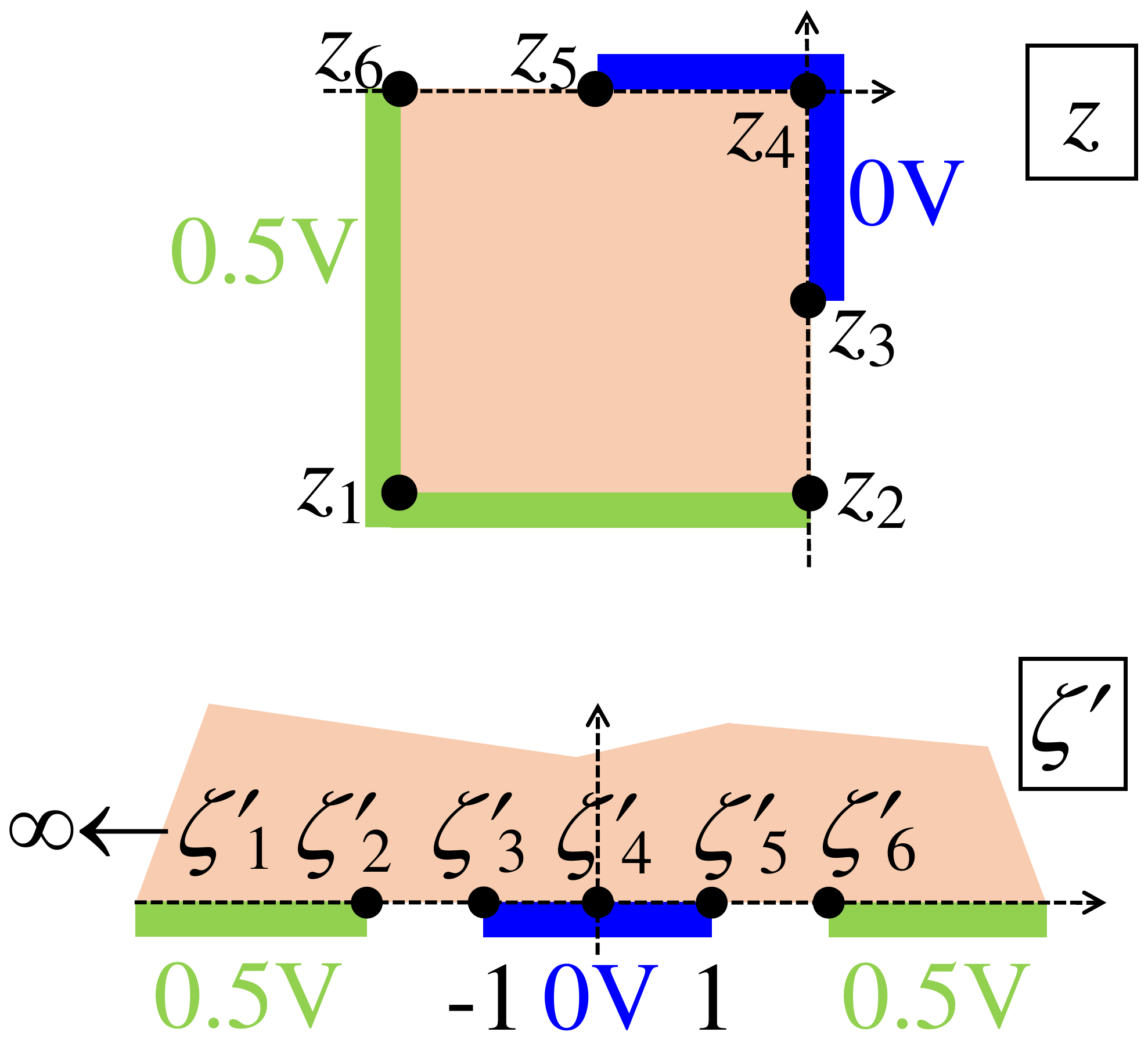}
                \caption{One quarter of Fig. \ref{fig:inv-Square_Rxq_z-plane1} and mapping to the $\zeta'$-plane. $\overline{R}_x$. }
                \label{fig:inv-Square_Rxq_z-plane2}
        \end{subfigure}
    \caption{Conformal mapping of a square with corner contacts to compute the cross resistance $\overline{R}_x$. }
   \label{fig:Square_with_corner-contacts_Rxq}
\end{figure}

A square Hall-plate can have mid-contacts as in Fig. \ref{fig:Square_Rx_z-plane1} or corner contacts as in Fig. \ref{fig:inv-Square_Rxq_z-plane1}. We call the Hall-plate with mid-contacts 'original device', and the Hall-plate with corner contacts 'complementary device'. Therefore, the potentials indicated in Fig. \ref{fig:Square_Rx_z-plane1} occur during the measurement of $R_x$, whereas Fig. \ref{fig:inv-Square_Rxq_z-plane1} shows the potentials at the contacts during a measurement of $\overline{R}_x$. A quarter of the complementary device in the $z$-plane is shown in Fig. \ref{fig:inv-Square_Rxq_z-plane2}. It is mapped to the upper half of the $\zeta'$-plane in the lower part of Fig. \ref{fig:inv-Square_Rxq_z-plane2}. The resistance between the two contacts in Fig. \ref{fig:inv-Square_Rxq_z-plane2} are also $\overline{R}_x$. The mapping is given by 
\begin{equation}\label{apx-Square1}
z = C' \!\!\int_0^{\zeta'}\!\! \frac{\mathrm{d}\,\zeta'}{\sqrt{\zeta'}\sqrt{{\zeta'}^2-{\zeta'_6}^2}} .
\end{equation}
The shape in the $\zeta'$-plane in Fig. \ref{fig:inv-Square_Rxq_z-plane2} is identical to Fig. \ref{fig:inv-Greek-cross_Rxq_z-plane2}. Thus, equations (\ref{apx-Greek-cross2}), (\ref{apx-Greek-cross3}), and (\ref{apx-Greek-cross4}) also hold,
\begin{equation}\label{apx-Square2}\begin{split}
\overline{\lambda}_x & = \frac{1}{2}\,\frac{K'}{K}\!\left(\frac{1}{\zeta'_6}\right) , \\
4\lambda_x & = \frac{K'}{K}\!\!\left(\!\frac{\zeta'_6\!-\!1}{\zeta'_6\!+\!1}\right) , \\
\frac{1}{\zeta'_6} & = \sin\!\left(2\theta\right) .
\end{split}\end{equation}
We define the filling factor $f_{SC}$ as the sum of the lengths of all contacts in Fig. \ref{fig:inv-Square_Rxq_z-plane1} over the entire perimeter. Then it holds 
\begin{equation}\label{apx-Square5}\begin{split}
f_{SC} & = \frac{2|z_4-z_5|}{2|z_4-z_5|+2|z_5-z_6|} \\
\Rightarrow \frac{1}{f_{SC}} & = 1+\frac{\int_1^{\zeta'_6} \frac{\mathrm{d}\zeta'}{\sqrt{\zeta'}\sqrt{{\zeta'_6}^2-{\zeta'}^2}}}{\int_0^1 \frac{\mathrm{d}\zeta'}{\sqrt{\zeta'}\sqrt{{\zeta'_6}^2-{\zeta'}^2}}} .
\end{split}\end{equation}
The integrals are tabulated in Refs.~\onlinecite{Prudnikov1.2.28,Prudnikov1.2.27} 
\begin{equation}\label{apx-Square5b}\begin{split}
\int_0^1 \frac{\mathrm{d}\,\zeta'}{\sqrt{\zeta'}\sqrt{{\zeta'_6}^2-{\zeta'}^2}} & = \sqrt{\frac{2}{\zeta'_6}}\; F\!\!\left(\sqrt{\frac{2}{1+\zeta'_6}},\frac{1}{\sqrt{2}}\right) , \\ 
\int_1^{\zeta'_6} \frac{\mathrm{d}\,\zeta'}{\sqrt{\zeta'}\sqrt{{\zeta'_6}^2-{\zeta'}^2}} & =  \sqrt{\frac{2}{\zeta'_6}} \;F\!\!\left(\sqrt{1-\frac{1}{\zeta'_6}},\frac{1}{\sqrt{2}}\right) .
\end{split}\end{equation}
This finally gives the filling factor $f_{SC}$ for a square Hall-plate with corner contacts from Fig. \ref{fig:inv-Square_Rxq_z-plane1} with identical properties like a disk-shaped Hall-plate with contacts of size $\epsilon$ from Fig. \ref{fig:regular-4C-Hall-disk_Rx_z-plane1},
\begin{equation}\label{apx-Square6}\begin{split}
\frac{1}{f_{SC}} = 1+\frac{F\!\!\left(\sqrt{1-\sin\left(\pi\frac{1+\epsilon}{4}\right)},\frac{1}{\sqrt{2}}\right)}{F\!\!\left(\sqrt{2-\frac{2}{1+\sin\left(\pi(1+\epsilon)/4\right)}},\frac{1}{\sqrt{2}}\right)} .
\end{split}\end{equation}
In (\ref{apx-Square6}) a disk with contacts size $\epsilon$ is equivalent to a square with corner contacts having the filling factor $f_{SC}(\epsilon)$. The complementary device is a disk with contacts size $-\epsilon$, which is equivalent to a square with mid-contacts with the filling factor $1-f_{SC}(\epsilon)$. Hence, a disk with contacts size $\epsilon$ corresponds to a square with mid-contacts of size $f_{SM}(\epsilon)=1-f_{SC}(-\epsilon)$, which gives 
\begin{equation}\label{apx-Square7}\begin{split}
\frac{1}{f_{SM}} = 1+\frac{F\!\!\left(\sqrt{2-\frac{2}{1+\sin\left(\pi(1-\epsilon)/4\right)}},\frac{1}{\sqrt{2}}\right)}{F\!\!\left(\sqrt{1-\sin\left(\pi\frac{1-\epsilon}{4}\right)},\frac{1}{\sqrt{2}}\right)} .
\end{split}\end{equation}

\begin{figure*}[t]
%\vspace{1mm}
  \centering
                \includegraphics[width=0.98\textwidth]{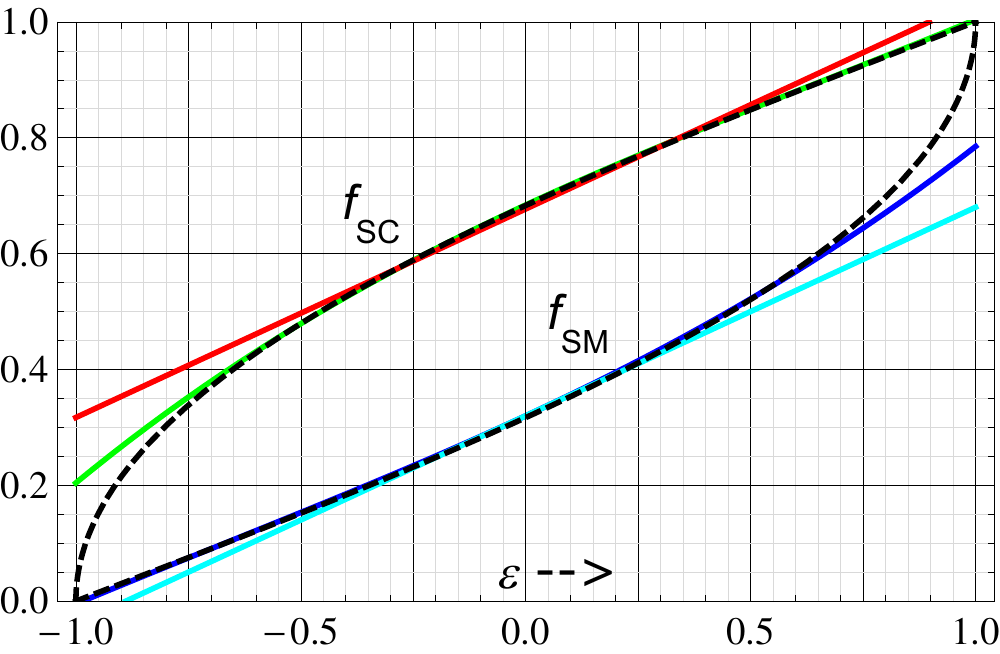}
    \caption{Plot of the filling factors of $90$° symmetric square Hall-plates with corner contacts, $f_{SC}$, and with mid-contacts, $f_{SM}$, versus the size of contacts of equivalent disk-shaped Hall-plates, $\epsilon$. The black dashed curves are the exact filling factors from (\ref{apx-Square6}), (\ref{apx-Square7}). The solid colored curves are the approximations from (\ref{apx-Square15}). }
   \label{fig:fSC_fSM_vs_eps}
\end{figure*}

For medium-sized contacts we can develop (\ref{apx-Square6}) and (\ref{apx-Square7}) into Taylor series in $\epsilon$, 
\begin{equation}\label{apx-Square15}\begin{split}
f_\mathrm{SC} & \approx 0.684+0.36\epsilon-0.08\epsilon^2+0.04\epsilon^3 \\ 
& \approx 0.677+0.36\epsilon , \\
f_\mathrm{SM} & \approx 0.32+0.36\epsilon+0.07\epsilon^2+0.035\epsilon^3 \\ 
& \approx 0.32+0.36\epsilon .
\end{split}\end{equation}
The relative errors of the linear approximations in (\ref{apx-Square15}) are less than $1\%$ for $-0.37<\epsilon<0.49$ for $f_{SC}$ and $-0.27<\epsilon<0.31$ for $f_{SM}$. For the cubic approximations the relative errors are less than $1\%$ for $-0.63<\epsilon\le1$ for $f_{SC}$ and $-0.76<\epsilon<0.64$ for $f_{SM}$. Fig. \ref{fig:fSC_fSM_vs_eps} shows a plot of (\ref{apx-Square6}), (\ref{apx-Square7}), and (\ref{apx-Square15}).

\begin{figure*}
%\vspace{1mm}
  \centering
        \begin{subfigure}[t]{0.62\textwidth}
                \centering
                \includegraphics[width=1.0\textwidth]{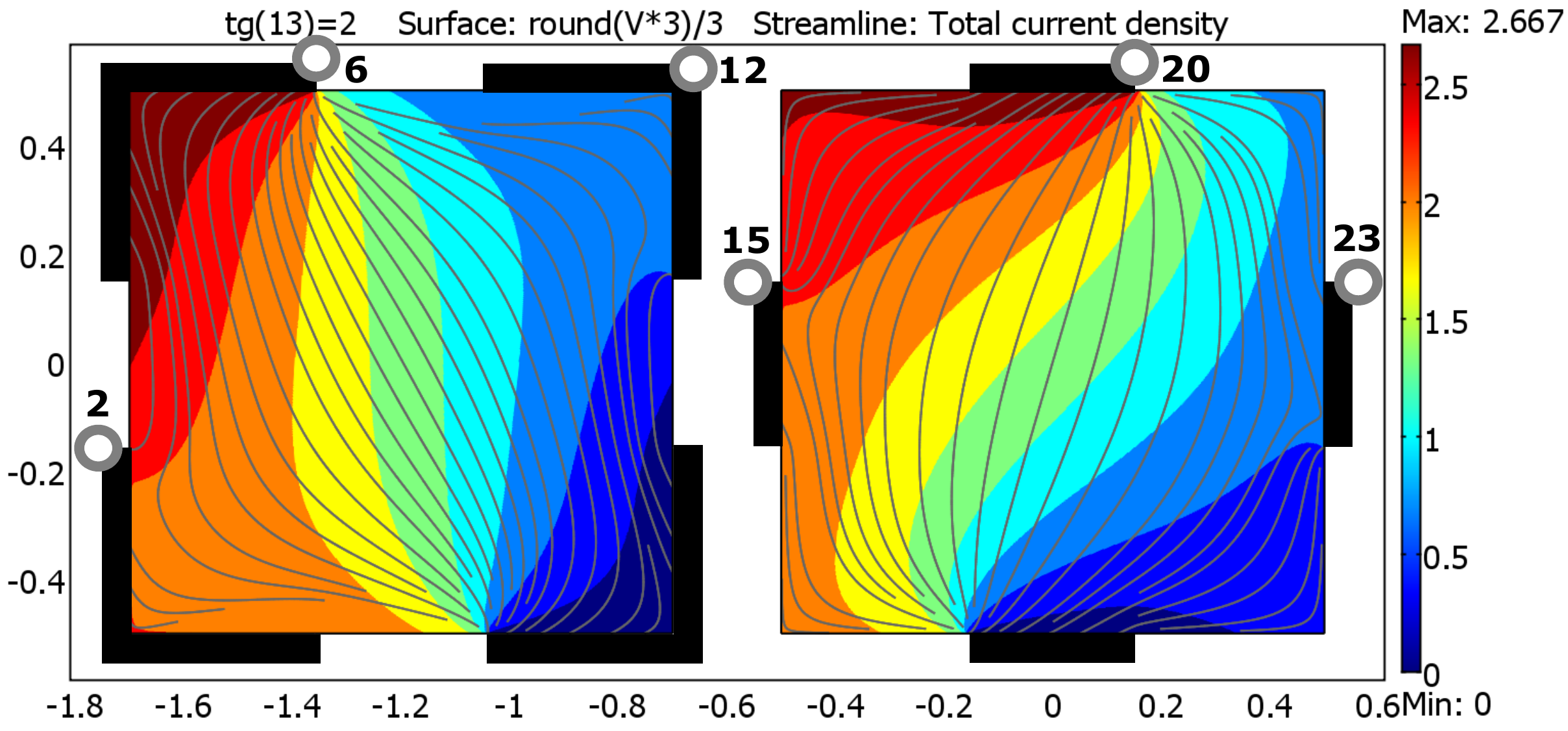}
                \caption{Current streamlines and potentials at $\tan(\theta_H)=2$ in square Hall-plates. The left Hall-plate has corner contacts with the filling factor $f_{SC}=0.683113$ and the right Hall-plate has mid-contacts with the filling factor $f_{SM}=0.316887$. Both are equivalent to a disk-shaped Hall-plate with contacts size of $\epsilon=0$. }
                \label{fig:streamlines_tg_eq_2_eps_eq_0_Squares}
        \end{subfigure}
        \hfill  % An dieser Stelle kann ein zusätzlicher Zwischenraum eingebunden werden: ~, \quad, \qquad, \hfill usw.
          % Eine leere Zeile erzwingt, dass die zweite Grafik darunter erscheint.
        \begin{subfigure}[t]{0.36\textwidth}
                \centering
                \includegraphics[width=1.0\textwidth]{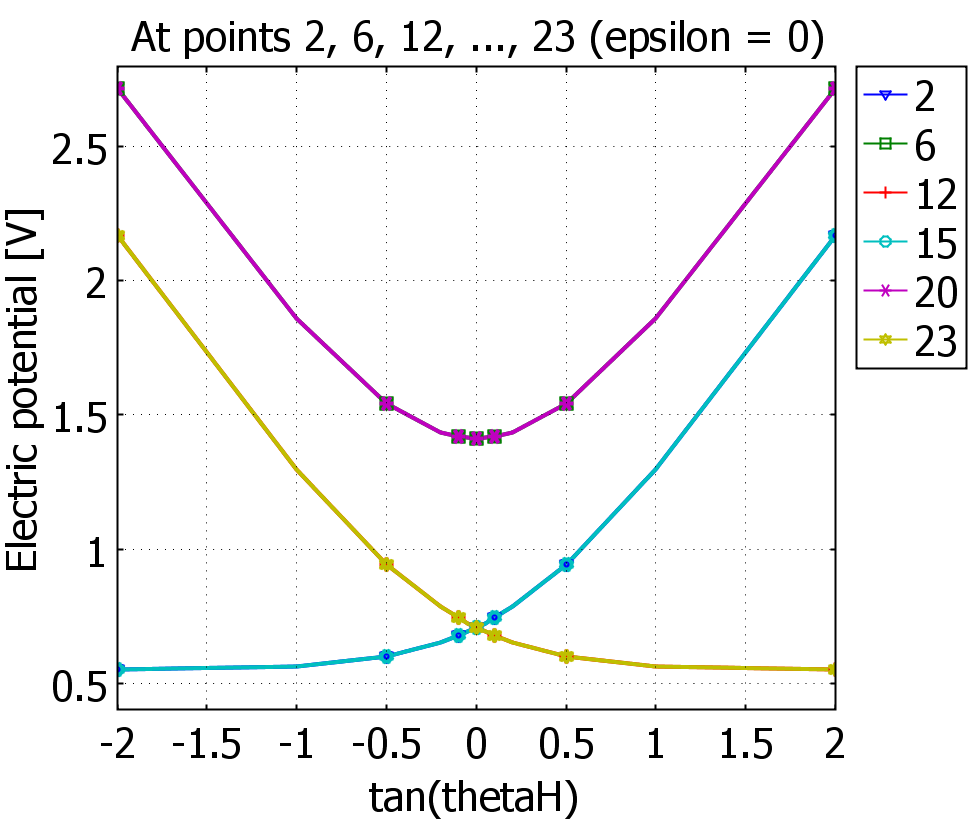}
                \caption{Potentials at the contacts versus $\tan(\theta_H)$ for the Hall-plates of Fig. \ref{fig:streamlines_tg_eq_2_eps_eq_0_Squares}. These pairs of curves are nearly identical: $2-15,6-20,12-23$. They are also identical to the ones in Figs. \ref{fig:eps_eq_0_rounded-Greek-crosses} and \ref{fig:eps_eq_0_Octagon}. }
                \label{fig:eps_eq_0_Squares}
        \end{subfigure}
	 \\
        \begin{subfigure}[t]{0.62\textwidth}
                \centering
                \includegraphics[width=1.0\textwidth]{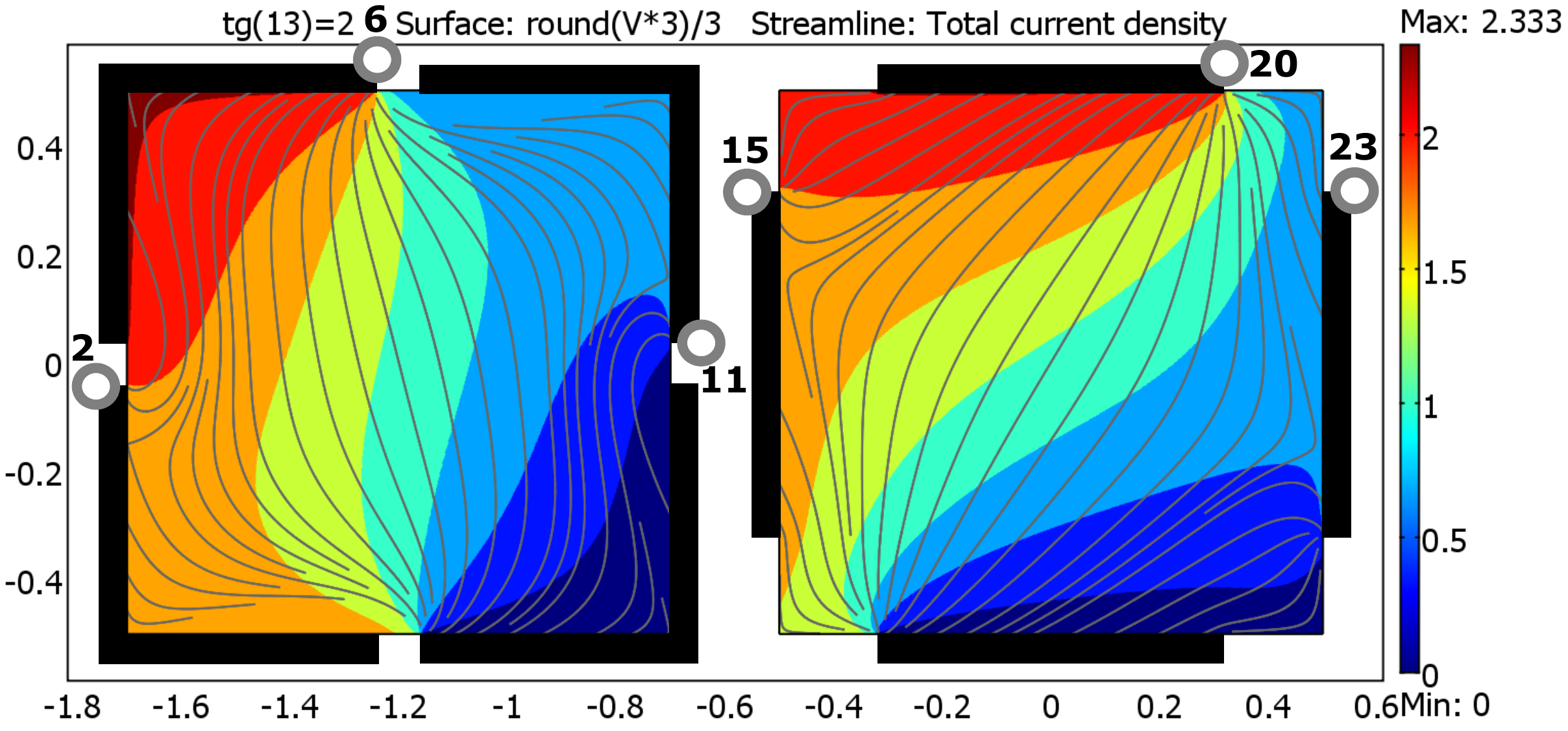}
                \caption{Current streamlines and potentials at $\tan(\theta_H)=2$ in square Hall-plates. The left Hall-plate has corner contacts with the filling factor $f_{SC}=0.914056$ and the right Hall-plate has mid-contacts with the filling factor $f_{SM}=0.638371$. Both are equivalent to a disk-shaped Hall-plate with contacts size of $\epsilon=5/7$ from Fig. \ref{fig:streamlines_tg_eq_2_eps_eq_5durch7_inv-Greek-crosses6}. }
                \label{fig:streamlines_tg_eq_2_eps_eq_5durch7_Squares}
        \end{subfigure}
        \hfill  
        \begin{subfigure}[t]{0.36\textwidth}
                \centering
                \includegraphics[width=1.0\textwidth]{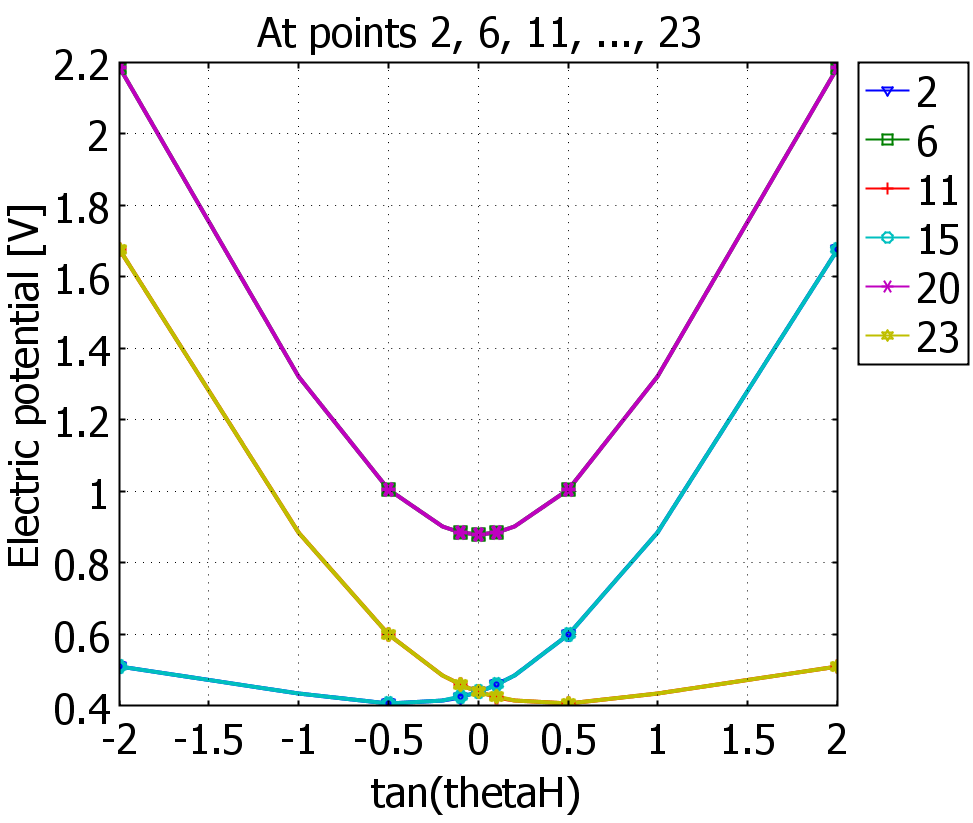}
                \caption{Potentials at the contacts versus $\tan(\theta_H)$ for the Hall-plates of Fig. \ref{fig:streamlines_tg_eq_2_eps_eq_5durch7_Squares}. These pairs of curves are nearly identical: $2-15,6-20,11-23$. They are also identical to the ones in Fig. \ref{fig:eps_eq_5durch7_inv-Greek-crosses6}. }
                \label{fig:eps_eq_5durch7_Squares}
        \end{subfigure}
    \caption{Results of FEM-simulations of square Hall-plates with corner contacts and mid-contacts. Sizes of contacts are chosen such that the square Hall-plates are equivalent to disk-shaped Hall-plates with $\epsilon=0$ and $\epsilon=5/7$. Contacts without any labelled circle are grounded. A current of $1$ A enters contacts comprising circles 6 or 20. %They are diametrically opposite to the grounded contacts. 
$R_\mathrm{sheet}=1\;\Omega$. The differences in the potentials of corresponding points $2-15,6-20,11\text{ or }12-23$ vanish apart from numerical inaccuracies. }
   \label{fig:Squares-checks}
\end{figure*}

Equations (\ref{apx-Square6}) and (\ref{apx-Square7}) give the following filling factors for three cases,  
\begin{equation}\label{apx-Square19}
\begin{array}{c|cccc} \mathrm{case} & \epsilon & \theta & f_\mathrm{SC} & f_\mathrm{SM} \\ \hline
1 & -1/4 & 3\pi/32 & 0.588850 & 0.231577 \\
2 & 0 & \pi/8 & 0.683113 & 0.316887 \\
3 & 5/7 & 3\pi/14 & 0.914056 & 0.638371 \end{array}
\end{equation}

Fig. \ref{fig:Squares-checks} shows the results of a numerical check with finite element simulation. The potentials of corresponding contacts of left and right Hall-plates are identical within the limits of numerical accuracy.

\clearpage
\subsection{Rectangular Hall-plates}
\label{sec:Rectangles}

\begin{figure}
%\vspace{1mm}
  \centering
        \begin{subfigure}[t]{0.24\textwidth}
                \centering
                \includegraphics[width=1.0\textwidth]{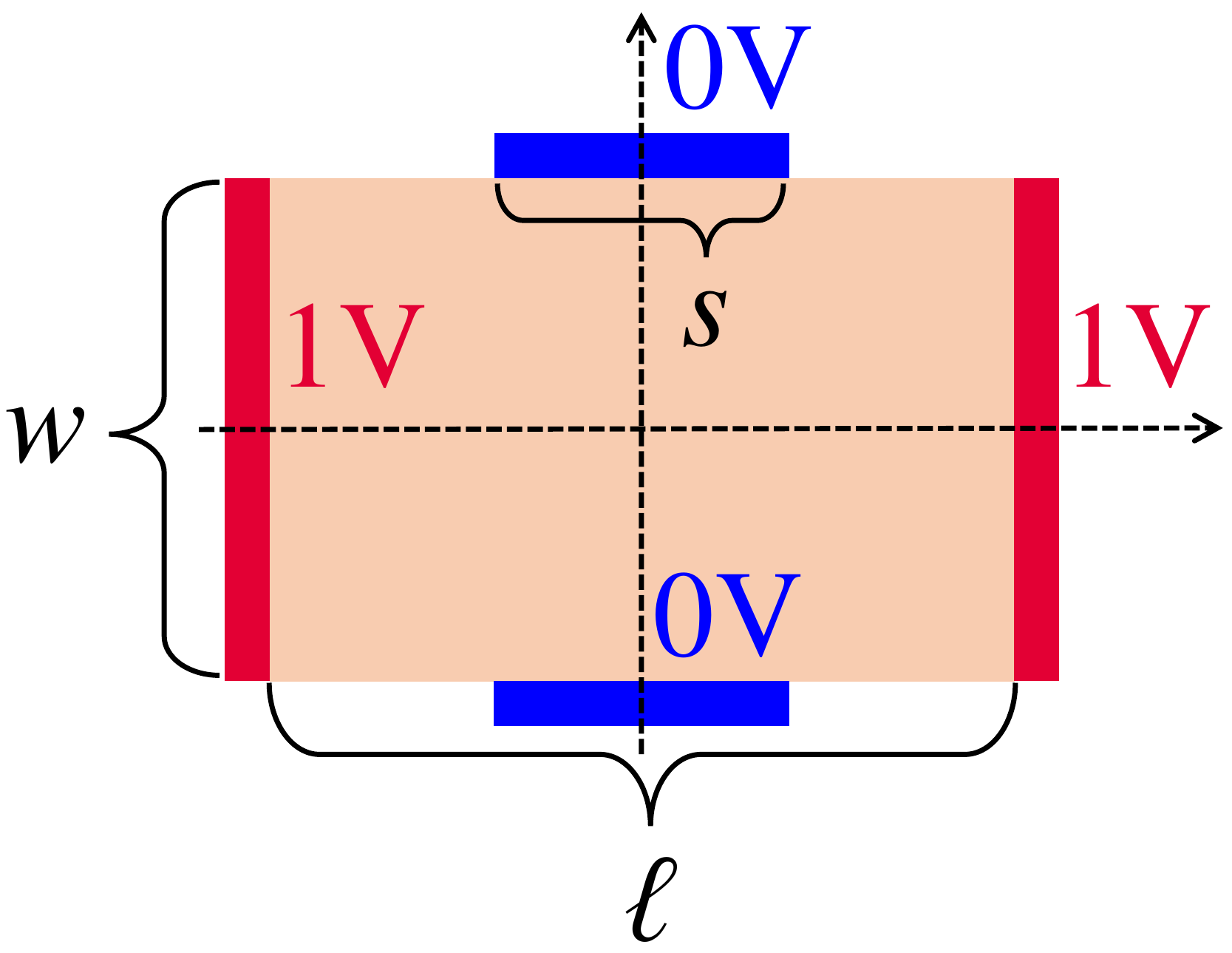}
                \caption{Rectangular Hall-plate with even symmetry. Resistance $=R_x$. }
                \label{fig:rectangle_Rx_z-plane}
        \end{subfigure}
	\hfill
        \begin{subfigure}[t]{0.23\textwidth}
                \centering
                \includegraphics[width=1.0\textwidth]{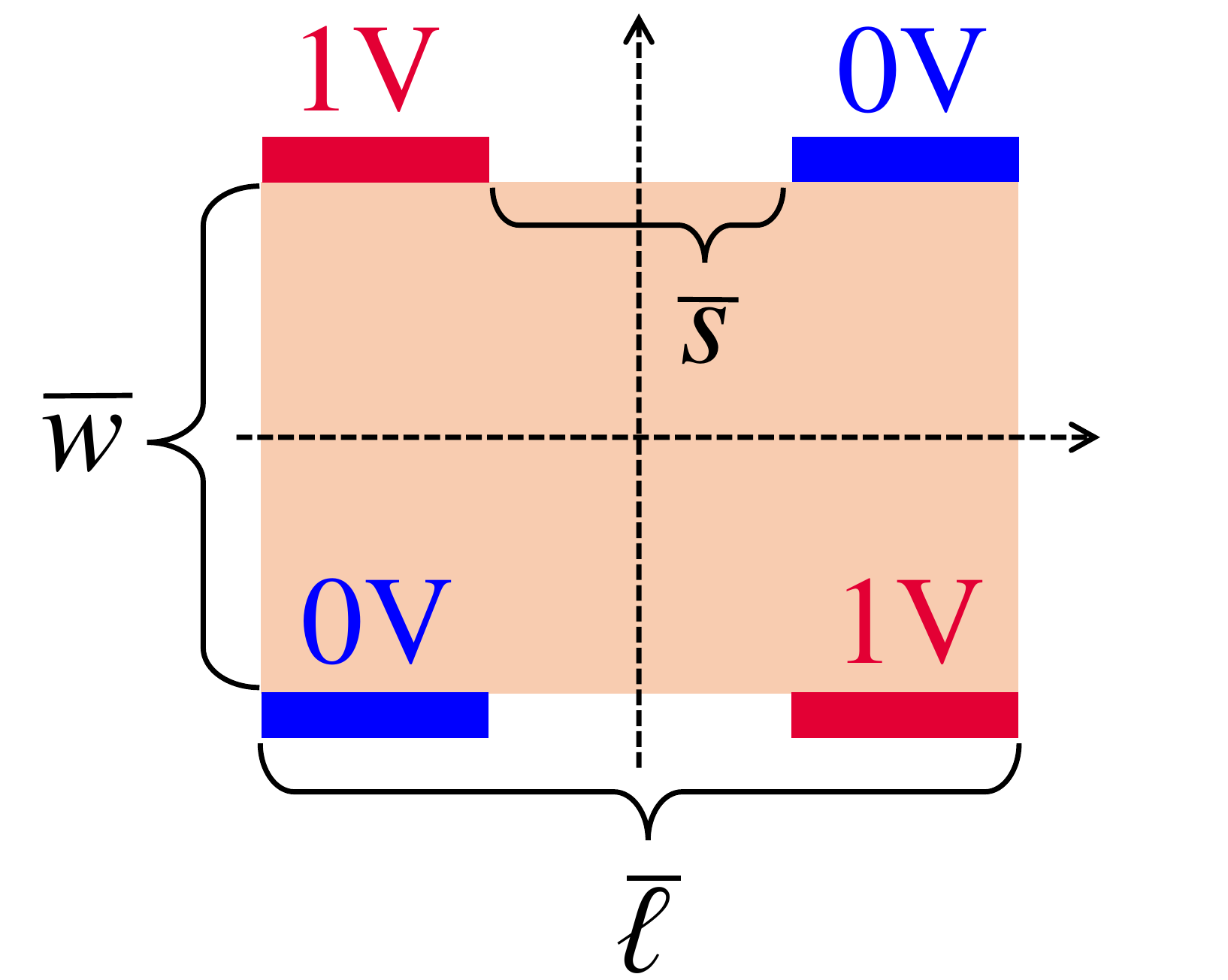}
                \caption{Complementary rectangular Hall-plate with odd symmetry. Resistance $=\overline{R}_x$. }
                \label{fig:compl_rectangle_Rxq_z-plane}
        \end{subfigure}
    \caption{Rectangular Hall-plates. }
   \label{fig:rectangle}
\end{figure}

In general, the rectangular Hall-plate of Fig. \ref{fig:rectangle_Rx_z-plane} is not regular symmetric. Yet, for \cite{AusserlechnerSNR2017}
\begin{equation}\label{apx-rect10}\begin{split}
& \frac{s}{\ell} = \frac{F\;\left(w_3,k\right)}{2K(k)} +\frac{1}{2}\quad\text{with }\;\;w_3=\frac{1}{k}-\frac{8}{1+4k-k^2}, \\ 
&\text{with } k=\sqrt{L\left(2\frac{w}{\ell}\right)} \;\;\leftrightarrow \;\;\frac{\ell}{w}=2\frac{K(k)}{K'(k)},\;\; \text{ for }w\le\ell ,
\end{split}\end{equation}
the resistance between the flush contacts (with length $w$) is equal to the resistance between the partial contacts (with length $s$) and each pair of contacts is centered between the other pair of contacts. Then the rectangular Hall-plate can be mapped conformally to a regular symmetric one. %Hence, it is regular symmetric in an electrical sense. 
%TAYLOR-REIHEN FUER ell/w vs. k und s/ell

Next, we use Fig. A1 in Ref.~\onlinecite{AusserlechnerVdP2}, which shows the mapping of the unit cell of the electric potential during the measurement of $R_x$ to the upper infinite plane. Consequently, $\zeta_{2\mathrm{disk}}$ from Fig. \ref{fig:regular-4C-Hall-disk_Rx_t-and-zeta-plane} is identical to $1/x$ in Fig. A1(c) in Ref.~\onlinecite{AusserlechnerVdP2}, whereby we take $x$ from (A9) in Ref.~\onlinecite{AusserlechnerVdP2}. Therein, we replace $p$ by $f$ with $\sqrt{p}=(1-\sqrt{f})/(1+\sqrt{f})$ from (46) in Ref.~\onlinecite{ELEN}, and we replace $f$ by (9) in Ref.~\onlinecite{ELEN}, with $w_3$ from (\ref{apx-rect10}). This gives a relation between the sizes of the contacts of the disk-shaped Hall-plate, $\epsilon$, and the parameter $k$, which specifies the the aspect ratio of the rectangular Hall-plate of Fig. \ref{fig:rectangle_Rx_z-plane} and the size of its partial contacts, see (\ref{apx-rect10}),  
\begin{equation}\label{apx-rect15}\begin{split}
& k = \frac{1-\sin\left(\pi\frac{1+\epsilon}{8}\right)}{1+\sin\left(\pi\frac{1+\epsilon}{8}\right)}, \;\;3-2\sqrt{2}\le k\le 1 \quad\Leftrightarrow \\ 
& \epsilon = -1+\frac{4}{\pi}\,\mathrm{arctan}\left(\frac{4\sqrt{k}(1-k)}{6k-1-k^2}\right), \;\;-1\le\epsilon\le 1 .
\end{split}\end{equation}
%Hence, if we define $c_\mathrm{RE}$ as the ratio of the length of all four contacts of the rectangle from Fig. ZZZ2 over the length of the entire boundary, it holds 
%\begin{equation}\label{apx-rect20}\begin{split}
%c_\mathrm{RE} & = \frac{2s+2w}{2\ell+2w} = 1 - \frac{K(k)-F(w_3,k)}{2K(k)+K'(k)}\right) .
%\end{split}\end{equation}
\noindent If we have a disk-shaped Hall-plate with contacts size $\epsilon$, we use (\ref{apx-rect15}) to get $k$. With (\ref{apx-rect10}) we get the aspect ratio $\ell/w$ and the size of the partial contacts $s/\ell$ in Fig. \ref{fig:rectangle_Rx_z-plane}. 

\begin{figure*}
%\vspace{1mm}
  \centering
                \includegraphics[width=0.87\textwidth]{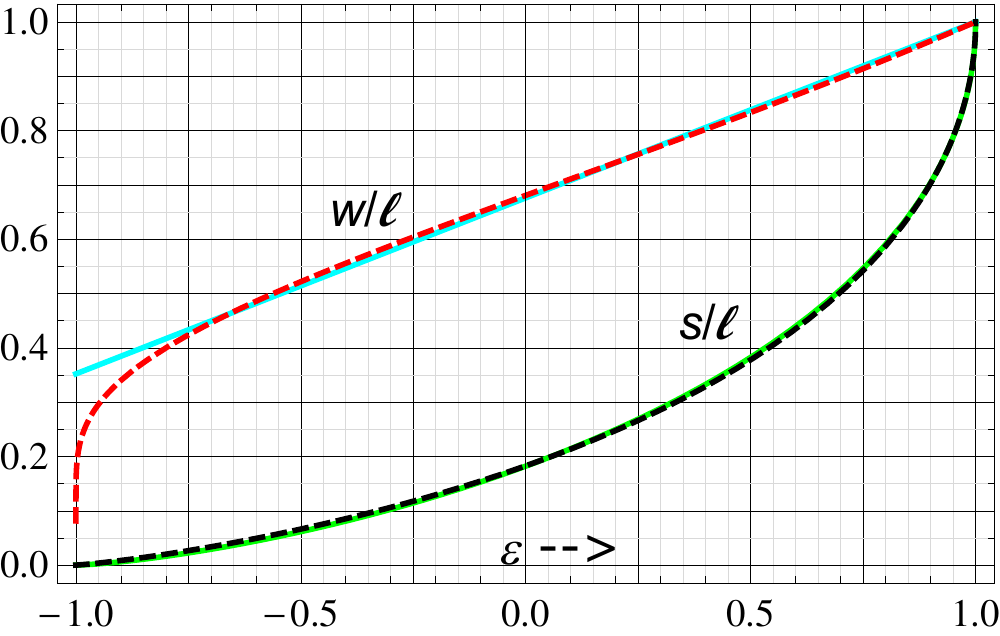}
    \caption{Plot of $s/\ell$ and $w/\ell$ of rectangular Hall-plates with even symmetry versus the size of the contacts of equivalent disk-shaped Hall-plates, $\epsilon$. The dashed curves are the exact formulae from (\ref{apx-rect10}), (\ref{apx-rect15}). The solid colored curves are the approximations from (\ref{apx-rect18}). }
   \label{fig:sdl_wdl_vs_eps}
\end{figure*}

\begin{figure*}
%\vspace{1mm}
  \centering
                \includegraphics[width=0.87\textwidth]{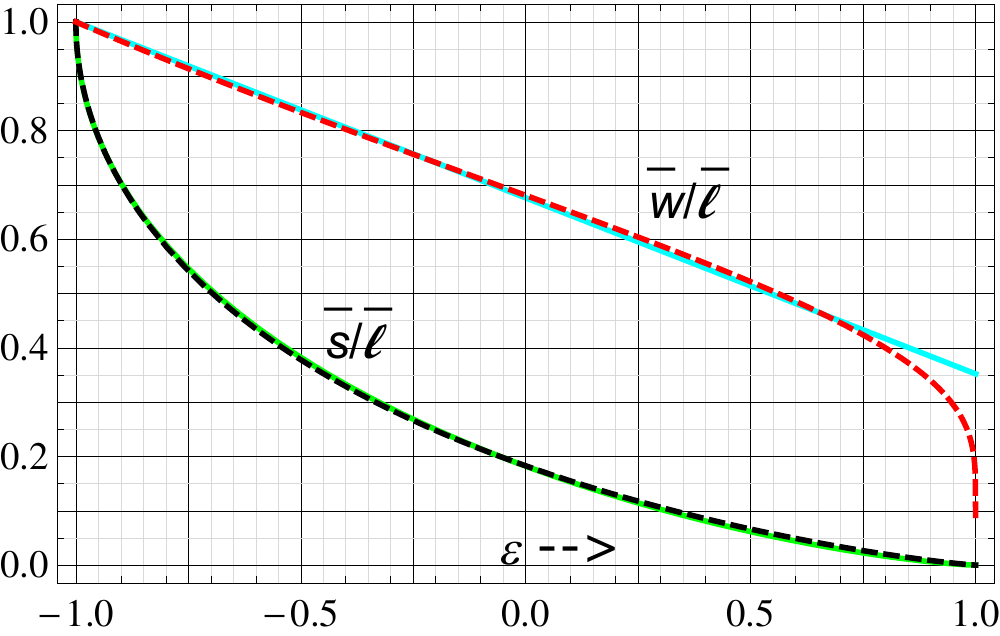}
    \caption{Plot of $\overline{s}/\overline{\ell}$ and $\overline{w}/\overline{\ell}$ of rectangular Hall-plates with odd symmetry versus the size of the contacts of equivalent disk-shaped Hall-plates, $\epsilon$. The dashed curves are the exact formulae from (\ref{apx-rect17}), (\ref{apx-rect16}). The solid colored curves are the approximations from (\ref{apx-rect18}). }
   \label{fig:sqdlq_wqdlq_vs_eps}
\end{figure*}

At the same time, the complementary disk has contacts size $-\epsilon$, and it corresponds to the complementary rectangle as shown in Fig. \ref{fig:compl_rectangle_Rxq_z-plane}. This rectangle has the same aspect ratio $\ell/w$ and the same distance-to-length-ratio $s/\ell$. Consequently, the rectangular Hall-plate of Fig. \ref{fig:compl_rectangle_Rxq_z-plane} is identical to a disk of Fig. \ref{fig:regular-4C-Hall-disk_Rx_z-plane1} for 
\begin{equation}\label{apx-rect16}\begin{split}
& \overline{k} = \frac{1-\sin\left(\pi\frac{1-\epsilon}{8}\right)}{1+\sin\left(\pi\frac{1-\epsilon}{8}\right)}, \;\;3-2\sqrt{2}\le \overline{k}\le 1 \quad\Leftrightarrow \\ 
& \epsilon = 1-\frac{4}{\pi}\,\mathrm{arctan}\left(\frac{4\sqrt{\overline{k}}(1-\overline{k})}{6\overline{k}-1-\overline{k}^2}\right), \;\;-1\le\epsilon\le 1 .
\end{split}\end{equation}
In (\ref{apx-rect16}) we used the overbar to distinguish it from (\ref{apx-rect15}): $k$ specifies the aspect ratio of a rectangular Hall-plate in Fig. \ref{fig:rectangle_Rx_z-plane} (= even symmetry), while $\overline{k}$ specifies the aspect ratio of a rectangular Hall-plate in Fig. \ref{fig:compl_rectangle_Rxq_z-plane} (= odd symmetry). We get the geometrical paramters of the odd-symmetric Hall-plate in Fig. \ref{fig:compl_rectangle_Rxq_z-plane} by inserting $\overline{k}$ into 
\begin{equation}\label{apx-rect17}\begin{split}
& \frac{\overline{s}}{\overline{\ell}} = \frac{F\;\left(\overline{w}_3,\overline{k}\right)}{2K(\overline{k})} +\frac{1}{2}\quad\text{with }\;\;\overline{w}_3=\frac{1}{\overline{k}}-\frac{8}{1+4\overline{k}-\overline{k}^2}, \\ 
&\text{with } \overline{k}=\sqrt{L\left(2\frac{\overline{w}}{\overline{\ell}}\right)} \;\;\leftrightarrow \;\;\frac{\overline{\ell}}{\overline{w}}=2\frac{K(\overline{k})}{K'(\overline{k})},\;\; \text{ for }\overline{w}\le\overline{\ell} .
\end{split}\end{equation}
Equation (\ref{apx-rect17}) is identical to (\ref{apx-rect10}) if we delete all overbars. There are approximations for  (\ref{apx-rect10}-\ref{apx-rect17}), 
\begin{equation}\label{apx-rect18}\begin{split}
\frac{s}{\ell} & \approx \left(1-\sqrt{\frac{1-\epsilon}{2}}\right)^{\!\!1.387} , \\
\frac{w}{\ell} & \approx 1-0.324*(1-\epsilon) , \\
\frac{\overline{s}}{\overline{\ell}} & \approx \left(1-\sqrt{\frac{1+\epsilon}{2}}\right)^{\!\!1.387} , \\
\frac{\overline{w}}{\overline{\ell}} & \approx 1-0.324*(1+\epsilon) ,
\end{split}\end{equation}
whereby $s/\ell$ and $\overline{s}/\overline{\ell}$ are accurate up to $\pm 0.5\%$ for $-1\le\epsilon\le 1$, $w/\ell$ is accurate up to $\pm 1\%$ for $\epsilon>-0.75$, and $\overline{w}/\overline{\ell}$ is accurate up to $\pm 1\%$ for $\epsilon<0.75$. Figs. \ref{fig:sdl_wdl_vs_eps} and \ref{fig:sqdlq_wqdlq_vs_eps} show plots of these exact and the approximate curves. 

Equations (\ref{apx-rect15}) and (\ref{apx-rect16}) give the following geometrical parameters of rectangular Hall-plates for three cases, 
\begin{equation}\label{apx-rect19}
\begin{array}{c|cccccc} \mathrm{case} & \epsilon & \theta & \ell/w & s/\ell & \overline{\ell}/\overline{w} & \overline{s}/\overline{\ell} \\ \hline
1 & \frac{-1}{4} & \frac{3\pi}{32} & 1.6561 & 0.11763 & 1.3223 & 0.26627 \\
2 & 0 & \frac{\pi}{8} &  1.4692 & 0.18272 & 1.4692 & 0.18272 \\
3 & \frac{5}{7} & \frac{3\pi}{14} &  1.1086 & 0.51341 & 2.2745 & 0.03157 \end{array}
\end{equation}
In (\ref{apx-rect19}) parameters without overbar refer to Fig. \ref{fig:rectangle_Rx_z-plane} (= even symmetry), and parameters with overbar refer to Fig. \ref{fig:compl_rectangle_Rxq_z-plane} (= odd symmetry). 

\begin{figure*}
%\vspace{1mm}
  \centering
        \begin{subfigure}[t]{0.62\textwidth}
                \centering
                \includegraphics[width=1.0\textwidth]{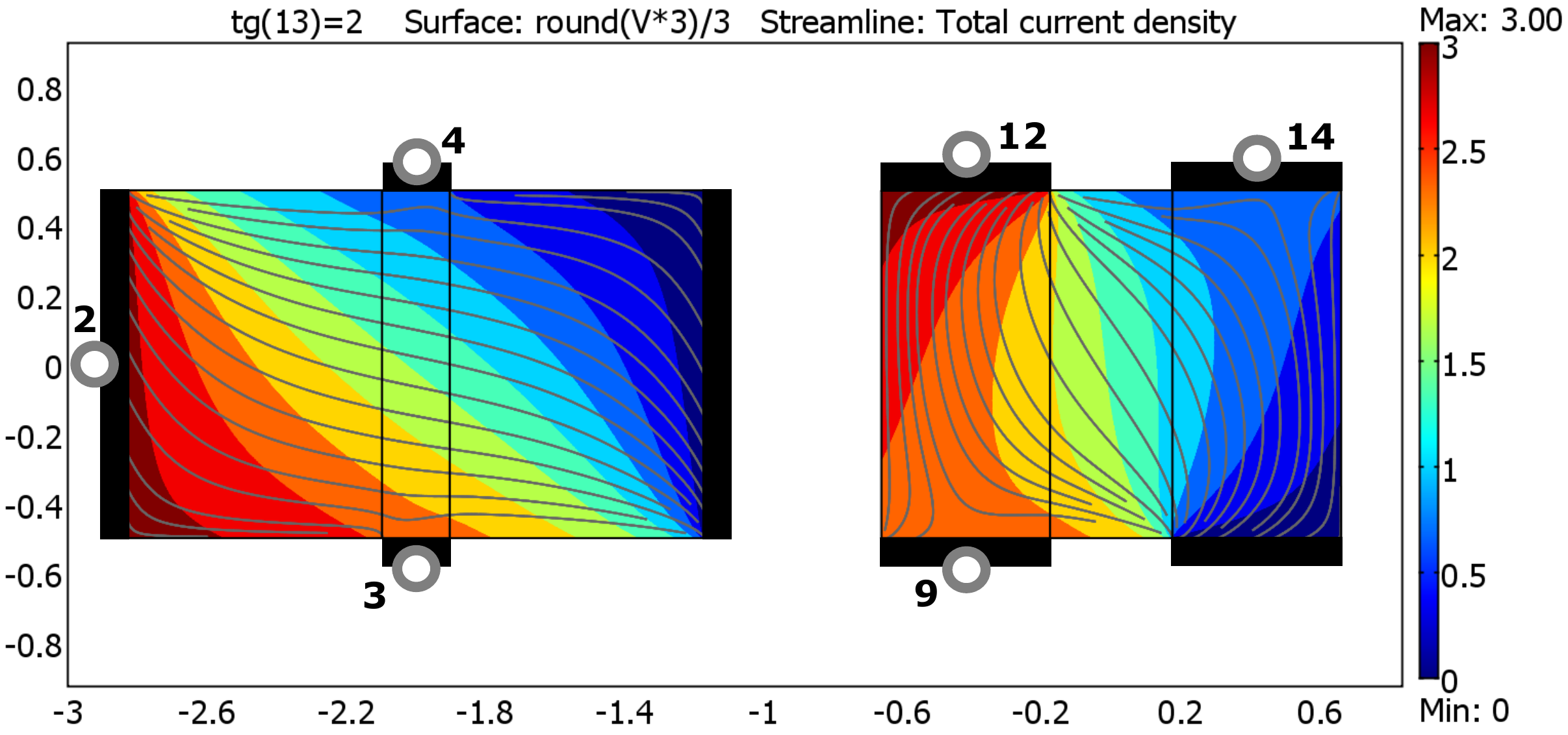}
                \caption{Current streamlines and potentials at $\tan(\theta_H)=2$ in rectangular Hall-plates. The left Hall-plate has even symmetry with $\ell/w=1.6561$ and $s/\ell=0.11763$, the right Hall-plate has odd symmetry with $\overline{\ell}/\overline{w}=1.3223$ and $\overline{s}/\overline{\ell}=0.26627$. Both are equivalent to a disk-shaped Hall-plate with contacts size of $\epsilon=-1/4$ from Fig. \ref{fig:streamlines_tg_eq_2_eps_eq_m250m_Greek-crosses1}. }
                \label{fig:streamlines_tg_eq_2_eps_eq_m250m_rectangle}
        \end{subfigure}
        \hfill  % An dieser Stelle kann ein zusätzlicher Zwischenraum eingebunden werden: ~, \quad, \qquad, \hfill usw.
          % Eine leere Zeile erzwingt, dass die zweite Grafik darunter erscheint.
        \begin{subfigure}[t]{0.36\textwidth}
                \centering
                \includegraphics[width=1.0\textwidth]{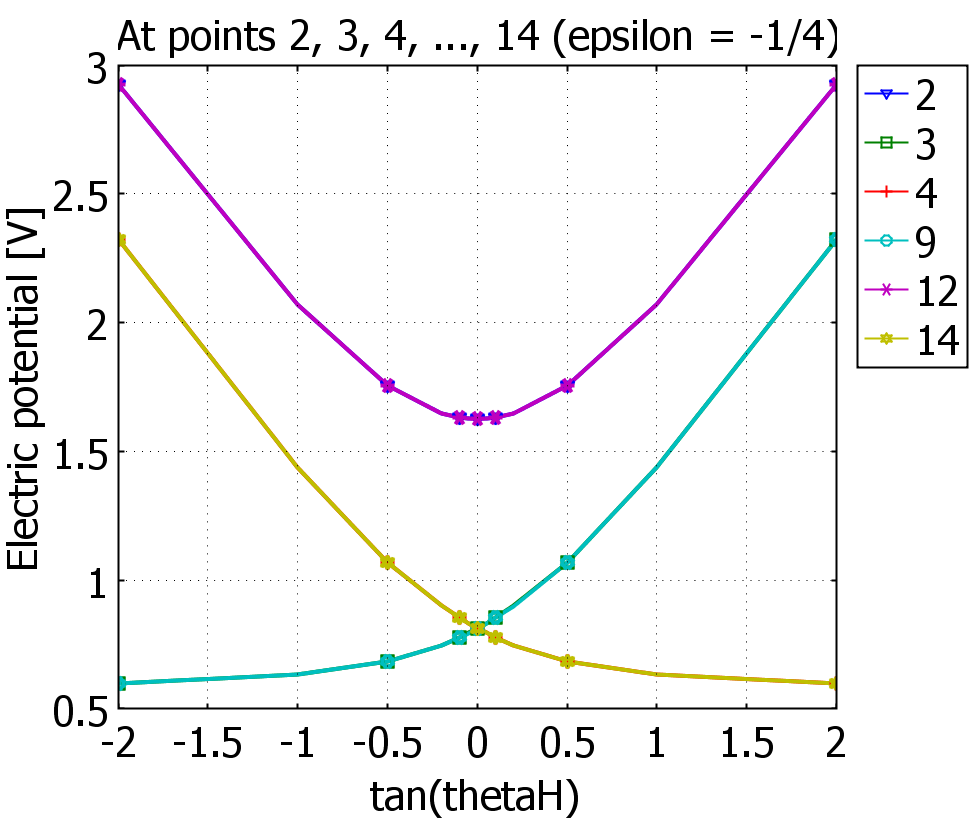}
                \caption{Potentials at the contacts versus $\tan(\theta_H)$ for the Hall-plates of Fig. \ref{fig:streamlines_tg_eq_2_eps_eq_m250m_rectangle}. These pairs of curves are nearly identical: $3-9,2-12,4-14$. They are also identical to the ones in Fig. \ref{fig:eps_eq_m250m_Greek-crosses1}. }
                \label{fig:eps_eq_m250m_rectangle}
        \end{subfigure}
	 \\
        \begin{subfigure}[t]{0.62\textwidth}
                \centering
                \includegraphics[width=1.0\textwidth]{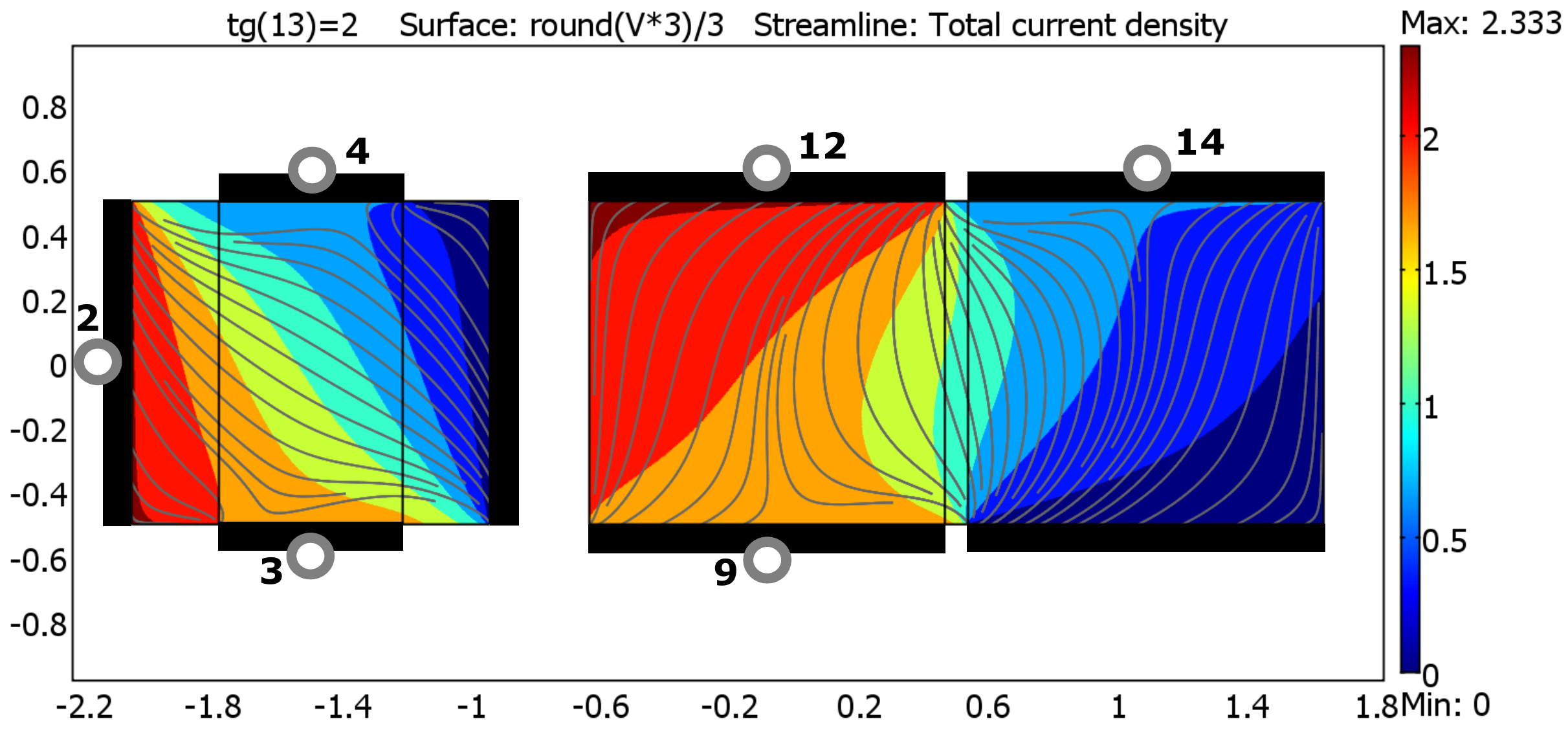}
                \caption{Current streamlines and potentials at $\tan(\theta_H)=2$ in rectangular Hall-plates. The left Hall-plate has even symmetry with $\ell/w=1.1086$ and $s/\ell=0.51341$, the right Hall-plate has odd symmetry with $\overline{\ell}/\overline{w}=2.2745$ and $\overline{s}/\overline{\ell}=0.03157$. Both are equivalent to a disk-shaped Hall-plate with contacts size of $\epsilon=5/7$ from Fig. \ref{fig:streamlines_tg_eq_2_eps_eq_5durch7_inv-Greek-crosses6}. }
                \label{fig:streamlines_tg_eq_2_eps_eq_5durch7_rectangle}
        \end{subfigure}
        \hfill  
        \begin{subfigure}[t]{0.36\textwidth}
                \centering
                \includegraphics[width=1.0\textwidth]{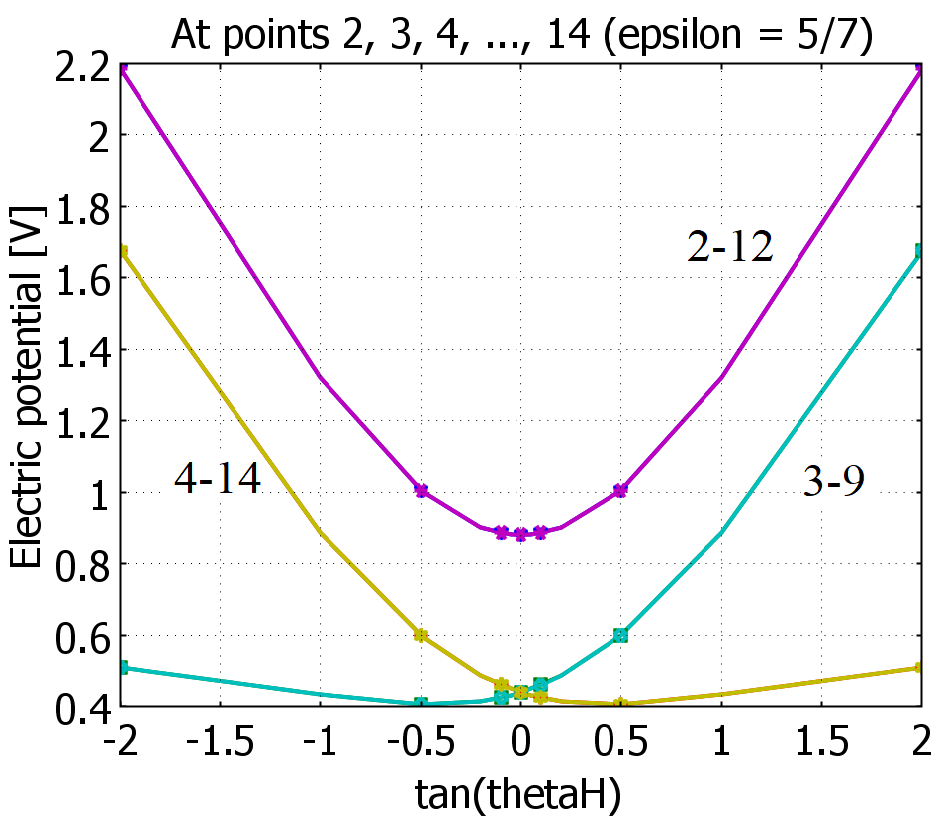}
                \caption{Potentials at the contacts versus $\tan(\theta_H)$ for the Hall-plates of Fig. \ref{fig:streamlines_tg_eq_2_eps_eq_5durch7_rectangle}. These pairs of curves are nearly identical: $3-9,2-12,4-14$. They are also identical to the ones in Fig. \ref{fig:eps_eq_5durch7_inv-Greek-crosses6}. }
                \label{fig:eps_eq_5durch7_rectangle}
        \end{subfigure}
    \caption{Results of FEM-simulations of rectangular Hall-plates of even and odd symmetry. Sizes of contacts are chosen such that the rectangular Hall-plates are equivalent to disk-shaped Hall-plates with $\epsilon=-1/4$ and $\epsilon=5/7$. Contacts without any labelled circle are grounded. A current of $1$ A enters contacts comprising circles 2 and 12. %They are diametrically opposite to the grounded contacts. 
$R_\mathrm{sheet}=1\;\Omega$. The differences in the potentials of corresponding points $3-9,2-12,4-14$ vanish apart from numerical inaccuracies. }
   \label{fig:rectangle-checks}
\end{figure*}

Numerical checks of (\ref{apx-rect15}) and (\ref{apx-rect16}), analogous to the ones in the preceding sections, are shown in Fig. \ref{fig:rectangle-checks}, where the plots of the potentials versus magnetic field are identical to the repsective ones in the preceding sections. The corresponding graphs for even and odd symmetric rectangles are also identical.

\newpage
%\nocite{*}
\bibliography{aipsamp}% Produces the bibliography via BibTeX.

\end{document}